\documentclass[
12pt, % Default font size is 10pt, can alternatively be 11pt or 12pt
a4paper, % Alternatively letterpaper for US letter
onecolumn,
oneside, %oneside,%twoside,% Alternatively onecolumn portrait % Alternatively portrait
hidelinks,%Remove boxes from hyperlink
table]{book}

\usepackage[columnsep=2cm,
left=1.575in,
right=0.78in,
top=1.18in,
bottom=.88in]{geometry} % Margins includeheadfoot, includefoot,

\usepackage[T1]{fontenc}
\usepackage[utf8]{inputenc}

\usepackage{titlesec}
\titleformat{\chapter}[display]
{\normalfont\huge\bfseries}{\chaptertitlename\ \thechapter}{0pt}{\Huge}

\titlespacing*{\section}{0pt}{1em}{0.3em}

\usepackage{sectsty}
\chapterfont{\setstretch{1.5}}
\usepackage[titles]{tocloft} % subfigure option only if using subfigure package

\DeclareUnicodeCharacter{0301}{*}

\usepackage{textcomp}
\usepackage{gensymb}
\usepackage{enumitem}
\usepackage{graphicx}
\usepackage[colorinlistoftodos]{todonotes}
\usepackage{fancyhdr}
\usepackage{setspace}
\usepackage{url}
\usepackage{verbatim}
\usepackage[bottom]{footmisc}
\usepackage[english]{babel}
\usepackage{multirow}

\usepackage{pifont}% http://ctan.org/pkg/pifont
\usepackage[nottoc, numbib]{tocbibind}

\usepackage{amsmath}
\usepackage{amsfonts}
\usepackage{amsthm}
\usepackage{amssymb}
\usepackage{longtable,tabularx}
\usepackage{csquotes}
\usepackage{siunitx}
\usepackage{booktabs}

\usepackage{moreverb}
\usepackage{bm}
\usepackage{algpseudocode}
\usepackage{algorithm}
\usepackage{subfig}
\usepackage{minitoc}
\usepackage{longtable}
\usepackage{multicol}
\usepackage{hyperref}
\usepackage{cleveref}
\usepackage{bookmark}
\usepackage{xcolor}
\usepackage{pdfpages}
\usepackage{tikz}
\usepackage[titletoc]{appendix}
\usepackage{mathtools}
\usepackage{etoolbox}
\usepackage[htt]{hyphenat}
\usepackage{soul}

\usepackage[acronym, nogroupskip, nopostdot]{glossaries} 
\newglossary[slg]{symbol}{sld}{sdn}{Nomenclature}
\makeglossaries
\glstoctrue
\usepackage{afterpage}

\usepackage{emptypage}

\usepackage{tcolorbox} \newcommand{\ciao}[1]{{\setlength\fboxrule{0pt}\fbox{\tcbox[colframe=black,colback=white,shrink tight,boxrule=0.2pt,extrude by=0.5mm]{\small #1}}}}

\let\Algorithm\algorithm
\renewcommand\algorithm[1][]{\Algorithm[#1]\setstretch{2}}

\usepackage{lscape}
\usepackage{adjustbox}
\usepackage{array}
\usepackage{multirow}
\usepackage{rotating}

\usepackage{balance} % For balanced columns on the last page
\usepackage{tabularx} % For full width table
\usepackage{xcolor,colortbl}
\usepackage{bbding}
\usepackage{dblfloatfix}
\usepackage{color}
\usepackage{pifont}% http://ctan.org/pkg/pifont

\usepackage{listings}

\PassOptionsToPackage{table,xcdraw}{xcolor}
\newcommand{\ie}{{\em i.e., }}
\newcommand{\eg}{{\em e.g., }}
\newcommand{\myverb}{\fontsize{9}{48}\usefont{OT1}{lmtt}{b}{n}\noindent }

\makeatletter
\newcommand\tabcaption{\def\@captype{table}\caption}
\newcommand\figcaption{\def\@captype{figure}\caption}
\makeatother

\usepackage{makecell}

\renewcommand{\texttt}[1]{%
	\begingroup
	\ttfamily
	\begingroup\lccode`~=`/\lowercase{\endgroup\def~}{/\discretionary{}{}{}}%
	\begingroup\lccode`~=`[\lowercase{\endgroup\def~}{[\discretionary{}{}{}}%
	\begingroup\lccode`~=`.\lowercase{\endgroup\def~}{.\discretionary{}{}{}}%
	\catcode`/=\active\catcode`[=\active\catcode`.=\active
	\scantokens{#1\noexpand}%
	\endgroup
}

\newcolumntype{L}[1]{>{\raggedright\let\newline\\\arraybackslash\hspace{0pt}}m{#1}}
\newcolumntype{C}[1]{>{\centering\let\newline\\\arraybackslash\hspace{0pt}}m{#1}}
\newcolumntype{R}[1]{>{\raggedleft\let\newline\\\arraybackslash\hspace{0pt}}m{#1}}

\makeatletter
\newcommand*\fsize{\dimexpr\f@size pt\relax}
\makeatother

\usepackage[backend=biber,
 giveninits=true,
 isbn=true,
 url=true,
 doi=true,
 style=ieee,
 citestyle=numeric-comp,
 dashed=false,
 maxnames = 5]{biblatex}
\newcommand{\RNum}[1]{\uppercase\expandafter{\romannumeral #1\relax}}

{%
  \fancyhf{}% clear all header and footer fields
  \fancyhead[R]{\nouppercase{\leftmark}}
  \fancyfoot[C]{\thepage} % except the center
  \renewcommand{\headrulewidth}{0pt}
  \renewcommand{\footrulewidth}{0pt}
}

\newacronym{iot}{IoT}{Internet of Things}
\newacronym{ip}{IP}{Internet Protocol}
\newacronym{http}{HTTP}{Hyper Text Transfer Protocol}
\newacronym{ietf}{IETF}{Internet Engineering Task Force}
\newacronym{ids}{IDS}{Intrusion Detection System}
\newacronym{pca}{PCA}{Principal Component Analysis}
\newacronym{svm}{SVM}{Support Vector Machines}
\newacronym{sdn}{SDN}{Software Defined Networking}
\newacronym{tcp}{TCP}{Transmission Control Protocol}
\newacronym{udp}{UDP}{User Datagram Protocol}
\newacronym{dns}{DNS}{Domain Name System}
\newacronym{tcam}{TCAM}{Ternary Content-Addressable Memory}
\newacronym{ml}{ML}{Machine Learning}
\newacronym{ftp}{FTP}{File Transfer Protocol}
\newacronym{tls}{TLS}{Transport Layer Security}
\newacronym{dos}{DoS}{Denial-of-Service}
\newacronym{ddos}{DDoS}{Distributed Denial-of-Service}
\newacronym{ssh}{SSH}{Secure Shell Protocol }
\newacronym{cnc}{C\&C}{Command and Control}
\newacronym{dga}{DGA}{Domain Generation Algorithm}
\newacronym{nxd}{NXD}{Non-eXistent-Domains}
\newacronym{rf}{RF}{Random Forest}
\newacronym{iF}{iForest}{isolation Forest}
\newacronym{https}{HTTPS}{Hyper Text Transfer Protocol Secured}
\newacronym{tld}{TLD}{Top Level Domain}
\newacronym{sld}{SLD}{Second Level Domain}
\newacronym{byod}{BYOD}{Bring Your Own Device}
\newacronym{Mitm}{MitM}{Man-in-the-Middle}
\newacronym{dnssec}{DNSSEC}{Domain Name System SECurity extenstions}
\newacronym{doh}{DoH}{DNS over HTTPS}
\newacronym{nids}{NIDS}{Network Intrusion Detection System}
\newacronym{kdd}{KDD}{Knowledge Discovery in Databases}
\newacronym{icmp}{ICMP}{Internet Control Message Protocol}
\newacronym{ttl}{TTL}{Time To Live}
\newacronym{ccdf}{CCDF}{Complementary Cumulative Distribution Function}
\newacronym{api}{API}{Application Programming Interface}
\newacronym{fqdn}{FQDN}{Fully Qualified Domain Name}
\newacronym{pcap}{PCAP}{Packet CAPture}
\newacronym{rbf}{RBF}{Radial Basis Function}
\newacronym{dram}{DRAM}{Dynamic Random Access Memory}
\newacronym{eif}{EiF}{Extended isolation Forest}
\newacronym{ieee}{IEEE}{Institute of Electrical and Electronics Engineers}
\newacronym{tnsm}{TNSM}{Transactions on Network and Service Management}
\newacronym{tnse}{TNSE}{Transactions on Network Science and Engineering}
\newacronym{knn}{KNN}{K-Nearest Neighbour}

\begin{document}
\frontmatter
\thispagestyle{empty}
\pagenumbering{alph}

\begin{titlepage}
	\thispagestyle{empty}
	% \pdfbookmark[chapter]{Titlepage: Computationally Efficient Non-linear Kalman Filters for On-board Space Vehicle Navigation}{Thesis}
	\begin{center}

		% Upper part of the page

		% Title
		
		\Huge \textbf{Monitoring Security of Enterprise Hosts via DNS Data Analysis}\\
		\vspace{1.5cm}
		\huge \textbf{Jawad Ahmed}\\
		\vspace{2cm}
		\large A dissertation submitted in fulfillment\\of the requirements for the degree of\\
		\vspace{0.5cm}
		\large \textbf{Doctor of Philosophy}\\
		\vspace{2cm}
		\includegraphics[width=0.4\columnwidth]{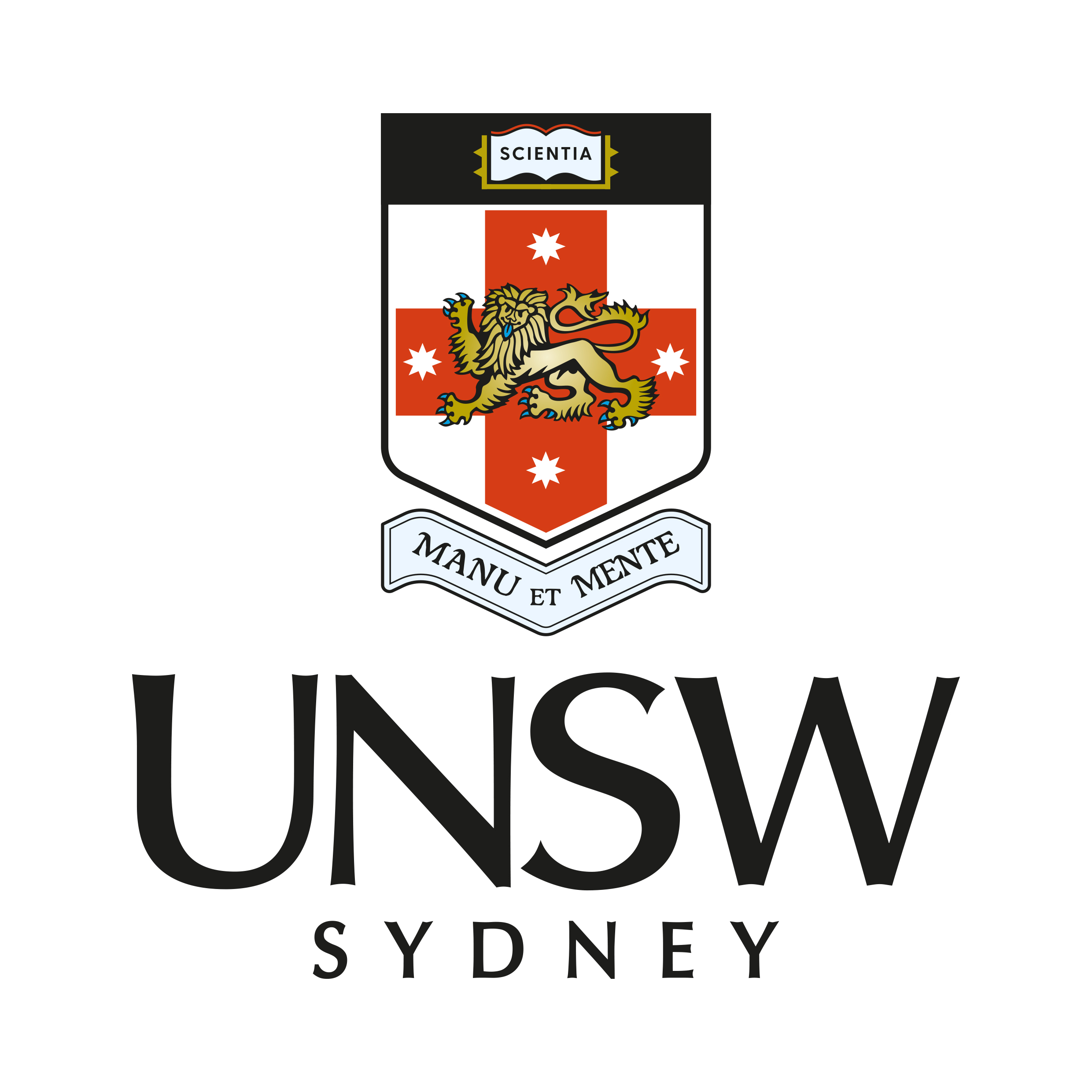}\\
		\vspace{0.5cm}
		\large School of Electrical Engineering and Telecommunications\\
		\vspace{0.5cm}
		\large The University of New South Wales\\
		\vspace{2cm}
		\large June 2021
	\end{center}
	
\end{titlepage}

\newpage
\cleardoublepage
%==========================SIGNED DECLARATION==========================================================================
\clearpage
\thispagestyle{empty}
\includepdf[pages={1}]{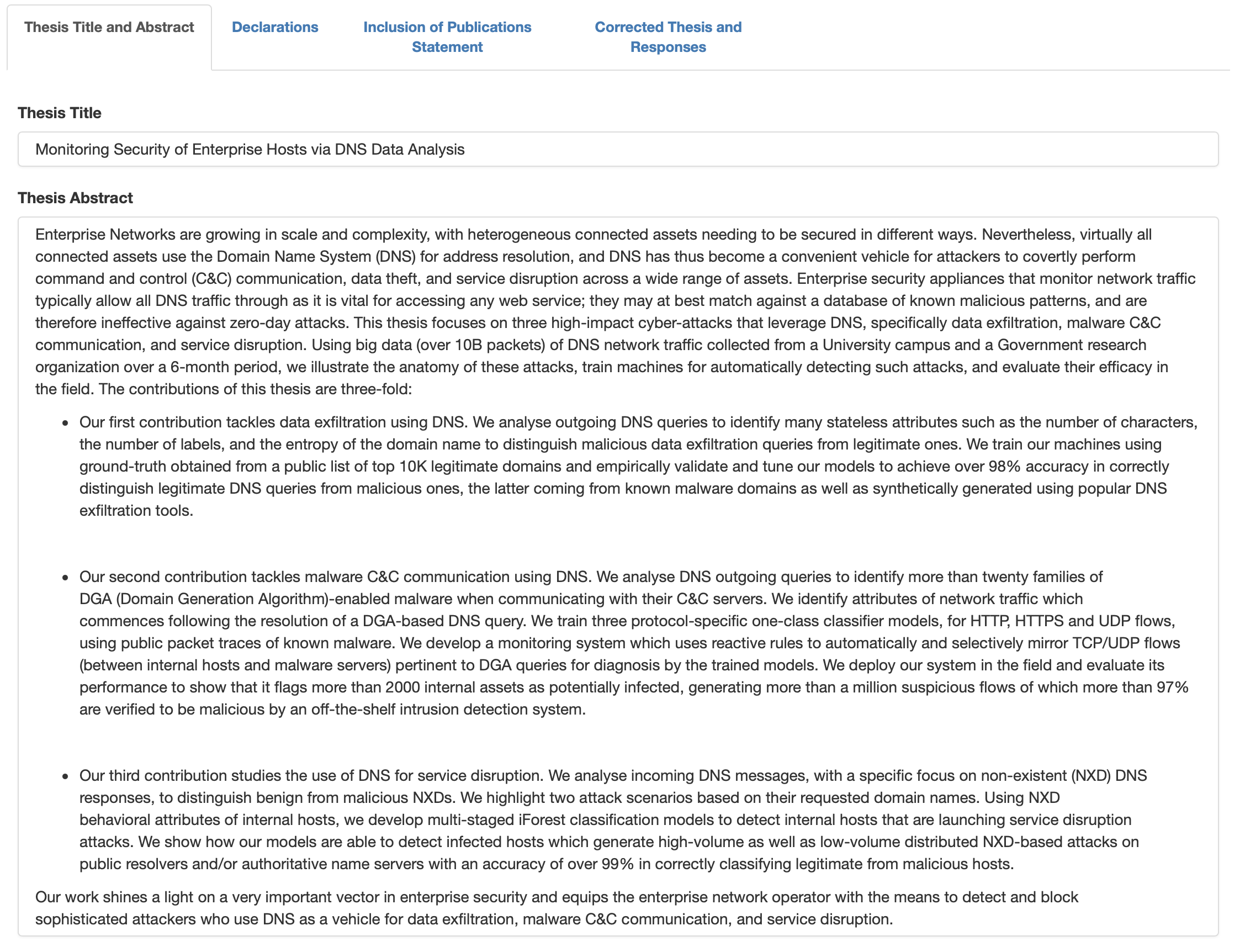}
\cleardoublepage

%\clearpage
%\thispagestyle{empty}
%\includepdf[pages={1}]{./originalitystatement}
%\cleardoublepage

\clearpage
\thispagestyle{empty}
\includepdf[pages={1}]{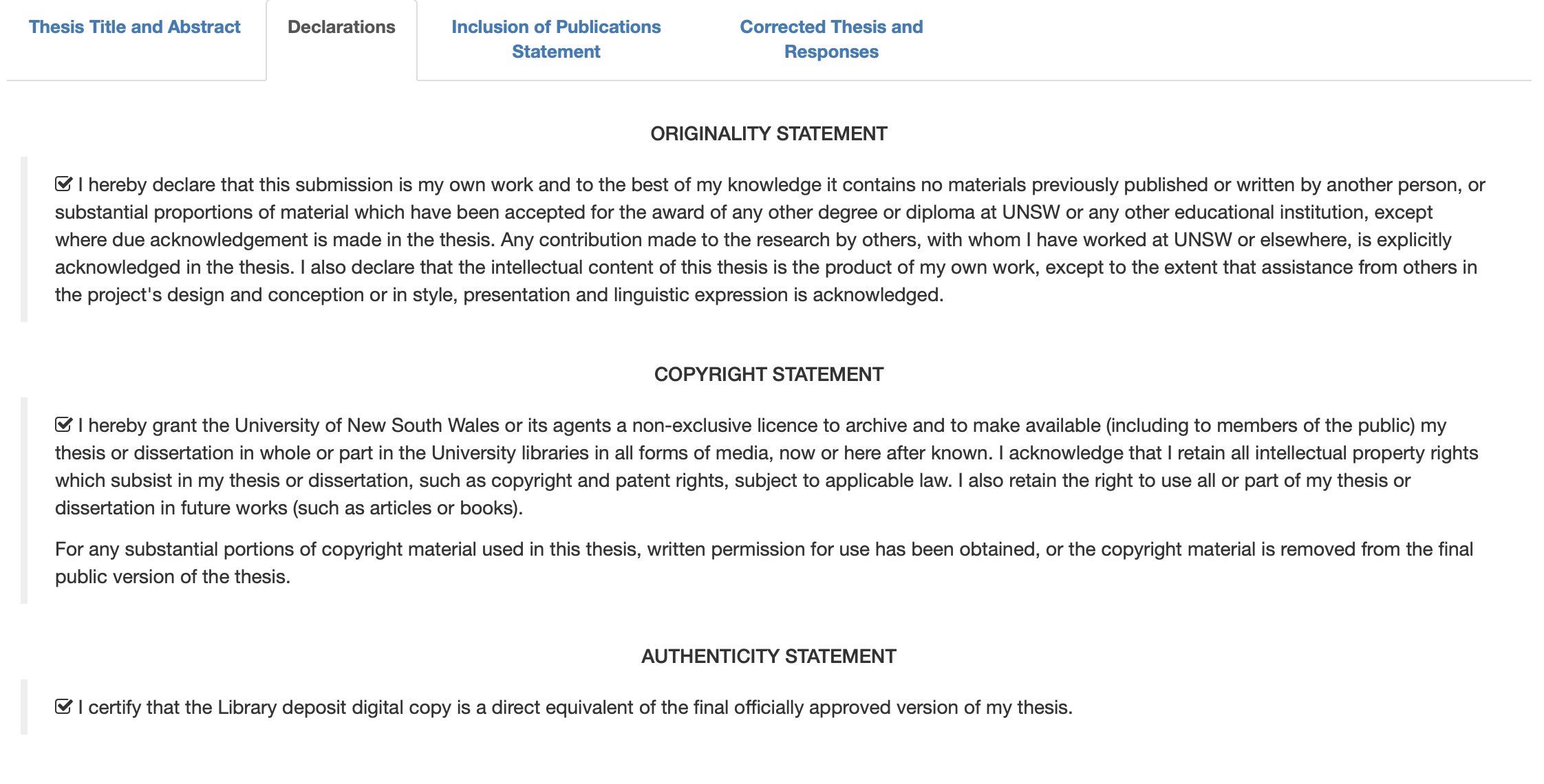}
\cleardoublepage

\chapter*{}
\thispagestyle{empty}
\vfil
\begin{center} To my family \end{center}
\vfil
\cleardoublepage

\pagenumbering{roman}
\setstretch{1.18}

\chapter*{Abstract}
%\mtcaddchapter[Abstract]
\bookmarksetupnext{level=section}
\addcontentsline{toc}{chapter}{Abstract}
\onehalfspace
\bookmarksetupnext{level=section}
\thispagestyle{plain}
\adjustmtc
\vspace{-1.6cm}

Enterprise Networks are growing in scale and complexity, with heterogeneous connected assets needing to be secured in different ways. Nevertheless, virtually all connected assets use the \gls{dns} for address resolution, and \gls{dns} has thus become a convenient vehicle for attackers to covertly perform \gls{cnc} communication, data theft, and service disruption across a wide range of assets. Enterprise security appliances that monitor network traffic typically allow all DNS traffic through as it is vital for accessing any web service; they may at best match against a database of known malicious patterns, and are therefore ineffective against zero-day attacks. 
This thesis focuses on three high-impact cyber-attacks that leverage DNS, specifically data exfiltration, malware C\&C communication, and service disruption. Using big data (over 10B packets) of DNS network traffic collected from a University campus and a Government research organization over a 6-month period, we illustrate the anatomy of these attacks, train machines for automatically detecting such attacks, and evaluate their efficacy in the field. The contributions of this thesis are three-fold:

\begin{itemize}
	\item Our first contribution tackles data exfiltration using DNS. We analyze outgoing DNS queries to identify many stateless attributes such as the number of characters, the number of labels, and the entropy of the domain name to distinguish malicious data exfiltration queries from legitimate ones. We train our machines using ground-truth obtained from a public list of top 10K legitimate domains and empirically validate and tune our models to achieve over 98\% accuracy in correctly distinguish legitimate DNS queries from malicious ones, the latter coming from known malware domains as well as synthetically generated using popular DNS exfiltration tools. 
	
	\item Our second contribution tackles malware C\&C communication using DNS. We analyze DNS outgoing queries to identify more than twenty families of DGA (Domain Generation Algorithm)-enabled malware when communicating with their C\&C servers. We identify attributes of network traffic which commences following the resolution of a DGA-based DNS query. We train three protocol-specific one-class classifier models, for HTTP, HTTPS and UDP flows, using public packet traces of known malware. We develop a monitoring system which uses reactive rules to automatically and selectively mirror TCP/UDP flows (between internal hosts and malware servers) pertinent to DGA queries for diagnosis by the trained models. We deploy our system in the field and evaluate its performance to show that it flags more than 2000 internal assets as potentially infected, generating more than a million suspicious flows of which more than 97\% are verified to be malicious by an off-the-shelf intrusion detection system. 
	
	\item Our third contribution studies the use of DNS for service disruption. We analyze incoming DNS messages, with a specific focus on non-existent (NXD) DNS responses, to distinguish benign from malicious NXDs. We highlight two attack scenarios based on their requested domain names. Using NXD behavioral attributes of internal hosts, we develop multi-staged iForest classification models to detect internal hosts that are launching service disruption attacks. We show how our models are able to detect infected hosts which generate high-volume as well as low-volume distributed NXD-based attacks on public resolvers and/or authoritative name servers with an accuracy of over 99\% in correctly classifying legitimate hosts. 
	
\end{itemize}

Our work shines a light on a very important vector in enterprise security and equips the enterprise network operator with the means to detect and block sophisticated attackers who use DNS as a vehicle for malware C\&C communication, data exfiltration, and service disruption.

\cleardoublepage

%\setstretch{1.5}

\chapter*{List of Publications} %
%\mtcaddchapter[List of Publications]
\bookmarksetupnext{level=section}
\addcontentsline{toc}{chapter}{List of Publications}
%\onehalfspace
\bookmarksetupnext{level=section}
\thispagestyle{plain}

\vspace{-5mm}
During the course of this thesis project, a number of publications have been made based on the work presented here and are listed below for reference.

\vspace{5mm}
\noindent{\underline{\large Journal Publications}}
\begin{enumerate}
	
	\item \textbf{J. Ahmed}, H. Habibi Gharakheili, and V. Sivaraman, ``Learning-Based Detection of Malicious Hosts by Analyzing Non-Existent DNS Responses,'' (under review at IEEE Globecom 2022) (Outcome of Chapter 5)

	\item \textbf{J. Ahmed}, H. Habibi Gharakheili, C. Russell and V. Sivaraman, ``Automatic Detection of DGA-Enabled Malware Using SDN and Traffic Behavioral Modeling,'' in IEEE Transactions on Network Science and Engineering TNSE, May 2022, doi: 10.1109/TNSE.2022.3173591 (Outcome of Chapter 4)

	\item \textbf{J. Ahmed}, H. Habibi Gharakheili, Q. Raza, C. Russell and V. Sivaraman, ``Monitoring Enterprise DNS Queries for Detecting Data Exfiltration From Internal Hosts," in IEEE Transactions on Network and Service Management, vol. 17, no. 1, pp. 265-279, March 2020, doi: 10.1109/TNSM.2019.2940735. (Outcome of Chapter 3)

\end{enumerate}

\vspace{5mm}
\noindent{\underline{\large Conference Publications}}
\begin{enumerate}
	\setcounter{enumi}{3}
	
	\item \textbf{J. Ahmed}, H. Habibi Gharakheili, Q. Raza, C. Russell and V. Sivaraman, ``Real-Time Detection of DNS Exfiltration and Tunneling from Enterprise Networks," 2019 IFIP/IEEE Symposium on Integrated Network and Service Management (IM), Arlington, VA, USA, 2019, pp. 649-653. (Outcome of Chapter 3)

	\item \textbf{J. Ahmed}, H. Habibi Gharakheili, Q. Raza, C. Russell and V. Sivaraman, ``Demo Abstract: A Tool to Detect and Visualize Malicious DNS Queries for Enterprise Networks," 2019 IFIP/IEEE Symposium on Integrated Network and Service Management (IM), Arlington, VA, USA, 2019, pp. 729-730. (Outcome of Chapter 3)

\end{enumerate}

\chapter*{Acknowledgment}
\bookmarksetupnext{level=section}
\addcontentsline{toc}{chapter}{Acknowledgment}
\onehalfspace
\bookmarksetupnext{level=section}

In the name of Allah Almighty, the most beneficent and the most merciful. All praise to Him, who bestowed upon me the wisdom and strength to accomplish this task gracefully. I owe acknowledgments and thanks to a number of people for their guidance, support, and prayers that enabled me to conclude this work successfully.

First and foremost, I would like to express my appreciation to my supervisor Vijay Sivaraman, had it not been for his constant guidance, instructions, encouragement, and occasional timely reminders, I would certainly never have successfully completed this thesis. His support has been extremely valuable. At the end of this long collaboration, I have got nothing but a lot of respect for him.
The second person I owe a great deal of thanks and respect to is my joint supervisor, Hassan Habibi Gharakheili, for his insightful and enlightening comments and opinions; he has been a reliable source of information and yet a meticulous critic. His expertise played a significant part in shaping the ideas presented in this thesis. We had many in-depth technical discussions which were especially valuable for my thesis. I strongly feel that this journey would not have been possible without the support and nurturing of Vijay and Hassan.
I would like to thank my supervisor at CSIRO/Data61 Craig Russell for his support and comments on my research work.

I would also like to express my great appreciation to the University of New South Wales (UNSW) and CSIRO/Data 61 for providing all the resources and a pleasant atmosphere for conducting research. Without their support, I would not have been able to embark on this. 

I would also like to thank my colleagues Minzhao, Ayyoob, Tara, Arunan, Chinaei, Iresha, Sharat, Arman, Lance, Qasim, and Asaf, Thank you all for your fruitful discussions and unwavering support that made this journey a wonderful one.

I have been extremely fortunate to have wonderful friends who were instrumental in making my journey to this point a memorable and pleasurable one. Tariq, Ali, Saad, Talal, Awais, Atif, Haq, Ihsan, Kaleem, Zubair, Akhtar, Bilal, Ahmed, Ali Raza, Usman, Saqib, Azhar, Yasir Rizwan, Yasir Malik, Badar, Babar and Sir Taj, thank you for being there for me.

I cannot thank enough my parents for all their selfless love. I would like to acknowledge the numerous sacrifices they have made for me starting from the day I was born and continuing to date. I would have been nothing without their love and compassionate support. I would also like to thank my brother and my sisters, for their love and affection. Last, but certainly not least, I am very grateful to my wife and my kids who have been very patient and supportive in the time I have been away from them. Needless to say without my wife's sacrifices, I would not have been able to complete work on my thesis successfully.

\onehalfspace

\vspace{-8mm}

\onehalfspace
\cleardoublepage

%\doublespace
\dominitoc
%\onehalfspace
\clearpage
\tableofcontents
\adjustmtc
\cleardoublepage

\clearpage
\listoffigures
\adjustmtc
%\mtcaddchapter[List of Figures]
\cleardoublepage

\clearpage
\listoftables
\adjustmtc
%\mtcaddchapter[List of Tables]

\doublespacing

\adjustmtc
\glsaddall
%\clearpage
%\printglossary[style=super, type= symbol , nonumberlist]
\printglossary[style=super, type=\acronymtype, nonumberlist]

\onehalfspace
\raggedbottom

%\afterpage{\blankpage}
%\afterpage{\blankpage}

\mainmatter
\doublespacing
\pagenumbering{arabic}
\pagestyle{fancy}{%
    \fancyhf{}
    \fancyhead[L,C]{}
    \fancyhead[R]{\nouppercase{\leftmark}}
    \fancyfoot[L]{}
    \fancyfoot[C]{\thepage}
    \fancyfoot[R]{}
    \renewcommand{\headrulewidth}{0pt}
    \renewcommand{\footrulewidth}{0pt}
}

\thispagestyle{fancy}
%\fancyfoot[L]{}
%\setstretch{1.5}

\chapter{Introduction}
\adjustmtc[2]
\mtcsetfeature{minitoc}{open}{\vspace{1em}}
\minitoc

Cyber attacks are becoming more frequent and sophisticated. As a result, enterprise networks constantly face the threat of valuable and sensitive data being stolen by cyber-attackers. Sophisticated attackers are increasingly exploiting the \gls{dns} service for malicious activities such as exfiltrating data as well as maintaining tunneled command and control communications for malware. This is because \gls{dns} traffic is usually allowed to pass through enterprise firewalls without deep inspection or state maintenance, thereby providing a covert channel for attackers to encode low volumes of data without fear of detection. Similarly, attackers are also exploiting \gls{dns} for service disruption such as to launch \gls{ddos} attacks on authoritative \gls{dns} servers and/or open resolvers - The victim is bombarded with random \gls{dns} to utilize their available resources. A famous example of this attack is Mirai attack on Dyn \gls{dns} architecture (in 2016). Unfortunately, Network operators lack the methods and tools to determine whether outgoing/incoming \gls{dns} traffic of an enterprise network is legitimate or cyber-breached.

\section{Problem Statement}

\gls{dns} is an essential protocol for the Internet that is used to resolve the domain name like www.domain.com to its corresponding \gls{ip}address, e.g. 10.0.0.2.  A host needs to ask its local recursive \gls{dns} resolver to resolve a domain. The \gls{dns} data is commonly collected at the recursive \gls{dns} resolvers. The \gls{dns} protocol restricts the length of the domain name (for outbound queries) to 255 bytes containing letters, digits, and hyphens. Also, since the \gls{dns} protocol is used mostly over the \gls{udp}, there is no guarantee that queries will be replied based on their order of arrival. 

\subsection{Data theft using Domain Name Queries}
One way for the attacker to exploit \gls{dns} is to register a domain (\eg {\fontsize{10}{48}\usefont{OT1}{lmtt}{b}{n}\noindent foo.com}) so that the attacker's malware in a host victim can then encode valuable private information (such as credit card numbers, login passwords or intellectual property) into a \gls{dns} request of the form {\fontsize{10}{48}\usefont{OT1}{lmtt}{b}{n}\noindent arbitrary-string.foo.com}. This \gls{dns} request gets forwarded by resolvers in the global domain name system to the authoritative server for the {\fontsize{10}{48}\usefont{OT1}{lmtt}{b}{n}\noindent foo.com} domain (under the attacker's control), which in turn sends a response to the host victim. This provides the attacker with a low-rate but covert two-way communication channel between a host victim and their \gls{cnc} center.

Interestingly, enterprise firewalls are typically configured to allow all packets on port 53 (used by \gls{dns}) since \gls{dns} is such a crucial service for virtually all applications. Some firewalls do offer enhanced \gls{dns} protection. Still, they require deep packet inspection of \gls{dns} messages to identify the covert channel and then isolate domains that contain encoded data. The significant resources needed for this capability \cite{infoblox2015} and the resulting impact on firewall forwarding performance usually results in enterprise network operators disabling such features. This ability to transit firewalls gives attackers a covert channel, albeit a low-rate one, by which to exfiltrate private data and to maintain communication with malware by tunneling other protocols (\eg \gls{ssh}, \gls{ftp} to command-and-control centers. As one example, the remote access trojan DNS Messenger \cite{sampleDNSMessenger2017} discovered in 2017 used \gls{dns} queries and responses to execute malicious powerShell commands on compromised hosts.

%Enterprise networks are often complex, with applications that rely on a mix of local and cloud-based services, and hence difficult to manage securely \cite{homelandSecReport2019}. Enterprise hosts often include powerful servers, personal computing devices, mobile phones, and unmanaged Internet of Things (IoTs). These devices may use a mixture of statically or dynamically assigned addresses from several public and private Internet Protocol (IP) address ranges. 
%Poorly administrated assets, like personal computers or unpatched servers \cite{unpatched2017}, are not only potential victims of cyber-attacks but are also sources of risk for other entities on the Internet. For example, hosts sitting behind the enterprise border firewall can be infected by malware from phishing emails, security holes in browser plugins, or other infected local devices. 

\subsection{Malware Command and Control Communication using DGAs}
Malware-infected machines, forming a botnet, are typically managed remotely by an adversary (aka botmaster) via a \gls{cnc} channel. Cyber-criminals primarily use a botnet for malicious activities such as stealing sensitive information, disseminating spam, or launching denial-of-service attacks. Therefore, law enforcement agencies routinely perform takedown operations on the blacklisted \gls{cnc} servers \cite{MIT2019}, disrupting their botnet activities. In response to these efforts, botmasters have developed innovative approaches to protect their infrastructure. The use of DGAs is one of the most effective techniques that has gained increasing popularity \cite{antonakakis2012throw}.

\gls{dga} make use of a ``seed'' (a random number that is accessible to both the botmaster and the malware agent on infected hosts) to generate a large number of custom domain names. Developing numerous time-dependent domain names and registering only the relevant one(s) ``just shortly'' before an attack allows a botnet to shift their \gls{cnc} domains on the fly and remain invisible for longer \cite{plohmann2016comprehensive}.
The botmaster waits for the malware to successfully resolve a Domain Name System (\gls{dns}) query for the registered domain, enabling the \gls{cnc} communications to take place. Note that even if a \gls{cnc} server is taken offline or blacklisted, this process can simply be restarted and a new server can come online. To date, more than 80 collections of DGA domains (each corresponding to a malware family) have been recorded by DGArchive \cite{dgarchive} and are publicly available. 

\subsection{Service Disruption using Random Domain Names}

gls{dns} works in such a way that if a query is being asked from the \gls{dns} authoritative name server or open resolver, it is an obligation on them to answer it even if the query is non-existent on their ecosystem. Non-existent domains are of two types: (a) Popular search engines and Anti-viruses utilize random domains to convey the one-time signal to their servers known as disposable domains (eg,``{\fontsize{10}{48}\usefont{OT1}{lmtt}{b}{n}\noindent elb.amazonaws.com.cn}'', ``{\fontsize{10}{48}\usefont{OT1}{lmtt}{b}{n}\noindent cloudfront.net}'', and `{\fontsize{10}{48}\usefont{OT1}{lmtt}{b}{n}\noindent avts.mcafee.com}'') - Benign domains also contain typographical mistakes for example a user accidentally write ``{\fontsize{10}{48}\usefont{OT1}{lmtt}{b}{n}\noindent googel.com}'' instead of ``{\fontsize{10}{48}\usefont{OT1}{lmtt}{b}{n}\noindent google.com}''; (b) Malicious NXDs to launch a type of DDoS attack, \gls{dns} water torture attack \cite{water2014} also known as random subdomain attack by dynamically generating random strings as the prefix of a victim domain. The \gls{dns} Water Torture Attack is a type of DDoS attack on \gls{dns} servers. This attack affects both authoritative servers and open/recursive resolvers, but mainly it targets the former.

\lhead{}

%Cyber actors use bots (compromised devices) to send a large number of randomly generated domain names on target servers (victims). The domain names contain the primary domain that is governed by its authoritative name server to tell the \gls{ip}address of that particular domain. During the attack, due to the high number of requests, the authoritative servers, and the recursive resolvers may have slow response to the queries being asked and potentially knock out the servers as well. Although the problem is well known since the last decade, it is mostly dealt concerning the servers perspective (identifying the victim) and by identifying the malicious queries. We see this as an opportunity to find out the regular hosts of an enterprise network to protect the reputation of the enterprise by identifying the hosts making non existent queries to outside world. 

Cyber actors use bots (compromised devices) to send many randomly generated domain names on their victim servers. The queried domain names relate to the primary domain governed by its authoritative name server to tell the IP address of that particular domain. Due to the high number of requests during the attack, the victim authoritative servers and/or the recursive resolvers may have slow responses to the queries being asked or potentially become unavailable. Although the problem has been well understood over the last decade, it is mostly dealt with from the perspective of the victim server by identifying the malicious queries. Therefore, we see this as an opportunity to detect potentially infected hosts of an enterprise that initiate non-existent queries to the outside world. 

%The common practice used for \gls{dns} security of enterprise networks relies on intrusion detection systems (IDS) that monitor network traffic to detect attacks. However, these solutions are not scalable at high data rates if they are software-based or costly if they are hardware-based. 

This thesis is structured into four parts to address the above-pointed issues. First, we review the current literature based on \gls{dns} security. We discuss the available solutions and identify the challenges and limitations in the state-of-the-art. We then develop tools and models to detect \gls{dns}-based attacks in enterprise networks. After a thorough analysis of real \gls{dns} traffic of our campus network, we extract numerous meaningful attributes that can distinguish malicious from legitimate queries. %We train and tune ML models to segregate benign and malicious queries. 
We highlight the prevalence and activity pattern of more than twenty families of \gls{dga}-enabled malware across internal hosts. %We identify malware traffic attributes and train three specialized one-class classifier models using behavioral attributes of malware HTTP, HTTPS, and \gls{udp} flows obtained from a malicious public dataset. 
We develop a monitoring system that uses \gls{sdn} reactive rules to automatically and selectively mirror \gls{tcp}/\gls{udp} flows pertinent to DGA queries (between internal hosts and malware servers) for diagnosis by the trained models. Finally, we draw insights into %high volume of incoming Non-Existent Domains (\gls{nxd}) responses and identify the difference between two scenarios of water torture attack (attack on authoritative \gls{dns} server and/or open resolver) using multistaged iForest models.
the use of \gls{dns} for service disruption.

\section{Research Contributions}
In the context \gls{dns} network security, the following can be considered as significant contributions made by this research: 

\begin{itemize}
	
	\item[1)] Our first contribution tackles data exfiltration using \gls{dns}. We analyze outgoing
	\gls{dns} queries to identify many stateless attributes such as the number of characters, the number of labels, and the entropy of the domain name to distinguish malicious data exfiltration queries from legitimate ones. We also develop, tune, and train a machine-learning algorithm to detect anomalies in \gls{dns} queries using a benign dataset of top-rank primary domains. To achieve this, we have used 14 days-worth of \gls{dns} traffic from each organization. We then implement our scheme on live 10 Gbps traffic streams from the network borders of the two organizations, inject more than three million malicious \gls{dns} queries generated by two exfiltration tools, and show that our solution can identify them with high accuracy. We compare our solution with the two-class classifier used in prior work. (Chapter 3)
	
	\item[2)] Our second contribution tackles malware \gls{cnc} communication using \gls{dns}. We analyze \gls{dns} outgoing queries to identify more than twenty families of DGA-enabled malware when communicating with their \gls{cnc} servers. We draw insights into the behavioral profile of DGA-enabled malware flows when communicating with \gls{cnc} servers by analyzing a Packet Capture (PCAP) trace (3.2B packets) collected during the peak hour from our campus network. We identify malware traffic attributes and train three specialized one-class classifier models using behavioral attributes of malware HTTP, HTTPS, and \gls{udp} flows obtained from a public dataset. We develop a monitoring system that uses \gls{sdn} reactive rules to automatically and selectively mirror \gls{tcp}/\gls{udp} flows pertinent to DGA queries (between internal hosts and malware servers) for diagnosis by the trained models.  (Chapter 4)
	
	\item[3)] Our third contribution studies the use of \gls{dns} for service disruption. We analyse incoming \gls{dns} messages, with a specific focus on \gls{nxd} \gls{dns} responses, to distinguish benign from malicious \gls{nxd}s. We highlight two attack scenarios based on their requested domain names. Using \gls{nxd} behavioral attributes of internal hosts, we develop multi-staged iForest classification models to detect internal hosts that are launching service disruption attacks. We show how our models can detect infected hosts which generate high-volume as well as low-volume distributed \gls{nxd}-based attacks on public resolvers and/or authoritative name servers with an accuracy of over 99\% incorrectly classifying legitimate hosts. (Chapter 5)

	For each of the schemes developed above, we demonstrate the feasibility and efficacy of our methods through the implementation in our campus network via simulation and practical results. We believe that our research will play an important role in securing the internal hosts of any enterprise network using its \gls{dns} traffic.

\end{itemize}

\section{Thesis Organization}
% I will refer to the chapters correctly with their specific labels later.

The rest of this thesis is organized as follows. First, chapter \ref{chap:ch2} surveys the landscape of \gls{dns} based network security and highlights the shortcomings in the current state-of-the-art, and discusses that there is still a need for thorough research based on \gls{dns} security in enterprise networks.
In Chapter \ref{chap:ch3}, we propose real-time detection of data exfiltration in enterprise networks. Chapter \ref{chap:ch4} gives the extensive data analysis of our campus network to draw insights into the prevalence of DGA domains. It also discusses how to identify the infected hosts that are communicating with \gls{cnc} servers by selectively mirrored their traffic using \gls{sdn}. In Chapter \ref{chap:ch5}, we propose an architecture using multi-staged iForest classification models to detect internal hosts that are launching service disruption attacks. Finally, we conclude the thesis in Chapter \ref{chap:ch6} with pointers to directions for future work.

\chapter[Security of DNS Infrastructure]{Security of Enterprise DNS Infrastructure} \label{chap:ch2}
\minitoc

With the exponential growth of networking devices in enterprise networks, cyber-attacks are becoming increasingly sophisticated and intense. DNS is an essential protocol used by every networking device for address resolution. Thus, attackers have used DNS to perform malicious activities such as data theft, C\&C communication, and launch DDoS attacks covertly. According to Forescout Research Labs \cite{forescout}, new DNS vulnerabilities have the potential to impact millions of devices. This chapter will give a thorough background of DNS infrastructure, vulnerabilities, and the existing detection mechanisms. Furthermore, we will identify the existing gaps in the current literature and why our research is essential to address some of the existing gaps.

\begin{figure}[t!]
	\centering
	\includegraphics[width=0.95\textwidth]{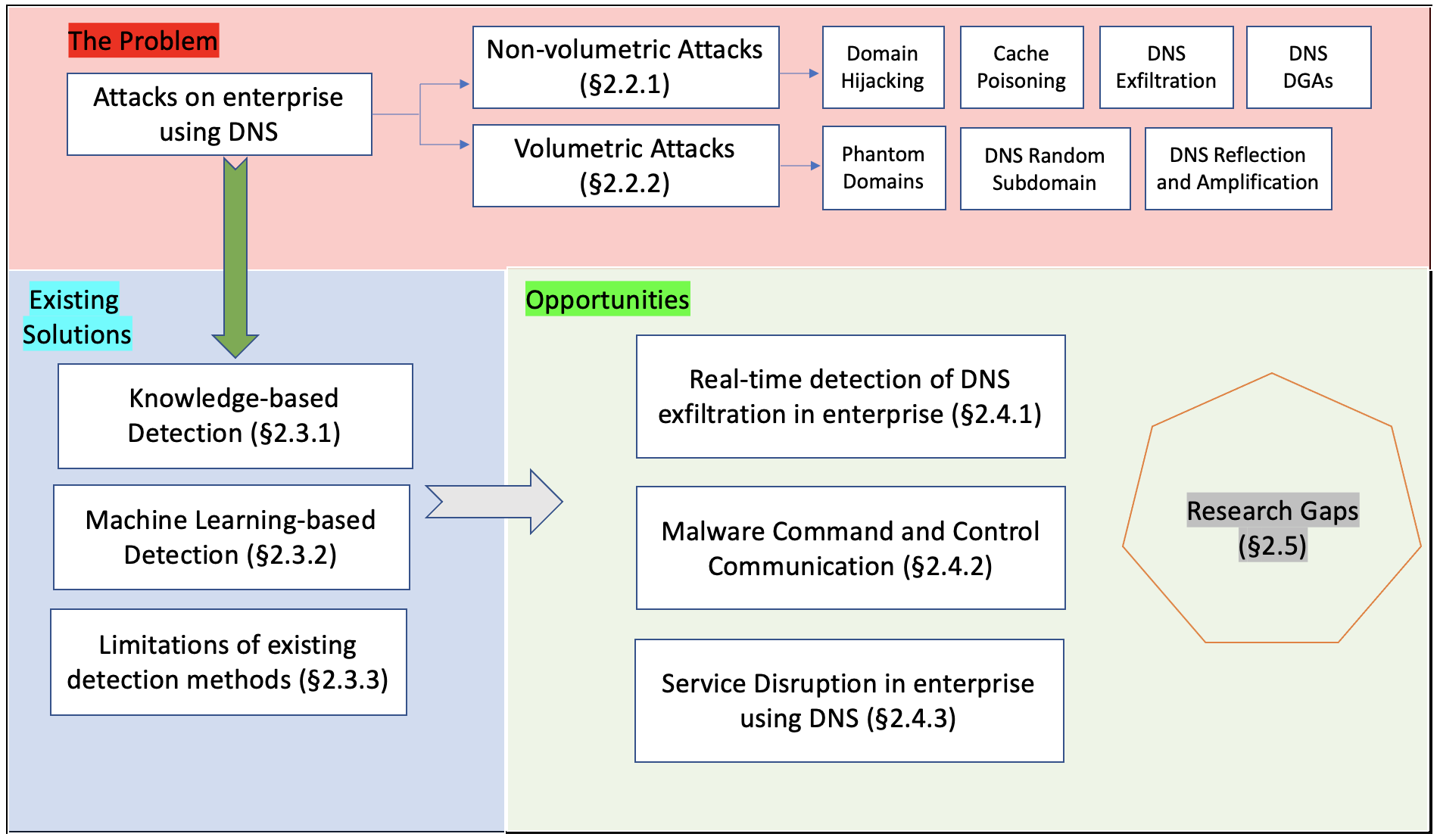}
	\caption{Key concepts covered in this chapter}
	\label{fig:DNS-Taxonomy}
\end{figure}

\section{DNS Infrastructure: Vulnerabilities and Challenges}

Before going into DNS security details, we first discuss the domain name space and steps involved in DNS address resolution. We then discuss the vulnerabilities and challenges face by the DNS infrastructure. Fig.~\ref{fig:DNS-Taxonomy} gives an overview of key topics covered in this chapter. 

\subsection{DNS Hierarchy}

\begin{figure}[t!]
	\centering
	\includegraphics[width=0.88\textwidth]{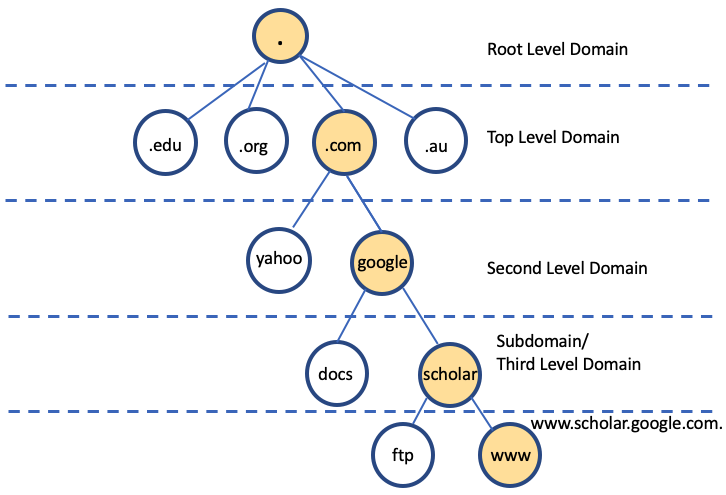}
	\caption{Domain Name Space.}
	\label{fig:DNS-hier}
\end{figure}

The basic function of DNS is to translate a human-readable domain name into an IP address and vice versa. To achieve this, DNS uses a distributed hierarchical domain server system as shown in Fig~\ref{fig:DNS-hier}. It consists of a tree-based structure with a maximum tree depth of 128. A tree is divided into zones.  On the top, we have a root domain containing all top-level domains (TLDs) or zones. The top-level domain is further divided into domain names such as google or yahoo, which we labeled as the second-level domain. Each domain is divided into subdomains, also known as the third-level domain, such as docs or scholar. Finally, we have hostnames such as ftp or www as highlighted in Fig~\ref{fig:DNS-hier}.

\subsection{Steps for DNS Resolution in an Enterprise Network}

\begin{figure}[t!]
	\centering
	\includegraphics[width=0.88\textwidth]{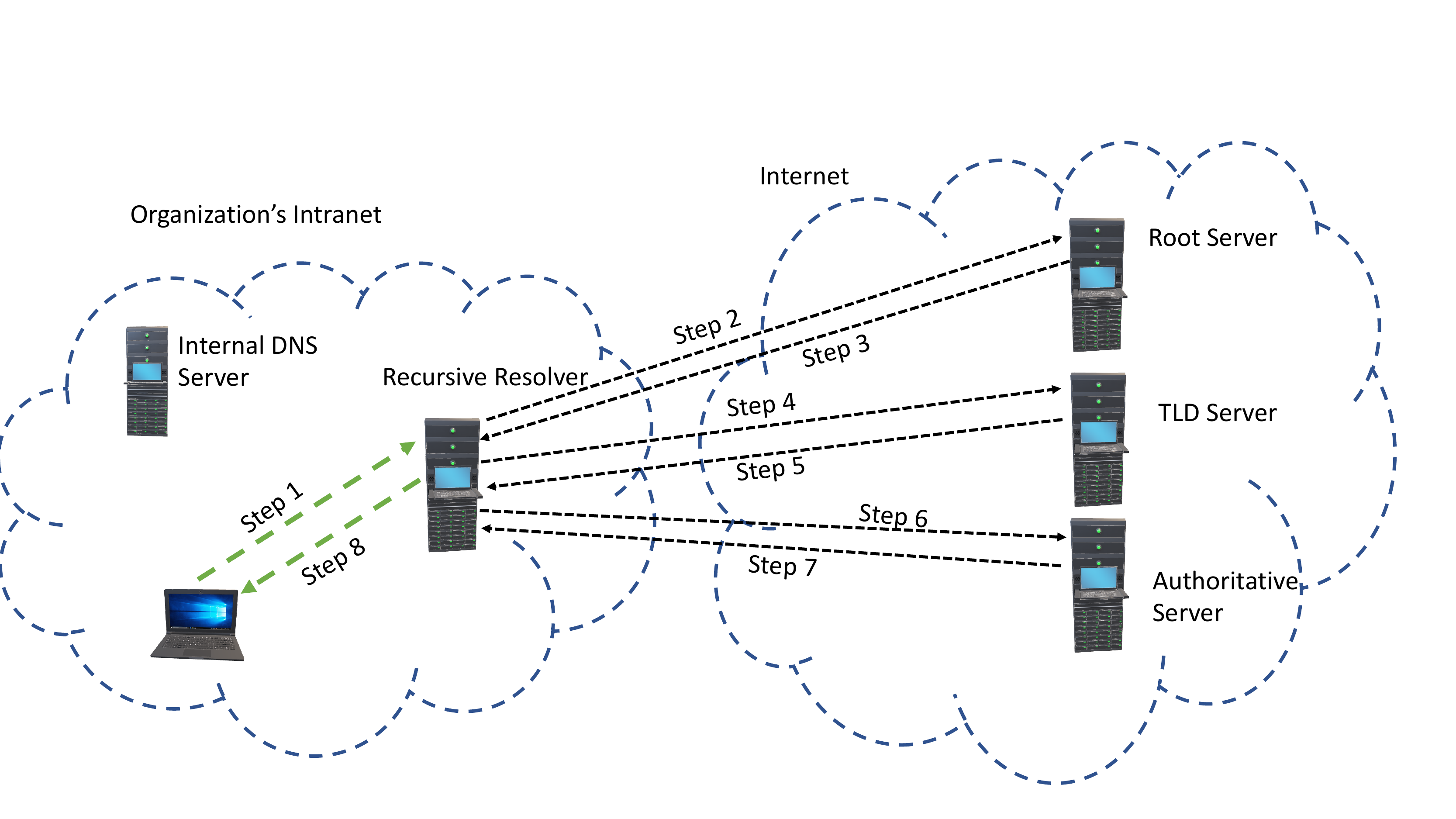}
	\caption{DNS Architecture.}
	\label{fig:dns_arch}
\end{figure}

This section gives all the necessary steps for DNS resolution in an enterprise network. Large enterprises usually focus on the security of their DNS infrastructure. Therefore they use recursive resolvers and their authoritative DNS servers. Fig. \ref{fig:dns_arch} illustrates a big picture of how a client obtains the IP address for a web server to get connected with the server within an enterprise network. The first step is that the client sends a DNS request of www.scholar.google.com to the recursive resolver of the enterprise. The recursive resolver has a local cache memory where it saves the responses with IP addresses for later use. It checks the current DNS request in its cache and forwards the request to the root server if the domain name is not appeared in its cache (step 2). 
The root servers have the resource details of all the top-level domain servers, such as in this case, it will provide the IP address of the .com server (step 3). The recursive resolver then contacts the com server for IP resolution of the second-level domain (SLD) \ie google (step 4). The TLD server (.com in this scenario) sends back the IP address of the name server of google (step 5). The recursive resolver then requests the IP address of www.scholar.google.com from the Google name server (step 6). The google.com server responds to the query of recursive resolver with the IP address of https://www.scholar.google.com (step 7). Finally, the recursive resolver provides the IP address to the client as a response (step 8), and the client gets connected to the www.scholar.google.com server.

\subsection{Vulnerabilities and Challenges of Existing DNS Architecture}

Some enterprises do not restrict their clients and let them configure their own devices, such as configure the IP address of their local DNS resolver and change it to any other publicly available open DNS resolver. Similarly, there is an increase in using personal devices due to the advancement in the Internet of Things (IoT) \ie the concept of Bring Your Own Device (BYOD) that is making a security loophole as far as DNS security infrastructure is concerned. The attackers take advantage by using DNS as a vehicle to perform malicious activities covertly. 

DNS servers and open resolvers can cache most domain names and their IP addresses and respond to anyone who may ask for the particular domain name. However, DNS caches can be easily spoofed and manipulated by the attackers with the help of Man-in-the-Middle (MitM) attacks such as DNS hijacking and DNS cache poisoning attacks (explained briefly in the next section).

To resolve some of the aforementioned issues, Internet Engineering Task Force (IETF) introduced the domain name system security extensions (DNSSEC) in 2005 \cite{arends2005dns} to enhance the data authentication and data integrity of existing DNS architecture using public-key cryptography (digital signing) for authentication. Similarly, DNS over TLS (DoT) \cite{rfc7858} and DNS over HTTPS (DoH) \cite{rfc8484} have been introduced by IETF in 2016 and 2018 respectively to encrypt the traditional DNS communication by using encrypted protocols \cite{nisenoffuser}.  However, this can incur an increased overhead (4 times more bytes in any DNS query or response) as compared to the traditional unencrypted DNS traffic \cite{singanamalla2020oblivious}. Moreover, the signing and verifying operations in DNSSEC take valuable CPU time and memory. Due to these drawbacks, the adoption rate of these encrypted DNS technologies is fairly low in enterprise networks.

\section{DNS-Based Attacks}

\begin{figure}[t!]
	\centering
	\includegraphics[width=0.95\textwidth]{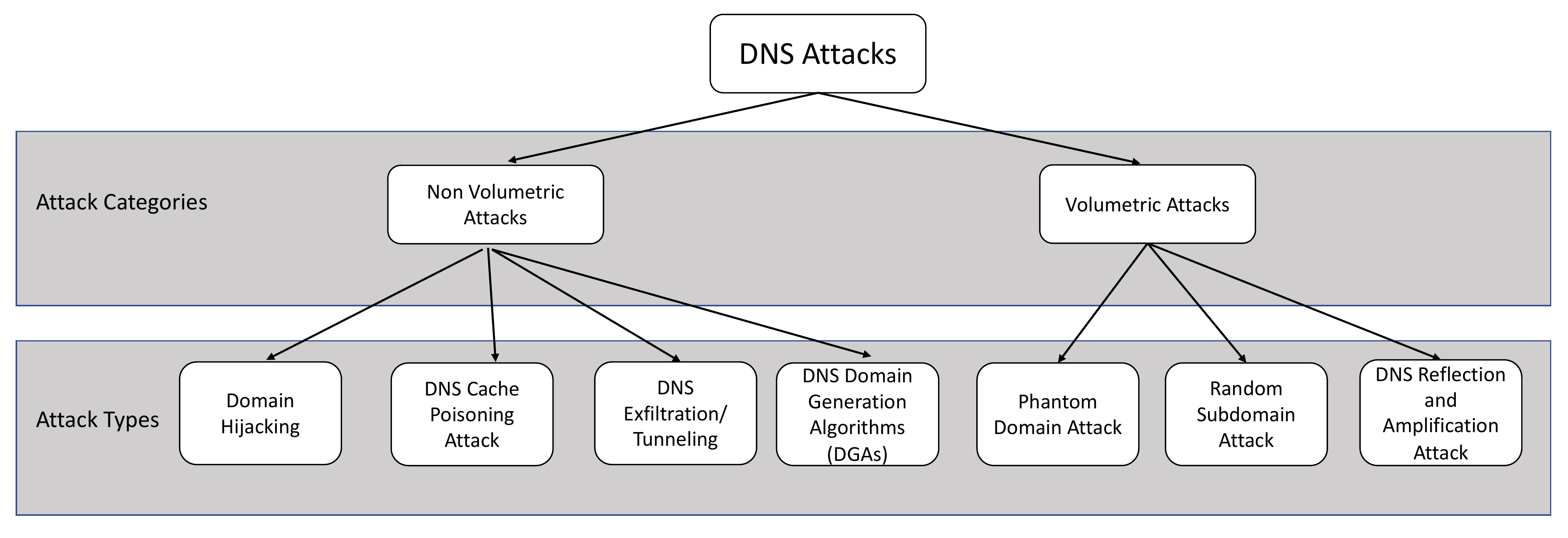}
	\caption{DNS Attack Categories.}
	\label{fig:dns_attacks}
\end{figure}

The world of the Internet and networking is exposed to an overwhelming number of DNS-based cyber-attacks and threats. 
In Fig. ~\ref{fig:dns_attacks}, we have classified DNS attacks into two main categories; \ie non-volumetric attacks and volumetric attacks.  
We will discuss the anatomy of these attacks below.

\subsection{Non Volumetric Attacks}
This section lists the non-volumetric attacks on DNS protocol that can affect any enterprise network with financial and reputational damages.

\textbf{Domain Hijacking:} Domain hijacking is a well-known type of DNS-based MitM attack in which queries are incorrectly resolved by the attackers by redirecting them to malicious websites. Attackers become successful in this attack type by: (a) installing malware on the victim's PC or routers and changing the DNS server information to redirect the DNS requests attacker's control sites; (b) eavesdropping on the local network and sending the spoofed response to the legitimate DNS request coming from the victim. In this scenario, when the legitimate response arrives, it is ignored by the local DNS server.

Another way to hijack a domain is to collect personal information about the actual owner of the domain to impersonate him and convince the domain registrar to update the information or transfer the domain to another registrar they control. 

\textbf{DNS Cache Poisoning Attack:} DNS cache poisoning attack aka DNS spoofing is another example of MitM attack that follows a similar concept as that of DNS hijacking \ie to redirect the DNS requests to attackers controlled sites or DNS servers. The anatomy of cache poisoning attack is to store incorrect and fake information against the legitimate domain names to trap innocent users into malicious activities. As a result, the internet traffic goes to the wrong places until the cached data is fixed. 

\begin{figure}[t!]
	\begin{center}
		\vspace{3mm}
		{\includegraphics[width=0.80\textwidth]{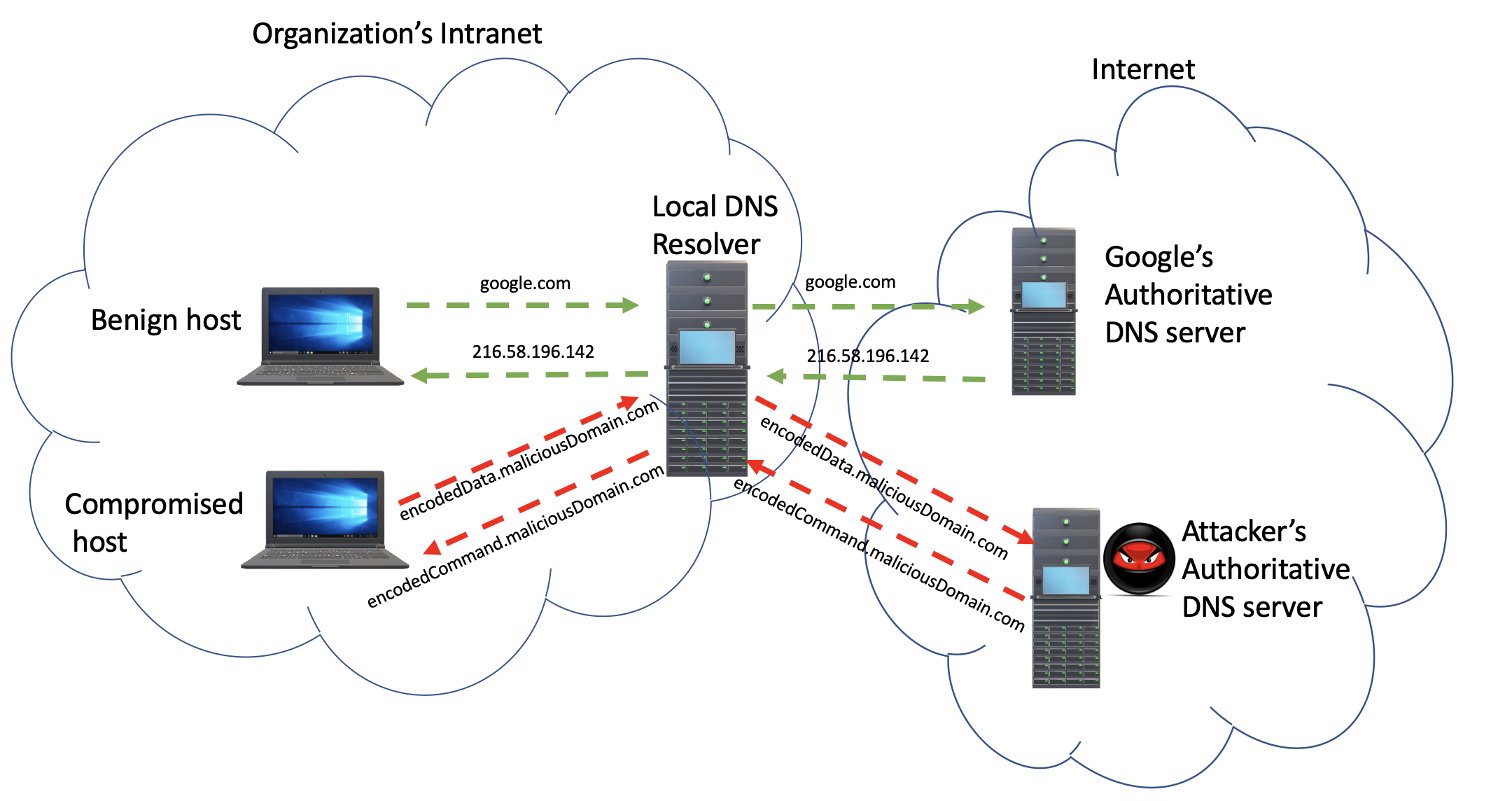}}
		\caption{DNS Exfiltration Attack - Basics.}
		\label{fig:DNSExfill}
	\end{center}
\end{figure}

\textbf{DNS Exfiltration Attack:} The anatomy of the DNS exfiltration attack is shown in Fig.\ref{fig:DNSExfill}. We see two hosts, one is benign, and the other is compromised with malware (such as backdoor). The malware then encodes the personal information (such as credit card data) in the subdomain of an attacker's controlled primary domain (\eg maliciousDomain.com).  The request then goes all the way from the local resolver of the organization to the attacker's domain. The attacker then decodes the encoded sensitive information and uses it for criminal activities.

\textbf{DNS Domain Generation Algorithms (DGAs):} DGA is a technique used by attackers to generate new domain names (based on a predefined algorithm) and IP addresses for malware’s C\&C servers. 
It uses a ``seed'' (a random number that is accessible to both the botmaster and the malware agent on infected hosts) to generate a large number of custom domain names. Generating numerous time-dependent domain names and registering only the relevant one(s) ``just shortly'' before an attack allows a botnet to shift their C\&C domains on the fly and remain invisible for longer \cite{plohmann2016comprehensive}. Some common examples of DGA-based malware include ``Gameover'', ``Suppobox'', ``Ramnit'' and ``Virut''.

The botmaster waits for the malware to successfully resolve a Domain Name System (DNS) query for the registered domain, enabling the C\&C communications to take place. Note that even if a C\&C server is taken offline or blacklisted, this process can simply be restarted, and a new server can come online. To date, more than 80 collections of DGA domains (each corresponding to a malware family) have been recorded by DGArchive \cite{dgarchive} and are publicly available. 

\subsection{Volumetric attacks}

In DNS data flooding attacks, a common example is the Denial of Service (DoS) attack that is a simple and powerful method to consume the available resources of a victim having the intentions to deny its services to legitimate users \cite{mirkovic2004taxonomy}. Most of the attacks on the Internet are performed by some of the hidden software. This software is commonly known as bots \cite{cooke2005zombie} \cite{freiling2005botnet} \cite{abu2006multifaceted}.

\textbf{Phantom Domain Attack:} Phantom domain attack is another interesting yet powerful attack that covertly uses DNS to target the victim DNS resolvers \cite{ramdas2019survey}. The anatomy of the phantom attack is that it uses phantom domains that are being controlled by the attackers, which provide a slow response or no response at all. The bots are sending thousands of requests to the victim DNS resolvers to respond to these phantom domains. The DNS server consumes resources while waiting for responses, eventually exhausting all the available compute and memory resources.

\textbf{Random Subdomain Attack:} Random subdomain attack aka DNS Water Torture attack or Slow Drip attack is a type of DDoS attack on authoritative DNS servers and/or open resolvers in which the victim is bombarded with random non-existent domain names (NXD) from attackers controlled machines such as bots. It requests subdomains or hosts that do not exist, which consumes memory and processing resources- eventually leading to degraded performance or failure. Mirai botnet is an example of this attack on Dyn DNS architecture in 2016 \cite{water2014, afek2020nxnsattack}.

\textbf{DNS Reflection and Amplification Attack:} DNS reflection and amplification attack is a major type of volumetric attack which makes use of the open DNS resolvers over the Internet to take part in the attack against a specific victim \cite{rozekrans2013defending}. The idea is to spoof the identity of the actual attacker so that the responses can go to the spoofed IP address (real victim) and increase the magnitude of the attack by asking for more than one type of DNS response which leads to increased response packet size. Malicious actors send thousands of spoofed DNS queries to the open DNS resolvers, which respond with the large-sized responses directed to the victim's IP address. The victim can be a stand-alone PC or an authoritative name server which can be overwhelmed by the unsolicited responses - can cause slow performance or outage depending on the available resources of the victim.

\section{Detection Systems in Cyber-Security}

Intrusion detection systems are the computer tools that are used to enhance the security of computer networks. IDS plays an essential role in identifying the network traffic and classifying it as benign or malicious. In addition, they can identify the intrusions from an external source as well as the internal intrusions. Detection systems are broadly classified into two main categories: Knowledge-based/Misuse-based detection and Anomaly-based detection.

\subsection{Knowledge-Based Detection}

The working principle of Knowledge-based (Misuse-based) detection is that it only classifies the attacks or the malicious content, and the traffic pattern is previously known. It is also known as signature-based detection because they depend on the signatures of the attacks provided by the expert in network security. Snort \cite{roesch1999snort} and Bro \cite{paxson1999bro} are the two open-source signature-based IDS that are widely used for this specific purpose.  The main advantage of using this type of intrusion detection is that it detects the known attacks without overwhelming the system with false-positive alarms \cite{buczak2016survey}.  However, to detect the latest attacks on the network, these systems require a continuous feed of updates in systems. Due to the increase in categories of attacks, it is challenging to update signatures of all attacks and vulnerabilities in the network. Another disadvantage of this approach is that it burns many resources in terms of updating the systems periodically.
%Existing up-to-date network intrusion detection systems such as Snort  \cite{roesch1999snort} and Suricata \cite{albin2012realistic} use signature-based matching and it uses deep packet inspection to detection attacks. These tools differentiate themselves by focusing on different areas; the former is lightweight while the latter provides deeper analysis \cite{day2011performance}. Snort compromises depth for lowering complexity which leads to a higher false-positive rate \cite{patton2001achilles}. 
Thus, there is a definite need for scalable intrusion detection systems to perform better for high-speed networks.

\subsection{Machine Learning-Based Detection: Anomaly Detection}

Big data analytics has become a hot topic in recent times in the field of industry and academia. Data has been increasing exponentially with time. Big data comes into practice where the data sets are extensive and are difficult to examine by traditional data processing methods \cite{cui2016big}. To analyze the big data, we use the tools and algorithms of Machine Learning (ML). 

The anomaly-based detection approach somehow bridges the gap of misuse-based detection by capturing the attacks that are not seen before and without using an attack signature. The essential criteria to find the anomaly is that the system is fed with normal traffic. Now, the system is aware of the behavior of normal traffic. So, it raises the alarm whenever it observes any traffic that diverges from this normal behavior. The main drawback of this technique is the high number of false alarms in some anomaly-based NIDS.  
The framework in \cite{tegeler2012botfinder} was designed to investigate the botnets in the network \cite{mirkovic2004taxonomy}. The authors in \cite{tegeler2012botfinder} presented a novel system to detect these bots in the overall network with the help of machine learning techniques by looking at the high-level information about the network traffic. The framework is based on a learning-based approach that automatically generates the bot detection models.

\subsection{Limitations of Existing Detection Mechanisms}
\label{subsec:Limitations-existing}

ML security applications range from identifying malicious activity within a network to predicting the attack at any period by observing the anomalies in the network traffic. The bulk of the research falls on one of two sides, offline analysis on a labeled dataset with multiple ML algorithms \cite{kayacik2005selecting} \cite{shrivas2014ensemble} or real-time analysis on lab-grown data with a single ML algorithm \cite{seufert2007machine}. The trade-off in this spectrum is between computation and timeliness. The former approach emphasizes mass data extraction (Deep Packet Inspection) but is not feasible due to the global trend towards end-to-end encryption \cite{felt2017measuring}. Attempts at using purely flow-level information have yielded mixed results depending on how the data was captured. \cite{giotis2014combining}. 

%In \cite{lakhina2004diagnosing}, the authors have proposed to identify the anomaly detection by using the ML algorithm of Support Vector Machines (SVM). The limitation in this research work is that the dataset used is inappropriate because it has a small subset of malicious traffic from largely benign traffic. Furthermore, anomaly detection relies heavily on network behavior being static and predictable; the false positive rate spikes when there is a slight deviation from the model \cite{thottan2003anomaly}. 

Decision tree classifiers are not suitable for large volumes of benign traffic. Most of the research works conducted in this area use disproportional datasets \cite{farid2010combining}. A typical dataset is the KDD99 which has an attack-to-traffic ratio of 4:1, which is acknowledged as a significant limitation by the research community \cite{koc2012network}. The advantage of using neural network methods is that they are not limited to known attacks or signatures \cite{shun2008network}. These trained models can classify unknown attacks with credible success \cite{poojitha2010intrusion}. The two major disadvantages of this model are the need for large volumes of labeled data and time, and the predictions are computationally expensive relative to other models.

\section{Review on Current State-of-the-Art DNS security}

This section will focus on the current state-of-the-art  DNS security, specifically command-and-control (C\&C) communication, data theft, and service disruption across a wide range of assets. 

\subsection{Monitoring DNS Queries of Enterprise Hosts for DNS Exfiltration}

\textbf{Malware's Perspective:}
From a security standpoint, the DNS protocol is an excellent covert channel. 
According to IDC 2021 Global DNS threat report, DNS-based malware is ranked second among commonly used attack vectors following phishing attacks  \cite{global2021}. Due to its crucial internet role, misconfiguration of the DNS can lead to network disconnects. Therefore, security policies rarely restrict it (e.g., allowing resolutions only to specific domain names). In addition, the DNS protocol is often less monitored than other Internet protocols (e.g., HTTP, FTP, and mail transfer protocols).  It follows that the use of the DNS protocol as a covert channel has been a part of previous cyber campaigns, including a 56M credit cards theft from Home Depot in 2014 \cite{HomeDepot2014}, and a 25k credit cards theft from Sally Beauty \cite{SallyBeauty2014}. During the Covid-19 pandemic, financial firms have been affected the most by DNS-based attacks with an average of over \$1 Million \cite{global2021}.

In an enterprise network, the users within the network can still make the DNS queries out, which can cause the data exfiltration over the DNS. For example, FireEye has discovered the new malware named Multigrain that has stolen the credit card data from the point of sale systems and send that information by exfiltrating data over the DNS \cite{Multigrain2016}.  BernhardPOS, NewposThings, and FrameworkPOS are some other examples.

Feederbot \cite{Feederbot2011} is one from the malware family which has used DNS for sending its command and control messages to the botmaster \cite{dietrich2011botnets}. Morto worm \cite{Morto2011} has been used by the attackers for sending the malware commands that were using the DNS protocol via sending the DNS queries with the Query type of TXT. Wekby pisloader \cite{wekby}is also from the bad actors that are using their DNS server to send and receive the command and control commands to and from the compromised hosts. The information is encoded in the DNS requests. A remote access trojan named DNSMessenger \cite{dnsmsg} was discovered in 2017, which uses DNS queries and responses to execute the malicious PowerShell commands on compromised hosts. It is a sophisticated attack as it does not require any file writing to the compromised hosts. The technique uses DNS TXT records to extract the PowerShell commands saved as DNS TXT records remotely.

\textbf{Malicious Domains' Perspective:}
DNS traffic has been analyzed to identify malicious network activities \cite{jawad2019DNS,PAM19}. Studies in \cite{surveyCovertCahnnel2007,surveyDNS2018} survey the available research literature on the misuse of DNS protocol for various attacks. Common malicious activities that utilize DNS include command and control (C$\&$C) traffic tunneled over DNS channel, circulating spam messages, transferring credit card numbers (or other sensitive information), and hosting scams and phishing websites  \cite{antonakakis2011detecting,antonakakis2012throw}. Therefore, it is important to profile and detect these malicious activities. Over the last decade, there has been an increasing amount of works \cite{hao2010internet,bilge2011exposure,reputationDNS2010,initDNSBehaviour2011,mowbray2014finding} on identifying these malicious activities mostly related to C$\&$C communications \cite{dietrich2011botnets,feily2009survey} and phishing \cite{khonji2013phishing}. 
%The primary focus of our work is to detect the queries involved in the exfiltration of the sensitive information to the attacker from the compromised host within the network or facilitate C$\&$C communications from an enterprise network. 

\textbf{Machine Learning Based Detection of DNS Exfiltration and Tunneling:}

Researchers have used three categories of methods for detecting DNS exfiltration and tunneling, namely statistical-based techniques \cite{paxson2013practical,cambiaso2016feature,NgViz2010,DNSfreqAnalys2010,qi2013bigram}, supervised multi-class classification \cite{liu2017detecting,das2017detection,buczak2016detection,schuppen2018fanci}, and unsupervised one-class classification \cite{engelstad2017detection,asaf}. 

Work in \cite{paxson2013practical} proposed a method to find maximum information that can be encoded in a sub-domain portion of a DNS query name to detect whether the query contains encoded data or not. The authors used an information-theoretic approach, namely the use of Kolmogorov complexity. The authors established an upper bound on the volume of surreptitious communication by investigating inter-query time and query record type. In \cite{cambiaso2016feature}, authors employed mutual information and principal component analysis for dimensionality reduction based on consecutive DNS request and response sizes. In \cite{NgViz2010,DNSfreqAnalys2010,qi2013bigram},  authors have proposed DNS tunnel detection using character frequency analysis. However, the detection criteria are based on the threshold value for which attackers can go undetected easily.

In \cite{liu2017detecting,das2017detection} the authors employed a supervised learning-based model with logistic regression to classify queries as either normal or exfiltration.
Buczak et al. \cite{buczak2016detection} used the Random Forest algorithm for the two-class classification of benign and malicious DNS queries. Similarly, Samuel et al. \cite{schuppen2018fanci} proposed a model to detect malicious DNS query names (generated by malware-infected machines) using Random Forest.
However, attributes used in prior works to train the model are stateful (such as tracking the inter-arrival time of DNS packets or the frequency of query type) or require both DNS query and response messages (such as response length) \cite{liu2017detecting,buczak2016detection}.  {Also, this body of work essentially trains a model with both benign and malicious instances (\ie a two-class classifier) and the accuracy of detecting malicious queries dropped when a new family of attack is introduced (\eg model accuracy varied from 27\% to 75\% depending upon model parameters in \cite{buczak2016detection})}.

\textbf{Deep Learning Based Detection of DNS Exfiltration}

Over the past two years, the interest of researchers on DNS exfiltration has shifted from ML-based detection systems to deep learning-based detection systems \cite{darem2021visualization,wang2019bidirectional,zhang2019dns,CHEN2021102095}. Work in \cite{CHEN2021102095} employs a combination of Convolutional Neural Network (CNN) and Long Short-Term Memory (LSTM). For LSTM, authors use one hidden layer and they assumed that the first 128 bytes/characters contain the actual message information; therefore, they set the length of the hidden layer to 128. Similarly, for CNN, the authors used three convolutional layers, two max-pooling layers, and one softmax layer. Their model is trained on two classes of FQDN, i.e., benign and malicious. Our model instead is trained on attributes of benign FQDNs, resulting in a one-class classifier. The key difference is its way of detecting anomalies and ability to detect zero-day attacks (i.e., it flags all anomalous domain names that deviate from the attributes of the benign domain names). Although deep learning-based approaches give a high percentage of accuracy, they require a huge amount of data (typically more than a million instances) for improved performance. Furthermore, Deep Learning-based methods demand an increased computing resources and add more complexity to operational networks \cite{luong2019applications,mohammed2019machine}.

%Let us compare the results of existing above three existing works with our work. We can see in Table.~\ref{tab:metrics}, that the accuracy of Ahmed's model is 90\% which is more significant than Chen's CNN method, however, we see a considerable improvement made by Chen's LSTM method with an accuracy of 99.3\%. We can see that our proposed model outperforms Chen's LSTM method with an accuracy of more than 99.75\%. shows the overall performance of all the models. The table also gives the comparison in values of other performance metrics. We can say that our novel DNN classifier distinguishes the DNS exfiltration attack domains from benign domains in better proportion. 
\subsection{Behavioral Analysis of DGA-Enabled Malware of Enterprise Hosts}

Malware behavioral analysis has been widely studied by many researchers \cite{yen2008traffic,anderson2016identifying,anderson2017machine,antonakakis2011detecting,antonakakis2012throw,tegeler2012botfinder,yadav2012detecting} using different tools and techniques \cite{talukder2020tools}. The most relevant works to ours can be divided into three categories: (a) detection of malicious traffic based on unusual DNS queries (predicting the presence of DGA domains) \cite{yadav2010detecting,antonakakis2011detecting,antonakakis2012throw,garcia2014empirical,schuppen2018fanci,yadav2012detecting,anderson2016deepdga, vinayakumar2018detecting,marin2020deepmal,choudhary2018algorithmically,spooren2019detection}, (b) network behavioral analysis of known malware and botnet by inspecting their traffic data and/or metadata \cite{gu2007bothunter, gu2008botminer,gu2008botsniffer,tegeler2012botfinder,anderson2016identifying,anderson2017machine}, (c) use of SDN  and/or programmable networking in detecting cyber-attacks \cite{gupta2018sonata,afek2017network,fayaz2015bohatei,zhangposeidon}.

\textbf{Malicious DNS Queries:} 
DNS traffic has been analyzed to identify malicious network activities \cite{PAM19, antonakakis2011detecting, schiavoni2014phoenix}. Over the past decade, there has been an increasing number of works \cite{hao2010internet,bilge2011exposure,reputationDNS2010,initDNSBehaviour2011} on detecting malicious network activities mostly related to DNS exfiltration, DNS tunneling, and C$\&$C communications \cite{dietrich2011botnets,feily2009survey,jawad2019DNS,ahmed2019monitoring, ahmed2019demo}. 

%To detect DNS exfiltration and tunneling attributes like length, entropy, and the number of labels for domain names of benign queries are considered to train a model \cite{jawad2019DNS,ahmed2019monitoring, ahmed2019demo}. 

In the past, blacklists were used to detect C\&C communications between servers and infected hosts. However, blacklisting has been defeated by attackers since they migrated from a static domain mapping to the use of algorithmically generated {domain names}.
In response to this change, researchers have attempted to automatically detect DGA domains using statistical modeling of DNS traffic \cite{yadav2010detecting,antonakakis2011detecting,antonakakis2012throw,garcia2014empirical,schuppen2018fanci,yadav2012detecting}, or machine/deep learning techniques \cite{anderson2016deepdga, vinayakumar2018detecting,marin2020deepmal,choudhary2018algorithmically,spooren2019detection, mcdole2020analyzing}. Antonakakis et al. \cite{antonakakis2012throw} develop a clustering-based method for detecting new (unknown) algorithmically generated domains (AGD) as well as classifying known AGDs using supervised learning.  {The authors evaluated their proposed solution in a large ISP network and found several new families (unseen before) of DGAs, operating on the network.} They used statistical attributes including entropy and n-grams measures as well as structural attributes such as length and label count in the domain name, extracted from NXD (non-existent domains) responses on a per-host basis. %Their benign dataset represents the top 10K Alexa domains whereas, for the malicious dataset, they used the 60K NXD responses gathered from four DGA-enabled malware families (\ie Bobax, Conficker, Murofet, and Sinowal).
Schuppen et al. \cite{schuppen2018fanci} {employ} manually-engineered features to train a binary classifier using Random Forest algorithm to determine whether domains in NXDomain-failed DNS queries (\ie queries to non-existent domain names) are benign or malicious. It remains unclear how this model performs in classifying unseen malicious domains. We note that generating a rich training dataset of benign NXDs is nontrivial, and a labeled benign dataset may get polluted by some new (but unknown) malicious NXDs. 

\textbf{Malware and Botnet Behavioral Analysis:} 
Flow-based analysis has been used to detect malware and botnet traffic. However, monitoring high-rate network traffic in large and complex enterprise environments is difficult and computationally expensive. Works in \cite{gu2007bothunter, gu2008botminer,gu2008botsniffer,tegeler2012botfinder, sivaguru2020inline} develop flow-based features like packet count and distribution of packet length by analyzing every packet of the network to model the patterns in encrypted traffic. 
Similarly, Anderson et al. \cite{anderson2016identifying,anderson2017machine} employed supervised learning algorithms to classify malware and benign traffic. Their model was trained by a variety of host-level attributes, including packet size, flows inter-arrival, DNS query (\eg TLD, TTL, domain rank in { Alexa}), HTTP data (such as server code, content type, Accept-language, and location), and TLS data (such as TLS cipher suites and TLS extension) that are measured over a while. 

Our behavioral modeling approach differs from prior works in three ways: (a) we only monitor the behavior of selected flows pertinent to certain servers (resolved by DGA responses) instead of monitoring all traffic of every host on the network, (b) we choose to extract statistical attributes of flows (encrypted and unencrypted) to train our models, without the need for inspecting payloads like HTTP content type or TLS handshake cipher keys, and (c) our models are built by one-class classification algorithms and hence become sensitive to changes in any attribute while multi-class models become sensitive to changes in only discriminative attributes.
Also, it is essential to note that we only use a database of known DGA domains and check domain queries against this database in real-time instead of classifying a domain as benign or malicious. Our objective is to detect infected hosts of enterprise networks by monitoring their selected flows. %We isolate the malware traffic with the help of reactive rule insertion and finding the efficacy of our system based on the one-class classifiers trained from the malware traffic of known malware-related { datasets}. Lastly, we validate our results with an existing intrusion detection system (IDS).

\textbf{Programmable Networking for Network Security:}

Software-Defined Networking (Programmable Networks) is an evolving paradigm to manage the networks in a better way. SDN is a technology that decoupled the control plane with a data plane (i.e. switches and routers). Now the network designs are not vendor-centric that was the case in traditional networks. Then, the network operators had to configure every device with low-level commands individually. 

Programmable networking has recently gained popularity among researchers, specifically for network security use-cases \cite{gupta2018sonata,afek2017network,fayaz2015bohatei,zhangposeidon}. Gupta et al. \cite{gupta2018sonata} {develop }scalable telemetry (\ie partitioning different types of traffic such as TCP and ICMP) that can be used to collect and analyze the network traffic in real-time using a programmable data-plane. Furthermore, the authors show that their approach reduces the workload of the overall system to be able to operate at a line rate.
Similarly, Zhang et al. \cite{zhangposeidon} proposed an approach that uses P4 programmable switches for a better defense mechanism against DDoS attacks. The authors considered the case of volumetric DDoS attacks. They provided the defense strategies in a modular fashion that can be adopted for each network and can be used for new defense strategies other than just the DDoS. Although these prior works primarily leverage the programmability features in the data plane, we instead employ the programmable control-plane available by Openflow-based SDN to dynamically select suspicious flows for diagnosis by trained machine learning models.

{
	In a relevant work, \cite{ceron2016mars}, Ceron et al. developed an automated system for offline malware analysis, recording the network behavior of a given malware in a controlled sandbox environment orchestrated by an SDN controller. {A known host on their sandbox is infected by malware (from a set), and the SDN controller inspects every packet from/to this infected host for taking required actions like rate-limiting, blocking, or re-configuring the topology, upon finding certain patterns in the packet payload (\ie regex signature) or headers (\eg contacting specific IP address and/or TCP/UDP port numbers)}. 
	
}

\subsection{Identifying Malicious Hosts by Analyzing DNS NXDs}

{\textbf{Identifying Malicious Queries:}} To resolve the NXD attacks, specifically water torture attacks, researchers come up with significant countermeasures such as rate limiting and IP address blocking. However, it can severely affect the legitimate users since the malicious queries can forward through the open resolvers or ISP cache servers. Therefore, the IP address of the cache server will be blocked, and the legitimate users will not have access to the ISP cache server. Similarly, if the volume of queries exceeds the limit set in rate-limiting, it will block all queries, even from legitimate users. Researchers \cite{10.1145/2534142.2534146,takeuchi2016detection} have also examined domain names to detect the malicious queries by focusing on NXD error responses only. Kazato et al. \cite{10.1145/2534142.2534146} predicted whether a domain name included random words via a score calculated by comparing bigrams of domain names of malicious domains and those of benign domain names.

{\textbf{Identifying Attack on DNS Servers:}} A group of researchers \cite{alonso2016mining, takeuchi2016detection} have identified NXD attacks on DNS servers by setting the threshold on the number of non-existent domains. The approaches can be fruitful for heavy volume attacks (such as bursty data) - moreover, choosing a threshold value would be challenging. However, this approach does not work efficiently for the lightweight and distributed NXDs, bypassing the threshold-based security systems.

\section{Research Gaps in Prior Works}

A considerable number of previous research investigations have focused on the detection of malicious domains \cite{hao2010internet, bilge2011exposure,reputationDNS2010, initDNSBehaviour2011}.  The techniques used are mostly based on passive DNS analysis. Leyla et al. \cite{bilge2011exposure} came up with the EXPOSURE system to detect the suspicious domains by extracting 15 features from the DNS traffic. The features are divided into four main categories i.e, DNS query-name based, time-based, answer based and TTL based. The authors have claimed that their system has been validated in a real-world dataset containing 100 billion DNS requests. The system captures the misused suspicious domains taking part in botnet command and control and spamming. Manos et al. \cite{reputationDNS2010} proposed a similar method by analyzing the passive DNS queries. They gave NOTOS as a name to their approach, a dynamic reputation system for the DNS domain names. They have extracted 41 features grouped into two main categories i.e, network-based features and zone-based features. The main idea is that malicious domains have unique characteristics that can easily be separated from legitimate domains based on their extracted features. Such approaches, however, have failed to address the information theft over the DNS queries, and it does not detect the malicious domains in real-time.

Manos et al. \cite{antonakakis2011detecting} proposed another system called Kopis, which is operating at the upper DNS hierarchy i.e, the authoritative name servers and Top-level domain servers, in contrast to NOTOS and EXPOSURE, which were operated at the local recursive DNS servers. This approach has enhanced the visibility of the DNS messages. The same process is followed by \cite{antonakakis2012throw} to capture the DNS traffic. However, the aim is to identify the DGA (Domain Generation Algorithms) bots by focusing only on the DNS queries with name error responses i.e, NXDomain responses. Although extensive research has been carried out on malware detection over DNS protocol at the upper DNS hierarchy, the features considered may not be that effective at the network level.

We believe that a two-class classification approach (\ie signature-based) is insufficient to address new and increasing types of attacks. Also, obtaining ``ground truth'' on a diverse set of malicious instances to train the classifier is difficult \cite{Sommer2010}.   
The authors of \cite{engelstad2017detection} employed unsupervised machine learning algorithms (\ie one-class support vector machine and k-means) to detect DNS tunneling. Their primary focus was to identify infected mobile devices using stateful attributes, including the time between a DNS query and its corresponding response and the size of individual devices' DNS query/response. {In \cite{homem2017harnessing}, Homem et al. benchmarked the performance of four algorithms (multi-class decision trees, support vector machine, K-nearest neighbors, and neural networks) in identifying tunneled traffic (\eg HTTP, HTTPS, and FTP) over DNS.} The authors used only three attributes of DNS packets, including the size of IP packet, length of query name, and entropy of query name.
Similar to our approach, Nadler et al. \cite{asaf} proposed an anomaly-based solution to detect low throughput data exfiltration over DNS.  This work evaluated the performance of isolation forest and support vector machine learning algorithms. However, the authors maintain several attributes for each primary domain over the last $n$ hours (\eg rate of A and AAAA records, the average length of query name). This makes it difficult to detect malicious queries in real-time.

{\subsection{Novelty of Our Approach}}	

To the best of our knowledge, our work in chapter \ref{chap:ch3} is the first that presents a thorough analysis of attributes for query names from operational enterprise networks. Our focus is on attributes of fully qualified domain names that can be extracted in ``real-time'', without a need for states (\ie ``stateless'') -- we assume that DNS  traffic is not encrypted over TLS. We also provide fascinating insights into the practical considerations of such a detection scheme. 
Our scheme can be extended by collecting states only for those hosts that generate anomalous queries and ultimately mitigate malicious DNS tunneling/exfiltration -- such mitigation is beyond the scope of this research.

Our work in chapter \ref{chap:ch4} develops an automatic SDN-based system for detecting malware-infected hosts in real-time by relying more on the behavioral activity profile of ``selected flows'' rather than the content of packets. Further, our SDN switch does not send any network packets to the controller ( allowing the solution to scale to high rates). Instead, packets that need to be inspected in the software are sent as copies on a separate interface of the switch, to which a software inspection engine is attached. 

In chapter \ref{chap:ch5}, we have developed a monitoring system at the source side that can detect the internal hosts generating high volume NXD attacks as well as distributed NXD attacks by using multi-level machine learning algorithms. Existing research focuses on detecting heavy volume NXD attacks at the reception side \ie the authoritative name server or the queried domain (primary domain). Instead, our work is to detect the enterprise hosts generating attack traffic to other networks to protect our campus network's reputation.

\section{Conclusion}

Domain Name System (DNS) is an essential application layer protocol used by every Internet-connected device to obtain the IP address of web servers. However, due to its prevalence and importance, DNS traffic is rarely inspected by intrusion detection systems, making this protocol vulnerable to various cyber-attacks. In this chapter, we have studied and characterized DNS-based attacks. We have also discussed the detection mechanisms and the underlying challenges and gaps in the existing research. We investigate some of these critical research problems in the rest of this thesis, beginning with data theft in enterprise networks.

\chapter[Monitoring DNS Queries for Detecting Data Exfiltration]{Monitoring Enterprise DNS Queries for Detecting Data Exfiltration from Internal Hosts}\label{chap:ch3}

\minitoc

In the previous chapter, we highlighted the importance of DNS security in enterprise networks. In this chapter, we analyze DNS data to draw insights into the characteristics of exfiltrated domains and then develop a machine learning-based method to detect the exfiltrated domains in real-time in two enterprise networks. Parts of this chapter have been published in \cite{jawad2019DNS,ahmed2019demo,ahmed2019monitoring}.

\section{Introduction}

Cyber-criminals have exploited DNS to maintain covert communication channels with compromised hosts. The resulting damages can be huge, amounting to several million dollars in a single attack \cite{efficientIP2017}. Based on a recent DNS security survey of Infoblox \cite{infoblox}, 46 percent of the businesses of North America and Europe have exploited by DNS exfiltration, and about 45 percent affected by DNS tunneling. Several high-profile data exfiltration breaches have been reported since 2014: the Sally Beauty breach (a theft of 25K credit cards) \cite{SallyBeauty2014} and FrameworkPOS malware (a theft of 56M credit cards from Home Depot) \cite{HomeDepot2014} in 2014, BernhardPOS malware \cite{sampleBernhardPOS2015} in 2015, MULTIGRAIN malware \cite{Multigrain2016} in 2016, Win32.Backdoor.Denis \cite{BackdoorDenis2017} in 2017, and UDPoS Malware \cite{UDPOS2018} in 2018. In addition, there have been several DNS tunneling incidents in which malware actors used their DNS servers to send and receive the command and control commands to and from compromised hosts. Examples include Feederbot \cite{Feederbot2011}, and botmaster \cite{dietrich2011botnets}, Morto worm \cite{Morto2011}, and Wekby pisloader \cite{wekby}.

This chapter develops and validates a mechanism for real-time detection of DNS exfiltration and tunneling in two operational networks -- a large University and a mid-sized Government Research Institute. Our {\bf first} contribution is to collect and conduct a thorough analysis of real DNS traffic from the two organizations over several days and extract stateless attributes of DNS messages, such as length, entropy, dots, numerics, uppercase characters, and the number of labels, that can distinguish malicious from legitimate queries. Our {\bf second} contribution is to develop, tune, and train a machine-learning algorithm to detect anomalous DNS queries based on the above attributes using a known dataset of benign domains as ground truth based on 14 days worth of DNS data from the two organizations. For our {\bf third} contribution, we implement our scheme on live 10 Gbps traffic streams from the network borders of the two organizations, inject more than three million malicious DNS queries generated using two exfiltration tools (our customized tool and an open-source tool) and show that our scheme can identify such malicious activity with high accuracy.
We also show our one-class classifier outperforms an existing two-class classifier in detecting unknown DNS exfiltration attacks.
We draw insights into anomalous DNS queries detected by our models, looking into their anomaly scores, tracking query counts in real-time, the number of enterprise hosts querying them, and investigating the TTL/Type fields of their corresponding responses. We make our tools and datasets available to the public to facilitate further Research into this area.

\vspace{-3mm}

\section{DNS Queries of Enterprise Hosts: Data Collection and Attributes Extraction} \label{sec:3.2}

In this section, we first analyze the characteristics of DNS traffic (with a specific focus on query names) collected from the border of two enterprise networks, a medium-sized research institute and a large University campus. In both instances, the IT department of the enterprise provisioned a full mirror (both inbound and outbound) of their Internet traffic (each on a 10 Gbps interface) to our data collection system from their border routers (outside of the firewall). We obtained appropriate ethics clearances for this study (UNSW Human Research Ethics Advisory Panel approval number HC17499, and CSIRO Data61 Ethics approval number 115/17). We extracted DNS packets from each enterprise Internet traffic stream in real-time by configuring rules to match incoming/outgoing IPv4 and IPv6 UDP packets on port 53 in an OpenFlow switch. The study here considers data collected over one week from 30-07-2018 to 05-08-2018. 
%First week's data is used for the analysis of the DNS queries of enterprise hosts and the second week's data is used for the evaluation and to draw insights to the real-time anomalous DNS queries which is discussed in Section \ref{sec:eval}.

\subsection{Our Dataset}\label{sec:DataSet}
Table~\ref{tab:datset} shows a summary of our dataset from each organization. We captured a total of 249M and 589M DNS packets from the border of the two networks and stored them in daily CSV files -- each row in our dataset represents a timestamped DNS packet including headers and payload. The data shows that 17\% of total DNS traffic is carried over IPv6 packets in both networks. Also, more than a third of our records correspond to outgoing DNS queries generated by enterprise hosts -- \ie 86.9M and 221M in the Research and University networks, respectively. We note that our dataset also contains queries for unqualified domain names (\ie 900K and 1.5M respectively in the Research and University networks) that are discarded in our analysis -- we use the cleaned dataset. Unqualified query names contain no delimiting dots (\eg
``{\fontsize{10}{48}\usefont{OT1}{lmtt}{b}{n}\noindent top\_10\_banks\_offering\_attractive\_home}'') 
or their top-level-domain is pure numeric (\eg ``{\fontsize{10}{48}\usefont{OT1}{lmtt}{b}{n}\noindent 129.178}''). After removing unqualified names, outgoing DNS queries in total span respectively 2.2M and 6.2M distinct fully qualified domain names (FQDN).

\begin{table}[t!]
	\centering
	\caption{Summary of our dataset.}
	\label{tab:datset}
	\begin{adjustbox}{max width=0.85\textwidth}
		\begin{tabular}{@{}lcc@{}}
			\toprule
			& \multicolumn{1}{l}{\textbf{Research}}   & \multicolumn{1}{l}{\textbf{University}} \\ \midrule
			Total DNS packets        & 249M               & 589M                 \\ \midrule
			~~IPv4 DNS packets      & 206M                  & 489M              \\ \midrule
			~~IPv6 DNS packets      & 43M                    & 100M              \\ \midrule
			~~DNS queries           & 142M                   & 341M                \\ \midrule
			~~DNS responses        & 107M                   & 248M              \\ \midrule
			Total Outgoing DNS queries & 86.9M             & 221M              \\ \midrule
			~~Outgoing DNS queries (IPv4) & 69.7M              & 177M              \\ \midrule
			~~Outgoing DNS queries (IPv6) & 17.2M              & 44M              \\ \midrule
			Outgoing DNS queries ({\color{teal}only qualified}) & 86M                     & 219.5M              \\ \midrule
			Unique query names (FQDN)         & 2.2M           & 6.2M             \\ \midrule
			%Unique query names (FQDN) with IPv6         & 607K          & 2.05M            \\ \midrule
			Unique primary domains & 397K                & 1.1M             \\ \midrule
			%Unique query names (FQDN) with IPv4         & 2.09M           & 6.07M             \\ \midrule
			%Unique query names (FQDN) with IPv6         & 607K          & 2.05M            \\ \midrule
			%Unique primary domains with IPv4 & 390K                & 1.3M             \\ \midrule
			%Unique primary domains with IPv6 & 191K               & 886K             \\ \midrule
		\end{tabular}
		%  \vspace{-3mm}
	\end{adjustbox}
\end{table}

\begin{figure}[t!]
	\centering
	\includegraphics[width=0.85\textwidth]{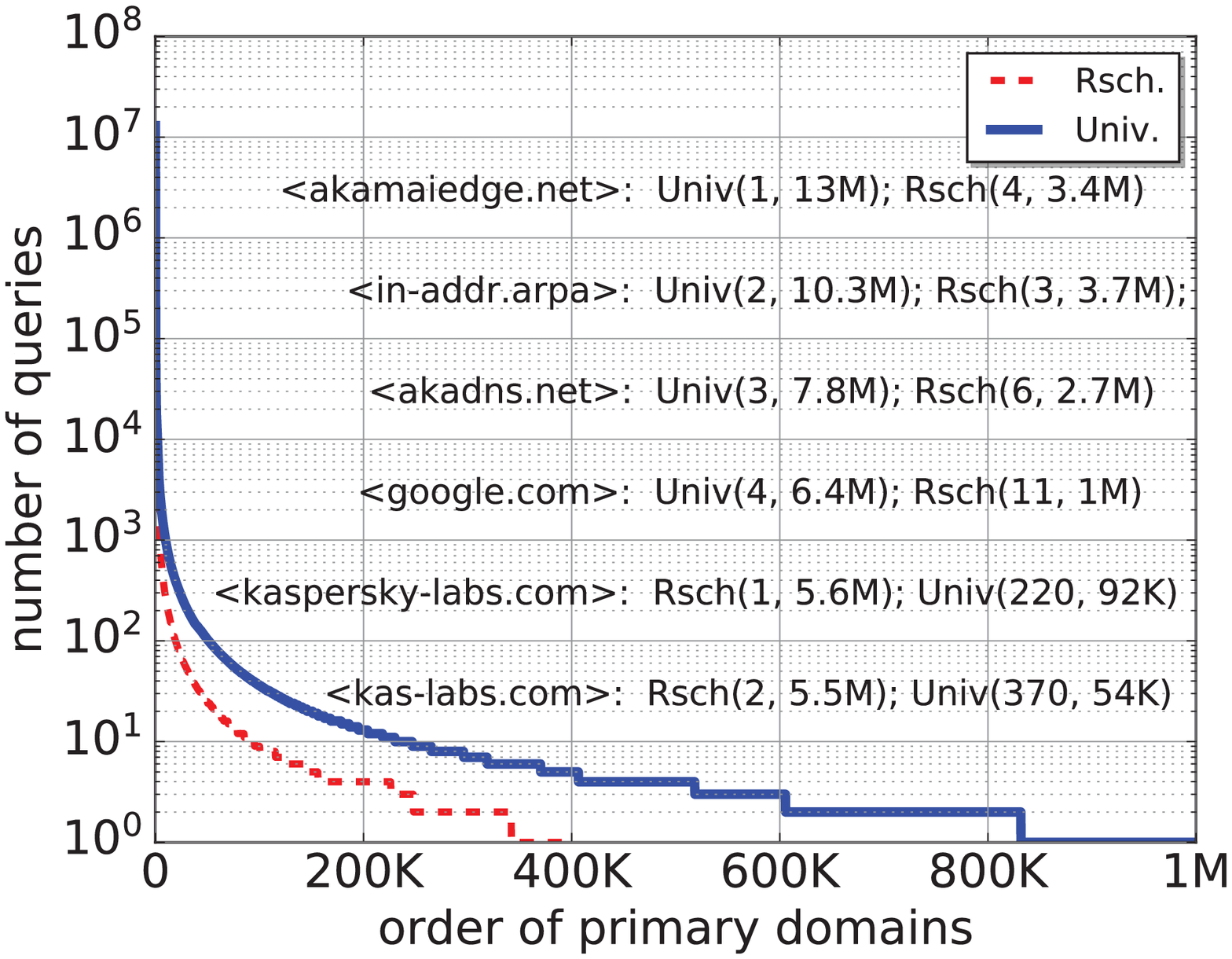}
	\vspace{-3mm}
	\caption{Number of queries per unique primary domain, over a week (Rsch: 397K, Univ: 1.1M).}
	\label{fig:freqDomain}
	\vspace{-5mm}
\end{figure}

These FQDNs are rooted themselves in 397K and 1.1M distinct \textit{primary domains} (\ie one level under ``{\fontsize{10}{48}\usefont{OT1}{lmtt}{b}{n}\noindent com}'' or ``{\fontsize{10}{48}\usefont{OT1}{lmtt}{b}{n}\noindent co.uk}''). 
Fig.~\ref{fig:freqDomain} shows the number of queries for each unique primary domain over the entire dataset, ordered from most queried on the left, to least queried on the right. There is a small number of domains on the left that predominate with very high query counts, followed by a long-tail of domains, all of which receive a fairly small number of queries (\ie less than 1000 over a week). It is seen that the top 4K (out of 397K) and 9K (out of 1.1M) domains respectively in the research institute and the University comprise the head in their respective curve. For example, only three domains namely 
``{\fontsize{10}{48}\usefont{OT1}{lmtt}{b}{n}\noindent akamaiedge.net}'', 
``{\fontsize{10}{48}\usefont{OT1}{lmtt}{b}{n}\noindent in-addr.arpa}'', and ``{\fontsize{10}{48}\usefont{OT1}{lmtt}{b}{n}\noindent akadns.net}'' contribute to 15\% of total queries generated by University hosts. 
In the research network, on the other hand, top three domains of
``{\fontsize{10}{48}\usefont{OT1}{lmtt}{b}{n}\noindent kaspersky-labs.com}'', 
``{\fontsize{10}{48}\usefont{OT1}{lmtt}{b}{n}\noindent kas-labs.com}'', and ``{\fontsize{10}{48}\usefont{OT1}{lmtt}{b}{n}\noindent in-addr.arpa}'' contribute to 17\% of total queries.
We note that queries for ``{\fontsize{10}{48}\usefont{OT1}{lmtt}{b}{n}\noindent in-addr.arpa}'' correspond to reverse DNS lookups which are commonly used by email servers to check and see if the message came from a valid server. Many email servers will reject messages from any server that does not support reverse lookups since spammers typically use invalid IP addresses.

In terms of queries ``reputation'', we used Majestic dataset \cite{Majestic} which is free and updates on a daily basis -- Majestic is a reverse search engine that computes the number and strength of links to a domain (it is a measure of trust instead of traffic estimates){\cite{rweyemamu2019clustering,kelkar2018analyzing}. To get a sense of reputation and probability of typical ranks, we show in Fig.~\ref{fig:ccdfRepuration} the complementary cumulative distribution function (CCDF) of the reputation rank for primary domains queried in both organizations. We can see that 44\% of total queries, in both organizations, are not listed in the top 1M domains of Majestic domains ranking (\ie CSV dataset released on 7-Aug-2018). Also, only 32\% and 34\% of queries in each network are among the top 10K most popular domains.
	In our Majestic dataset, ``{\fontsize{10}{48}\usefont{OT1}{lmtt}{b}{n}\noindent google.com}", 
	``{\fontsize{10}{48}\usefont{OT1}{lmtt}{b}{n}\noindent facebook.com}, and ``{\fontsize{10}{48}\usefont{OT1}{lmtt}{b}{n}\noindent youtube.com}" are top three ranked domains respectively. 
	
	\begin{figure*}[t!]
		\begin{center}
			%\vspace{-3mm}
			\mbox{
				%\hspace{-3mm}
				\subfloat[Research institute.]{
					\includegraphics[width=0.48\textwidth]{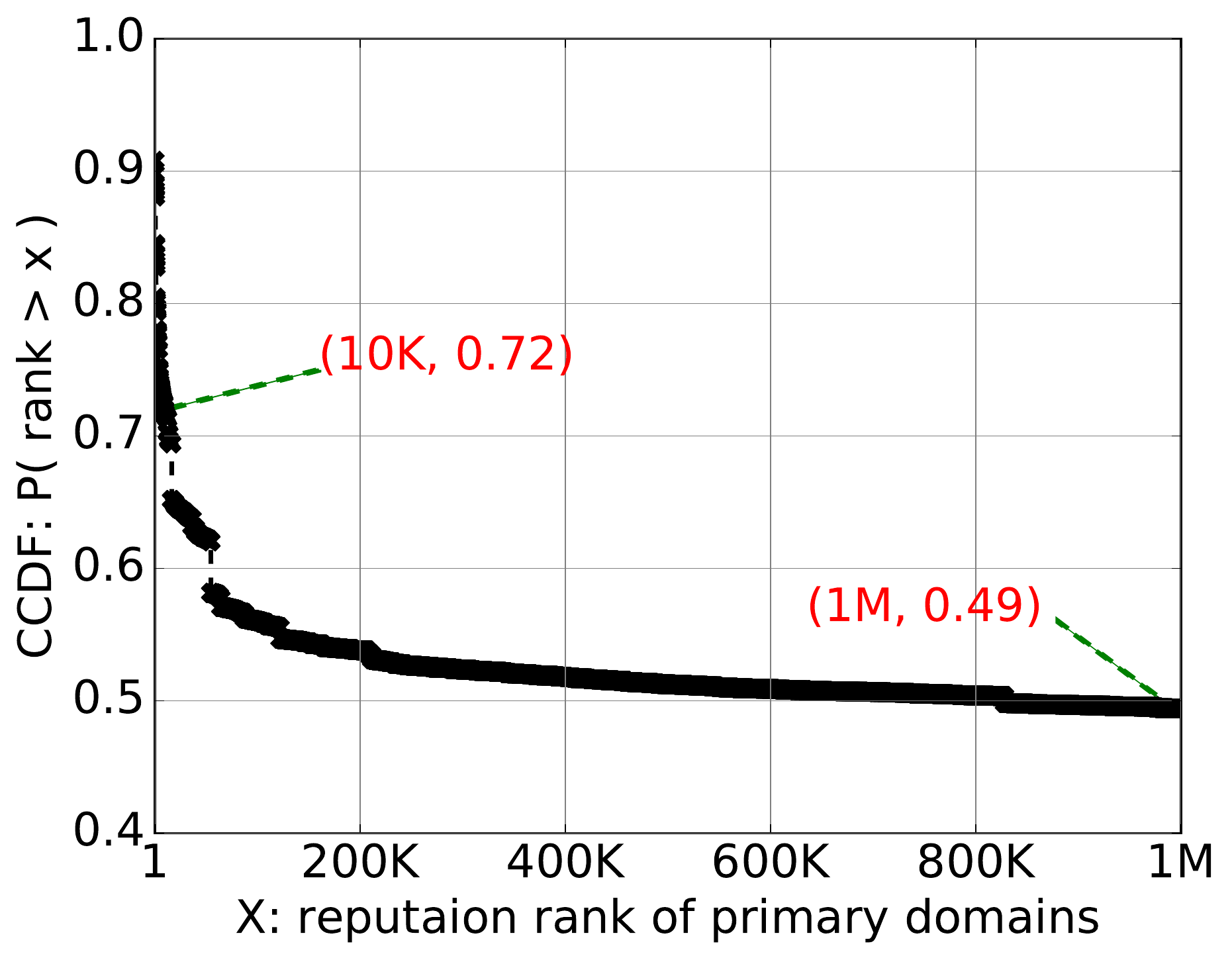}}\quad
				\label{fig:ccdfRepurationCSIRO}
				%\hspace{-5mm}
				\subfloat[University campus.]{
					\includegraphics[width=0.48\textwidth]{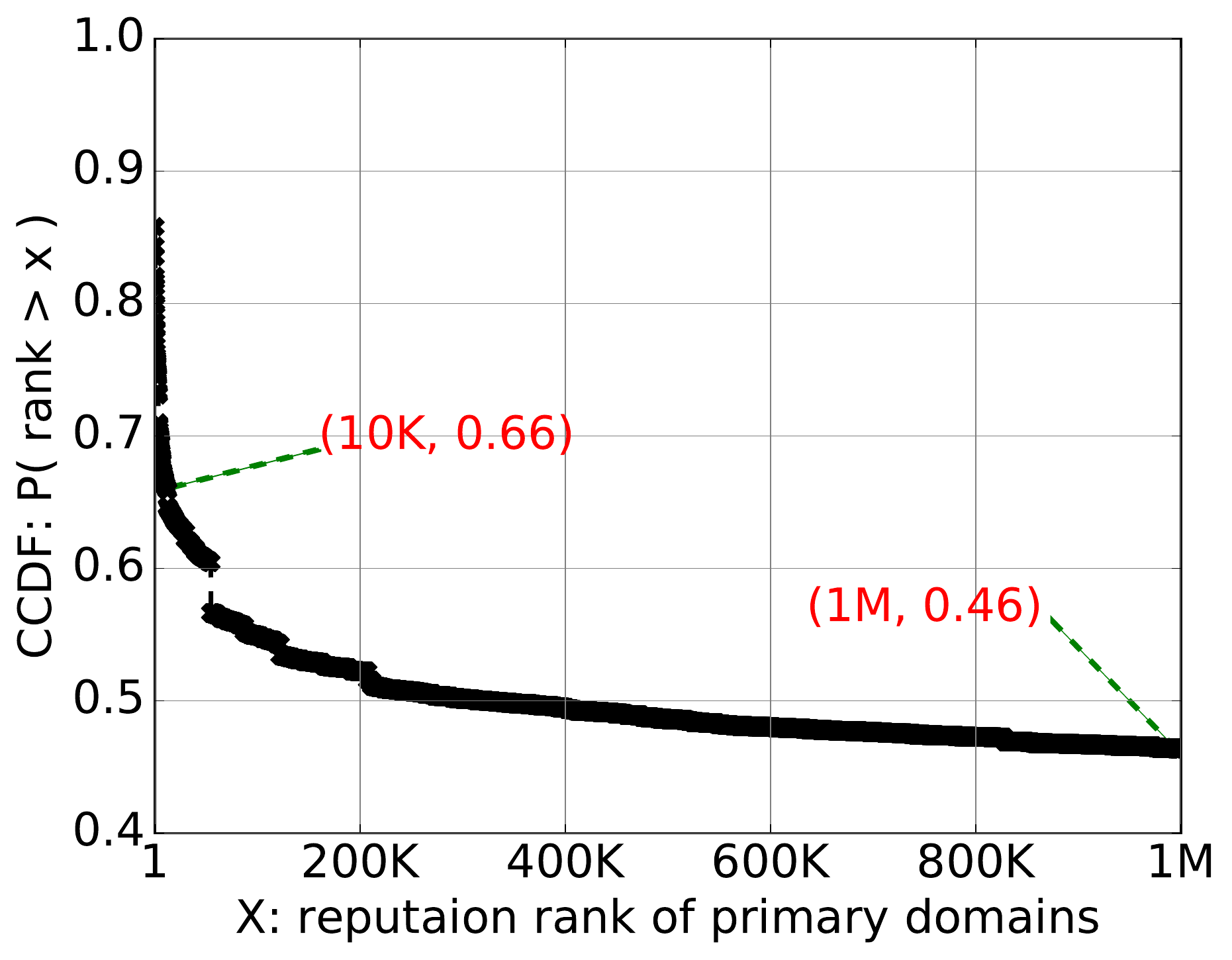}}\quad
				\label{fig:ccdfRepurationUNSW}
			}
			\caption{CCDF of reputation rank: (a) Research institute, and (b) University campus.}
			\label{fig:ccdfRepuration}
		\end{center}
		%\vspace{-6mm}
	\end{figure*}

	\begin{figure}[t!]
		\centering
		{
			\includegraphics[width=0.75\textwidth]{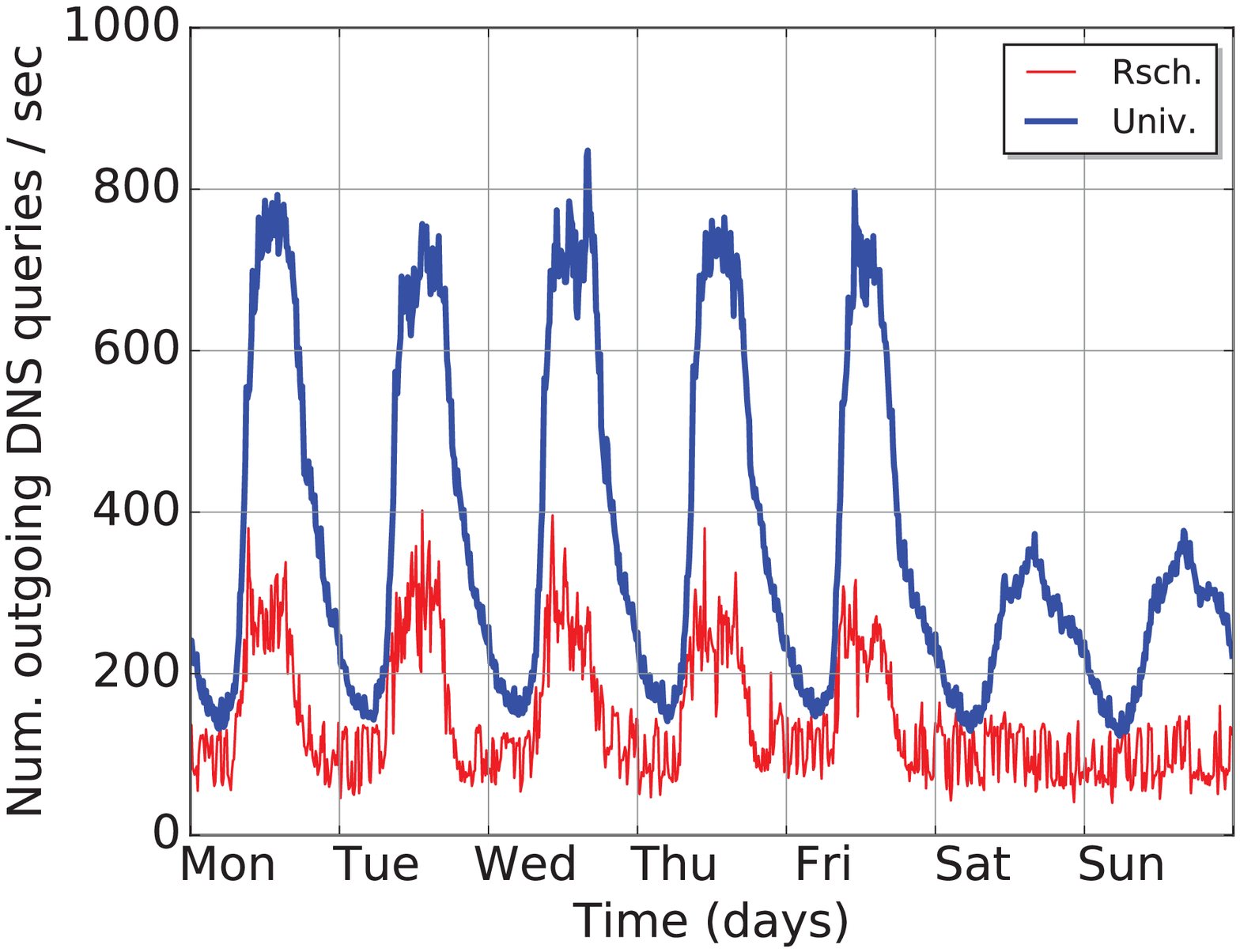}
			\caption{Real-time number of queries.}
			\label{fig:timetraceDNS}
		}
	\end{figure}

	Considering the load of DNS queries generated by enterprise hosts, shown in Fig.~\ref{fig:timetraceDNS}, we see that the number of packet-per-sec in the research network varies between 50 to 400 depending on the day of week and peak/off-peak hours. For the University network, on the other hand, a larger variation is observed -- \ie 150 to more than 800 pps. 
	
	\begin{table*} [ht]
		\centering
		\caption{A sample list of malicious and normal DNS queries with unusual length.}
		\label{tab:sampleDNS}
		% \resizebox{\columnwidth}{!}{  
		\begin{adjustbox}{max width=0.98\textwidth}   
			\renewcommand{\arraystretch}{1.5}
			\begin{tabular}{lc} %{\columnwidth}{cccccc}
				\toprule       \textbf{Query name (FQDN)}       &  \textbf{Security}   \\      \hline\hline
				\fontsize{10}{48}\usefont{OT1}{lmtt}{b}{n}
				\noindent
				6e517f3.grp10.ping.adm.cdd2e9cde9fee9cdc8.cdd0e8e9c8fce9d2e9fecdc4.c597f097ce87c5d3.ns.a23-33-37-54-deploy-akamaitechnologies.com& {\color{red}Malicious} \\      \hline
				\fontsize{10}{48}\usefont{OT1}{lmtt}{b}{n}
				\noindent 708001701462b7fae70d0a28432920436f70797269676874.20313938352d32303031204d696372.6f736f667420436f72702e0d0a0d0a0.433a5c54454d503e.cspg.pw& {\color{red}Malicious} \\   \hline
				\fontsize{10}{48}\usefont{OT1}{lmtt}{b}{n}
				\noindent 
				PzMnPiosOD4nOCwuOzomPS4nNjovPS8uOzsnNCstODkjOCwoMwAA.29a.de& {\color{red}Malicious} \\ \hline
				%\fontsize{10}{48}\usefont{OT1}{lmtt}{b}{n}
				%\noindent 
				%DIYNBPRYA0K5CVUWA.ns1.logitech-usa.com& {\color{red}Malicious} \\ \hline       
				\fontsize{10}{48}\usefont{OT1}{lmtt}{b}{n}
				\noindent             
				bwzm133h9gb3pp9s6l3mu7r73sh.arm2513pu79r9.1z19e1bgm1hwu8z6u2.9rzlkhbvi45gaag52t3rqtqd2t.p2gliv6gklwzvvlt2jp1z6li7v.avqs.mcafee.com& {\color{green!55!blue}Normal} \\ \hline
				\fontsize{10}{48}\usefont{OT1}{lmtt}{b}{n}
				\noindent 0.19.6ce.71c.444.25.41.0.0.0.4.27.0.0.0.0.0.0.0.0.0.9efc95e03d7f3a4ae446ecd0d049e5ae9e016ee33703c9cb3506cad4bbd98bc.b.f.00.s.sophosxl.net & {\color{green!55!blue}Normal}  \\ \hline
				%      \fontsize{10}{48}\usefont{OT1}{lmtt}{b}{n}
				%    \noindent         
				%  f4a55fc3f30keaayaayqivpqaggkbqggudp6hm-yacnusej1525121392-sonar.xy.fbcdn.net& {\color{green!55!blue}Normal} \\ \hline
				\fontsize{10}{48}\usefont{OT1}{lmtt}{b}{n}
				\noindent               
				p4-ces3lawazdkbw-qlrq5qalxdt7tycq-385202-i1-v6exp3.ds.metric.gstatic.com& {\color{green!55!blue}Normal} \\ \hline
				\fontsize{10}{48}\usefont{OT1}{lmtt}{b}{n}
				\noindent         
				\_ldap.\_tcp.AWS.\_sites.dc.\_msdcs.AD.us-east-1.ec2-utilities.amazonaws.com& {\color{green!55!blue}Normal} \\
				\bottomrule
			\end{tabular}
		\end{adjustbox}
		%}
	\end{table*}

	\subsection{Query Name Attributes Engineering}
	We now look at the attributes of the query name (FQDN) in each DNS query generated by enterprise hosts that are relevant to differentiating benign and malicious DNS queries traffic. Our aim is to use only ``Stateless'' attributes which can be derived from individual DNS query packets, independent of time-series characteristics of queried domains or hosts DNS activity -- there is no overhead in computing these attributes in real-time. {Our attributes are inspired by various prior works (referred against each attribute).}
	
	According to RFC 1035 \cite{rfc1035}, the total length of a domain name (dots included) is restricted to 255 characters, and domain names are represented as a sequence of ``labels'' separated by dots. The maximum length of a label is 63 characters. 
	It has been shown that DNS can be used for malicious purposes in the form of DNS tunneling or exfiltration in which valuable information (\eg credentials, credit card, or control messages) is embedded in the sub-domain portion of a query name. 
	Malware applications typically embed stolen data \cite{talos2016} into the subdomain part of a DNS query for a domain where the name server is under control of an attacker. A DNS query for ``{\fontsize{10}{48}\usefont{OT1}{lmtt}{b}{n}\noindent exfiltrated-data.example.com}'' would be forwarded to the name server of ``{\fontsize{10}{48}\usefont{OT1}{lmtt}{b}{n}\noindent example.com}'', which would record ``exfiltrated-data'' and decode and decrypt the sensitive information from that subdomain field.
	
	%The sub-domain strings are typically encoded and/or encrypted. For example, in PzMnPiosOD4nOCwuOzomPS4nNjovPS8uOzsnNCstODkjOCwoMwAA.badman.com, the string before 2LD is the subdomain containing the encoded and encrypted information which is sending to the dedicated name server of badman.com. Upon receiving the query for the badman domain, the malicious actor extracts the subdomain of that particular site and decodes it and decrypt it to get back the credit card or other sensitive information.

	%Breaching the confidentiality of an organization from a compromised host within the organization to an attacker outside over the DNS queries is commonly known as DNS exfiltration. These queries are generally encoded and encrypted with an efficient encryption scheme. Usually, the encrypted data is in the subdomain of the DNS query. For example, in PzMnPiosOD4nOCwuOzomPS4nNjovPS8uOzsnNCstODkjOCwoM\\
	%wAA.badman.com, the string before 2LD is the subdomain containing the encoded and encrypted information which is sending to the dedicated name server of badman.com. Upon receiving the query for the badman domain, the malicious actor extracts the subdomain of that particular site and decodes it and decrypt it to get back the credit card or other sensitive information.
	
	% {\color{red} Introduce the anatomy of DNS tunneling/exfiltration.}
	Table~\ref{tab:sampleDNS} lists samples of malicious \cite{sampleBernhardPOS2015,sampleFrameworkPOS2016,sampleDNSMessenger2017} and benign query names with ``unusual'' length and string pattern. For example, the top two malicious query names in this list respectively contain 129 and 136 characters. We note that the sub-domain portion of these query names comprises random-looking strings with a significant number of upper-case and numerical characters, and is fairly long. For example, the second malicious query name from the top (\ie for ``{\fontsize{10}{48}\usefont{OT1}{lmtt}{b}{n}\noindent cspg.pw}") contains 38 numeric characters (\ie 28\%), and the third malicious query name (\ie for ``{\fontsize{10}{48}\usefont{OT1}{lmtt}{b}{n}\noindent 29a.de}") contains 38 numeric characters (\ie 28\%) contains 23 uppercase letters (\ie 39\%). Given these observations, we define our attributes by three main categories namely characters count, entropy (an indication of randomness) of string, and length of discrete labels in the query name.

	\begin{figure}[t!]
		\begin{center}
			% \vspace{-3mm}
			\mbox{
				% \hspace{-3mm}
				\subfloat[Research institute.]{
					{\includegraphics[width=0.48\textwidth]{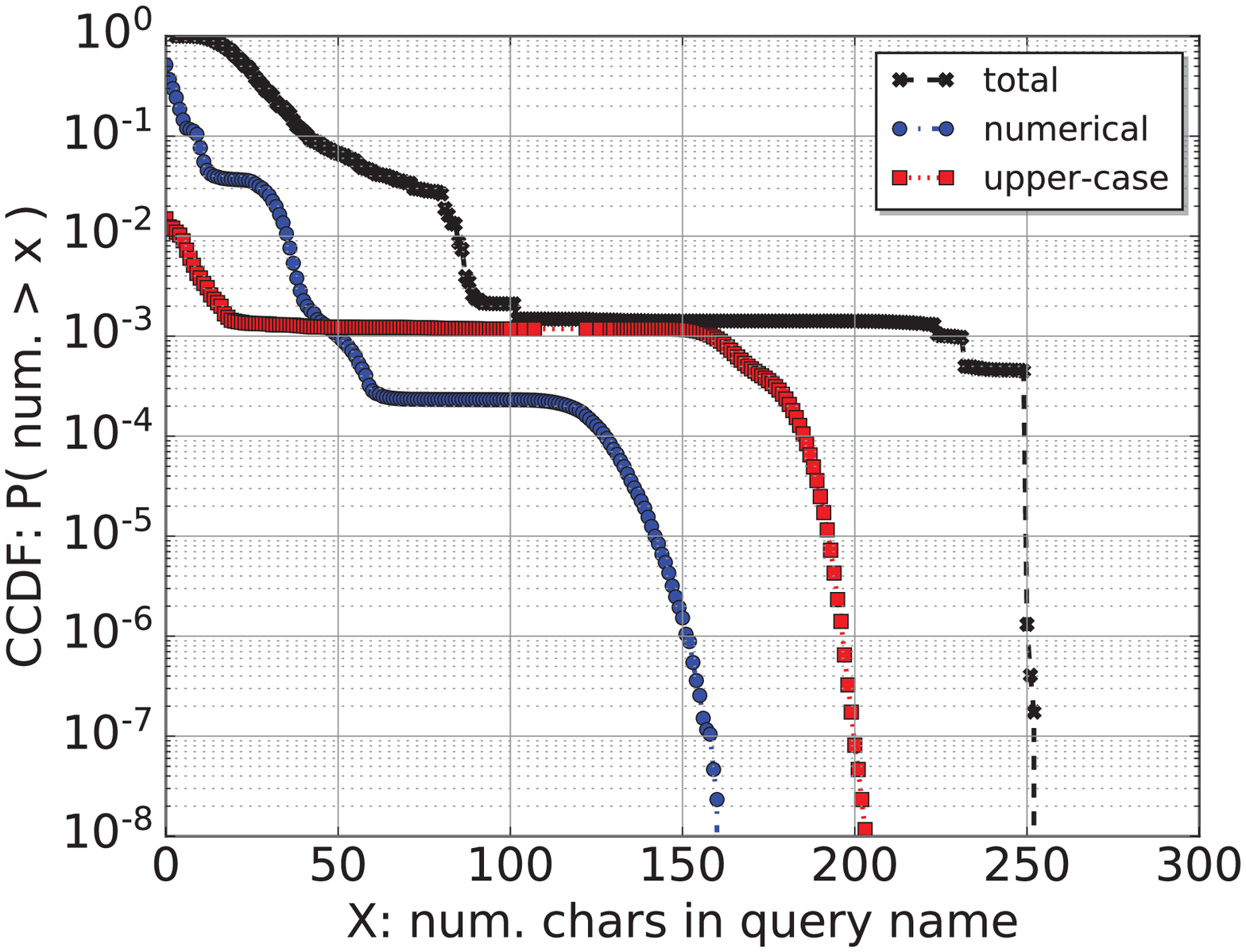}}\quad
					\label{fig:ccdfLengthCSIRO}
				}
				\subfloat[University campus.]{
					{\includegraphics[width=0.48\textwidth]{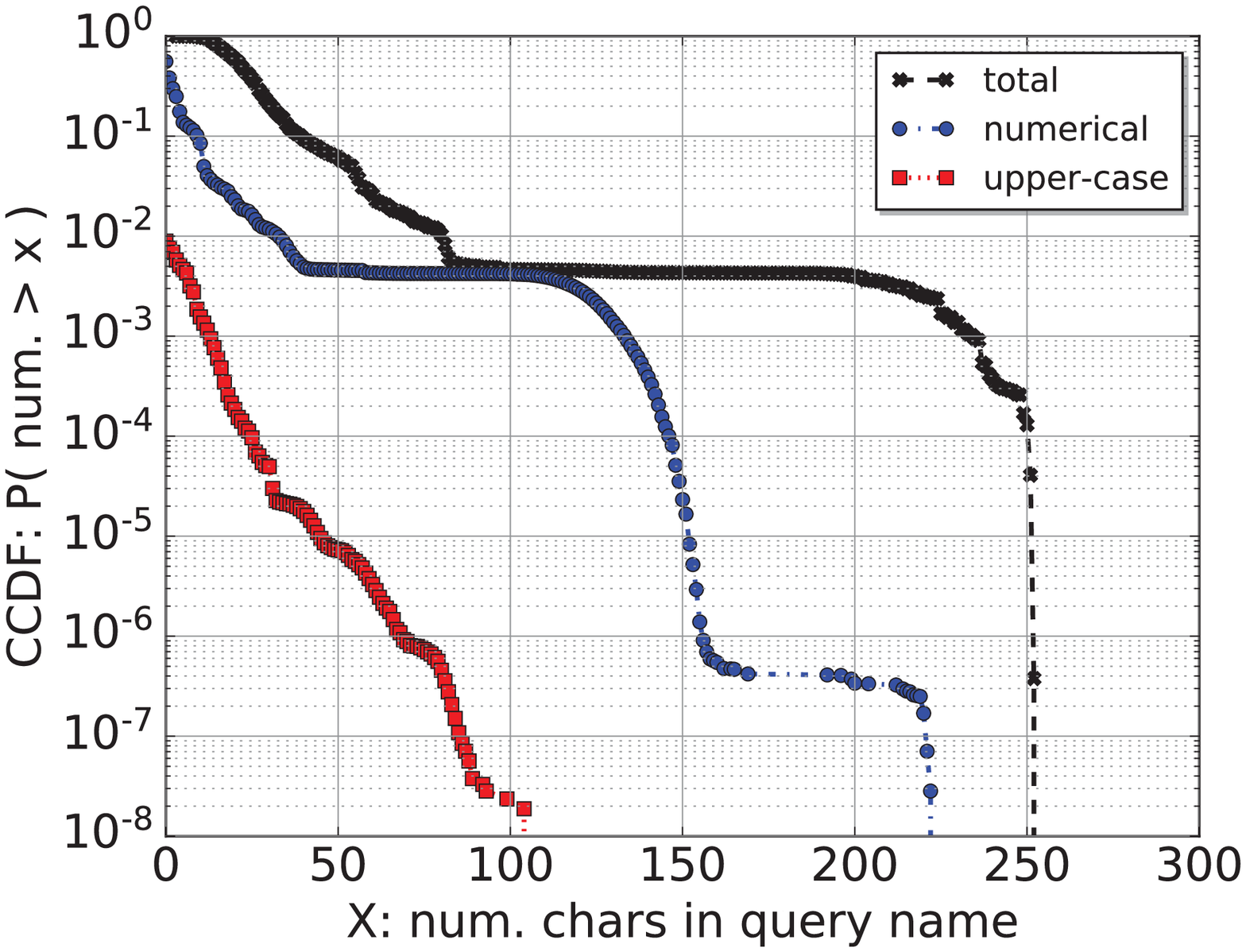}}\quad
					\label{fig:ccdfLengthUNSW}
				}
			}
			\caption{CCDF of number of characters in query name for: (a) Research institute, and (b) University campus.}
			\label{fig:ccdfLength}
		\end{center}
	\end{figure}

	\begin{figure*}[t!]
		\begin{center}
			% \vspace{-3mm}
			\mbox{
				% \hspace{-3mm}
				\subfloat[Research institute.]{
					{\includegraphics[width=0.48\textwidth]{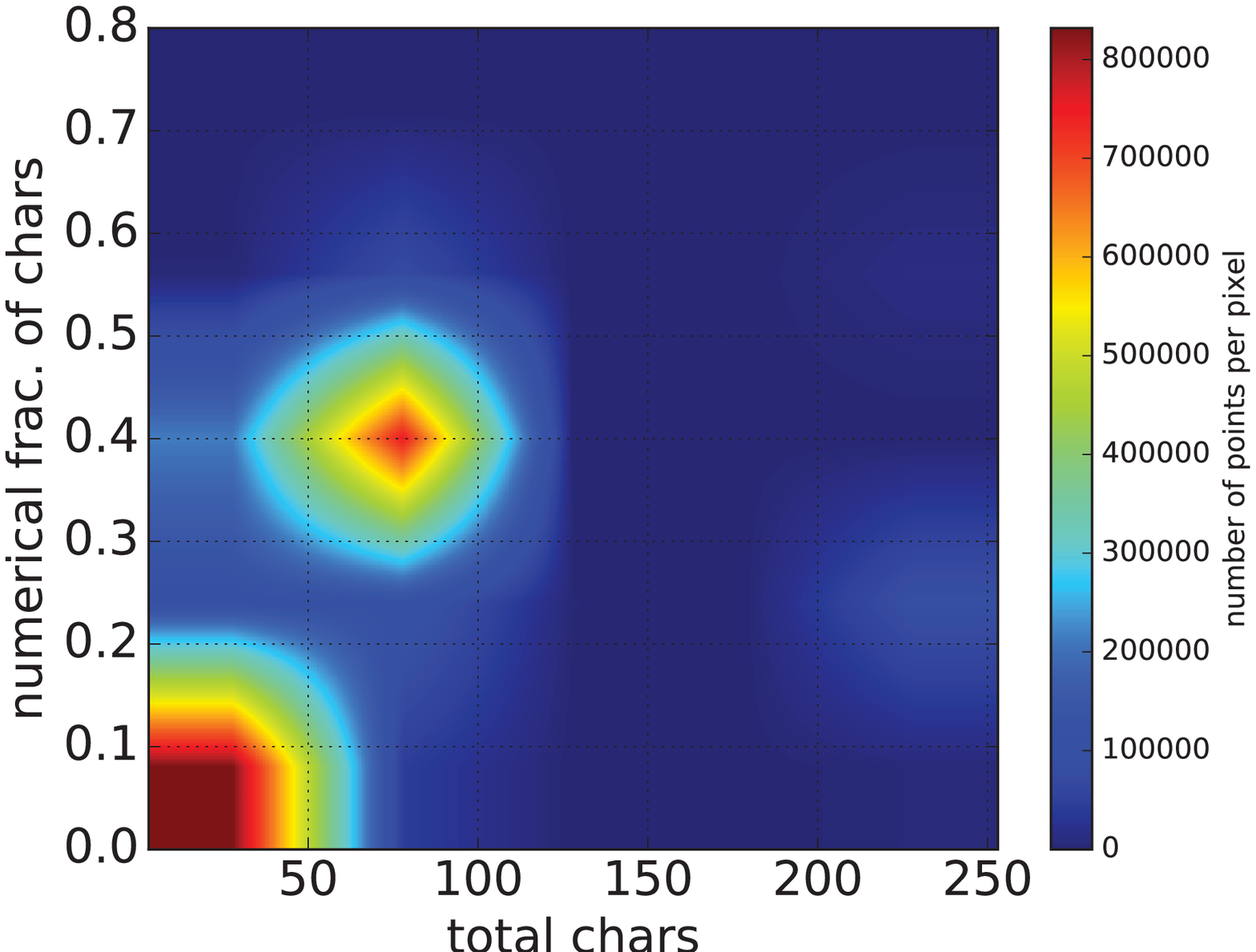}}\quad
					\label{fig:scatterNumercialCSIRO}
				}
				%\hspace{-5mm}
				\subfloat[University campus.]{
					{\includegraphics[width=0.48\textwidth]{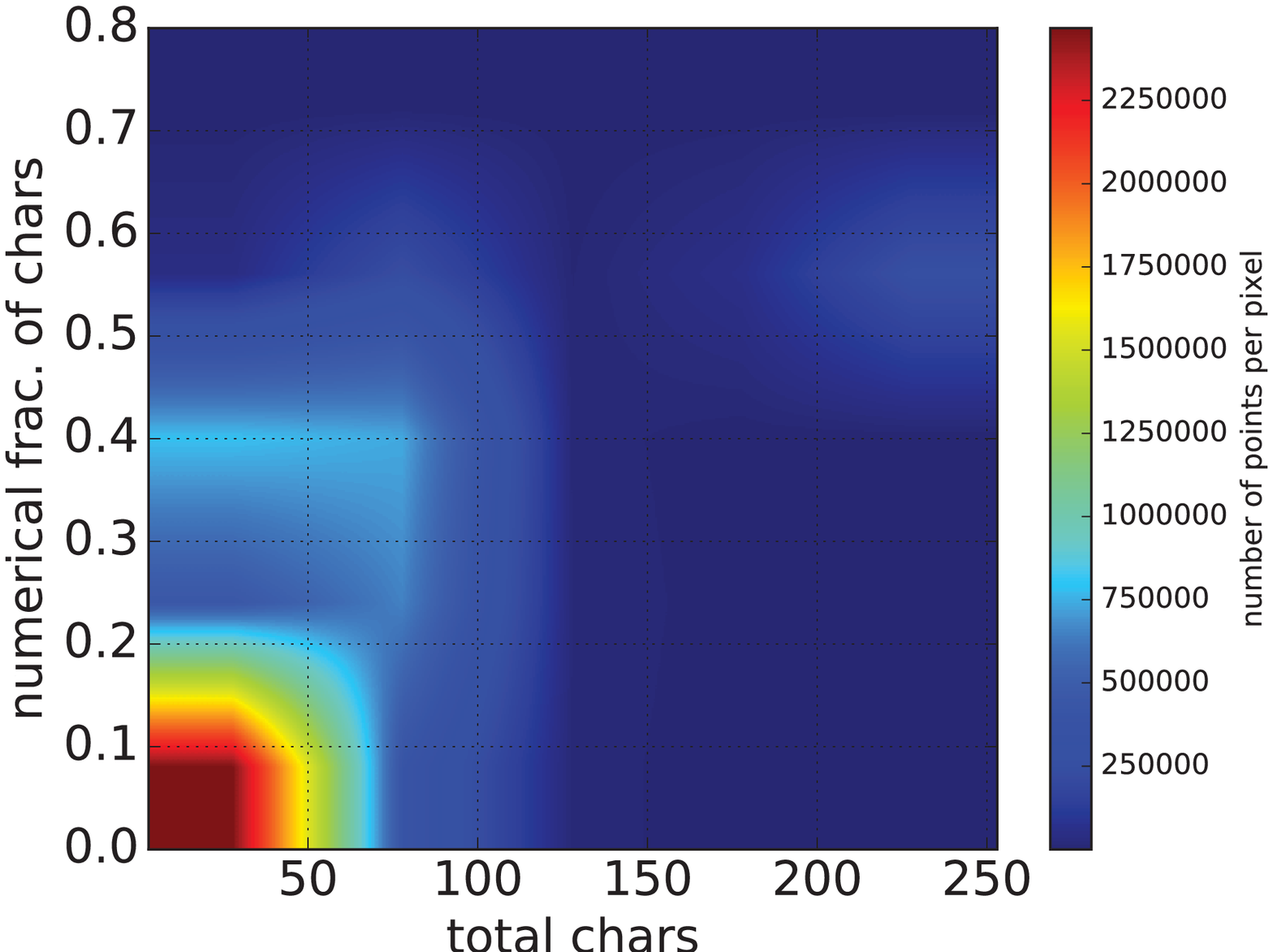}}\quad
					\label{fig:scatterNumercialUNSW}
				}
			}
			\caption{Scatter density map of numerical fraction of characters vs. total length of query name for: (a) Research institute, and (b) University campus.}
			\label{fig:scatterNumercial}
		\end{center}
	\end{figure*}

	\subsubsection{Count of Characters}
	The total number of characters is an important attribute since more characters imply that the query name probably carries embedded information for an outside host. 
	In Fig.~\ref{fig:ccdfLength}, we plot the distribution of character count for query names in our dataset to understand the typical value of these attributes.
	%A typical DNS query name is fairly short (\textbf{\color{red}Any ref?}). But malware tends to embed precious information into their query. Followings are the key attributes related to the number of characters. 
	
	%\begin{itemize}
	\textbf{Total count of characters in FQDN:}{\cite{homem2017harnessing}} We can see that more than 99\% of host queries in both organizations contain less than 80 characters, as shown by black cross markers in Fig.~\ref{fig:ccdfLength}. Only a very small fraction of query names (\ie about 0.3\%) are really long, each with more than 100 characters.  
	%We begin with the total number (\ie length) of characters in query names. We are sniffing the fully qualified domain names from the DNS traffic. As the information is likely to be transmitted in the queried domain in an encrypted fashion. The length of the domain is an important feature to look at. We have analyzed the distribution of the length of the domain by plotting the CCDF in figure \ref{fig:ccdfLength}. It is interesting to observe from the plot that only 1\% of the domains have a domain length of greater than 80 bytes. 
	It is important to note that antivirus tools tend to exchange legitimate data (\ie for signature lookup) over DNS \cite{asaf}. For example, in Table~\ref{tab:sampleDNS} the first two ``normal'' query names correspond to ``McAfee'' and ``Sophos'' antivirus. Interestingly, primary domains ``{\fontsize{10}{48}\usefont{OT1}{lmtt}{b}{n}\noindent mcafee.com}" with 1.9M queries (average query length of 84 characters), and ``{\fontsize{10}{48}\usefont{OT1}{lmtt}{b}{n}\noindent sophosxl.net}" with 145K queries (average query length of 106 characters) are among top ten frequent domains seen in our dataset from the Research institute and the University network respectively as shown in Table~\ref{tab:numChars}.
	Since the exfiltrated (or Command \& Control) message is carried by the sub-domain portion of an FQDN, we use the \textbf{count of characters in sub-domain} {\cite{liu2017detecting}} as our second attribute.

	\begin{figure*}[t!]
		\begin{center}
			%\vspace{-3mm}
			\mbox{
				%\hspace{-3mm}
				\subfloat[Research institute.]{
					{\includegraphics[width=0.48\textwidth]{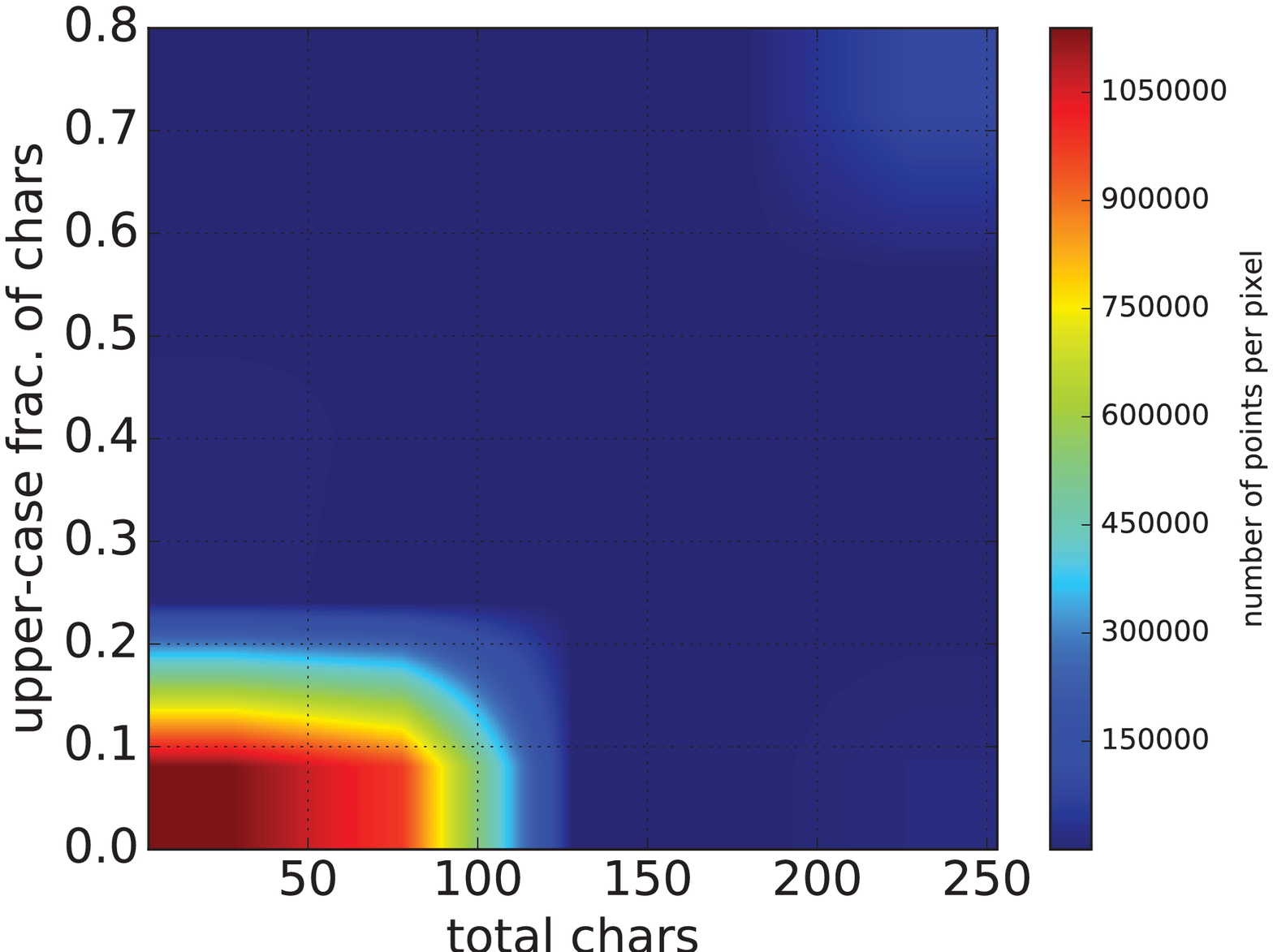}}\quad
					\label{fig:scatterUppercaseCSIRO}
				}
				% \hspace{-4mm}
				\subfloat[University campus.]{
					{\includegraphics[width=0.48\textwidth]{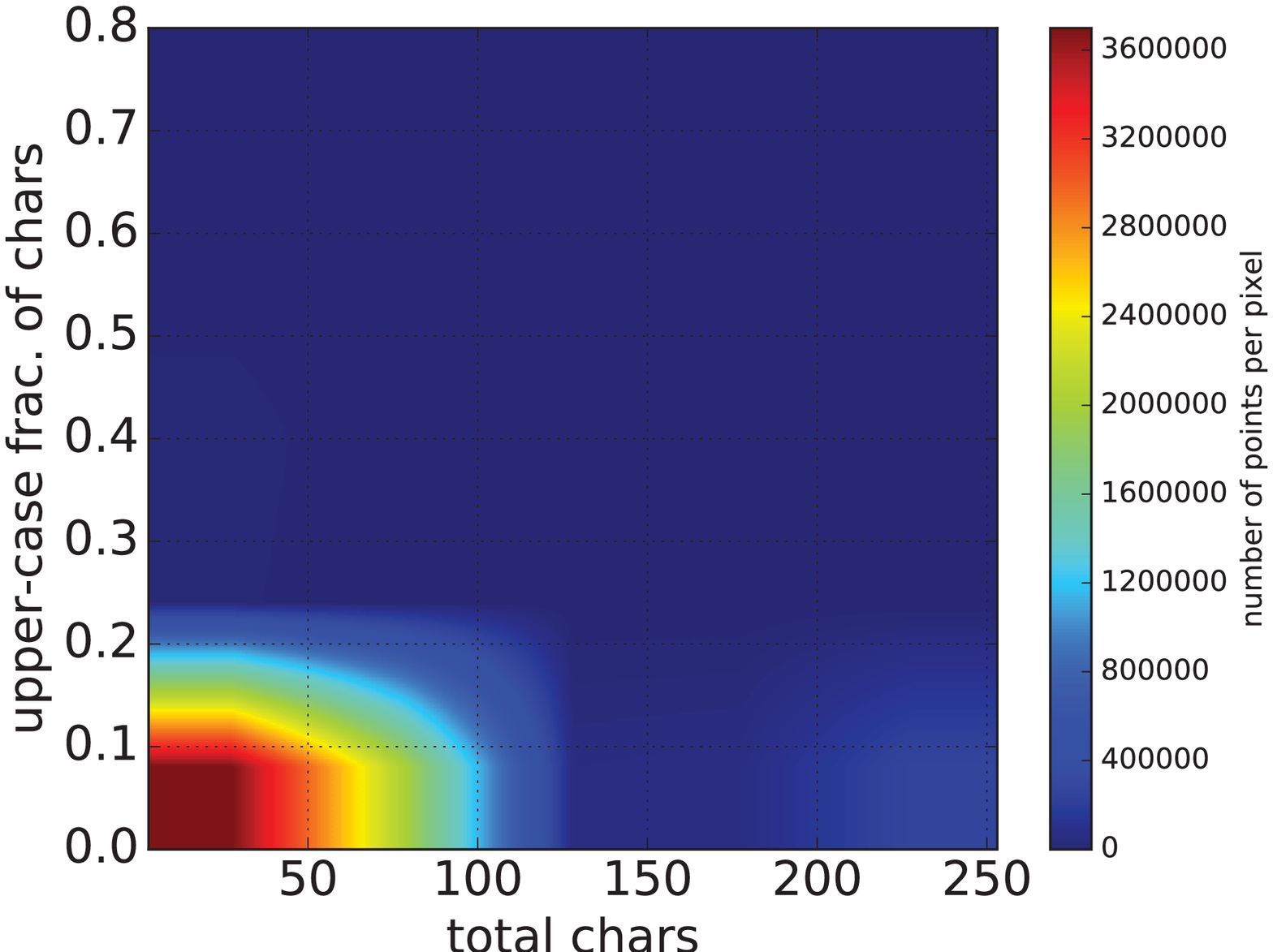}}\quad
					\label{fig:scatterUppercaseUNSW}
				}
			}
			\caption{Scatter density map of upper-case fraction of characters vs. total length of query name for: (a) Research institute, and (b) University campus.}
			\label{fig:scatterUppercase}
		\end{center}
		%\vspace{-1mm}
	\end{figure*}

	\begin{table}[t!]
		\centering
		\caption{Number of characters in FQDN for selected domains \protect\linebreak in our dataset.}
		%\vspace{-3mm}
		\label{tab:numChars}
		\begin{adjustbox}{max width=0.85\textwidth}
			\begin{tabular}{@{}lccccc@{}}
				\toprule
				\textbf{primary domain} & \textbf{\# FQDN}  & \textbf{\# unique FQDN} & \textbf{frac. Numerical (\%)} & \textbf{frac. Uppercase (\%)}  & \textbf{avg. Length} \\ \midrule
				\fontsize{10}{48}\usefont{OT1}{lmtt}{b}{n}
				\noindent       
				mcafee.com& 1.9M  & 571K  & 39.4  & 0.31e-3  & 84.01 \\ \midrule
				\fontsize{10}{48}\usefont{OT1}{lmtt}{b}{n}
				\noindent       
				sophosxl.net& 145K  & 41K  & 47.5  & 0.11  & 106.7 \\ \midrule
				\fontsize{10}{48}\usefont{OT1}{lmtt}{b}{n}
				\noindent       
				spotify.com& 84K  & 819  & 7  & 0         & 41.7 \\ \midrule
				\fontsize{10}{48}\usefont{OT1}{lmtt}{b}{n}
				\noindent       
				cnr.io& 121K  & 113K  & 19.97  & 70.08  & 209.8 \\ \midrule
				\fontsize{10}{48}\usefont{OT1}{lmtt}{b}{n}
				\noindent       
				e5.sk& 66K  & 131  & 13.8  & 0.15      & 129.06\\ \bottomrule       
			\end{tabular}
			%  \vspace{-3mm}
		\end{adjustbox}
	\end{table}

	Additionally, we use the \textbf{count of uppercase characters}{\cite{das2017detection}} and \textbf{count of numerical characters}{\cite{das2017detection}} in a query name to determine if it is benign or malicious. This is because the fraction of uppercase and numerical characters becomes high in encrypted/ encoded data 
	\cite{das2017detection} -- however, not all encrypted data is malicious.
	In Fig.~\ref{fig:ccdfLength}, it is seen that only about 1\% of all queries in each organization contain more than 30 numerical characters. Unsurprisingly, the upper-case character is very rare in domain names generated by hosts in both enterprise networks -- at least 98\% of queries contain no upper-case character, and less than 0.2\% of queries contain more than 10 capitals.
	
	To better understand the distribution of various characters in query names, we plot the scatter density maps of total characters count versus numerical fraction in Fig.~\ref{fig:scatterNumercial}, and total characters count versus uppercase fraction in Fig.~\ref{fig:scatterUppercase} -- dark red areas depict higher density of points and dark blue areas highlight the lower density of points. In Fig.~\ref{fig:scatterNumercial}, it can be seen that the numerical fraction of a query name typically stays below 20\% (mostly less than 10\%) when the query name has less than 60 letters (\ie dark red area on the left bottom of plots). Interestingly, for the Research Institute shown in Fig.~\ref{fig:scatterNumercialCSIRO}, we observe a crowded region around 40\% of numerical letters when the FQDN length is between 60 to 80 characters.
	In Fig.~\ref{fig:scatterUppercase}, we see that the fraction of uppercase letters is below 10\% for short query names (\ie less than 80 characters for the Research Institute and less than 50 characters for the University), and it tends to zero when query names get longer.

	\subsubsection{Entropy}
	Random (``not-readable'') sub-domains are common in DNS exfiltration/tunneling queries due to use of encryption and/or encoding \cite{asaf}. \textbf{Entropy} {\cite{homem2017harnessing}} is a measure to determine the degree of non-readability (or strength of encryption) and uncertainty in a string. We use Shannon entropy \cite{shannon2001mathematical} which takes a discrete random variable $X$ as input (\ie DNS query name in our case), and mathematically is given by: 
	
	\vspace{-2mm}
	\begin{equation}\label{eq:eq1}
		{
			H(X) = -\sum_{k=1}^N P(x_k) \log_2 P(x_k)
		}
		%\vspace{-2mm}
	\end{equation}

	where $P(x_k)$ is the probability of the k-th symbol (\ie lower-case/upper-case letter, numerical, dot, or hyphen) in the input string $X$ containing various characters where $N$ is the total number of unique characters. We note that only specific letters can be used in a valid DNS query name \cite{rfc1035} (\ie 52 alphabetic and ten numeric characters, a hyphen, and dot, thus $N =$ 64). This means that the entropy value of a query name will take a value between 0 and $\log_2 (64) = 6 $ \cite{e19080422}. Table \ref{table:entropyDNS} shows the entropy value for a sample list of query names, both benign and malicious. For example, the entropy of a simple query name such as ``{\fontsize{10}{48}\usefont{OT1}{lmtt}{b}{n}\noindent www.google.com}" equals to $2.84$, and it gets a higher value for a more random string such as the last entry in Table \ref{table:entropyDNS}, a query for ``{\fontsize{10}{48}\usefont{OT1}{lmtt}{b}{n}\noindent googleapis.com}" whose entropy value is $5.27$. We also observe that the entropy value of malicious queries (highlighted in bold text) varies and is not necessarily higher than that of benign queries. In Fig.~\ref{fig:ccdfEntropy}, we plot the CCDF of entropy for all FQDNs queried by hosts of the two organizations during a week. It can be seen that the entropy value for more than 90\% of query names is less than $4$ in both networks, and having an entropy greater than $5$ is less likely (\ie lower than $0.1$\%).
	
	\begin{table*} [ht]
		\centering
		\caption{Entropy value for a sample list of query names.}
		\label{table:entropyDNS}
		% \resizebox{\columnwidth}{!}{  
		\begin{adjustbox}{max width=0.99\textwidth}
			\renewcommand{\arraystretch}{2}       
			\begin{tabular}{lc} %{\columnwidth}{cccccc}
				\toprule       \textbf{Query name (FQDN)}       &  \textbf{Entropy}   \\      \hline\hline
				\fontsize{10}{48}\usefont{OT1}{lmtt}{b}{n}
				\noindent
				www.google.com & \textit{2.84} \\ \hline   
				\fontsize{10}{48}\usefont{OT1}{lmtt}{b}{n}
				\noindent       
				202.135.201.205.23000000000012.sb-adfe2ko9.senderbase.org& \textit{3.75} \\ \hline
				\fontsize{10}{48}\usefont{OT1}{lmtt}{b}{n}
				\noindent       
				708001701462b7fae70d0a28432920436f70797269676874.20313938352d32303031204d696372.6f736f667420436f72702e0d0a0d0a0.433a5c54454d503e.cspg.pw& \textbf{3.92} \\  \hline
				%\fontsize{10}{48}\usefont{OT1}{lmtt}{b}{n}
				%\noindent             
				%r1.sn-n8v7snek.googlevideo.com& \textit{3.96} \\ \hline
				\fontsize{10}{48}\usefont{OT1}{lmtt}{b}{n}
				\noindent       
				0.19.6ce.71c.444.25.41.0.0.0.4.27.0.0.0.0.0.0.0.0.0.9efc95e03d7f3a4ae446ecd0d049e5ae9e016ee33703c9cb3506cad4bbd98bc.b.f.00.s.sophosxl.net& \textit{3.98}  \\ \hline       
				%\verb|jkksuaacaakaclj5aeaaaagsaqaaajqbfu66anvmm5bmf777aeaaaabqaaaabmi.6kn5qeaaoaafqbhefojitozrtghtsxdz5o5acettcnutep75dfwg2tsnv4nffnq.62seg6duz2lzfwljqb.a.j.e5.sk| & {\color{black}4.51} \\ \hline                   
				\fontsize{10}{48}\usefont{OT1}{lmtt}{b}{n}
				\noindent
				6e517f3.grp10.ping.adm.cdd2e9cde9fee9cdc8.cdd0e8e9c8fce9d2e9fecdc4.c597f097ce87c5d3.ns.a23-33-37-54-deploy-akamaitechnologies.com& \textbf{4.50} \\  \hline   
				\fontsize{10}{48}\usefont{OT1}{lmtt}{b}{n}
				\noindent       
				PzMnPiosOD4nOCwuOzomPS4nNjovPS8uOzsnNCstODkjOCwoMwAA.29a.de& \textbf{4.59} \\  \hline
				\fontsize{10}{48}\usefont{OT1}{lmtt}{b}{n}
				\noindent           
				f4a55fc3f30keaayaayqivpqaggkbqggudp6hm-yacnusej1525121392-sonar.xy.fbcdn.net& \textit{4.78} \\  \hline
				\fontsize{10}{48}\usefont{OT1}{lmtt}{b}{n}
				\noindent         
				DIYNBPRYA0K5CVUWA.ns1.logitech-usa.com& \textbf{4.86} \\  \hline         
				%\fontsize{10}{48}\usefont{OT1}{lmtt}{b}{n}
				%\noindent   
				%			%bwzm133h9gb3pp9s6l3mu7r73sh.arm2513pu79r9.1z19e1bgm1hwu8z6u2.9rzlkhbvi45gaag52t3rqtqd2t.p2gliv6gklwzvvlt2jp1z6li7v.avqs.mcafee.com& \textit{4.85} \\ \hline         
				\fontsize{10}{48}\usefont{OT1}{lmtt}{b}{n}
				\noindent   
				0ca7d.1.288.WYB52Q2ZPIU2SEUTDDDGEJDQFAO6F2C53AVC6IVAZZLR2PJHEWQWRFG6Z2NPQ3J.CQ4888.1d19d9c4.cnr.io& {\color{black}5.10} \\  \hline
				\fontsize{10}{48}\usefont{OT1}{lmtt}{b}{n}
				\noindent   
				X2AR6GEQVHCSMXKFUNVIZU67PVMD5EF3N74E4TLOEOYK47WEXKMQ.hash.rocketeer.ct.googleapis.com& \textit{5.27} \\    
				\bottomrule
			\end{tabular}
		\end{adjustbox}
		%}
	\end{table*}

	\begin{figure}[t!]
		\begin{center}
			\mbox{
				\subfloat[Research institute.]{
					{\includegraphics[width=0.48\textwidth]{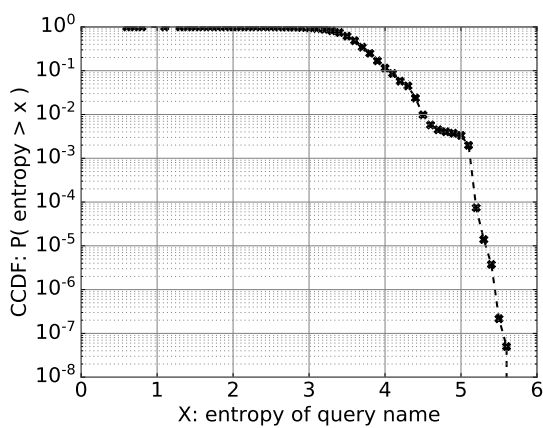}}\quad
					\label{fig:ccdfEntropyCSIRO}
				}
				%  \vspace{-3mm}
				\subfloat[University campus.]{
					{\includegraphics[width=0.48\textwidth]{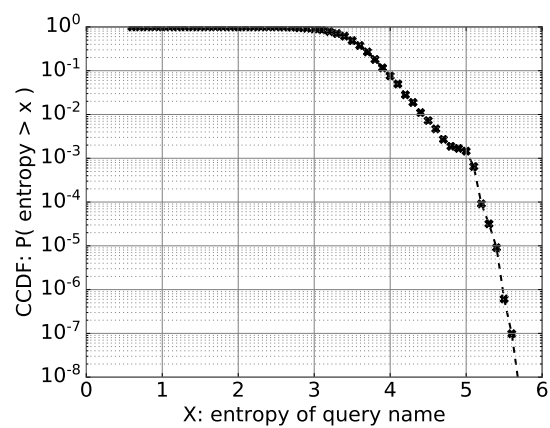}}\quad
					\label{fig:ccdfEntropyUNSW}
				}
			}
			\caption{CCDF of entropy of query name for: (a) Research institute, and (b) University campus.}
			\label{fig:ccdfEntropy}
		\end{center}
	\end{figure}

	\subsubsection{Labels}
	
	This category comprises two attributes of labels inside a FQDN. For example, in the query name ``{\fontsize{10}{48}\usefont{OT1}{lmtt}{b}{n}\noindent www.scholar.google.com}", there are four labels separated by dots.
	%namely {\fontsize{10}{48}\usefont{OT1}{lmtt}{b}{n}\noindent www}, {\fontsize{10}{48}\usefont{OT1}{lmtt}{b}{n}\noindent scholar}, {\fontsize{10}{48}\usefont{OT1}{lmtt}{b}{n}\noindent google}, and {\fontsize{10}{48}\usefont{OT1}{lmtt}{b}{n}\noindent com}, separated by dots. 
	We use the \textbf{number of labels} {\cite{buczak2016detection}} as our sixth attribute. This is because DNS exfiltration/tunneling traffic tends to use certain patterns of labels in their query names. Table \ref{tab:ExampleDomains} shows the label patterns for five selected domains in our dataset from the Research institute network. We abstract a label pattern by an array (samples are shown in the second column) whose elements indicate the length (\ie character count) of the corresponding label in the query name -- \eg the pattern for ``{\fontsize{10}{48}\usefont{OT1}{lmtt}{b}{n}\noindent www.scholar.google.com}" is represented by {(3,7,6,3)}.
	We see that queries for each of the primary domains, listed in Table \ref{tab:ExampleDomains}, appear in various number of patterns -- 
	{the primary domain is obtained by combining the top level domain (TLD) and the second level domain (2LD) (\eg in ``{\fontsize{10}{48}\usefont{OT1}{lmtt}{b}{n}\noindent www.scholar.google.com}", the primary domain is ``{\fontsize{10}{48}\usefont{OT1}{lmtt}{b}{n}\noindent google.com}")}.
	For example, the domain ``{\fontsize{10}{48}\usefont{OT1}{lmtt}{b}{n}\noindent sophosxl.net}"
	is queried by 2208 distinct label patterns during one-week period of our dataset, and each pattern is seen in 66 queries on average. For ``{\fontsize{10}{48}\usefont{OT1}{lmtt}{b}{n}\noindent e5.sk}" domain, on the other hand, we observe only 10 unique patterns, each repeats more than 600 times.

	{
		%\vspace{-5mm}
		\begin{table}[t!]
			\centering
			{
				\caption{Labels pattern in query names for selected domains.}
				%\vspace{-3mm}
				\label{tab:ExampleDomains}
				\begin{adjustbox}{max width=0.85\textwidth}
					\begin{tabular}{@{}lccc@{}}
						\toprule
						\textbf{primary domain} & \textbf{sample patterns} & \textbf{\# unique patterns}  & \textbf{avg \# queries / pattern} \\ \midrule
						\fontsize{10}{48}\usefont{OT1}{lmtt}{b}{n}
						\noindent       
						sophosxl.net& (1, 63, 63, 36, 16, 1, 2, 1, 8, 3)   & 2208  & 66  \\ 
						& (1, 63, 63, 18, 40, 1, 2, 1, 8, 3)                 &     &  \\ \midrule
						\fontsize{10}{48}\usefont{OT1}{lmtt}{b}{n}
						\noindent       
						mcafee.com& (3, 11, 7, 4, 4, 4, 3, 1, 26, 4, 6, 3)  & 316  & 6K \\ 
						& (3, 11, 1, 4, 4, 4, 3, 1, 26, 4, 6, 3)              &     &  \\ \midrule
						\fontsize{10}{48}\usefont{OT1}{lmtt}{b}{n}
						\noindent       
						spotify.com& (48, 48, 48, 48, 16, 2, 7, 3)    & 43   & 1.9K   \\ 
						& (23, 2, 7, 3)                &     &  \\ \midrule
						
						\fontsize{10}{48}\usefont{OT1}{lmtt}{b}{n}
						\noindent       
						cnr.io& (5, 1, 3, 63, 63, 63, 30, 8, 3, 2)   & 46    & 2.6K \\ 
						&  (5, 1, 3, 63, 63, 63, 8, 8, 3, 2)               &     &  \\ \midrule
						\fontsize{10}{48}\usefont{OT1}{lmtt}{b}{n}
						\noindent       
						e5.sk& (63, 63, 63, 24, 1, 1, 2, 2)    & 10  & 660 \\ 
						&  (63, 63, 18, 1, 1, 2, 2)              &     &  \\ \bottomrule       
					\end{tabular}
					%\vspace{-3mm}
				\end{adjustbox}
			}
		\end{table}
	}

	\begin{figure}[t!]
		\begin{center}
			\mbox{
				\subfloat[Research institute.]{
					{\includegraphics[width=0.80\textwidth]{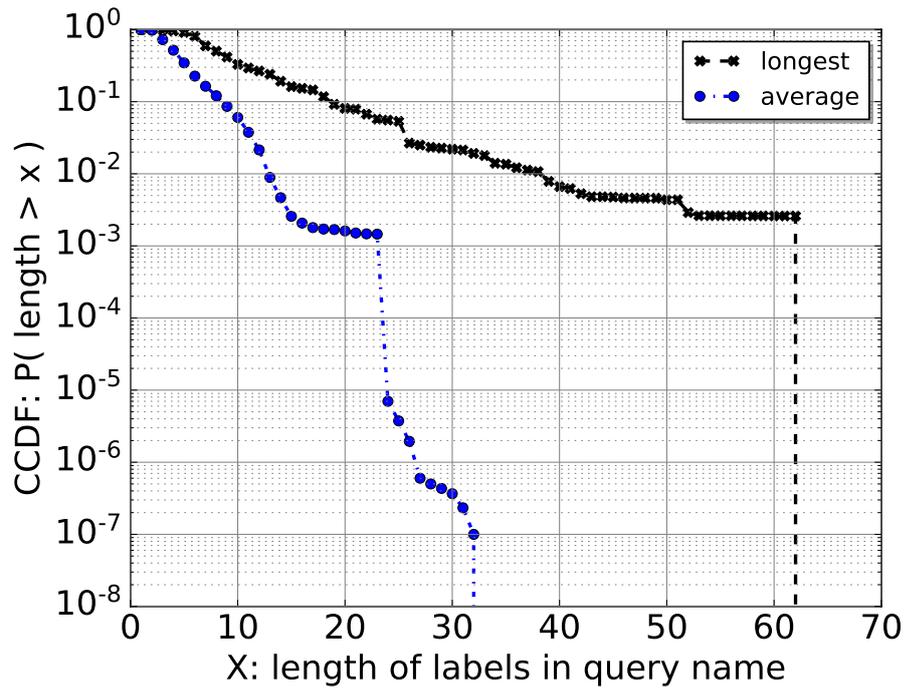}}\quad
					\label{fig:ccdfLongestLabelCSIRO}
				}
			}
			\mbox{
				\subfloat[University campus.]{
					{\includegraphics[width=0.80\textwidth]{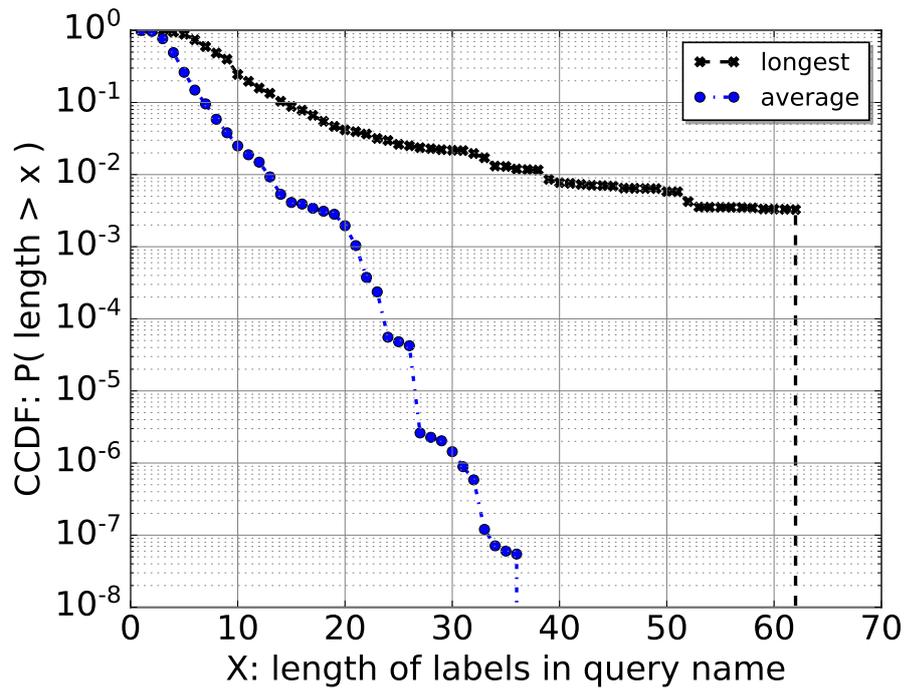}}\quad
					\label{fig:ccdfLongestLabelUNSW}
				}
			}
			\caption{CCDF of length of labels in query name for: (a) Research institute, and (b) University campus.}
			%\vspace{-5mm}
			\label{fig:ccdfLongestLabel}
		\end{center}
	\end{figure}

	Another interesting observation is that queries for three domains namely ``{\fontsize{10}{48}\usefont{OT1}{lmtt}{b}{n}\noindent sophosxl.net}", ``{\fontsize{10}{48}\usefont{OT1}{lmtt}{b}{n}\noindent cnr.io}", and ``{\fontsize{10}{48}\usefont{OT1}{lmtt}{b}{n}\noindent e5.sk}" have several labels with 63 characters (\ie the max limit according to RFC), whereas queries for ``{\fontsize{10}{48}\usefont{OT1}{lmtt}{b}{n}\noindent spotify.com}" and ``{\fontsize{10}{48}\usefont{OT1}{lmtt}{b}{n}\noindent mcafee.com}" do not use label length greater than 48 and 26 characters respectively. For our last two attributes, we use \textbf{maximum label length} {\cite{buczak2016detection}} and \textbf{average label length} {\cite{buczak2016detection}} in a query name. 
	Fig. \ref{fig:ccdfLongestLabel} depicts the CCDF of the longest, and the average label length for FQDNs observed in the two organizations. It is seen that for 90\% of queries, their longest label does not exceed 20 characters, and their average label length is ten characters (or less) in both networks. On the other hand, only about 1\% of queries have the longest label of more than 40 characters in the Research and the University networks, respectively. 
	
	{
		\textbf{Summary:} Our main achievement in this section is to identify and capture eight attributes (from the query name section of each outgoing DNS request packet) that collectively have strong predictive power in determining whether the query name is normal or malicious. The attributes include: \textbf{(1)} Total count of characters in FQDN, \textbf{(2)} count of characters in sub-domain, \textbf{(3)} count of uppercase characters, \textbf{(4)} count of numerical characters, \textbf{(5)} entropy, \textbf{(6)} number of labels, \textbf{(7)} maximum label length, and  \textbf{(8)} average label length.
	}

	%Traditionally, the attackers have performed the DNS exfiltration with long FQDNs. However, an intelligent attacker may reduce the length of the FQDN to transfer sensitive information without being detected. \textbf{Longest label within the domain} and \textbf{Average label within the domain} features are significant in the above scenario to detect the malicious activity over the domain name. 
	
	%Moreover, it can also be observed that only 0.01\% of the queries are using the average label length of 20 characters.  
	
	\begin{table}[t!]
		\centering
		{
			\caption{Summary of additional dataset (days 8-14) used for evaluation.}
			\label{tab:datset2}
			\begin{adjustbox}{max width=0.85\textwidth}
				\begin{tabular}{@{}lcc@{}}
					\toprule
					
					& \multicolumn{1}{l}{{\textbf{Research}}}   & \multicolumn{1}{l}{{\textbf{University}}} \\ \midrule
					%				Total DNS packets        & 243M              & 597M              \\ \midrule
					%				~~IPv4 DNS packets     & 199M             & 493M             \\ \midrule
					%				~~IPv6 DNS packets    & 44M             & 104M          \\ \midrule
					%				~~DNS queries          & 140M               & 340M               \\ \midrule
					%				~~DNS responses      & 103M                & 256M             \\ \midrule
					Total Outgoing DNS queries & 79.6M            & 228M         \\ \midrule
					~~Outgoing DNS queries (IPv4) & 62.6M             & 182M              \\ \midrule
					~~Outgoing DNS queries (IPv6)& 17.0M              & 46M              \\ \midrule
					Outgoing DNS queries ({\color{teal}only qualified}) & 78.8M                & 217M             \\ \midrule
					Unique query names (FQDN)        & 2.1M           & 6.1M            \\ \midrule
					%Unique query names (FQDN) with IPv6         & 607K          & 2.05M            \\ \midrule
					Unique primary domains & 382K                & 1.15M            \\ \midrule
					%Unique query names (FQDN) with IPv4         & 2.09M           & 6.07M             \\ \midrule
					%Unique query names (FQDN) with IPv6         & 607K          & 2.05M            \\ \midrule
					%Unique primary domains with IPv4 & 390K                & 1.3M             \\ \midrule
					%Unique primary domains with IPv6 & 191K               & 886K             \\ \midrule
				\end{tabular}
			\end{adjustbox}
		}
	\end{table}

	\section{Detection of Anomalous Queries}\label{sec:ML}
	We now develop a machine learning technique to determine if a DNS query of an enterprise host is normal or not
	(\ie ``anomaly detection''). {By training a model with only normal query names we aim to detect new/unknown malicious attacks (\ie anomalous queries) which can be missed by the two-class classifier.} 
	The machine is invoked with the eight attributes of each DNS query explained in the previous section.
	{To validate the efficacy of our models, in this section, we extend our dataset by including additional records (78.8 M and 217 M qualified DNS queries from the Research Institute and the University campus networks, respectively) collected over the one week of 6-Aug-2018 to 12-Aug-2018 - a summary of this additional dataset (\ie days 8-14) is shown in Table~\ref{tab:datset2}. In total, we analyze 14 days worth of DNS queries from the two enterprises.}

	\subsection{Machine Training}
	
	We train our anomaly detection machine with benign data from four days of our dataset -- we keep the remaining ten days worth of data for testing. {Ground truth of benign domains in the literature is primarily drawn from highly ranked popular domains \cite{surveyDNS2018}. For example, Alexa top-ranked domains are commonly used -- Alexa no longer publishes free top one million sites. However, we note that Alexa ranking is based on the browsing behavior of Internet users (\ie estimate of global traffic to a domain). As a result, some malicious domains may appear among top \textit{K} Alexa domains due to a burst of requests from a high number of infected clients querying them \cite{survey2018}. 
		We, therefore, use an alternative, Majestic Million \cite{Majestic} that releases a free dataset of top 1M domains and updates it daily. Majestic ranks sites by the number of subnets linking to that site -- it is a measure of trust instead of traffic estimates \cite{rweyemamu2019clustering,kelkar2018analyzing}. For the benign training instances, we only use the top 10,000 primary domains in the Majestic list. We also include FQDNs for ``{\fontsize{10}{48}\usefont{OT1}{lmtt}{b}{n}\noindent sophosxl.net}'' domain which is not among the top 10K Majestic dataset -- the Majestic dataset is used as a reference of domain reputation to determine whether a queried domain is benign or not.}
	
	%I assume that you are using the majestic data set to validate if a domain from your captured traffic is benign or not? This could be made much clearer to the reader. 
	
	%domainsToCheck = ['senderbase.', 'mcafee.' , 'spotify.', 'sophosxl.', 'spamhaus.', 'googleapis.', 'gstatic.']
	
	%if ((int(fields[9]) < 10000) or any(dom in queryName for dom in domainsToCheck)):

	\subsection{Algorithms and Tuning Parameters}\label{sec:MLtuning}
	%We employ two popular anomaly detection algorithms: SVM and IsolationForest from the SciKit-learn machine learning library.
	The objective is to maximize the detection of anomalous queries while reducing the rate of false alarms (\ie incorrectly detecting a normal query as anomalous or vice versa).
	Many supervised machine-learning algorithms for detecting anomalies such as one-class SVM and Replicator Neural Network suffer from high false alarms. They are optimized for profiling the inlier behavior rather than detecting anomalies. 
	We employ ``Isolation Forest (\textit{iForest})'' \cite{liu2008isolation} which is an effective algorithm in detecting anomalous instances in high-dimensional datasets with minimal memory and time complexities. 
	
	{The iForest algorithm {\cite{liu2008isolation}} works based on the concept of isolation without employing any distance or density measure. This algorithm aims to isolate test instances by randomly selecting a feature and then randomly selecting a split value from a range (within min and max obtained from training) values of the selected feature. Then, the score is calculated as the number of conditions (path length) to check for isolating a test instance. Note that isolating normal instances require more conditions. The process is repeated several times to avoid issues due to randomness, and the average path length is calculated and normalized.}
	
	%The iForest algorithm {\cite{liu2008isolation}} ``isolates'' observations by randomly selecting an attribute and then randomly selecting a split value in the range of values (\ie between min and max) for the selected attribute. Since recursive partitioning can be represented by a tree structure, the number of splittings required to isolate an instance is equivalent to the path length from the root node to the terminating node. This path length averaged over a forest of such random trees, measures normality and the decision function (\ie normal instances are expected to have a reasonably large path length in random partitioning). %Random partitioning produces noticeably shorter paths for anomalies. 
	%Therefore, when a forest of random trees collectively produces shorter path lengths for a particular instance, it is highly likely to be an anomaly. %{\color{red}Note that to understand more about iForest, the authors in \cite{liu2008isolation} explain it in detail along with the examples. }

	\textbf{Algorithm Tuning:}  
	We used {\fontsize{10}{48}\usefont{OT1}{lmtt}{b}{n}\noindent scikit-learn} and its APIs, an open-source machine-learning package written in Python, to train and test our machine. We have used three tuning parameters for iForest during the training phase, namely the number of trees (\textit{n\_estimators}), height limit of trees (\textit{max\_samples}), and contamination rate. We tune the value of each parameter while fixing the other two parameters and validate the accuracy of our machine for both benign and malicious instances (that we have the ground truth)in both organizations. The default value for the number of trees is 100, the height limit of trees is set to ``auto'' (implying 8 given the size of our dataset), and the contamination rate is 10\%.
	
	{To tune the algorithm, we require ground truth for both benign and malicious instances. Our ground-truth for benign instances are chosen based on the top 10K domains of the Majestic list (\S\ref{sec:DataSet})) -- {we have 1.7 M instances for the research organization and 4.8 M for the university campus network}). For the ground-truth of malicious instances, we generated DNS exfiltration queries with our open-source tool, forked from an open-source project called ``DNS Exfiltration Toolkit'' (DET) \cite{DET}.} We ran our tool on a machine inside the University network that exfiltrates the content of a CSV file containing 1000 samples of random credit card details (obtained from \cite{creditCard}) to an authoritative name server
	% (\ie {\fontsize{10}{48}\usefont{OT1}{lmtt}{b}{n}\noindent nozzle.centie.net.au}) 
	under our control located in the Research network. DET employs AES-256 encryption and uses two tuning parameters: the max length of the query name (\ie 30 to 218 characters) and the max length of labels (\ie 30 to 63 characters) to diversify our synthetic malicious queries. We generated a total of 1.4M exfiltration queries that are publicly available at \cite{ourData} in the form of a CSV file. 
	
	\begin{table}[t!]
		\centering
		\caption{Detection accuracy of ground-truth instances after tuning.}
		\label{table:DetectionAccuracy}
		\begin{adjustbox}{max width=0.85\textwidth}     
			\renewcommand{\arraystretch}{1.1}           	
			\begin{tabular}{lcc}
				\toprule
				& \textbf{Benign} & \textbf{Malicious} \\ \hline
				{\textbf{Research Institute}}  & $98.44$\% & $95.07$\% \\ \hline     
				{\textbf{University Campus}} & $97.99$\% & $98.49$\% \\
				\bottomrule   
			\end{tabular}
		\end{adjustbox}
	\end{table}

	We found that setting the number of trees equal to 2 results in high accuracy of more than $91$\% for benign and $63$\% for malicious instances -- increasing this parameter does not enhance the accuracy but increases the model size and prediction time. 
	Having fixed the number of trees to 2 and the contamination rate to 10\%, we varied the height of trees from 1 to 20. The detection performance rises by increasing the height limit of trees and gets stabilized at the value of 18 with the best accuracy of more than $90$\% and $98$\% for ground-truth benign and malicious instances, respectively. We then fixed the number of trees to 2 and the height limit of isolation trees to 18 to quantify the impact of contamination rate. Decreasing the contamination rate from 10\% to 2\% improved the performance of our model for both organizations as shown in Table \ref{table:DetectionAccuracy}, with the accuracy of more than $97$\% for benign instances and more than $95$\% for malicious instances. %as the contamination rate varies the threshold value for the classification of an instance based on the value of the decision function. 

	To summarize, we found the optimal value of tuning parameters equal to 2, 18, and 2\% respectively for the number of trees, the height limit of trees, and the contamination rate. Furthermore, for optimal tuning parameters, the iForest algorithm sets the threshold value of anomaly score to $0.54$, distinguishing normal and anomalous instances.
	
	Table \ref{tab:perfBenign} shows the performance of our machine (after tunning) for selected benign instances -- for cross-validation. It can be seen that the rate of false alarms is mostly less than $5$\% in both organizations, though we see a higher false rate (\ie more than $10$\%) for ``{\fontsize{10}{48}\usefont{OT1}{lmtt}{b}{n}\noindent in-addr.arpa}" and ``{\fontsize{10}{48}\usefont{OT1}{lmtt}{b}{n}\noindent sophosxl.net}" domains in the University network. We found out that the attributes of some of FQDNs of primary domains ``{\fontsize{10}{48}\usefont{OT1}{lmtt}{b}{n}\noindent in-addr.arpa}" and ``{\fontsize{10}{48}\usefont{OT1}{lmtt}{b}{n}\noindent sophosxl.net}" are similar to those of exfiltration FQDNs (possibly benign DNS exfiltration). In the next section, we will pre-filter instances for these domains that are highly trusted (\ie certainly benign) without passing them to the anomaly detection machine.  
	
	%Then we fixed the number of trees and the contamination rate and analyze the effect of increasing the tree height limit of the isolation trees in isolation forest algorithm. The result shows that the accuracy of the algorithm improves as we increase the amount of tree height from 1 to n, and the performance of IsoForest is the best at tree height equals eight. 

	%As discussed previously that we have to train our machine with the benign domains, so we started by employing \textbf{Support Vector Machines (SVM) } that is a type of classification technique which is used to split the data into the best possible way. 
	%Suppose that we have two classes of symbols, we call it support vectors. 
	%The basic principle of SVM is to find the widest margin (hyperplane) between the support vectors in such a way that the distance between the two groups is as far as possible. We aim to build a machine by training it with only the benign domains, One-class SVM is used to find the outliers and anomaly in the normal trained data. 
	
	%We then employed \textbf{Isolation Forest} algorithm by training the machine with the attributes of only benign domains for all the 4 days of enterprise. 

	%\vspace{6mm}
	\begin{table}[t!]
		\centering
		{
			\caption{Anomaly detection for Research institute.}
			\label{table:perfRsch}
			\begin{adjustbox}{max width=0.85\textwidth}  
				\renewcommand{\arraystretch}{1.3}    
				\begin{tabular}{lccc}
					\toprule
					{\textbf{Input}}  & {\textbf{Output}}    & \textbf{Days 1-4} & \textbf{Days 5-14} \\ \hline\hline
					\multirow{2}{*}{Benign domains (top 10K)} & normal    & $98.44$\%     & $98.35$\% \\
					& anomalous & \textbf{1.56\%}       & \textbf{1.65\%} \\ \cline{1-4}
					\multirow{2}{*}{Others (beyond top 10K)}         & normal    & $78.43$\%      & $77.35$\% \\
					& anomalous & $21.57$\%       & $22.65$\% \\  
					\bottomrule   
				\end{tabular}
			\end{adjustbox}
		}
	\end{table}

	\begin{table}[t!]
		\centering
		{
			\caption{Anomaly detection for University campus.}
			\label{table:perfUniv}
			\begin{adjustbox}{max width=0.85\textwidth}   
				\renewcommand{\arraystretch}{1.3}   		  
				\begin{tabular}{lccc}
					\toprule
					{\textbf{Input}}  & {\textbf{Output}}    & \textbf{Days 1-4} & \textbf{Days 5-14} \\ \hline\hline
					\multirow{2}{*}{Benign domains (top 10K)} & normal  & $97.99$\%     & $97.83$\% \\
					& anomalous & \textbf{2.01\%}   & \textbf{2.17\%} \\ \cline{1-4}
					\multirow{2}{*}{Others (beyond top 10K)}          & normal  & $70.57$\%     & $63.38$\% \\
					& anomalous & $29.43$\%      & $36.62$\% \\  
					\bottomrule   
				\end{tabular}
			\end{adjustbox}
		}
	\end{table}

	\begin{table*}[t!]
		\centering
		\caption{Performance of our machine for trusted domains.}
		\label{tab:perfBenign}
		\begin{adjustbox}{max width=0.95\textwidth}
			\renewcommand{\arraystretch}{1.5}   	
			\begin{tabular}{@{}lcccc|cccc@{}}
				\toprule
				& \multicolumn{4}{c|}{\textbf{Research institute}}   & \multicolumn{4}{c}{\textbf{University campus}} \\ \midrule
				\textbf{primary domain} & \textbf{normal}  & \textbf{anomalous} & \textbf{Avg. query length} & \textbf{false-rate (\%)} & \textbf{normal}  & \textbf{anomalous} & \textbf{Avg. query length}& \textbf{false-rate (\%)}  \\ \midrule
				
				\fontsize{10}{48}\usefont{OT1}{lmtt}{b}{n}
				\noindent 
				akadns.net&2.6M  &24K  &38&  0.91 & 7.6M  &  191K &  38&  2.4 \\ \midrule
				\fontsize{10}{48}\usefont{OT1}{lmtt}{b}{n}
				\noindent 
				googleapis.com& 165K  &  1.6K   &76 &  0.96 &  526K  &  15K   &76  & 2.7 \\ \midrule
				\fontsize{10}{48}\usefont{OT1}{lmtt}{b}{n}
				\noindent 
				gstatic.com&207K&    362&  69&  0.17  & 835K    &  986   &76  & 0.11 \\ \midrule
				\fontsize{10}{48}\usefont{OT1}{lmtt}{b}{n}
				\noindent 
				in-addr.arpa&3.7M    &49K  &26&  1.32& 9.2M &1.1M  &26&  10.7 \\ \midrule
				%      \fontsize{10}{48}\usefont{OT1}{lmtt}{b}{n}
				%      \noindent 
				%      ip6.arpa&52    &200K  & 71&  99.97&168    &80K&  71&  99.78 \\ \midrule
				\fontsize{10}{48}\usefont{OT1}{lmtt}{b}{n}
				\noindent 
				mcafee.com&1.9M&    735  &84  &0.03&635K    & 13K&   88&  2.01 \\ \midrule
				\fontsize{10}{48}\usefont{OT1}{lmtt}{b}{n}
				\noindent 
				onmicrosoft.com&22K&    1.6K&   51  &6.55&201K  &  1537&   53&  0.75 \\ \midrule
				\fontsize{10}{48}\usefont{OT1}{lmtt}{b}{n}
				\noindent 
				senderbase.org&1.1M&    14K  &66  &1.32&  2.2M    & 2816  & 66  & 0.12 \\ \midrule
				\fontsize{10}{48}\usefont{OT1}{lmtt}{b}{n}
				\noindent 
				sophosxl.net& 138K   &6.5K  &103   & 4.44&2.5M  &   394K&   119&   13.7 \\ \midrule
				\fontsize{10}{48}\usefont{OT1}{lmtt}{b}{n}
				\noindent 
				spamhaus.org&12K  &  597&  31&  4.7& 947K   &   7.7K &  32 &  0.81 \\ \midrule
				\fontsize{10}{48}\usefont{OT1}{lmtt}{b}{n}
				\noindent 
				spotify.com&579  &31  &45&  5.08& 468K  &  1.2K&  168&   0.25 \\ \midrule
				Top 100 domains  & 7.9M &  135K&   20&  1.68& 24M    & 351K  & 20  & 1.41 \\ 
				(\eg {\fontsize{10}{48}\usefont{OT1}{lmtt}{b}{n}\noindent google}, {\fontsize{10}{48}\usefont{OT1}{lmtt}{b}{n}\noindent apple})  &   &   & &  & &  & &    \\ \bottomrule       
			\end{tabular}
			%  \vspace{-3mm}
		\end{adjustbox}
	\end{table*}

	\section{Performance Evaluation}\label{sec:eval}
	{ In this section, we evaluate the efficacy of our scheme by: (a) cross-validating and testing the accuracy of the trained model for benign instances and quantifying the performance in real-time on live 10 Gbps traffic streams from the two organizations, (b) testing the detection rate for malicious DNS queries that we generate using our customized tool (\ie DET \cite{DET}) and an open-source tool (\ie Iodine \cite{iodine}), (c) comparing our one-class classifier with a two-class classifier, and (d) drawing insights into the top three anomalous domains for which malicious DNS queries are made in the Research and University networks.}
	Note that our proposed approach is generic and hence can be readily used in different organizations, but the model needs to be trained by the specific data of each organization.
	
	{
		
		\subsection{Performance Metrics} \label{sec:performance}
		We begin with three performance metrics, namely accuracy, anomaly score, and responsiveness of our models.}

	\textbf{Accuracy:} As mentioned in the previous section, we trained our model with benign instances from 4 days' worth of our data (\ie Days 1-4), and tested with all instances from Days 5-14 in addition to remaining instances from Days 1-4 that were not used for training (\ie ``Others''). Tables \ref{table:perfRsch} and \ref{table:perfUniv} show the rate of detection (\ie normal versus anomalous) for the benign and Others instances in the two networks -- { instances in the Benign category are among the top 10K of the Majestic ranking list, and instances in the Others category are beyond 10K}. It can be seen that $98$\% of benign instances are correctly detected as normal during both cross-validation (\ie Days 1-4) and testing (\ie Days 5-14) phases. We note that our machine raises a false alarm for about $2$\% of benign domains, as highlighted in bold text.

	To address this, we populate a whitelist of domains that are highly trusted. Our whitelist comprises only the top 100 domains from the Majestic ranking dataset (\eg ``{\fontsize{10}{48}\usefont{OT1}{lmtt}{b}{n}\noindent google.com}'', ``{\fontsize{10}{48}\usefont{OT1}{lmtt}{b}{n}\noindent bbc.com}'', ``{\fontsize{10}{48}\usefont{OT1}{lmtt}{b}{n}\noindent amazonaws.com}'') as well as popular legitimate (\eg ``{\fontsize{10}{48}\usefont{OT1}{lmtt}{b}{n}\noindent akadns.net}'', ``{\fontsize{10}{48}\usefont{OT1}{lmtt}{b}{n}\noindent in-addr.arpa}'', ``{\fontsize{10}{48}\usefont{OT1}{lmtt}{b}{n}\noindent spotify.com}'') and security services (\eg ``{\fontsize{10}{48}\usefont{OT1}{lmtt}{b}{n}\noindent spamhaus.org}'', 		``{\fontsize{10}{48}\usefont{OT1}{lmtt}{b}{n}\noindent senderbase.org}''). Note that these security services are using disposable domains (\ie ``single-time use'') for the purpose of signaling over DNS queries (\eg ``{\fontsize{10}{48}\usefont{OT1}{lmtt}{b}{n}\noindent 0.0.0.0.1.0.0.4e.135jg5e1pd7s4735ftrqweufm5.avqs.mcafee.com}'' \cite{chen2014dns}).

	\begin{table}[t!]
		\centering
		{
			\caption{Anomaly detection combined with whitelisting for Research institute.}
			\label{table:perfFilteringRsch}
			\begin{adjustbox}{max width=0.85\textwidth}     
				\renewcommand{\arraystretch}{1.3}    		
				\begin{tabular}{lccc}
					\toprule
					{\textbf{Input}}  & {\textbf{Output}}    & \textbf{Days 1-4} & \textbf{Days 5-14} \\ \hline\hline
					\multirow{2}{*}{Benign domains (top 10K)} & normal    & $98.92$\%     & $99.20$\% \\
					& anomalous & \textbf{1.08\%}       & \textbf{0.80\%} \\ \cline{1-4}
					\multirow{2}{*}{Others (beyond top 10K)}         & normal    & $83.50$\%      & $87.48$\% \\
					& anomalous & $16.50$\%       & $12.52$\% \\  
					\bottomrule   
				\end{tabular}
			\end{adjustbox}
		}
	\end{table}

	%\vspace{-3mm}
	\begin{table}[t!]
		\centering
		{
			\caption{Anomaly detection combined with whitelisting for University campus.}
			\label{table:perfFilteringUniv}
			\begin{adjustbox}{max width=0.85\textwidth}    
				\renewcommand{\arraystretch}{1.3}    		 
				\begin{tabular}{lccc}
					\toprule
					{\textbf{Input}}  & {\textbf{Output}}    & \textbf{Days 1-4} & \textbf{Days 5-14} \\ \hline\hline
					\multirow{2}{*}{Benign domains (top 10K)} & normal  & $98.92$\%     & $98.90$\% \\
					& anomalous & \textbf{1.08\%}      & \textbf{1.10\%} \\ \cline{1-4}
					\multirow{2}{*}{Others (beyond top 10K)}          & normal  & $90.98$\%     & $81.74$\% \\
					& anomalous & $9.02$\%      & $18.26$\% \\  
					\bottomrule   
				\end{tabular}
			\end{adjustbox}
		}
	\end{table}

	Employing whitelisted domains would slightly enhance detection. Our refined results are shown in Tables \ref{table:perfFilteringRsch} and \ref{table:perfFilteringUniv}. We can see a slight reduction in the rate of false alarms for benign domains -- it is now capped at $1.20$\% for both networks, as highlighted in bold text. 
	We note there are a total of 10K (out of 923K) and 15K (out of 1.4M) false alarms for benign instances in the Research and University network, respectively. Although the rate of false positives is about 1\%, network operators can further reduce this by employing off-the-shelf intrusion detection systems (IDSs) such as Zeek on flagged instances. With their signatures of known malicious traffic, Tools like Zeek can filter out those flagged instances that are indeed benign. 
	Note that quantifying the rate of false alarms for detected anomalies under ``Others'' is non-trivial due to the lack of ground-truth labels.

	%{\color{red} Note that there are 12 to 30\% of the instances in the Others category that are detected as anomalous. This is because most of the times dynamic network services instances are classified as exfiltration, \eg ``{\fontsize{10}{48}\usefont{OT1}{lmtt}{b}{n}\noindent 0138-0-syd108a6a2cf5a54f4e577d7e5bc3e03b647896fa\\c935.beacon.rum.dynapis.info}'' which significantly contributes towards the instances classified as anomalous. }

	\begin{table}[t!]
		\centering
		\caption{Avg. anomaly score for research institute.}
		%\vspace{-3mm}
		\label{table:anomalyRsch}
		\begin{adjustbox}{max width=0.85\textwidth}	   
			\renewcommand{\arraystretch}{1.3}    		
			\begin{tabular}{lccc}
				\toprule
				{\textbf{Input}}  & {\textbf{Output}}    & \textbf{Days 1-4} & \textbf{Days 5-14} \\ \hline\hline
				\multirow{2}{*}{Benign domains} & normal  & $0.36$     & $0.36$ \\
				& anomalous & $0.59$      & $0.61$ \\ \cline{1-4}
				\multirow{2}{*}{Others}      		 & normal  & $0.44$       & $0.43$\\
				& anomalous & $0.64$      & $0.65$ \\  
				\bottomrule	
			\end{tabular}
		\end{adjustbox}
	\end{table}

	\begin{table}[t!]
		\centering
		\caption{Avg. anomaly score for University campus.}
		%\vspace{-3mm}
		\label{table:anomalyUniv}
		\begin{adjustbox}{max width=0.95\textwidth}   
			\renewcommand{\arraystretch}{1.3}    			
			\begin{tabular}{lccc}
				\toprule
				{\textbf{Input}}  & {\textbf{Output}}    & \textbf{Days 1-4} & \textbf{Days 5-14} \\ \hline\hline
				\multirow{2}{*}{Benign domains} & normal  & $0.39$     & $0.39$ \\ 										& anomalous & $0.57$      & $0.58$ \\ \cline{1-4}
				\multirow{2}{*}{Others}      		 & normal  & $0.43$       & $0.43$\\									& anomalous & $0.63$      & $0.62$ \\  
				\bottomrule	
			\end{tabular}
		\end{adjustbox}
	\end{table}

	\textbf{Anomaly Score:} 
	Anomaly detection algorithms use this score to determine if an instance is classified as normal or abnormal. For the iForest algorithm, the anomaly score varies from 0 to 1, where 0 means purely normal and 1 indicates a definite anomaly. A value of an anomaly score of less than 0.5 is reasonable enough to be interpreted as normal \cite{liu2008isolation}. 
	%{We have captured the anomaly score of each instance and took the average of the instance after separating the classified instances as normal or abnormal.({\color{red} HASSAN: the last sentence does not read well.})} 

	%Our model generates a label (\ie benign or anomalous) along with an score (\ie anomaly score). 
	Tables \ref{table:anomalyRsch} and \ref{table:anomalyUniv} show the average anomaly score (\ie normal versus anomalous) for the benign and Others instances in the two networks. It can be seen that the average anomaly score of benign instances during the cross-validation phase (\ie Days 1-4) is $0.36$ and $0.39$ which is well below the threshold value of $0.54$ (obtained during model tuning in \S\ref{sec:MLtuning}) for the Research Institute and University networks respectively. Similarly, the average anomaly score of benign instances during the testing phase (\ie Days 5-14) is $0.36$ and $0.40$ for the Research Institute and University networks respectively.

	\begin{table}[t!]
		\centering
		\caption{Avg. time complexity of our scheme.}
		%\vspace{-3mm}
		\label{table:realTime}
		\begin{adjustbox}{max width=0.85\textwidth}     
			\renewcommand{\arraystretch}{1.3}           	
			\begin{tabular}{lc}
				\toprule
				{\textbf{~~~~extracting attributes}}  & 54 $\mu$sec \\ \hline     
				{\textbf{~~~~detecting anomalies}} & 746 $\mu$sec\\ \hline
				{\textbf{Total time per each query name}} & 800 $\mu$sec\\ 
				\bottomrule   
			\end{tabular}
		\end{adjustbox}
	\end{table}
	
	{\textbf{Responsiveness:} } In terms of responsiveness, we have quantified the average time for extracting eight attributes and anomaly detection (via running prediction against the trained model) by testing more than 300 million DNS queries in our dataset from the two enterprise networks. Our attributes extraction and anomaly detection engines run on a virtual machine using 4 CPU cores, 6GB of memory, and storage of 50GB. As shown in Table~\ref{table:realTime}, on average it takes $800$ $\mu$sec to determine if a DNS query is normal or not. This indicates that our scheme can process approximately 1250 DNS queries per second, well above the actual rate of DNS queries in both organizations where the peak value is 800 DNS queries per second, as shown in Fig.~\ref{fig:timetraceDNS}.

	\begin{figure*}[ht]
		{
			\begin{center}
				% \vspace{-3mm}
				\mbox{
					% \hspace{-3mm}
					\subfloat[DET tool.]{
						{\includegraphics[width=0.85\textwidth,height=0.40\textwidth]{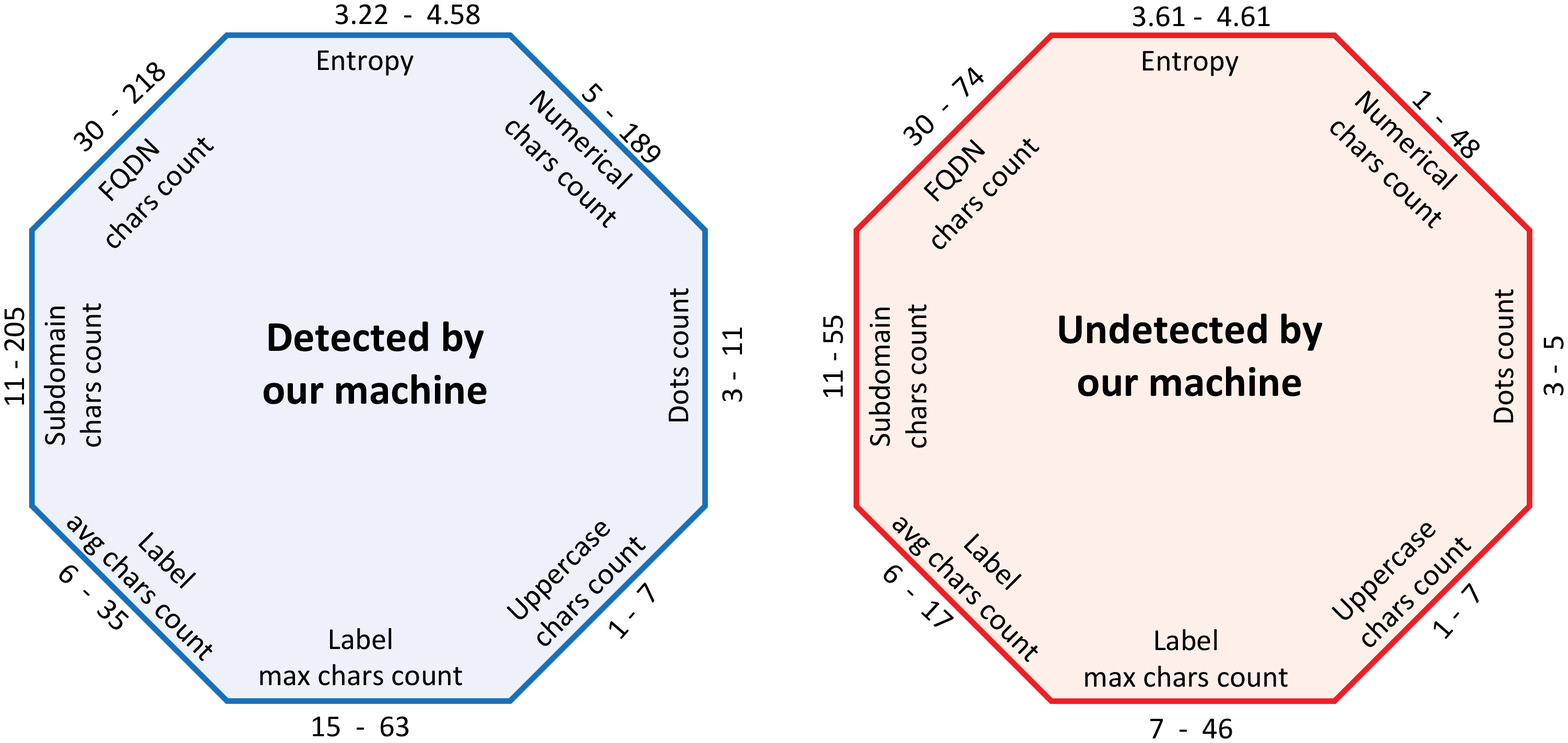}} \quad
						\label{fig:maliciousProfileDETUNSW}
					}
				}
				
				\mbox{
					
					%\hspace{-5mm}
					\subfloat[Iodine tool.]{
						{\includegraphics[width=0.85\textwidth,height=0.40\textwidth]{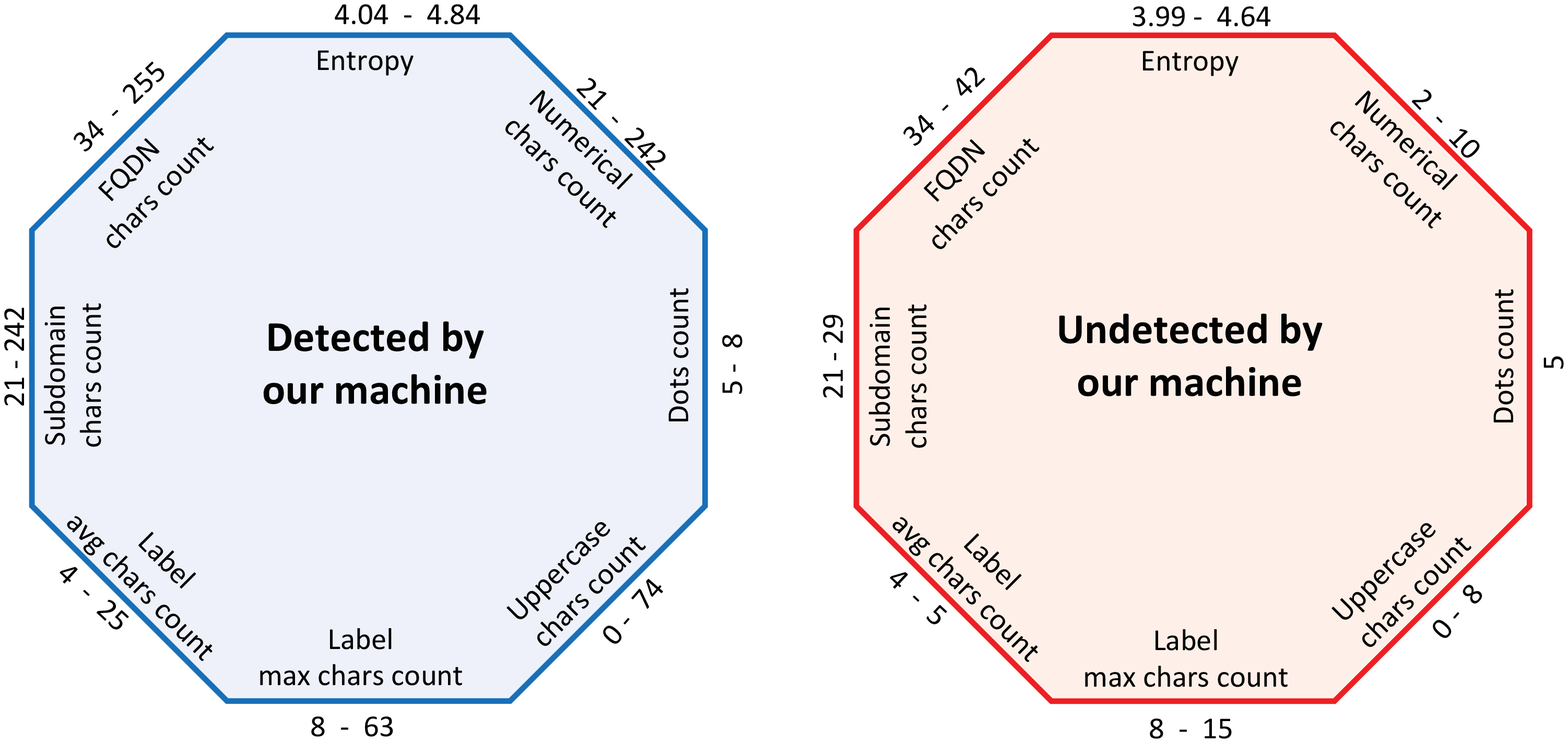}}\quad
						\label{fig:maliciousProfileIodineUNSW}
					}
				}
				%    \vspace{-2mm}
				\caption{Attributes of DNS exfiltration query names of: (a) DET tool, and (b) Iodine tool, detected vs. undetected by the University model.}
				\label{fig:maliciousProfile}
			\end{center}
		}
	\end{figure*}

	{\subsection{Evaluating Models using Known DNS Exfiltration Data}
		In this subsection, we evaluate the efficacy of our detection scheme using DNS exfiltration data (\ie ground-truth) including two large sets generated by our customized DET tool and the open-source Iodine tool, and a small set collected from publicly reported real malicious DNS queries.}
	
	\textbf{Our DET Tool:} 
	%We now evaluate the efficacy of our scheme with known (\ie ground-truth) DNS exfiltration queries. 
	We showed previously in Table \ref{table:DetectionAccuracy} that our models for the Research Institute and the University campus respectively were able to correctly detect $95.07$\% and $98.49$\% of exfiltration queries (generated by our DET tool) as anomalous instances.

	{
		In Fig.~\ref{fig:maliciousProfileDETUNSW}, we show the value of attributes for detected instances (blue octagon on the left) versus undetected instances (red octagon on the right) using the model generated from data of the university campus.} Even though undetected instances were shorter both in total length and average label length, it is important to note that there is a fair overlap of value range comparing detected (\ie classified as anomalous) with undetected instances (\ie classified as normal) across all attributes, suggesting that attributes collectively would determine a fairly accurate output of our model. To explain it further, we look at the attributes of two pairs of FQDNs generated from our DET tool (one classified as normal and one classified as anomalous by the model of the research institute), as shown in Table \ref{table:SampleMaliciousWithFeatures} -- normal and anomalous classified FQDNs are shown in bold and italic fonts respectively. This is because we have obfuscated the actual primary domain used in the DET tool for privacy reasons. For each pair, we investigate distinguishing factors given some identical (or close) attributes, highlighted in bold in Table \ref{table:SampleMaliciousWithFeatures}. We see six common attributes (character count, numerical character count, number of dots, maximum label length, average length of labels, and sub-domain character count). However, entropy and upper-case attributes have relatively larger values in the detected instance (\ie italic text). Moving to the second example, where entropy, number of dots, and upper-case characters count are very close in two instances, the query length becomes an essential factor for the model detecting or missing a malicious instance.

	\begin{table*}[t!]
		\centering
		{
			\caption{Samples of malicious queries (DET) along with their attributes detected/undetected by the model of the research institute.}
			\label{table:SampleMaliciousWithFeatures}
			\begin{adjustbox}{max width=0.995\textwidth}  
				\renewcommand{\arraystretch}{2}         
				\begin{tabular}{lcccccccc}
					\toprule
					{\textbf{ Sample FQDNs}} & {\textbf{\# chars}} & {\textbf{entropy}} & {\textbf{\# numeric}} & {\textbf{\# dots}} & {\textbf{\# upp.}} & {\textbf{max label}}. & {\textbf{avg label}} & {\textbf{\# chars subdom. }} \\ \hline\hline
					\textbf{s6wIrxk.363937356263363563663038333865.maliciousDomain.com} & \textbf{58} & 3.98 & \textbf{31} & \textbf{3} & 2 & \textbf{30} & \textbf{13} & \textbf{39} \\
					\textit {ClBQxLW.656661356534343938623539393265.maliciousDomain.com} & \textbf{58} & 4.24 & \textbf{30} & \textbf{3} & 6 & \textbf{30} & \textbf{13} & \textbf{39} \\ 
					\textbf{J8tngo1.53061393230646235636634326137656436.maliciousDomain.com}	& 63	& \textbf{4.15} & 37 & \textbf{3} &\textbf{2} & 35	& 15	& 44 \\ % \cline{1-9} 
					\textit{dZrlKkg.645a726c4b6b677c217c337c217c3732613830656235373830.3333386331643631393837.maliciousDomain.com} & 101 & \textbf{ 4.15} &	64 &\textbf{4}	& \textbf{3}	& 50 & 	19 & 	82 \\
					\bottomrule   
				\end{tabular} 
			\end{adjustbox}
		}
		%\vspace{-3mm}
	\end{table*}

	{
		\textbf{Iodine Tool:}
		To further evaluate the efficacy of our scheme, we used the Iodine tool \cite{iodine} to generate an additional dataset of malicious DNS queries. First, we exfiltrated the same CSV file of 1000 samples of random credit card details similar to our DET tool. It took approximately $8$ seconds to transfer the entire CSV file. Next, we wrote a Python script to repeat the process with a delay (between runs) uniformly distributed between 20 and 40 seconds. We ran the script for three days. As a result, we captured more than 2.2 million unique instances -- our Iodine dataset is also made publicly available \cite{ourData}. 
		Note that, unlike our customized DET tool, we only used iodine with its default settings (\ie no variation of parameters in DNS queries).
		When this dataset was tested with our iForest model of the University campus, except for 275 instances all others were correctly detected as anomalous. The average anomaly score was $0.86$ and $0.49$ for correctly and incorrectly classified instances, respectively. We also tested against the model of the Research Institute. We found that a small number of malicious instances (1837 out of 2.2 million) were missed -- the average anomaly score was $0.67$ and $0.45$ (lower scores than the university model) for correctly and incorrectly classified instances, respectively.
		
		%out of those 2.2 million instances only 275 instances were miss-classified by our one-class classifier with an average anomaly score of $0.49$ for those miss-classified instances. 
		%The average anomaly score for the instances classified as malicious is $0.86$ which is well above the threshold value of $0.54$ (obtained during model tuning in \S\ref{sec:MLtuning}) for the Research Institute and University networks respectively. 
		In Fig.~\ref{fig:maliciousProfileIodineUNSW}, we show the value of attributes for Iodine instances (detected versus undetected) when tested against the university in the same way as we did in Fig.~\ref{fig:maliciousProfileDETUNSW} for DET instances. We can see that undetected instances (red octagon on the right) have fewer numerical characters in their query name -- 2 to 10 numerical chars versus 21 to 242 in detected instances. Additionally, it is observed that missed instances are relatively short (total chars count of 34-42), with a few uppercase chars (up to 8), and contain short labels (average about 5). Note that the length of DNS queries generated by iodine is typically longer (average of 207 chars). Still, we intentionally diversified the query length (30 to 218 with an average of 64) in our custom DET tool, resulting in a slightly higher percentage of missed instances.

		\begin{table*}[t!]
			\centering
			\caption{Anomaly score of queries publicly reported as DNS Exfiltration.}
			\label{table:knownDNSExfil}
			\begin{adjustbox}{max width=0.98\textwidth}     
				\renewcommand{\arraystretch}{2}           	
				\begin{tabular}{lc}
					\toprule
					{\textbf{Known Malicious FQDN}}  & {\textbf{Anomaly Score}} \\ \hline     
					\fontsize{10}{48}\usefont{OT1}{lmtt}{b}{n}
					\noindent {708001701462b7fae70d0a28432920436f70797269676874.20313938352d32303031204d696372.6f736f667420436f72702e0d0a0d0a0.433a5c54454d503e.cspg.pw} & 0.75 \\ \hline
					\fontsize{10}{48}\usefont{OT1}{lmtt}{b}{n}
					\noindent {9ad9ca2.grp10.tt1.dcd2fed0d2fefecdc8d2c4c8c8fecdde.e3e29f9a9ff9cbc79fdae3fcc4 d2c8c4cdd0feded295e9e9e9e9e9e9feea.e9e9e9e9e9e9e9e9e9e9e9e9e9e} & \\		       			\fontsize{10}{48}\usefont{OT1}{lmtt}{b}{n}
					\noindent {9e9e9e9e9e9e9e9e9e9e9e9e9e9e9e9e9.e9e9e9e9e9e9e9e9e9e9.ns.a23-33-37-54-deploy-akamaitechnologies.com} & 0.70 \\ \hline
					\fontsize{10}{48}\usefont{OT1}{lmtt}{b}{n}
					\noindent  {PzMnPiosOD4nOCwuOzomPS4nNjovPS8uOzsnNCstODkjOCwoMwAA.29a.de} & 0.68 \\ \hline
					\fontsize{10}{48}\usefont{OT1}{lmtt}{b}{n}
					\noindent {9ad9ca2.grp10.tt2.dcc8c8d0c8fccdd2fcd0dcdec8c8cdc8.e6dcc8c8d0c8fccdd2fcd0dcdec8c8cdc8e9dcdcdec8ded2feded0d2c8fc.ns.a23-33-37-54-deploy-akamaitechnologies.com} & 0.67 \\ \hline
					\fontsize{10}{48}\usefont{OT1}{lmtt}{b}{n}
					\noindent {ZTEZGKDFA0KNGUCQI.ns1.logitech-usa.com}  & 0.65 \\ \hline     
					\fontsize{10}{48}\usefont{OT1}{lmtt}{b}{n}
					\noindent  {WQPKBPRYA0IVDUQWI.ns1.logitech-usa.com} & 0.65 \\ \hline
					\fontsize{10}{48}\usefont{OT1}{lmtt}{b}{n}
					\noindent  {QRBJBPRYA0JBKUGVI.ns1.logitech-usa.com} & 0.65 \\ \hline
					\fontsize{10}{48}\usefont{OT1}{lmtt}{b}{n}
					\noindent  {SXLXBPRYA0IVDUKTI.ns1.logitech-usa.com} & 0.65 \\ \hline 
					\fontsize{10}{48}\usefont{OT1}{lmtt}{b}{n}
					\noindent  {SXLXBPRYA0IVDUKTI.ns1.logitech-usa.com} & 0.65 \\ \hline
					\fontsize{10}{48}\usefont{OT1}{lmtt}{b}{n}
					\noindent {6e517f3.grp10.ping.adm.cdd2e9cde9fee9cdc8.cdd0e8e9c8fce9d2e9fecdc4.c597f097ce87c5d3.ns.a23-33-37-54-deploy-akamaitechnologies.com} & 0.59 \\ \hline
					{\fontsize{10}{48}\usefont{OT1}{lmtt}{b}{n}
						\noindent {RoyNGBDVIAA0.0ffice36o.com}} & {0.58} \\ \hline
					{\fontsize{10}{48}\usefont{OT1}{lmtt}{b}{n}
						\noindent {iucCGJDVIBDSNF3GK000.0ffice36o.com} }& {0.58} \\ \hline
					{\fontsize{10}{48}\usefont{OT1}{lmtt}{b}{n}
						\noindent {viLxGJDVIBJAIMQGQ000.0ffice36o.com}} & {0.58} \\ \hline
					{\fontsize{10}{48}\usefont{OT1}{lmtt}{b}{n}
						\noindent {gLtAGJDVIAJAKZXWY000.0ffice36o.com}} & {0.58} \\ \hline
					{\fontsize{10}{48}\usefont{OT1}{lmtt}{b}{n}
						\noindent {TwGHGJDVIATVNVSSA000.0ffice36o.com}} & {0.58} \\ \hline
					{\fontsize{10}{48}\usefont{OT1}{lmtt}{b}{n}
						\noindent {1QMUGJDVIA3JNYQGI000.0ffice36o.com} }& {0.57} \\ \hline
					{\fontsize{10}{48}\usefont{OT1}{lmtt}{b}{n}
						\noindent {t0qIGBDVIAI0.0ffice36o.com}} & {0.57} \\ 			
					
					\bottomrule   
				\end{tabular}
			\end{adjustbox}
		\end{table*}
		
		\textbf{Real malicious DNS queries:}
		Additionally, we tested $17$ samples of DNS queries from known real malware reported on various forums \cite{sampleBernhardPOS2015,sampleFrameworkPOS2016,sampleDNSMessenger2017,sampleDNSpionage2018}. The top ten instances correspond to POS malware, and the bottom eight instances were recently found as part of a new attack targeting networks of a private airline company \cite{sampleDNSpionage2018}. Our trained model was able to detect all of them as abnormal instances. Table \ref{table:knownDNSExfil} gives the anomaly score of these known malicious domains. It can be seen that values are well above the average anomaly score of benign instances shown in Table \ref{table:anomalyRsch} and \ref{table:anomalyUniv}.

		\subsection{Comparing Multi-Class Classifier with One-Class Classifier}
		Existing proposals have predominantly used stateful attributes (\eg mean/variance of time-interval between a pair of DNS query/response, frequency of A, AAAA, TXT types resource records, or time between two DNS responses from a given domain) { along with a combination of stateless attributes} to detect data theft over DNS protocol that is fundamentally different from our approach (\ie using stateless attributes only). Hence, we cannot compare the performance of our model with prior research work. However, we compare the efficacy of our proposed one-class classifier (\ie iForest anomaly detector) with a two-class classification model (\ie Random Forest in \cite{allard2011tunneling,buczak2016detection}) using the stateless attributes considered in this chapter.  
		%Therefore, we have compared our approach by using the same stateless attributes (that we used to train iForest) to train a multi-class machine learning algorithm that is used extensively by the research community \cite{allard2011tunneling,buczak2016detection} (i.e, Random Forest) instead of one class classifier.

		\begin{table*}[t!]
			\centering
			\caption{Detecting wild malicious DNS queries from two enterprises.}
			\label{tab:wild}
			\begin{adjustbox}{max width=0.98\textwidth}  
				\renewcommand{\arraystretch}{1.5}         
				\begin{tabular}{lcccccc}
					\toprule
					& {\textbf{Anomalous domains}} & {\textbf{Daily avg. \# queries}} & {\textbf{Daily avg. \# enterprise hosts}} & {\textbf{Distinct types}} & {\textbf{Avg. TTL (sec)}} \\ \hline\hline
					\multirow{4}{*}{Research Institute} & {\fontsize{10}{48}\usefont{OT1}{lmtt}{b}{n}\noindent imrworldwide.com} & 20.9K & 12 & A [64\%], AAAA [36\%] & 1250 \\
					& {\fontsize{10}{48}\usefont{OT1}{lmtt}{b}{n}\noindent cnr.io} & 17.4K & 4 & TXT [99.9\%], AAAA [0.01\%] & 0 \\
					& {\fontsize{10}{48}\usefont{OT1}{lmtt}{b}{n}\noindent 1rx.io}  & 2.5 K & 10& AAAA [58.9\%], A [41.1\%] & 45 \\
					& {\fontsize{10}{48}\usefont{OT1}{lmtt}{b}{n}\noindent 2o7.net} & 1.2K & 8 & A[53.4\%], AAAA [46.6\%] & 144 \\ \cline{1-6} 
					
					\multirow{3}{*}{University Campus} & {\fontsize{10}{48}\usefont{OT1}{lmtt}{b}{n}\noindent imrworldwide.com} & 80.4K & 203 & A[98\%], AAAA [1.8\%], CNAME[0.02\%] & 1781 \\
					& {\fontsize{10}{48}\usefont{OT1}{lmtt}{b}{n}\noindent 360.cn} & 69K& 122 & A[97.5\%], AAAA [2.5\%] & 3139 \\  
					& {\fontsize{10}{48}\usefont{OT1}{lmtt}{b}{n}\noindent adlooxtracking.com} & 3K & 51 & A [88.1\%], AAAA [11.9\%] & 903 \\
					\bottomrule   
				\end{tabular}
			\end{adjustbox}
		\end{table*}
		
		%cnr 17441.14286	38.71428571
		%imr 21639.28571	34.85714286
		%1rx 2191.857143	15.28571429
		%207 1182.428571	21.42857143 

		%360.cn 69063.57143	60.57142857
		%imr 80437.28571	22.14285714
		%adloox 3038.285714	71.14285714

		We build a new dataset for two-class classification using 4-days worth of top 10K majestic million as benign (same as for the iForest model) and 1.4 million records generated by our DET tool as malicious instances.
		%To achieve this, we took our benign domains (\ie top 10K majestic million primary domains) as normal and ground-truth malicious instances as an exfiltration attack.
		%We train a Random Forest model with 4-days worth of benign instances (same as for the iForest model) as well as 1.4 million instances generated by our DET tool. 
		{
			We split this dataset to 60\% for training and 40\% for testing. The accuracy of validation (on training data) and testing is about 99\% and 97\% respectively, with the default parameters of the Random Forest classifier. Also, we obtain the confidence-level of the Random Forest model to assess its confidence in making decision for test instances (benign or malicious). The model displays an average confidence of 99.9\% for correctly classified benign instances -- this measure is 76\% for misclassified benign instances. Similarly, when we present malicious queries from the DET tool, the average confidence is 99.9\% and 70\% for correct and incorrect predictions, respectively.
			
			To further evaluate the efficacy of Random Forest, we tested instances from the Iodine dataset (not used in model training). The model correctly detected only 0.001\% of Iodine instances (2543 out of 2.2M) as malicious. Therefore, we tuned three parameters of the model, namely the number of trees (1 to 200), the number of selected attributes for each tree (1 to 8), and depth of the tree (1 to 20). We used the same method described in IV and found the optimal values equal to 8, 5, and 9 respectively for the number of trees, the number of attributes, the depth. With these optimized parameters, the rate of correctly detecting Iodine queries as malicious improved to 54.8\%. However, the average confidence of the tuned model for correctly and incorrectly classified instances was 75\% and 52\% respectively -- the model confidence becomes lower than its performance for the trained DET instances. 
			
			Moreover, we presented the 17 samples of publicly reported malicious DNS queries (listed in Table~\ref{table:knownDNSExfil}) to the tuned Random Forest model and found that 13 out of 17 instances were misclassified. This again proves that multi-class classifiers display poor resilience to morphed attacks that deviate from known attack signatures.}
		%Then we checked the performance of our trained machine and the accuracy was 100\% for trained instances and 99.99\% for the testing instances. The results show that it performed very well for the known traffic, however, when we tested it with the publicly reported 17 samples of DNS queries from known malware given in \ref{table:knownDNSExfil}, all of the instances were classified as normal instead of malicious (i.e, classified as malicious by our proposed system). Furthermore, we have also tested this approach with the additional malicious dataset captured in the previous subsection, the result shows that the random forest classifier has the accuracy of 0.0004\% of correctly detecting bad as bad (i.e, more than 99\% times it is classifying malicious as normal). This shows that the existing two-class classification approach is not suitable for finding the novel exfiltration attacks. 

		{\subsection{Malicious DNS Queries from Enterprise Networks}
			We now look at real DNS queries of the two enterprises that were detected as anomalies in \S\ref{sec:performance}. Focusing on ``Others" in Tables~\ref{table:perfFilteringRsch} and \ref{table:perfFilteringUniv} (\ie domains that are not among top 10K Majestic ranking list), we see about $12$\% and $18$\% of instances in the Research and University networks respectively that are flagged as anomalous DNS queries.} We have further analyzed these instances to gain insights into their primary domains. We found that only a few primary domains contribute more than $80$\% of anomalous instances in both organizations. Table~\ref{tab:wild} lists these top domains. Interestingly, domain ``{\fontsize{10}{48}\usefont{OT1}{lmtt}{b}{n}\noindent imrworldwide.com}'' is seen in both the Research and University networks -- this domain is known spyware that tracks web activity of victim hosts \cite{imr}. We also see ``{\fontsize{10}{48}\usefont{OT1}{lmtt}{b}{n}\noindent adlooxtracking.com}'' the third top domain detected in the University network which is a notorious domain that redirects web users to phishing/unsafe webpages, resulting in freeware downloads \cite{adloox}. In addition, our model detects suspicious domains ``{\fontsize{10}{48}\usefont{OT1}{lmtt}{b}{n}\noindent 2o7.net}'', ``{\fontsize{10}{48}\usefont{OT1}{lmtt}{b}{n}\noindent cnr.io}'', ``{\fontsize{10}{48}\usefont{OT1}{lmtt}{b}{n}\noindent 1rx.io}'', and ``{\fontsize{10}{48}\usefont{OT1}{lmtt}{b}{n}\noindent 360.cn}'' with a fairly significant number of queries. However, we cannot verify that they are malware or spyware -- this may need further investigations into end-hosts that generate these queries.

		{ \textbf{Insights:} }
		To better understand these DNS queries detected as malicious, we have further analyzed their corresponding DNS responses -- note that DNS responses are exclusively used in this section for drawing further insights into anomalous queries. As mentioned above, Table~\ref{tab:wild} lists the top malicious primary domains along with their statistics, including the daily average number of DNS queries generated for each domain, the daily average number of enterprise hosts querying for each domain, distinct DNS types with their distribution, and the average of TTL values (specified in their corresponding response). %Note that the average is taken from the whole week of DNS dataset (6th Aug to 12th Aug 2018).  
		For the research institute, we can see that ``{\fontsize{10}{48}\usefont{OT1}{lmtt}{b}{n}\noindent imrworldwide.com}'' is queried more than 20,000 times a day (on average) and only 12 unique hosts (\ie IP addresses) make these queries. Analyzing these IP addresses, we found (by reverse lookup) that five of them are recursive resolvers of the research institute -- having recursive resolvers as querying hosts is also observed for other anomalous domains. Focusing on seven hosts which are regular clients, four were found actively making anomalous queries on four days, while the other three hosts do not display malicious behavior over the rest of the week of our analysis (though they are present on the network). This observation suggests that those four regular hosts are possibly infected by malware or spyware. 
		Moreover, we found that of those four regular clients, three generated queries to all top malicious domains, except ``{\fontsize{10}{48}\usefont{OT1}{lmtt}{b}{n}\noindent cnr.io}'' over the entire week in the research network.
		Consistently generating anomalous queries over the week is seen in two regular clients for ``{\fontsize{10}{48}\usefont{OT1}{lmtt}{b}{n}\noindent 1rx.io}'', and one regular client for ``{\fontsize{10}{48}\usefont{OT1}{lmtt}{b}{n}\noindent 2o7.net}''. 
		
		%Considering a consistent behaviour for anomalous domains five regular clients that query for ``{\fontsize{10}{48}\usefont{OT1}{lmtt}{b}{n}\noindent 1rx.io}'', two of them are present over the entire week. Also, for {\fontsize{10}{48}\usefont{OT1}{lmtt}{b}{n}\noindent 2o7.net}, one out of three regular hosts are making queries to this domain for the whole week.
		Similarly, for the university campus, we see on average 80,000 daily queries for ``{\fontsize{10}{48}\usefont{OT1}{lmtt}{b}{n}\noindent imrworldwide.com}'' from an average of 203 unique enterprise hosts. By reverse lookup of host IP addresses, we found that seven of them are recursive resolvers and the remaining 196 hosts are regular clients -- 150 of these clients consistently send queries for ``{\fontsize{10}{48}\usefont{OT1}{lmtt}{b}{n}\noindent imrworldwide.com}'' during the entire week. Interestingly 130 of them fall under one subnet of size /24.
		%Out of these regular clients, 150 clients are making queries for the whole week. By looking at the IP addresses, we found out that 130 clients are from the same subnet.  
		Furthermore, we found that a total of 290 university hosts generate queries for at least one of the top three malicious domains (\ie ``{\fontsize{10}{48}\usefont{OT1}{lmtt}{b}{n}\noindent imrworldwide.com}'' or ``{\fontsize{10}{48}\usefont{OT1}{lmtt}{b}{n}\noindent adlooxtracking.com}'' or ``{\fontsize{10}{48}\usefont{OT1}{lmtt}{b}{n}\noindent 360.cn}'') -- of these hosts, 35 make queries for all of these top three malicious domains. By reverse lookup we discovered that 6 of them are recursive resolvers and the remaining 29 hosts are from the same subnet of size /24, indicating that this particular subnet might be infected by malware or spyware.
		
		% ({\color{red} what is the relation between 203 and 290?}) {\color{blue} 203 is the average number of enterprise hosts making queries to imrworldwide and this 290 is the total number of unique enterprise hosts making queries to imrworldwide or adloox or 360.cn domains.} unique hosts, 35 regular clients are making queries to all the malicious domains and all of them are from the same subnet 
		
		%This means that there are probably more internal hosts making those queries at first place. Similarly, for {\fontsize{10}{48}\usefont{OT1}{lmtt}{b}{n}\noindent cnr.io} domain, there are 4 enterprise hosts making more than 17K daily average number of queries. By reverse lookup we came to know that all those four hosts are recursive resolvers. Same is the case with remaining two domains of research institute \ie for {\fontsize{10}{48}\usefont{OT1}{lmtt}{b}{n}\noindent 1rx.io} 5 out of 10 domains are recursive resolvers and for {\fontsize{10}{48}\usefont{OT1}{lmtt}{b}{n}\noindent 2o7.net} 5 out of 8 domains are recursive resolvers.
		
		We further investigated the type field in DNS queries for these frequent malicious domains. ``{\fontsize{10}{48}\usefont{OT1}{lmtt}{b}{n}\noindent A}''-type and ``{\fontsize{10}{48}\usefont{OT1}{lmtt}{b}{n}\noindent AAAA}''-type records map domains to IPv4 and IPv6 addresses respectively. Our first observation by looking at distinct types of anomalous queries in Table~\ref{tab:wild} is that there is a much greater percentage use of IPv6 in the Research Institute than in the University network. Secondly, we observe that ``{\fontsize{10}{48}\usefont{OT1}{lmtt}{b}{n}\noindent TXT}'' strongly dominates the type of DNS queries for ``{\fontsize{10}{48}\usefont{OT1}{lmtt}{b}{n}\noindent cnr.io}'' in the Research Institute which clearly indicates a data exfiltration/tunneling over DNS \cite{iotbd16}. Note that sophisticated attackers tend to use other types (\ie``{\fontsize{10}{48}\usefont{OT1}{lmtt}{b}{n}\noindent A}'', ``{\fontsize{10}{48}\usefont{OT1}{lmtt}{b}{n}\noindent AAAA}'', ``{\fontsize{10}{48}\usefont{OT1}{lmtt}{b}{n}\noindent CNAME}'', ``{\fontsize{10}{48}\usefont{OT1}{lmtt}{b}{n}\noindent NS}'' and ``{\fontsize{10}{48}\usefont{OT1}{lmtt}{b}{n}\noindent MX}'' instead of ``{\fontsize{10}{48}\usefont{OT1}{lmtt}{b}{n}\noindent TXT}") to hide their malicious activities over DNS.

		DNS responses contain a Time-To-Live (TTL) field in seconds, indicating the duration of a DNS resource record to be cached on the host machine. According to RFC 1033 \cite{rfc1033}, it is important to set an appropriate TTL value since very low values result in overloading the DNS server, and very high values may limit the flexibility of changing resource records in real-time. According to RFC 1912 \cite{rfc1912}, it is recommended to set the TTL value between one to five days. But, CDN (Content Distribution Network) services tend to use smaller TTLs for fast reaction to dynamic resource changes. Unfortunately, malicious entities also use small TTLs for minimizing their footprints and becoming more resistant against DNS blacklisting \cite{bilge2011exposure}. In Table~\ref{tab:wild}, we compute the average TTL for each of the top malicious domains. We observe that malicious domains use relatively smaller TTLs (\ie less than an hour), for example it is set to 0 in all of the DNS responses for ``{\fontsize{10}{48}\usefont{OT1}{lmtt}{b}{n}\noindent cnr.io}''. Another example is ``{\fontsize{10}{48}\usefont{OT1}{lmtt}{b}{n}\noindent 1rx.io}'' for which the average TTL is 45 seconds.

		%Upon further investigation, we observed that some enterprise hosts are also making queries to {\fontsize{10}{48}\usefont{OT1}{lmtt}{b}{n}\noindent imrworldwide.com}. Out of 25 distinct hosts, we found out that ten enterprise hosts are making queries to this domain for four days out of the whole week.
		
		%in both the networks, interestingly it behaves quite different for example, two-third of the queries are of type A in research institute whereas 98$\%$ of the DNS queries are of type A. {\color{red} You are supposed to draw insights here. what is the meaning of these values? why is it important?} Also, the average TTL value its 167K sec in research and only 3K sec in the university campus. 
		%Another interesting finding is the TTL value of {\fontsize{10}{48}\usefont{OT1}{lmtt}{b}{n}\noindent cnr.io} which is always 0 sec. Similarly, average TTL value of {\fontsize{10}{48}\usefont{OT1}{lmtt}{b}{n}\noindent 1rx.io} is 45 sec which is also less than a minute. We have found out in the literature that the attacker uses a small value of TTL \cite{bilge2011exposure}. 
		
		%Moreover, we look into the number of enterprise hosts that generate these malicious DNS queries. %and we have found out that a very small number of hosts are sending queries to these domains and by digging more into the analysis, 
		%We found that more than 70\% of the hosts {\color{red}are the same in both the organizations!!!} making queries to all these malicious domains. {\color{red} HASSAN: where do you get this 70\%? how does this relate to column 3 in Table X?}

		\begin{figure*}[t!]
			\begin{center}
				% \vspace{-3mm}
				\mbox{
					% \hspace{-3mm}
					\subfloat[Research institute.]{
						{\includegraphics[width=0.85\textwidth,height=0.49\textwidth]{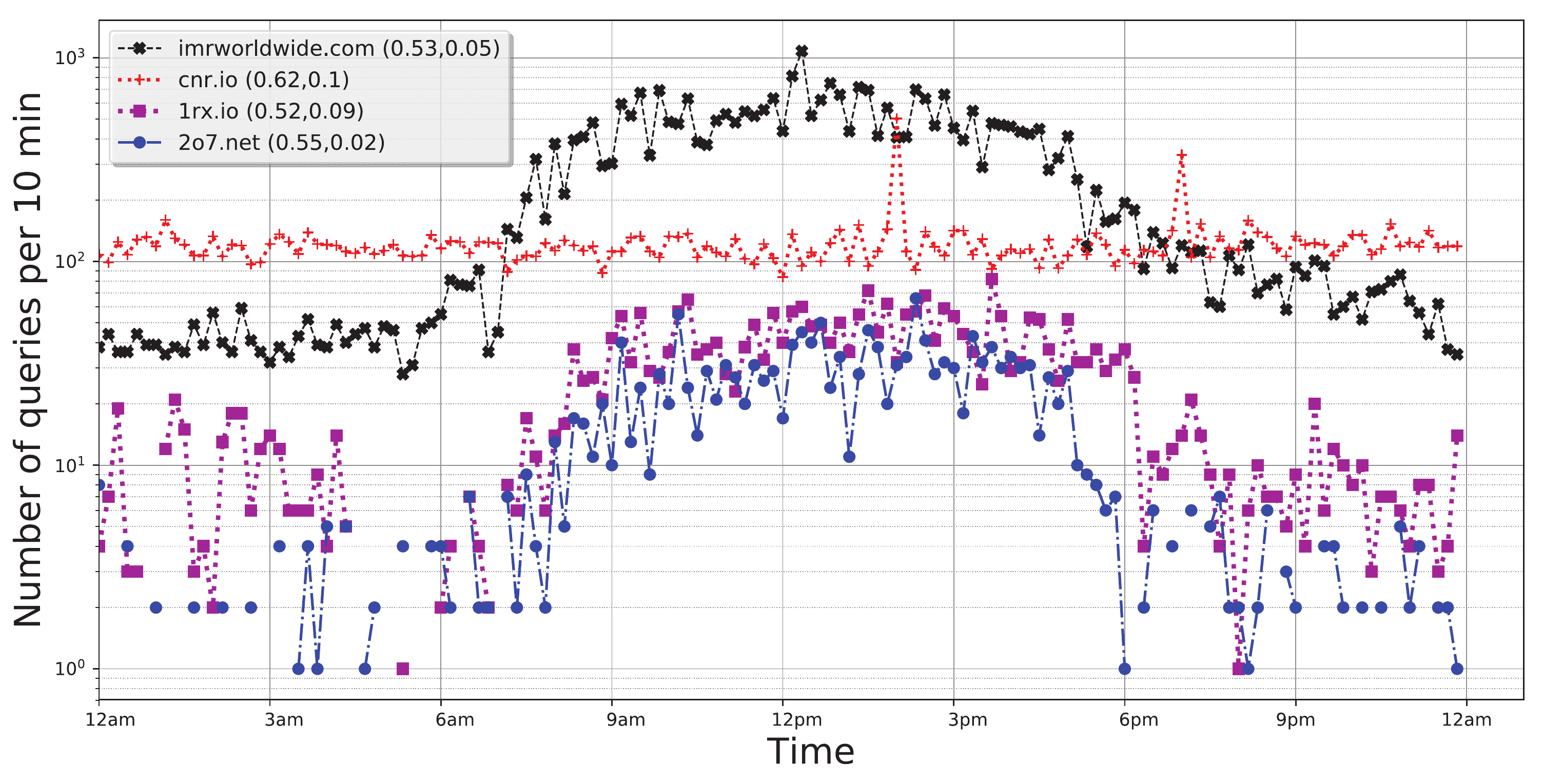}}\quad
						\label{fig:RealMaliciousCSIRO}
					}
				}
				\mbox{
					
					%\hspace{-5mm}
					\subfloat[University campus.]{
						{\includegraphics[width=0.85\textwidth,height=0.49\textwidth]{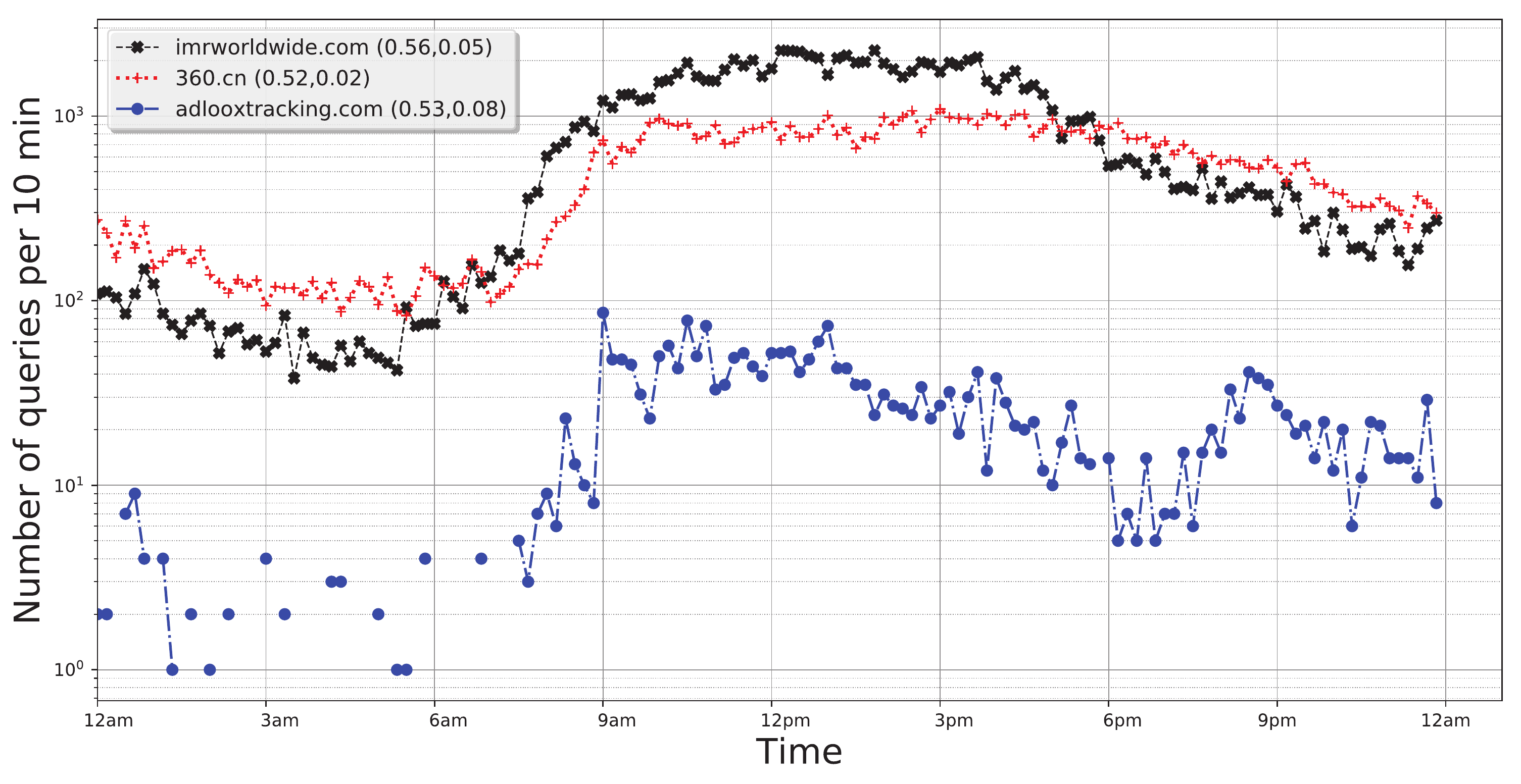}}\quad
						\label{fig:RealMaliciousUNSW}
					}
				}
				\vspace{-2mm}
				\caption{Number of DNS queries for top malicious domains over a day.}
				\label{fig:RealMalicious}
			\end{center}
		\end{figure*}

		We plot in Fig.~\ref{fig:RealMalicious} the query count (computed every 10 minutes) of top anomalous primary domains on day 6-Aug-2018, as an example -- missing points in this figure correspond to zero query count over those 10-min epochs. Note that the mean and standard deviation of the anomaly score is shown next to each domain name in the legend. Our first observation is that the query count for all primary domains %(\eg {\fontsize{10}{48}\usefont{OT1}{lmtt}{b}{n}\noindent imrworldwide.com} shown by dashed black lines and cross markers) 
		is higher during working hours (\ie increasing an order of magnitude at about 8 am, staying at a certain level, and falling back at about 5 pm), though the primary domain ``{\fontsize{10}{48}\usefont{OT1}{lmtt}{b}{n}\noindent cnr.io}'' in Fig.~\ref{fig:RealMaliciousCSIRO} displays a fairly consistent pattern of query count over a day (except one spike at around 2 pm). 
		%{\fontsize{10}{48}\usefont{OT1}{lmtt}{b}{n}\noindent imrworldwide.com} domain can be observed as a dominant one with respect to the other malicious domains with a peak value of more than 1000 DNS queries per 10 minutes interval at around 1 pm, whereas {\fontsize{10}{48}\usefont{OT1}{lmtt}{b}{n}\noindent cnr.io} domain can be seen as continuously occurring with more or less same number of DNS queries for whole day. {\color{red} the message of this paragraph is not so clear.} {\color{blue} Purpose of this plot and paragraph is to give readers an idea that how frequent these domains have been queried in our enterprise networks and are they following any pattern?}
		
		% As we have already mentioned above that  {\fontsize{10}{48}\usefont{OT1}{lmtt}{b}{n}\noindent cnr.io} domain is only queried by recursive resolvers and its pattern is also quite similar for overall day whereas {\fontsize{10}{48}\usefont{OT1}{lmtt}{b}{n}\noindent imrworldwide.com} and {\fontsize{10}{48}\usefont{OT1}{lmtt}{b}{n}\noindent 360.cn} domains are queried from the most number of end users/enterprise hosts comparing to the other malicious domains so they are active in the working hours of both the enterprise networks.
		Finally, looking at the anomaly score of queries for these selected malicious domains (as shown in the legend of Fig. \ref{fig:RealMalicious}), the primary domain {\fontsize{10}{48}\usefont{OT1}{lmtt}{b}{n}\noindent cnr.io} has the largest mean value $0.62$ (the closer to $1$ means more anomalous). The average score for other malicious domains in both networks varies between $0.52$ to $0.56$ which is well above the average score for benign instances (\ie less than $0.40$) reported in Table \ref{table:anomalyRsch} and \ref{table:anomalyUniv}. We acknowledge that the anomaly score is just an indicator, that can be used to flag suspicious domains. The network operators can then perform further investigation into these identified domains.

		\section{Real-Time Deployment in Campus Network}
		
		This section presents our web-tool to access and visualize real-time detection of malicious DNS queries for an enterprise network of a large university campus in Sydney, Australia. We showcase how to visualize our real-time learning-based detection engine operational on 10 Gbps traffic streams from the network border of the university campus.

		\begin{figure}[t!]
			\centering
			\includegraphics[width=0.85\textwidth,height=0.4\textwidth]{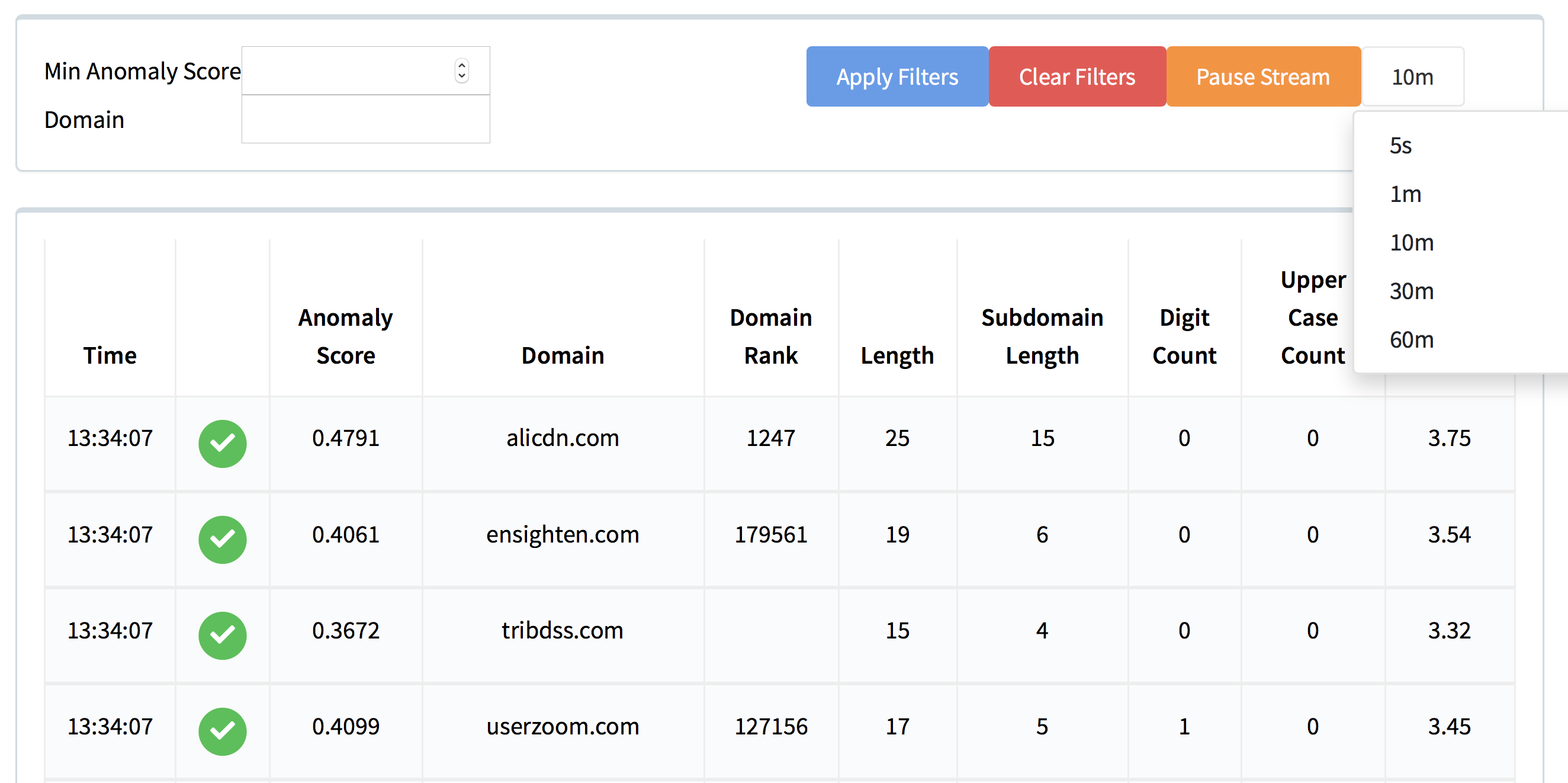}
			\caption{Web-UI of our real-time DNS exfiltration and tunneling detector.}
			\label{fig:arch1}
		\end{figure}

		\begin{figure}[t!]
			\centering
			\includegraphics[width=0.85\textwidth,height=0.4\textwidth]{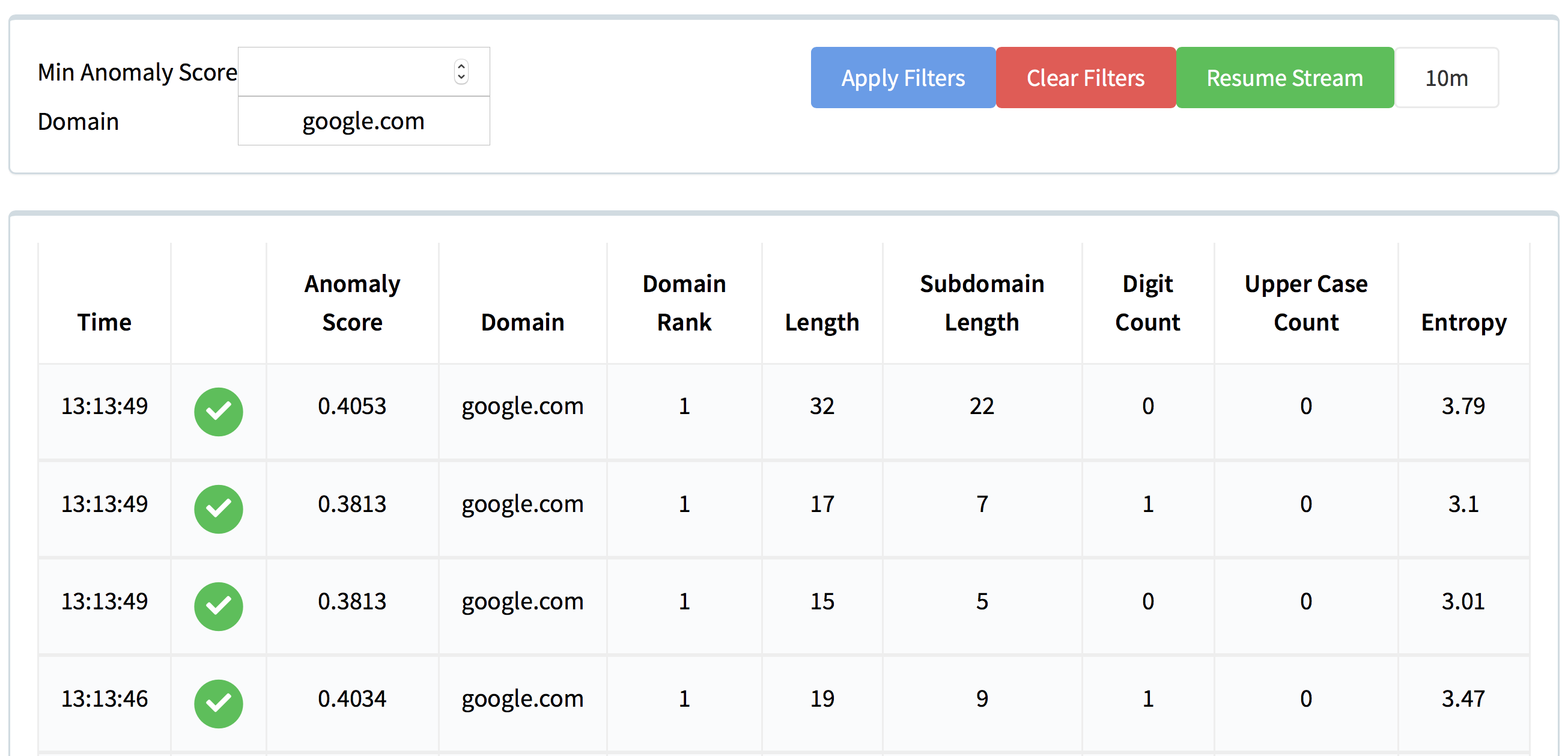}
			\caption{Filtering queries for a specific primary domain name.}
			\label{fig:arch2}
		\end{figure}

		\begin{figure}[t!]
			\centering
			\includegraphics[width=0.85\textwidth,height=0.4\textwidth]{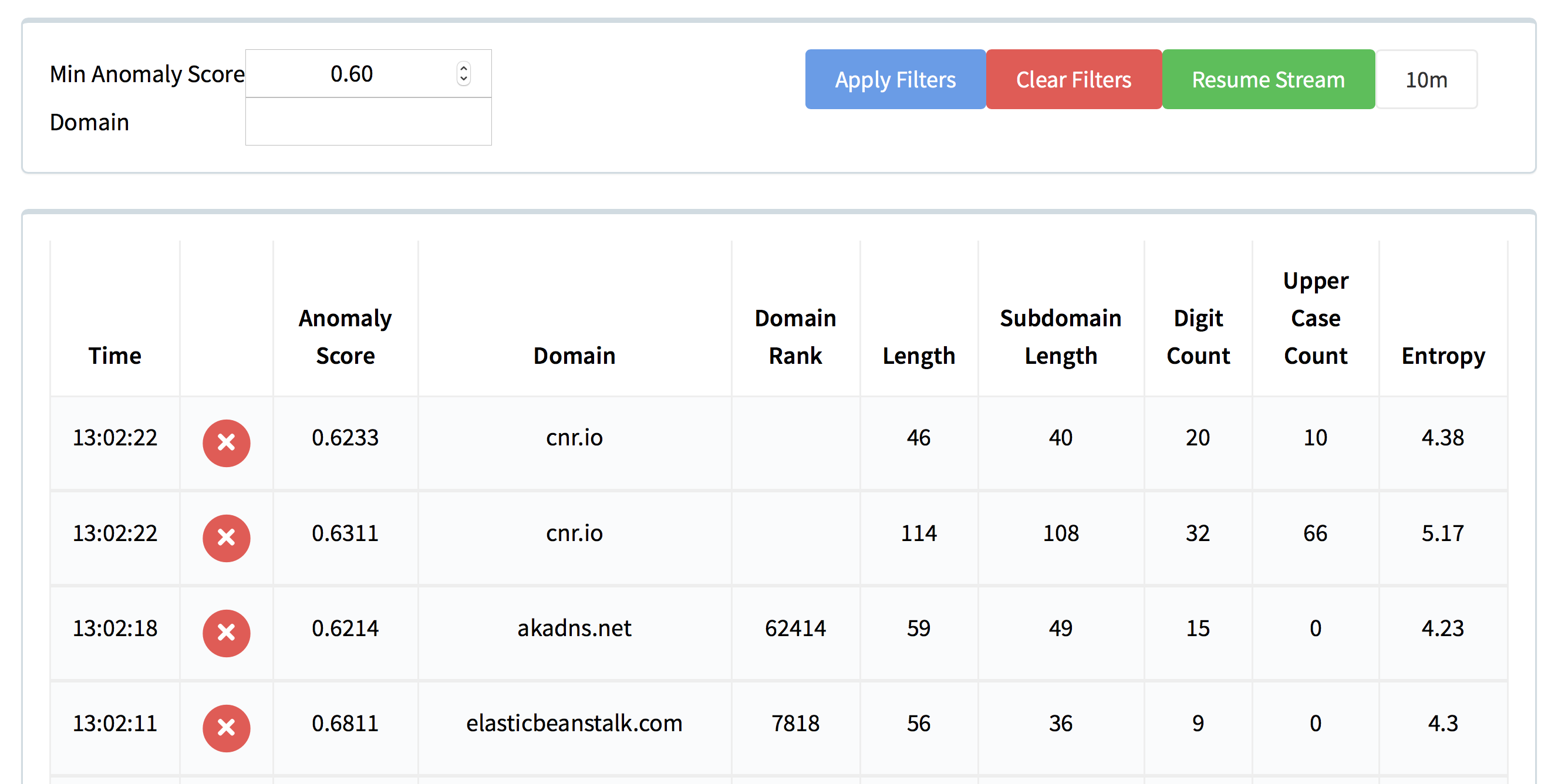}
			% \vspace{-3mm}
			\caption{Filtering queries with a minimum anomaly score.}
			%\vspace{-4mm}
			\label{fig:arch3}
		\end{figure}
		
		%We developed a tool to provide an intuitive user-interface for real-time malicious DNS queries going out from an enterprise network, using ReactJS, a JavaScript framework. Our web-tool is publicly available at \cite{ourData}.
		\vspace{-3mm}
		\subsection{Visualizing Detection of Malicious DNS Queries}
		We developed a tool to provide an intuitive user-interface for real-time monitoring of outgoing DNS queries in enterprise networks using ReactJS. Our web-tool is publicly available at \cite{ourData}. 
		
		This tool allows the user to visualize timestamped queries in real-time at various time scales, including last 5 sec, 1 min, 10 min, 30 min, and 60 min, as shown in Fig.~ \ref{fig:arch1}. It also shows selected attributes for each query and the output of our model. 
		%The tool gives the user an opportunity to look at the malicious domains within the range of last n seconds or minutes where n provides the different level of time granularity like 5sec, 1, 10, 30 and 60 mins in real-time as shown in Figure \ref{fig:arch1}. 
		Records marked by a red cross correspond to DNS queries detected as malicious by our machine learning model, while green ticks mark benign queries -- for privacy reasons, we do not show the FQDNs of live traffic in our web-tool.
		The ``'domain rank'' shows the reputation of the primary domain obtained from Majestic list \cite{jawad2019DNS} -- if a primary domain is not found in top million Majestic list then its rank is shown empty, for example {\fontsize{10}{48}\usefont{OT1}{lmtt}{b}{n}\noindent tribdss.com} in Fig.~\ref{fig:arch1}. 
		
		%Some of the primary domains are not ranked as we are using majestic million file for ranking of the primary domain. Therefore, if the primary domain is not in top million list than the rank column field does not contain any number.
		
		The user can enter a specific primary domain name, filtering all records (in real-time) associated with that primary domain. For example, in Fig.~\ref{fig:arch2} the user has selected {\fontsize{10}{48}\usefont{OT1}{lmtt}{b}{n}\noindent google.com}, the tool is showing the results specific to this filtered domain. Furthermore, our tool is capable of filtering query records based on anomaly score (computed by our model), as shown in Fig.~\ref{fig:arch3} which shows all queries with the anomaly score of 0.60 or more  -- anomaly score varies from 0 to 1, where 0 is least anomalous and 1 is the most anomalous. 
		This value is calculated by the machine learning algorithm used while classifying a query name. Additionally, our tool has a function to pause/resume the stream. If the user notices a malicious domain in real-time, then (s)he can pause the visualization stream to analyze the attributes of the malicious domain. %Overall, the benefit of using our tool is to visualize the outgoing malicious DNS queries from an enterprise network border router in real-time. 

		\section{Conclusion}\label{conclusion}
		Enterprise networks are potential targets of cyber-attackers for stealing valuable and sensitive data over DNS channels.  We have developed and validated a mechanism for real-time detection of DNS exfiltration and tunneling from enterprise networks. By analyzing DNS traffic from two organizations, we have identified attributes of DNS query names that can be extracted efficiently in real-time distinguishing legitimate from malicious queries. We then developed, tuned, and trained a machine-learning algorithm to detect anomalies in DNS queries using a known dataset of benign domains as ground truth. 
		Lastly, we evaluated the efficacy of our scheme on live 10 Gbps traffic streams from the borders of two enterprise campus networks by injecting more than three million malicious DNS queries via DET and Iodine tools -- our tools and datasets are publicly available. We showed that our solution outperforms the two-class classifier in detecting new malicious DNS queries.
		We have drawn insights into anomalous DNS queries by their anomaly scores, the trace of query count over time, enterprise hosts querying them, and TTL and Type fields of their corresponding responses.
		Lastly, we demonstrated our web tool for visualizing our learning based on the detection of anomalous DNS queries (DNS exfiltration and tunneling from enterprise networks) operational in our campus network.

		%This chapter dealt with the threat of sensitive information theft that is facing by the enterprise networks due to least security checks on DNS protocol by the firewalls.  We have analyzed the DNS data of one week from two organizations and extracted the stateless attributes. We gave the insights of the attributes by showing the distribution of several attribute for both of the enterprise networks. We then trained the machine with the attributes computed from the top ranked benign primary domains and tuned the machine to detect the anomalous DNS queries going out of the enterprise. We validated our solution by implementing it on the live traffic of 10 Gbps and injected more than a million malicious DNS queries. Our results showed that our system was highly accurate in identifying these malicious DNS queries. Furthermore, our trained machine has the capability to detect 1.6K packets per second, whereas we have the peak point of 800 pps over a week as shown in Table \ref{table:featureExtractionTestingTime}. Our tools and datasets are publicly available for validation and further research.

		\chapter[Automatic Detection of DGA-Enabled Malware ]{Automatic Detection of DGA-Enabled Malware Using SDN and Traffic Behavioral Modeling}
		\label{chap:ch4}
		\minitoc
		
		In the previous chapter, we analyzed outgoing DNS queries to tackle the data exfiltration over DNS. 
		This chapter combines Software Defined Networking (SDN) and machine learning to develop an accurate, cost-effective, and scalable system for detecting infected hosts communicating with external C\&C servers, subsequent to the resolution of DGA query names. Our solution dynamically selects network flows for diagnosis by trained models in real-time and relies more on the behavioral traffic profile than packet content. Parts of this chapter are published in IEEE Transactions on Network Science and Engineering (TNSE) \cite{9772333}.
		
		Our contributions are threefold: (1) We analyze full DNS traffic collected from the border of our university campus network (2.4B records for 75 days) to highlight the prevalence and activity pattern of more than twenty families of DGA-enabled malware across internal hosts. We draw insights into the behavioral profile of DGA-enabled malware flows when communicating with C\&C servers by analyzing a Packet Capture (PCAP) trace (3.2B packets) collected during the peak hour from our campus network;
		(2) We identify malware traffic attributes and train three specialized one-class classifier models using behavioral attributes of malicious HTTP, HTTPS, and UDP flows obtained from a public dataset. We develop a monitoring system that uses SDN reactive rules to automatically and selectively mirror TCP/UDP flows pertinent to DGA queries (between internal hosts and malware servers) for diagnosis by the trained models; and (3) We evaluate the efficacy of our approach by testing suspicious traffic flows (selectively recorded by SDN reactive rules over a 50-day trial) against our trained models, identify infected hosts from suspicious flows, and verify our detection with an off-the-shelf Intrusion Detection System (IDS) software tool. 
		
		\section{Introduction}
		
		% risk and scale of cyber-attacks
		Cyber threats and data breaches continue to increase in both frequency and complexity, placing businesses and individuals at constant risk. 
		According to Cybersecurity Ventures \cite{CyVentures20}, cybercrime damages will cost the world \$6 trillion annually by 2021. Enterprises, small and large, remain among the top lucrative targets of automated attacks \cite{Accenture2019,smallBizRisk20}. 
		% state of enterprise networks
		Enterprise networks are often complex, with applications that rely on a mix of local and cloud-based services, and hence difficult to manage securely \cite{homelandSecReport2019}. Enterprise hosts often include powerful servers, personal computing devices, mobile phones, and unmanaged Internet of Things (IoTs). These devices may use a mixture of statically or dynamically assigned addresses from several public and private Internet Protocol (IP) address ranges. 
		Poorly administrated assets, like personal computers or unpatched servers \cite{unpatched2017}, are not only potential victims of cyber-attacks but are also sources of risk for other entities on the Internet. For example, hosts sitting behind the enterprise border firewall can be infected by malware from phishing emails, security holes in browser plugins, or other infected local devices. 
		
		Malware-infected machines forming a botnet are typically managed remotely by an adversary (aka botmaster) via a C\&C channel. Cyber-criminals primarily use a botnet for malicious activities such as stealing sensitive information, disseminating spam, or launching denial-of-service attacks. Therefore, law enforcement agencies routinely perform takedown operations on the blacklisted C\&C servers \cite{MIT2019}, disrupting their botnet activities. In response to these efforts, botmasters have developed innovative approaches to protect their infrastructure. The use of DGAs is one of the most effective techniques that has gained increasing popularity \cite{antonakakis2012throw}.
		
		DGAs use a ``seed'' (a random number accessible to both the botmaster and the malware agent on infected hosts) to generate a large number of custom domain names. Generating numerous time-dependent domain names and registering only the relevant one(s) ``just shortly'' before an attack allows a botnet to shift their C\&C domains on the fly and remain invisible for longer \cite{plohmann2016comprehensive}.
		The botmaster waits for the malware to successfully resolve a Domain Name System (DNS) query for the registered domain, enabling the C\&C communications to take place. Note that even if a C\&C server is taken offline or blacklisted, this process can simply be restarted, and a new server can come online. To date, more than 80 collections of DGA domains (each corresponding to a malware family) have been recorded by DGArchive \cite{dgarchive} and are publicly available. 
		
		There exist a number of research works \cite{hao2010internet,bilge2011exposure,reputationDNS2010,initDNSBehaviour2011, jawad2019DNS,PAM19, antonakakis2011detecting} that analyze DNS traces to identify malicious activities, detecting C\&C servers, infected hosts, or malicious domains. However, their proposed methods largely require the extraction of information from DNS packets, correlating queries and responses, and maintaining many states over a reasonably long duration. All of these processing steps collectively demand heavy compute resources and hence make it difficult to scale cost-effectively. On the other hand, existing firewalls and intrusion detection systems rely primarily on inspecting every packet traversing the network, which makes them expensive. Further, correlating malicious DNS queries with subsequent C\&C communication flows would significantly impact their inferencing accuracy and cost.

		{
			In this chapter, we employ the SDN paradigm to judiciously combine selective packet inspection (only DNS proactively) and flow behavioral analysis (reactively) to intelligently detect malware-infected hosts on the network. 
			The novelty of this chapter arises from the dynamic inter-relation of SDN and machine learning technologies for a sophisticated yet cost-effective cyber-security solution. Commodity SDN switches today can forward data very cost-effectively at Terabits-per-second. When combined with intelligent machine learning algorithms in software, it provides the flexibility and agility to deal with existing and emerging threats.
			{This is ideal for dynamically selecting flows (after malicious DNS queries), and inferring their behavioral health using trained models.} We use public data of malware families to develop our machine learning models.
			
			Our \textbf{first} contribution highlights the prevalence and activity pattern of more than twenty DGA-enabled malware families on internal hosts of a university campus network by analyzing entire DNS traffic (consisting of 2.4B records) collected over 75 days from the network border (outside of the firewall). We also analyze a PCAP trace of full campus Internet traffic (collected during the peak hour with a total load of about 10Gbps) to draw insights into the behavioral pattern of DGA-enabled malware flows.
			For our \textbf{second} contribution, we identify key traffic attributes of malware, and train one-class classifier specialized models by attributes of malicious HTTP, HTTPS, and UDP flows obtained from a public dataset. We then develop a monitoring system that uses SDN reactive rules to automatically and selectively mirror TCP/UDP flows (between enterprise hosts and malware servers pertinent to DGA queries) for making inferences by the trained models. 
			\textbf{Finally}, we evaluate the efficacy of our proposed model by testing suspicious traffic flows (mirrored by SDN reactive rules and recorded over a 50-day trial) against our trained models, identify infected hosts from suspicious flows, and verify our detection with an off-the-shelf IDS software tool.  Also, we compare the performance of our one-class and multi-class models.
		}

		\section[Analyzing Network Traffic Data]{Analyzing Network Traffic Data: Prevalence of DGA-Enabled Query Names and Network Behavior of Malware}
		\label{sec:Insights}
		In this section, we begin by analyzing the DNS traffic of a campus network to demonstrate the prevalence of DGA-enabled domain names (obtained from a public dataset) which are found in DNS queries of internal hosts. We, then, analyze a one-hour PCAP trace of the entire campus traffic (in/out) to understand the network behaviors of internal hosts when they communicate with Internet-based servers following their DGA-related DNS query.

		\subsection{Our Datasets} \label{subsec:ourData}
		In this work, we use four different datasets including (a) 75 daily DNS PCAPs collected from the border of a university campus network, (b) a one-hour PCAP trace of the entire traffic of the university campus network to/from the Internet, (c) 82 archived files containing more than 65 million domain names used by DGA-related malware families, and (d) { public network traces (PCAPs and NetFlow records) of known malware and benign traffic}.
		
		%\subsubsection{PCAP Traces of Campus Network DNS Traffic}
		
		\textbf{PCAP Traces of DNS Traffic:}
		We collected daily DNS PCAP traces from the border of the University campus network. Each PCAP has a size of about 15 GB on average. The IT department of the campus network provisioned a full mirror (both inbound and outbound) of its Internet traffic (on a 10 Gbps interface each) to our data collection system from its border router (outside of the firewall). We obtained appropriate ethics clearance (Human Research Ethics Advisory Panel approval number HC17499) for this study. % (approval number will be disclosed when this chapter is de-anonymized). % 
		We extracted DNS packets from each enterprise Internet traffic stream in real-time by configuring rules to match incoming/outgoing IPv4 and IPv6 UDP packets on port 53 in an OpenFlow switch. This work analyzes data collected over 75 days from 16-Sept-2019 to 1-Dec-2019. %We have the full dump of DNS packets every day with a separate PCAP file. 
		Our detailed analysis is described later in this section (\S\ref{subsec:insightsMalwareFamilies}, \S\ref{subsec:insightsDailyPattern}, and \S\ref{subsec:insightsInfectedHosts}).
		
		{
			\textbf{One-Hour PCAP Trace of Full Campus Traffic:}
			In addition to DNS packet traces, we recorded all incoming/outgoing packets (only the first 96 bytes of each packet) of the campus network using {\myverb{tcpdump}} tool during the peak hour (2pm-3pm) of a weekday on 31st May 2019.  This PCAP trace, with over 250 GB, consisting of 3.2B packets, represents traffic of large-scale enterprise networks. {Fig.~\ref{fig:Aggload} shows the aggregate load (packet rate and bit rate, moving averaged over 30-sec intervals) on the Internet link of the campus network.} It can be seen that on average more than a million packets per second are exchanged between internal hosts and the Internet, resulting in an aggregate load of 10 Gbps. %These number shows that it is costly to keep all the campus data and to analyze the data is very difficult to even for an hour
			We will use this relatively large dataset in \S\ref{subsec:1hourPCAP} to highlight the behavioral profile of malware flows (needles in the haystack of enterprise network traffic) pertinent to DGA queries.
		}
		
		%428716085
		
		%\subsubsection{DGArchive}
		\textbf{DGArchive:}
		A group of researchers conducted an extensive study \cite{plohmann2016comprehensive} on several families of DGA-based malware. Authors made their dataset (``DGArchive'') available to the public \cite{dgarchive}, and they have been consistently expanding their database over time. We have used the latest version of the database (uploaded on 7th Jan 2019), which contains domain names from 86 DGA families. We excluded four families for our study because they overlap with legitimate domains like {\myverb{github.com}}, {\myverb{itunes.com}}, or {\myverb{doodle.com}}. Such overlaps are seen because their algorithms are restricted to generate only six-letter strings that can lead to the generation of existing legitimate domains. Note that datasets of these four families contain more than 20 million records of domain names, and hence it was infeasible to check all domains, filtering out the legitimate ones manually. Therefore, we used the files corresponding to the remaining 82 malware families and developed our database of about 83 Million records of domain names in total (65 Million unique records). Note that some families share several records with other families -- the details of the overlap across various families can be found in \cite{plohmann2016comprehensive}. 
		%{\color{magenta} Another approach is to examine the registration date of the domain name when they get registered to find out whether they are malicious or benign. We believe that it is quite challenging in “real-time” prior to commencement of TCP/UDP flow and the associated delay could result in lost malicious flow. In addition, diverting “selected flows” which are often relatively short and small volume would not incur {a }significant cost of computing and/or TCAM entry.}
		We will use the DGArchive dataset in \S\ref{subsec:insightsMalwareFamilies}, \S\ref{subsec:insightsDailyPattern}, and \S\ref{subsec:insightsInfectedHosts}. We acknowledge that our detection method primarily relies upon the knowledge-base of DGArchive and hence may miss some ``novel'' malicious query names that are not captured by this database. Note that this limitation is inherent to any signature-based detection method (ours included). This public database is actively updated at a frequency of weeks to months. Therefore, in practice, one can check this public repository daily or weekly to obtain the latest signatures (domain names used by latest malware families).
		
		{
			\textbf{Network Traces of Known Malware and Benign Traffic:}
			Authors of \cite{garcia2014empirical} released a public dataset called ``CTU-13'' that contains packet traces of malicious traffic as well as labeled NetFlow records of benign traffic.
			Malicious traffic traces consist of 13 PCAPs (76.8M packets) from network activities of seven real botnets including Menti \cite{menti}, Murlo \cite{murlo},  Neris \cite{neris}, NSIS.ay \cite{nsis}, Rbot \cite{rbot}, Sogou \cite{sogou}, and Virut \cite{virut} on Windows operating systems --  executable binary files of these malware were installed on lab computers \cite{ctu-13} at the CTU University, Czech Republic, in 2011. %{ It also contains labeled NetFlow files, which contain normal flow characteristics as well. }%The goal of the dataset was to have a large capture of real botnet traffic along with normal traffic and background traffic. The CTU-13 dataset consists of thirteen PCAP captures (called scenarios) of different botnet samples. 
			%In total, the size of the PCAPs is 72 Gigabytes. 
			Benign traces contain 3.6M NetFlow records of normal traffic (matching certain conditions \cite{garcia2014empirical}) from a controlled and known set of computers on a testbed.
			We will use this dataset in \S\ref{sec:SystemDesign} to train our classifier models, validate, and test their performance in distinguishing malicious flows from benign ones.
			Our analysis is motivated by evidence \cite{NSDI2010malware} that Web-based ``reusable'' tools for remote command of malware are available for sale on the Internet. Also, malware writers may generate a large number of polymorphic variants of the same malware using executable code obfuscation techniques, however, these variants will ultimately display similar activity patterns when executed \cite{AiSec2014malware}.
		}
		
		\begin{table*} [t!]
			\centering
			\caption{DGA-related domain families found in the campus network (from 16-Sep-2019 to 1-Dec-2019). }
			%\vspace{-2mm}
			\label{tab:ListMalwares}
			\begin{adjustbox}{max width=0.98\textwidth}   
				\renewcommand{\arraystretch}{1.5}
				\begin{tabular}{|l|c|c|c|c|c|c|c|c|c|c|c|c|c|c|c|c|c|c|c|c|c|c|c|c|c|c|} %{\columnwidth}{cccccc}
					\hline      \textbf{Malware family}       &  \rotatebox{90}{ModPack}  & \rotatebox{90}{Tinba} & \rotatebox{90}{Gameover} & \rotatebox{90}{Nymaim} & \rotatebox{90}{Bamital} & \rotatebox{90}{Suppobox} & \rotatebox{90}{Ranbyus} & \rotatebox{90}{Matsnu} & \rotatebox{90}{Ramnit} & \rotatebox{90}{Cryptolocker} & \rotatebox{90}{Gozi} & \rotatebox{90}{Tsifiri} & \rotatebox{90}{Simda} & \rotatebox{90}{Urlzone} & \rotatebox{90}{Vawtrak} & \rotatebox{90}{Banjori} & \rotatebox{90}{Downloader} & \rotatebox{90}{Necurs} & \rotatebox{90}{Dyre} & \rotatebox{90}{Locky} & \rotatebox{90}{Bobax} & \rotatebox{90}{Ccleaner} & \rotatebox{90}{Pandabanker} & \rotatebox{90}{Mirai} & \rotatebox{90}{Murofet} & \rotatebox{90}{Dircrypt} \\ \hline
					{{\multirow{2}{*}{DNS queries count}}}  & 552,877 & 22,284 & 9,419 & 1,338 & 842 & 768 & 399 & 324 & 260 & 123 & 94 & 83 & 72 & 69 & 34 & 30 & 28 & 12 & 11 & 8 & 7 & 6 & 4 & 2 & 2 & 2  \\   \cline{2-27}
					& $93.85$\% & $3.78$\% & $1.60$\%& \multicolumn{23}{c|}{$0.77$\%}\\ \hline
					Unique domain names & 6 & 327 & 305 & 38 & 396 & 175 & 13 & 15 &  77 & 7 & 13 & 2 & 6 & 1 & 8 & 5 & 1 & 3 & 4 & 2 & 4 & 3 & 2 & 1 & 1 & 1 \\
					\hline
				\end{tabular}
			\end{adjustbox}
			%}
		\end{table*}
		
		\subsection{DGA-Fueled Malware Families} \label{subsec:insightsMalwareFamilies}
		We begin by analyzing DGA-based DNS queries found in the campus network traffic. Out of 2.4B DNS queries made (during 75 days) by internal hosts of the campus network, about 589K were found in DGArchive, and hence considered as DGA-based queries belonging to a total of 26 known families. Table~\ref{tab:ListMalwares} shows the breakdown of query count across these families. It is seen that ``ModPack'' heavily dominates ($93.85$\%), followed by ``Tinba'' and ``Gameover'' respectively, contributing to  $3.78$\% and $1.60$\% of total DGA-based queries -- other 23 families are not very frequent (their collective contribution is less than one percent). %In terms of the domain names used by these DGA-enabled malware queries, 
		
		\begin{figure}[t!]
			\begin{center}
				%\vspace{-3mm}
				\mbox{
					%\hspace{-3mm}
					\subfloat[Aggregate packet rate.]{\includegraphics[width=0.48\textwidth]{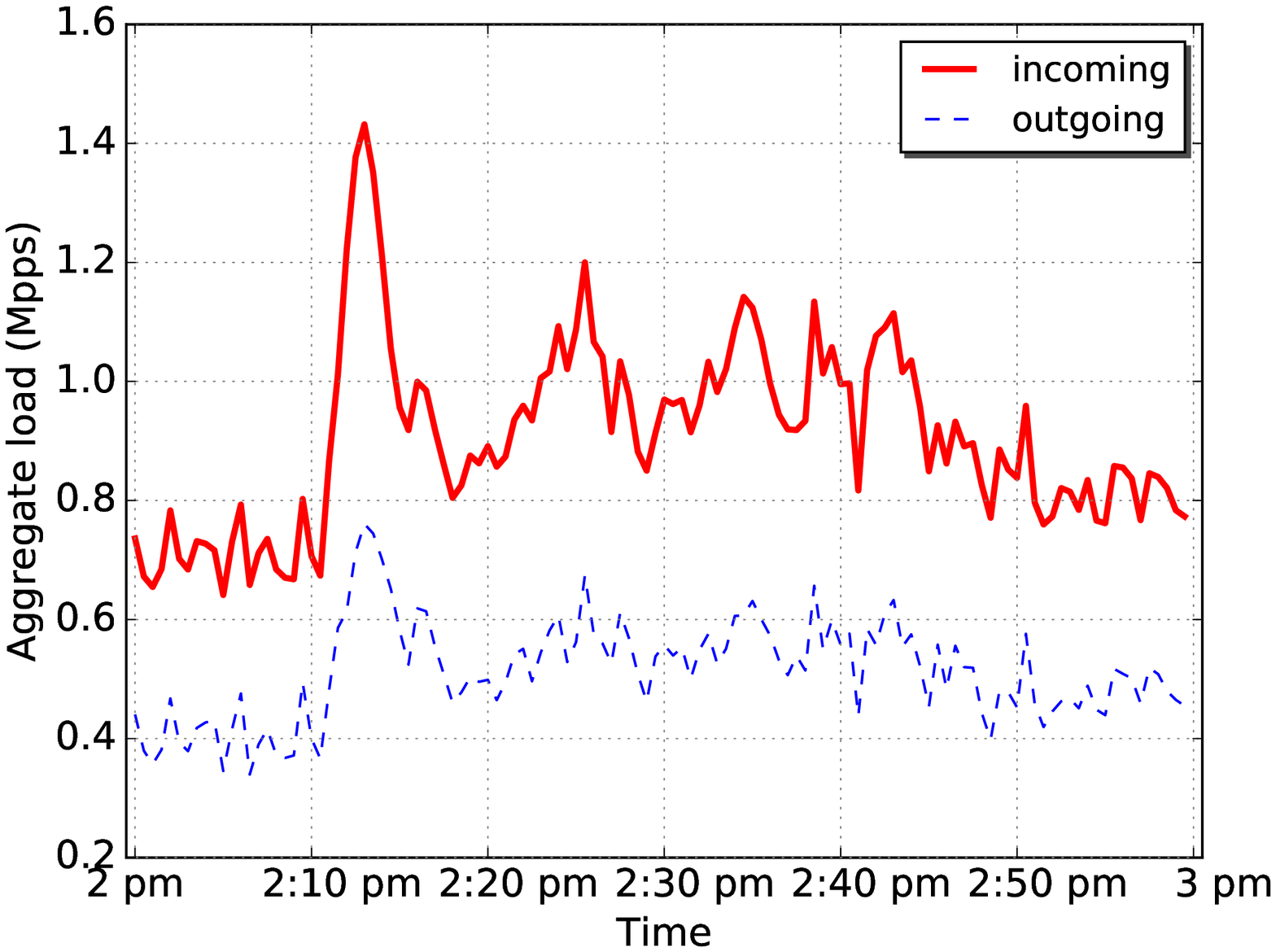}
						\label{fig:bitrate}
						\quad
					}
					%\hspace{-5mm}
					\subfloat[Aggregate bit rate.]{
						\includegraphics[width=0.48\textwidth]{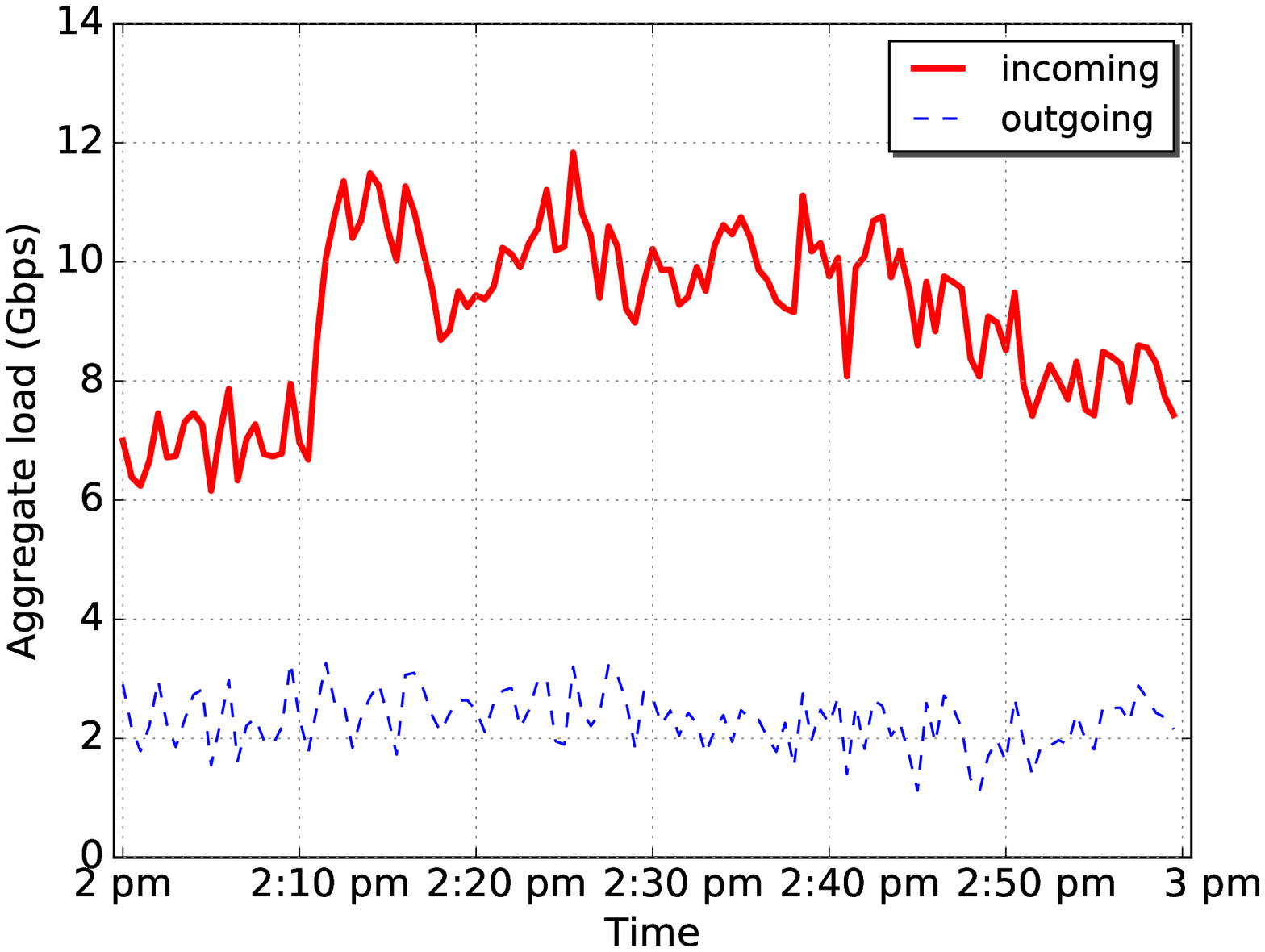}    
						\label{fig:pktrate}
						\quad
					}
				}
				
				\caption{Aggregate load: (a) {packet rate}, and (b) {bit rate}, during peak hour (2pm-3pm) of a weekday (31-May-2019).}
				
				\label{fig:Aggload}
			\end{center}
		\end{figure}

		\begin{table}[t!]
			\centering
			\caption{Top ten most frequently used DGA domain names found $\break$in the campus network.}
			\label{tab:domainNames}
			\vspace{-3mm}
			\begin{adjustbox}{max width=0.88\textwidth}   
				\renewcommand{\arraystretch}{1.3}
				\begin{tabular}{|l|c|c|}
					\hline      \textbf{Domain name}  & \textbf{DGA family} & \textbf{\# occurrences}         \\ \hline
					\myverb{gvaq70s7he[.]ru} &	ModPack	& 530,647
					\\ \hline
					\myverb{76236osm1[.]ru} &	ModPack	& 22,151
					\\ \hline
					\myverb{vqponckshykx[.]in}	& Tinba	& 301
					\\ \hline
					\myverb{uecrbipuperq[.]online} &	Tinba	& 284\\
					\hline
					\myverb{qipnhdggsteb[.]org} &	Tinba	& 284
					\\ \hline
					\myverb{xllqwgtppipp[.]info} &	Tinba	& 269
					\\ \hline
					\myverb{edyrsdetxwnu[.]info} &	Tinba	& 189
					\\ \hline
					\myverb{rkcrurklbstr[.]in}	& Tinba	& 185
					\\ \hline
					\myverb{jdlrshfmxkqdeprhypbejn[.]org} &	Gameover &	179
					\\ \hline
					\myverb{vwxwvcmicwnu[.]org} &	Tinba	& 177 \\ \hline
					
				\end{tabular}
			\end{adjustbox}
		\end{table}

		These 26 families in total generated 1416 unique domain names. Table \ref{tab:domainNames} lists the top ten most frequently used domain names found in DGA queries and their corresponding family. Surprisingly,  only two domain names (\ie {\myverb{gvaq70s7he[.]ru}} and {\myverb{76236osm1[.]ru}}) dominate almost all the queries for the ModPack family. %used only two domains (\ie {\myverb{gvaq70s7he[.]ru}} and {\myverb{76236osm1[.]ru}}) for all of its 55K DNS queries. 
		Other families like Tinba, however, use a variety of domain names in their DNS queries.

		\begin{figure}[t!]
			\centering
			\includegraphics[width=0.88\textwidth]{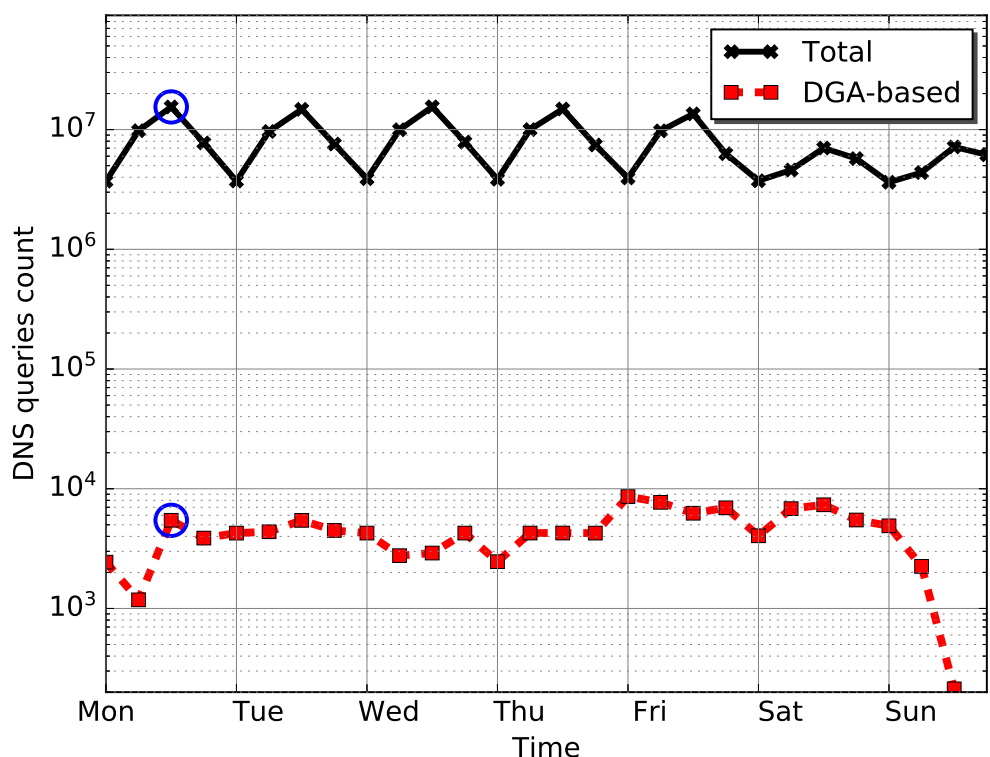}
			\vspace{-5mm}
			\caption{A weekly trace of DNS queries count: total versus DGA-based queries (from 25-Nov-2019 to 1-Dec-2019) -- blue circles highlight representative points to illustrate that count of DGA-enabled queries is at least three orders of magnitude less than count of total DNS queries.}
			\label{fig: comparisonTotalvsMalwareQueries}
			\vspace{-5mm}
		\end{figure}

		Various malware types participate in various malicious actions  \cite{plohmann2016comprehensive} such as unauthorized access to victim machines, stealing personal information, or actively taking part in denial-of-service (DOS) attacks. ``ModPack'' \cite{dgarchive} was found by the Canadian Centre for Cyber Security (CCIRC) \cite{ccirc} which potentially relates to Andromeda \cite{andromeda}. 
		Andromeda is malware that infected millions of computers across the world \cite{andromedaEvidence} to perform its botnet activities (\ie to steal, to destroy websites, or to spread malicious code). ``Tinba'' (Tiny banker was first discovered in 2012), is a malware program, targeting banking websites to steal online banking data \cite{tinba}. Similarly, ``Gameover'' Zeus looks for personal and sensitive data \eg banking information, customer data, and secret corporate information \cite{gameover}.

		We note that the occurrence of DGA-based queries is at least three orders of magnitude less than typical DNS queries made by internal hosts, as shown by a weekly trace in Fig.~\ref{fig: comparisonTotalvsMalwareQueries}. Each data-point in this plot represents the number of DNS queries over a 6-hour window. For example, it can be seen that on Monday at 12pm (as highlighted by blue circles), out of 15.4M DNS queries sent out of the campus network, only 5.4K are DGA-enabled. Such low-profile activity allows various malware families to go undetected in large enterprise networks \cite{lowNoise}. %This creates an opportunity for network administrators to monitor the health of internal hosts by further suspicious traffic and infected  focus

		\subsection{Daily Activity Pattern of DGA-Based Domains} \label{subsec:insightsDailyPattern}

		\begin{figure}[t!]
			\centering
			\includegraphics[width=0.88\textwidth]{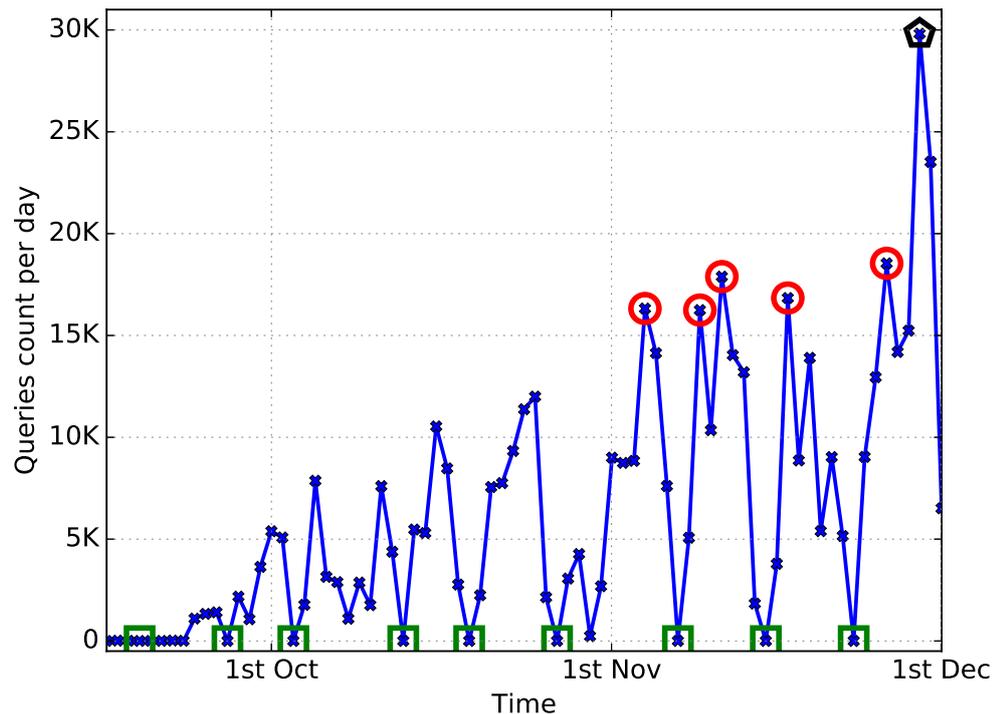}
			\vspace{-3mm}
			\caption{Time-trace of DGA-based DNS queries across 75 days (between 16-Sep-2019 and 1-Dec-2019) -- green squares represent no DGA queries, red circles represent more than 15K DGA queries, and the black pentagon highlights DGA queries daily count peaks at 30K.}
			\label{fig:3monthsMalware}
			\vspace{-3mm}
		\end{figure}

		Let us now focus on the temporal activity pattern of various DGA-enabled malware families.
		%In this subsection, we give insights into the daily pattern of the DGA domains and characterize the DGA-based families into different groups based on their temporal patterns. 
		Fig.~\ref{fig:3monthsMalware} illustrates the daily count of DGA queries during 75 days \ie from 16-Sep-2019 to 1-Dec-2019. 
		It can be seen that there is no specific pattern of daily activity at an aggregate level. For certain days, \ie mostly Thursdays, Fridays, and Saturdays, they become completely inactive (zero queries as highlighted by green squares). Some other days they are heavily active (more than 15K queries per day as highlighted by red circles). 
		We observe a growing trend in the queries count daily over this period peaking at 30K towards the end of Nov 2019 (as highlighted by the black pentagon at the top right of the plot). Focusing on individual families, we found that almost all families (except Tsifiri, Downloader, Pandabanker, and Mirai) became active on the same day, resulting in significant peaks between 28th Nov 2019 and 1st Dec 2019.

		To better understand the time-of-day activity of DGA-enabled malware, we plot in Fig.~\ref{fig:hist} the hourly histogram of DNS queries (overall versus malicious) count during the 75-day study. %It can be seen that the DGA-enabled DNS queries found in the campus network over a day with {a } maximum peak of 26.5K queries over an hour. 
		Starting from total load in Fig.~\ref{fig:histDNS}, it can be seen that the distribution of overall DNS queries reflects the daily activity pattern of users -- it starts rising in the morning (8-9am) when students and staff come to the university campus, peaks at around noon (12-1pm) when most of the users go online during their lunch break, and starts falling in the afternoon (4-5pm) when the users leave the campus network at the end of working hours.
		Moving to the distribution of DGA queries in Fig.~\ref{fig:histDGA}, we observe that the probability of finding DGA queries on the campus network is relatively higher from noon to midnight, and it is lower between post-midnight and pre-noon (1am-11am). We note that the temporal activity pattern of DGA queries does not correlate with that of the overall network traffic (\ie mostly benign), { especially} during afternoon and evening hours. This increasing trend in malware activity starting from mid-day could be due to waking times hard-coded in their software, possibly configured in a time zone different from ours. One may rightly ask \textit{how many} internal hosts, of \textit{which} type, and from \textit{where} in the network are these hosts making DGA-based DNS queries? We will provide insights into possible infected hosts later in Table~\ref{tab:infecteddistro} (briefly) and Section \ref{sec:SystemDesign} (in detail).

		\begin{figure}[t!]
			\begin{center}
				%\vspace{-3mm}
				\mbox{
					%\hspace{-3mm}
					\subfloat[Total DNS queries.]{
						\includegraphics[width=0.47\textwidth,height=0.30\textwidth]{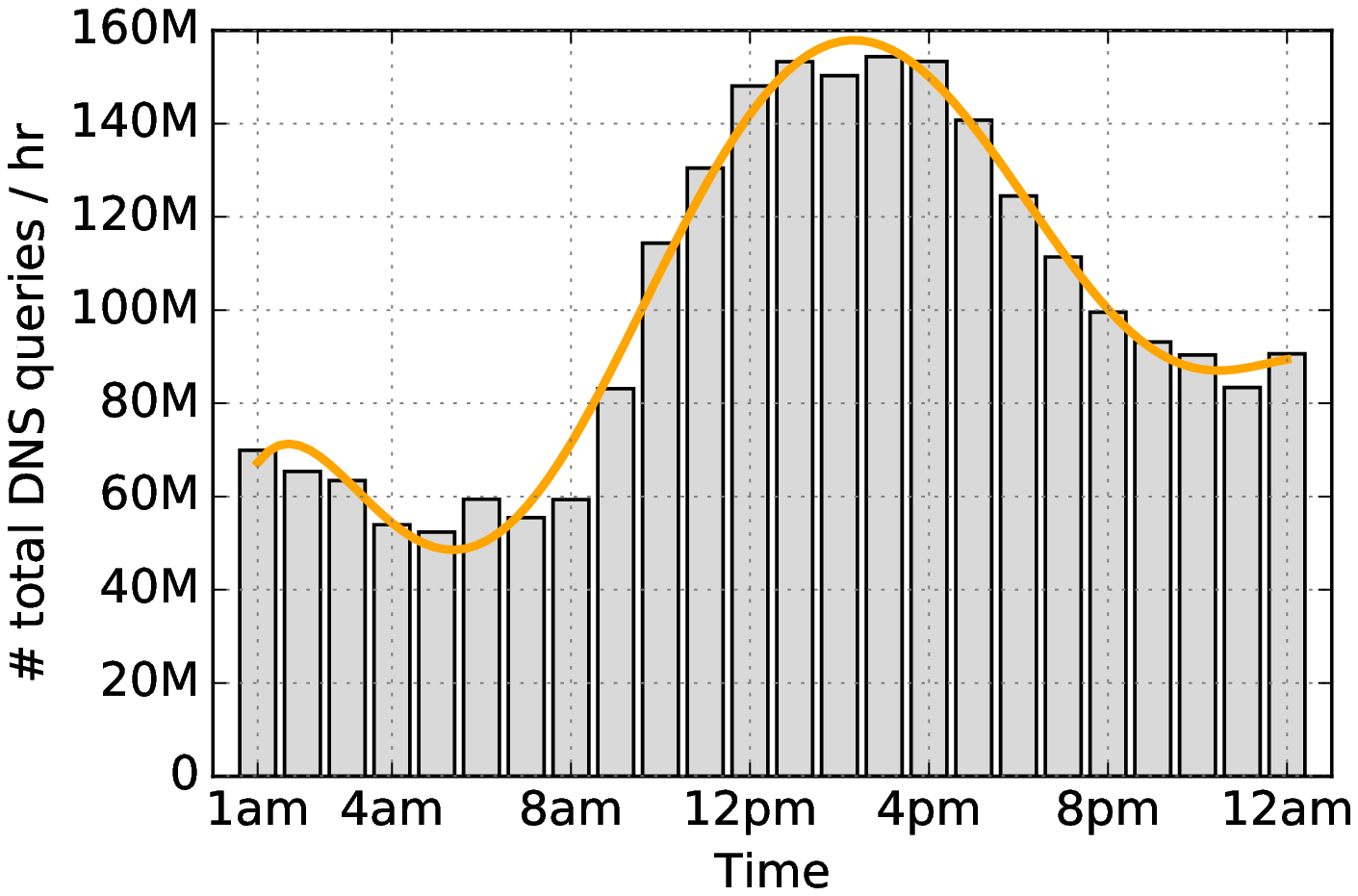}    
						\label{fig:histDNS}
						\quad
					}
					\subfloat[DGA-based DNS queries.]{ \includegraphics[width=0.47\textwidth,height=0.30\textwidth]{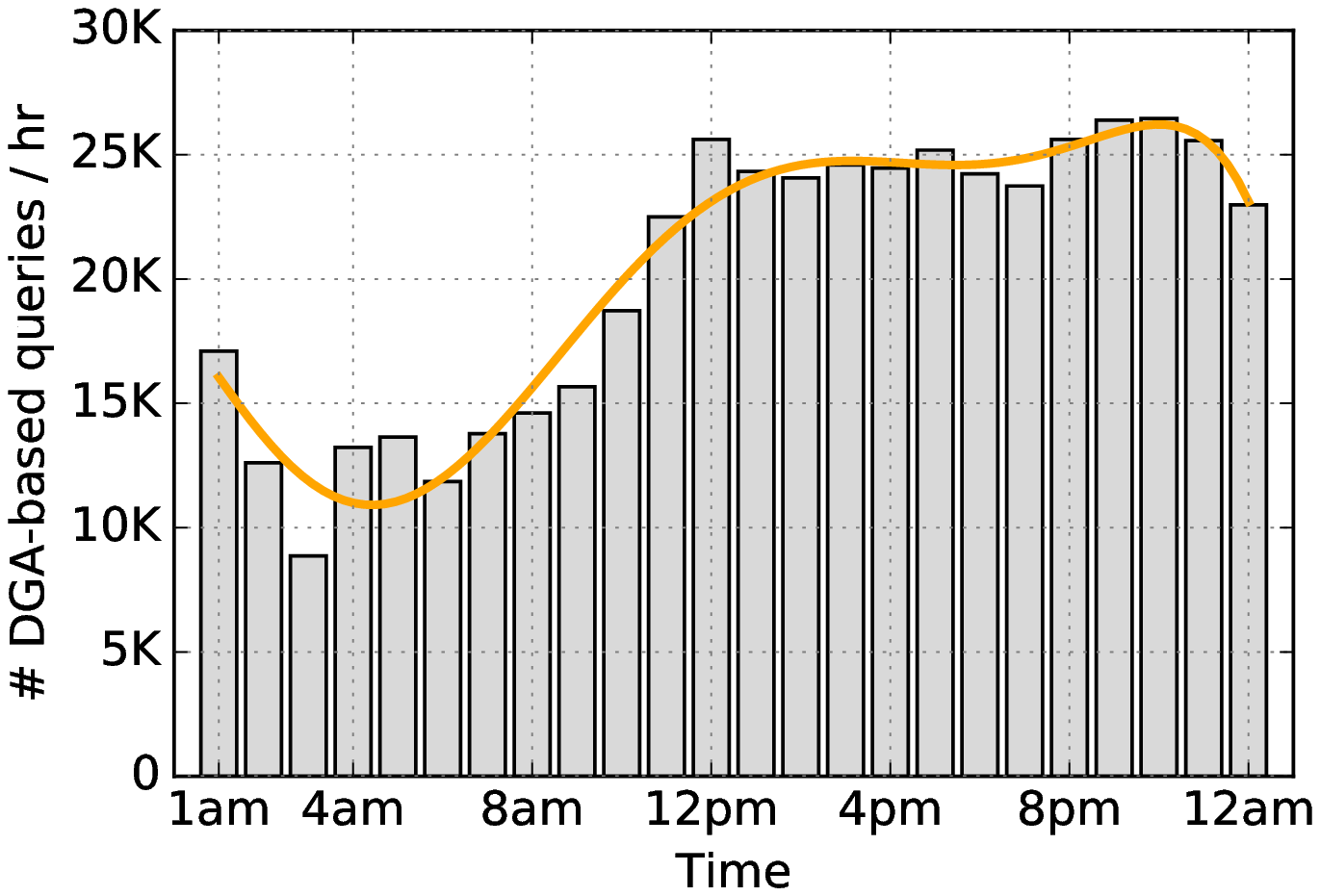}
						\label{fig:histDGA}
						\quad
					}
					%            \hspace{-7mm}
					%            \hspace{-7mm}        
				}
				\vspace{-4mm}
				\caption{Hourly histogram of: (a) total DNS queries, and (b) DGA-based DNS queries, across 75 days (between 16-Sep-2019 and 1-Dec-2019). }
				%        \vspace{-5mm}
				\label{fig:hist}
			\end{center}
			%\vspace{-6mm}
		\end{figure}

		\begin{figure*}[ht!]
			\begin{center}
				%\vspace{-3mm}
				\mbox{
					%\hspace{-3mm}
					\subfloat[ModPack (frequent and heavy).]{ \includegraphics[width=0.47\textwidth,height=0.30\textwidth]{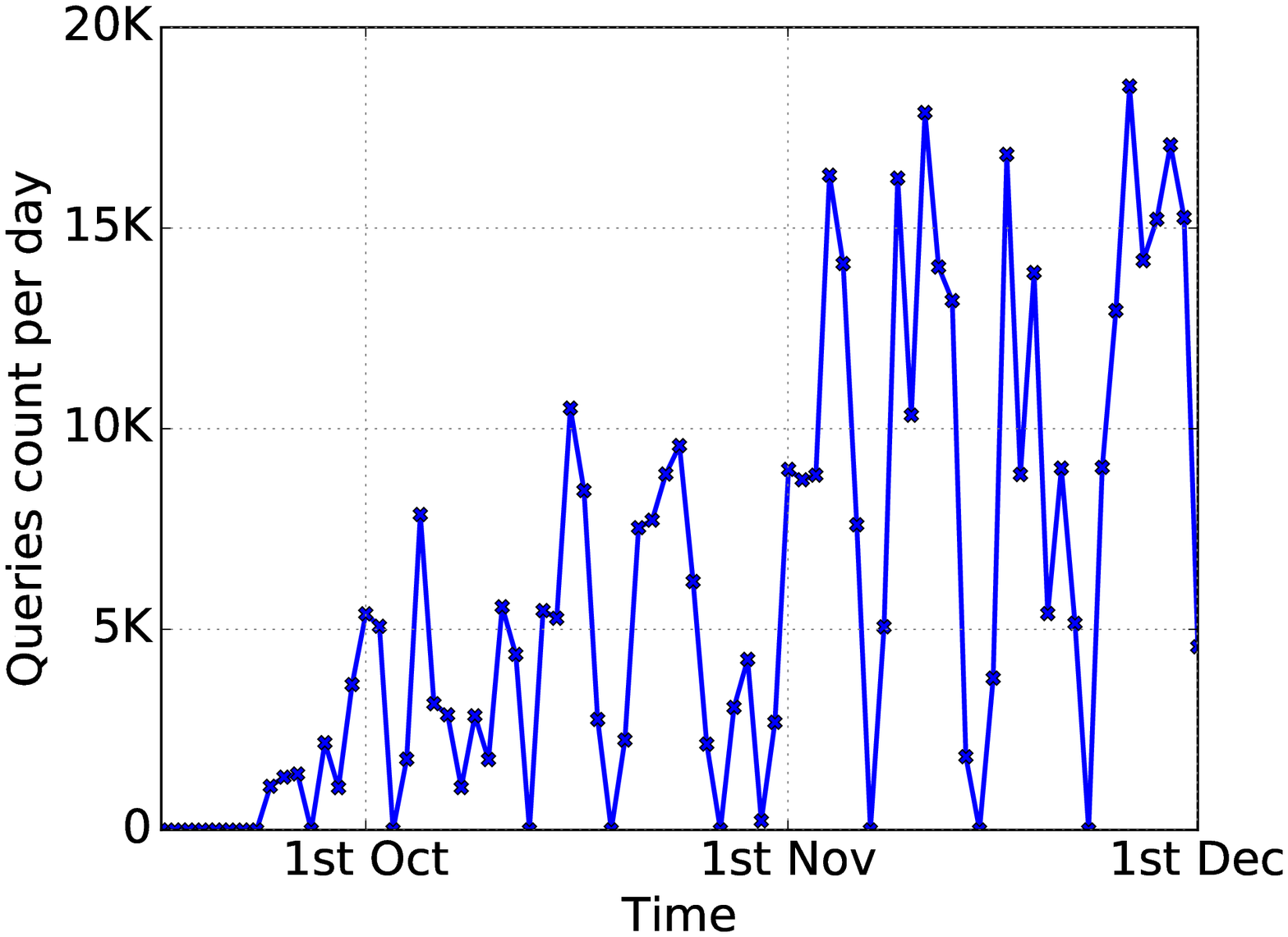}
						\quad
						\label{fig:modpack}
					}
					
					\subfloat[Suppobox (frequent and light).]{
						\includegraphics[width=0.47\textwidth,height=0.30\textwidth]{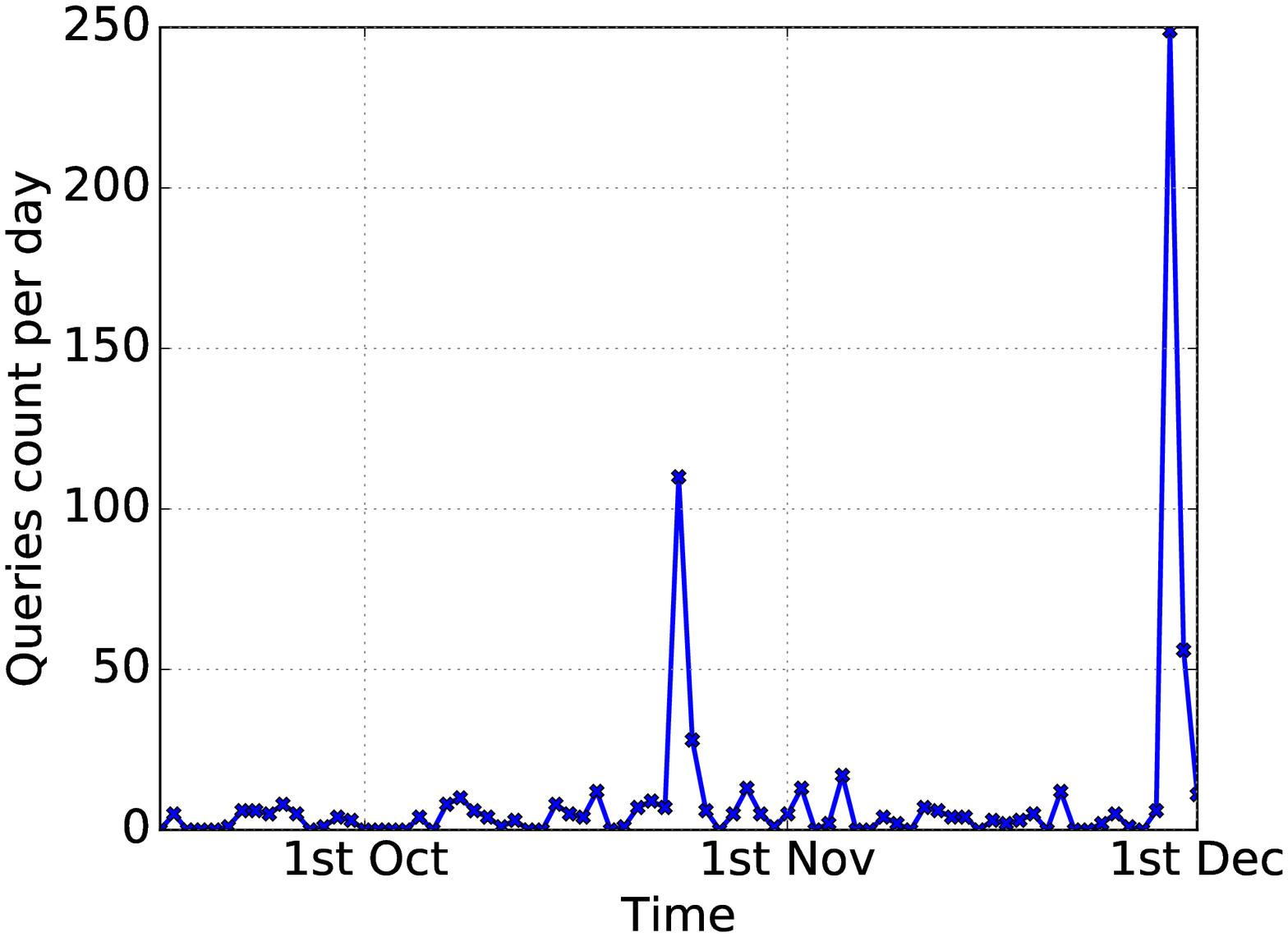}
						\quad
						\label{fig:suppobox}
					}    
				}
				\mbox{
					\subfloat[Ramnit (bursty).]{
						\includegraphics[width=0.47\textwidth,height=0.30\textwidth]{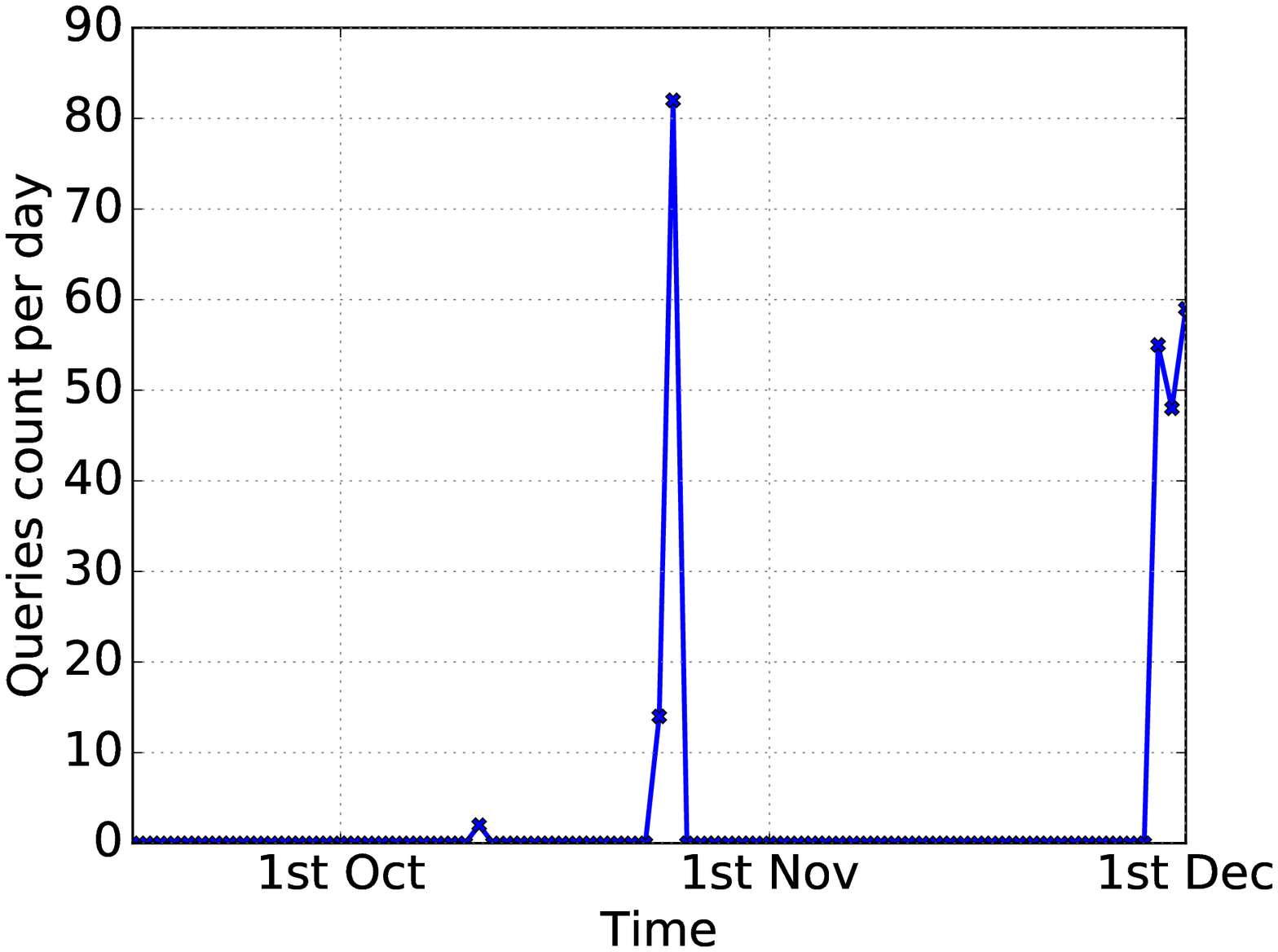} 
						\quad
						\label{fig:ramnit}
					}
					
				}
				\caption{Time-trace of daily DNS queries count for various DGA-enabled malware families: (a) ModPack, (b) Suppobox, and (c) Ramnit. }
				%        \vspace{-5mm}
				\label{fig:pattern_different_DGAs}
			\end{center}
		\end{figure*}

		By analyzing the DNS activity pattern of various DGA families, we categorize them into three groups, namely (a) Frequent and Heavy, (b) Frequent and Light, and (c) Bursty. In Fig.~\ref{fig:pattern_different_DGAs}, for each of these three categories, we illustrate the activity pattern of their representative family. Fig.~\ref{fig:modpack} corresponds to the most frequent and the heaviest family, the ``ModPack''. We observe that ModPack is highly active (thousands of queries) during most of the days (starting from 1st October), except for nine days on which it becomes completely inactive.
		Moving to the ``Suppobox" family in Fig.~\ref{fig:suppobox} representing a frequent and light family, it is seen that the daily queries count is fairly low (less than 20), but it rarely goes inactive on a day. 
		Lastly, as a bursty pattern family shown in Fig.~\ref{fig:ramnit}, we see that `Ramnit" displays two bursts (end of October and end of November) during the entire period of our analysis, and it remains completely inactive otherwise. 
		{It is important to note that these activities could be due to either a large number of infected hosts or certain infected hosts make a large number of DGA queries -- we will discuss in \S\ref{subsec:insightsInfectedHosts} the challenge of identifying infected hosts purely based on DNS traffic.}
		We also found some forms of coordination across various DGA families in terms of their activity. For example, Suppobox and Ramnit simultaneously became heavily active on 24th Oct as well as towards the end of November (\ie from 28-Nov to 01-Dec) as shown in Figures~\ref{fig:suppobox} and ~\ref{fig:ramnit}. We will highlight some examples of such coordination in \S\ref{subsec:1hourPCAP}.

		\begin{figure}[t!]
			\centering
			\includegraphics[width=0.88\textwidth,height=0.47\textwidth]{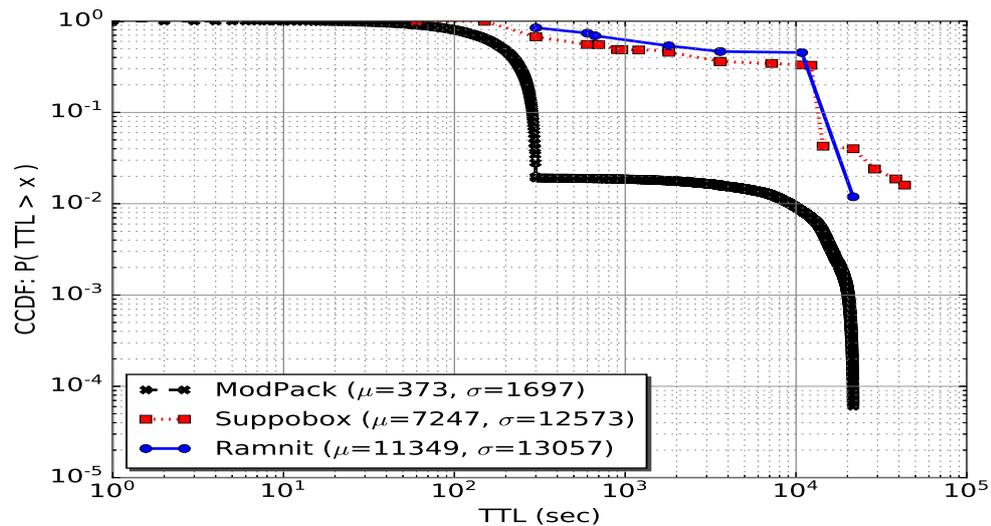}
			\vspace{-3mm}
			\caption{CCDF of TTL value in DGA-related DNS responses across representative malware families.}
			\label{fig:3monthsCombinedCCDF}
			\vspace{-3mm}
		\end{figure}

		To further investigate the periodicity of these three representative families, we extracted the TTL value from their DNS responses.  Fig.~\ref{fig:3monthsCombinedCCDF} shows the CCDF of TTL values for each family. It is seen that the ModPack family (dashed black lines with cross markers) tends to use fairly short TTLs with an average of 373 seconds -- $90$\% of ModPack DNS responses will live less than three minutes. 
		The TTL values in the Suppobox family are relatively longer, averaging at 7247 seconds, while $50$\% are more than 15 minutes. Lastly, the Ramnit family has the longest TTL values with an average of 11349 seconds. Similar to the Suppobox family, Ramnit also has $50$\% of its responses with a TTL greater than 15 minutes. Overall, the distribution of TTL values in various families explains (to a great extent) their frequency of occurrence -- the shorter the TTL, the more frequent they become. {We will use (in \S\ref{subsec: SDN-SYSTEM}) TTL values for timeout setting of reactive flow entries installed by the SDN controller to enable scalable management of switch TCAM table.}

		\subsection{Infected Enterprise Hosts} \label{subsec:insightsInfectedHosts}
		We now look at enterprise hosts that make DGA-enabled queries.  
		%In this subsection, we give an overview of how the DGA-enabled queries are distributed amongst the enterprise hosts. 
		Table~\ref{tab:infecteddistro} shows that a very high majority ($93.8$\%) of DGA queries are sourced from DNS recursive resolvers, %(552K out of 589K queries), 
		and hence the original querying end-hosts are invisible, except a limited number of hosts (8 enterprise servers and 13 regular hosts) which are probably configured to use public DNS resolvers directly (\eg Google {\myverb{8.8.8.8}}). We also note that $3.7$\% of DGA queries are generated by 19 WiFi NAT gateways -- the identity of infected hosts (WiFi clients) that generate these queries remains unknown, as they reside behind NAT gateways inside the network (we collect data from the border of the network). 
		
		{Obviously, enterprise recursive resolvers and NAT gateways hide the identity of infected hosts, and hence host analysis purely based on DNS traffic will not suffice for identifying infected hosts.
			To better understand the network activity of (possibly) infected hosts following their suspicious DNS queries (found in DGArchive), we analyze (in what follows) a 1-hour PCAP trace of full traffic dump from the campus network. We focus on subsequent TCP/UDP traffic exchanged between Internet-based C\&C servers (following the response of DGA-based DNS queries) and their respective enterprise hosts -- we will show in \S\ref{sec:results} how our SDN-based method in conjunction with trained models (\S\ref{sec:SystemDesign}) will be able to identify infected enterprise hosts.
			
			One may choose to inspect packets of enterprise DNS servers to gain richer insights into the health of internal hosts purely based on DNS traffic. However, obtaining all (incoming/outgoing) packets of various DNS servers requires significant changes to the distributed infrastructure of large enterprise networks \cite{PAM19}. 
			Our solution (\S\ref{sec:SystemDesign}), instead, is designed to be a ``bump-in-the-wire'' on the Internet link of the enterprise network and provides different visibility (a judicious combination of DNS and selected subsequent TCP/UDP flows) into activities of connected hosts from a different perspective. Furthermore, our detection system is transparent to the network, requires minimal change to existing infrastructure, is easy to deploy, and does not modify packets in any way.
		}

		\begin{table}[t!]
			\centering
			\caption{Enterprise hosts making DGA-enabled queries.}
			\vspace{-2mm}
			\label{tab:infecteddistro}
			\begin{adjustbox}{max width=0.88\textwidth}
				\begin{tabular}{@{}lr@{}}
					\toprule
					\textbf{Querying hosts (\# hosts)} & \multicolumn{1}{r}{\textbf{~~~~~~~~~~~~~~~~~~~\# DGA queries [\%]}}      \\ \midrule
					%Total              & 589,098                                                             \\ \midrule
					Recursive resolvers (3)            & 552,669  [93.82\%]                                        \\ \midrule
					Enterprise servers (8)                     & 14,217  [2.41\%]                                                     \\ \midrule
					End hosts (13)              & 283  [0.05\%]                                                       \\ \midrule
					WiFi NAT gateways (19)         & 21,929  [3.72\%]                                                        \\ \midrule
				\end{tabular}
				%    \vspace{-3mm}
			\end{adjustbox}
			\vspace{-3mm}
		\end{table}

		\subsection{Network Behavior of DGA-Fueled Malware} \label{subsec:1hourPCAP}
		{
			As explained in the previous section, purely monitoring DNS traffic does not lead to finding the hosts that are indeed suspected of malware infection. Furthermore, to draw insights into the behavior of suspected hosts (in terms of services used and/or any possible coordination across infected hosts while communicating with their corresponding C\&C servers), we analyze a concise but fairly active period of the full campus traffic dump.
			{This full dump of the entire Internet traffic contains 3.2B packets of which only 2M packets are DNS -- less than $0.1$\%.}

			Table \ref{tab:overview_1hourFindings} summarizes our findings from this analysis. During this one hour, eight DGA-enabled malware families were found in the DNS packets of this 1-hour PCAP trace. Also, a total of 14 unique C\&C servers were identified from DNS responses, and only five of these C\&C servers (corresponding to two DGA families) exchanged TCP traffic with enterprise hosts following their DGA-based DNS resolution. %Furthermore, we found that only two of eight DGA families were involved in communicating with the C\&C server. 
			Analyzing these follow-up TCP flows, we found 17 hosts (we call them ``suspicious hosts'') communicated with the five C\&C servers. By reverse DNS lookup, we verified that 8 of the suspicious hosts were regular end-hosts, 2 were enterprise servers, and 7 were WiFi NAT gateways.	
			{These 17 suspicious hosts accessed HTTP and HTTPS services offered by their corresponding C\&C servers, generating 33 suspicious HTTP flows ($75$\%) and 11 suspicious HTTPS flows ($25$\%) that collectively exchange a total of only 365 packets which contribute to a tiny fraction ($10^{-7}$) of total packets (3.2 B) recorded in this one-hour PCAP trace from campus Internet traffic.}

			\begin{table}[t!]
				\centering
				\caption{DGA-enabled malware and infected campus hosts found by analysis of one-hour PCAP of full campus traffic.}
				\vspace{-2mm}
				\label{tab:overview_1hourFindings}
				\begin{adjustbox}{max width=0.88\textwidth}
					\begin{tabular}{@{}lr@{}}
						\toprule
						\textbf{} & \multicolumn{1}{r}{\textbf{Count}}      \\ \midrule                    
						DGA-enabled families found in DNS queries of campus hosts &  8 \\ \midrule
						C\&C servers identified in responses to DGA queries   & 14 \\ \midrule
						C\&C servers exchanged traffic with internal hosts                & 5 \\ \midrule
						DGA-enabled families involved in C\&C traffic exchange                 & 2 \\ \midrule
						
						Num. of malware-infected hosts exchanged traffic with C\&C servers & 17        \\ \midrule
						~~~~~~End hosts & 8                                                        \\ \midrule 
						~~~~~~WiFi NAT gateways & 7   \\ \midrule
						~~~~~~Enterprise servers  & 2  \\ \midrule
						
					\end{tabular}
					
				\end{adjustbox}
				
			\end{table}

			\begin{figure*}[t!]
				\begin{center}
					\mbox{
						\subfloat[Sequence of flows.]{
							{\includegraphics[width=0.65\textwidth]{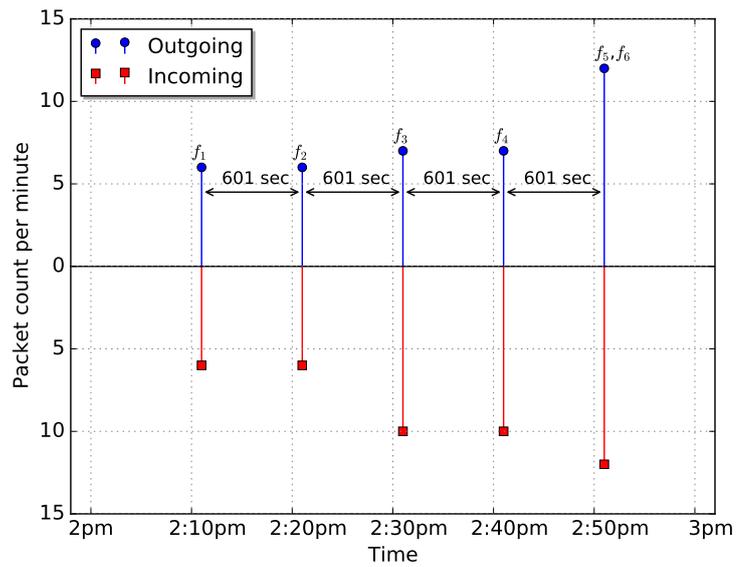}}
							\quad
							\label{fig:patternHostFlows-1}
						}
					}
					\mbox{
						\subfloat[Sequence of packets in flow ``$f_{2}$'', for example.]{
							{\includegraphics[width=0.65\textwidth]{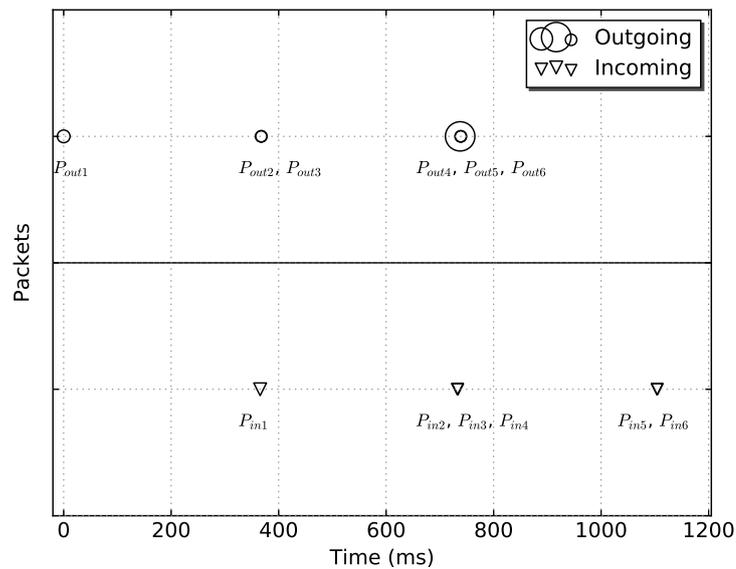}}\quad
							\label{fig:patternHostPkts-1}
						}
					}
					
					\caption{Communication pattern of a suspicious host with its C\&C server of Ramnit family: (a) sequence of  HTTPS flows, and (b) sequence of packets in a selected HTTPS flow.}
					
					\label{fig:patternHost-1}
				\end{center}
			\end{figure*}

			Additionally, we analyzed the behavior of these suspicious flows to highlight their activity patterns or possibly find coordination across suspicious hosts \cite{assadhan2009periodic}. Fig.~\ref{fig:patternHostFlows-1} shows the pattern of communications (on a flow basis) between a suspicious host (a WiFi NAT gateway in this case) and a C\&C server of Ramnit family with IP address ``{\myverb{89.185.44.100}}'' resolved by a query for domain name ``{\myverb{lvxlicygng.com}}" (with the response TTL value of 300 sec) -- we verified this address is blacklisted \cite{virusRamnit}. We observe that this WiFi gateway (possibly on behalf of a number NATed hosts) initiates six HTTPS flows (each with a duration of less than a few seconds) to the server in the time period between 2:10pm and 2:50pm. Note that two flows $f_5$ and $f_6$ are established concurrently. The height of each flow in the plot (Fig.~\ref{fig:patternHostFlows-1}) indicates the number of packets sent and received (outgoing direction shown by blue circles and incoming direction shown by red squares). Interestingly, there is a clear periodicity in the arrival of these flows -- they are well spaced by 601 sec ($\approx$10 minutes). Some of these flows (\eg $f_1$ and $f_2$) are symmetric in terms of incoming and outgoing packet count, and some are asymmetric (\eg $f_3$ and $f_4$).

			To further understand the network activity of these suspicious flows, we zoom in on the arrival and size of individual packets within each flow per direction. As an example, we show in Fig.~ \ref{fig:patternHostPkts-1} the time trace of packets in the flow $f_2$. The x-axis is time (in ms), and the y-axis indicates the direction (top row corresponds to outgoing packets and the bottom row corresponds to incoming packets). Also, each marker represents a packet (circles for outgoing and triangles for incoming), and the markers size indicates the relative length of the corresponding packet. The duration of this flow is about 1100 ms over which six packets are sent, and six packets are received. In this flow, all incoming packets from the server have the same size of 60 bytes.
			Within less than 400ms from the commencement of the flow, the three-way handshake is completed, \ie SYN ($P_{out1}$) $\rightarrow$ SYN-ACK ($P_{in1}$) $\rightarrow$ ACK ($P_{out2}$). Right after establishing the TCP connection, the host sends 60 bytes data over SSL ($P_{out3}$). In response, the server sends SSL data of 60 bytes ($P_{in2}$) followed by FIN-ACK ($P_{in3}$) and TCP-RST ($P_{in4}$) packets. Next, the host sends SSL data of 139 bytes ($P_{out4}$), followed by ACK ($P_{out5}$) and FIN-ACK ($P_{out6}$) packets. Lastly, the server sends two more TCP RST packets ($P_{in5}$ and $P_{in6}$) back-to-back. 
			{Observing such activity patterns will help us (in \S\ref{subsec:CTU-modeling}) identify flow-level attributes needed for modeling malware traffic behavior. As stated earlier in Chapter \ref{chap:ch2}, our main objective in this chapter is to develop a ``cost-effective'', yet ``accurate'' solution for detecting malicious flows and infected hosts in ``large-scale enterprise networks''. 	We will use flow-level attributes (as opposed to computationally expensive packet-level attributes) to diagnose whether selected suspicious traffic is malicious, or not.}

			\begin{figure}[t!]
				\centering
				\includegraphics[width=0.88\textwidth]{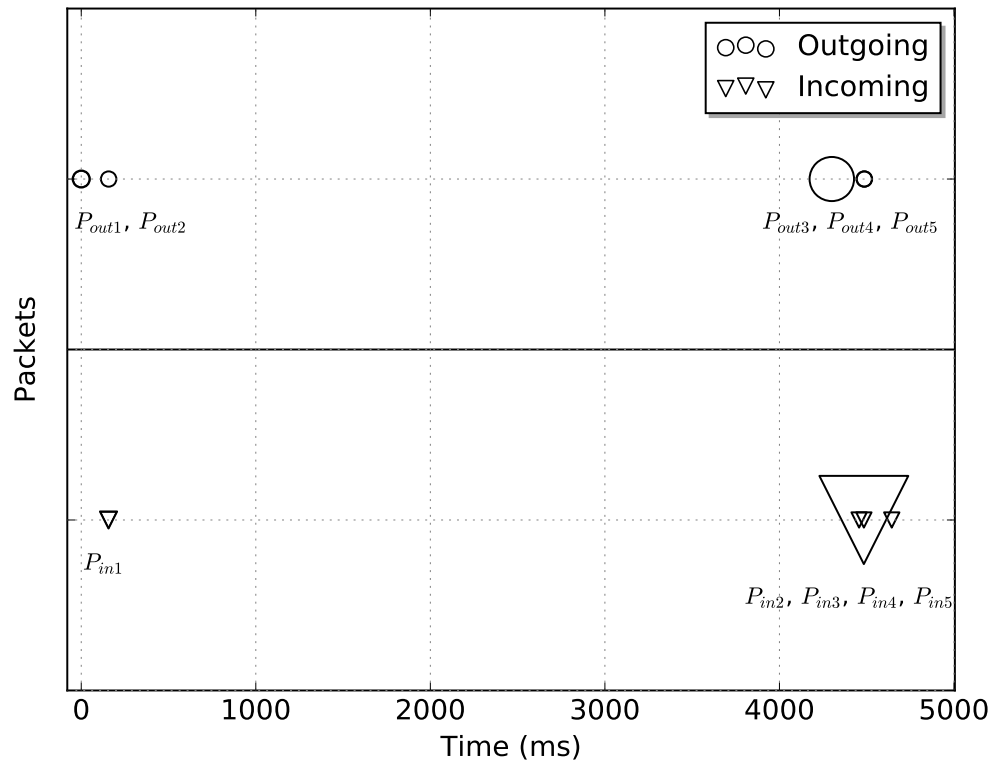}
				\vspace{-4mm}
				\caption{Time trace of packets (outgoing/incoming) in a selected HTTP flow.}
				\label{fig:http_pktBased}
			\end{figure}
			
			As another example, we show in Fig.~\ref{fig:http_pktBased} the flow activity of an end-host establishing a suspicious HTTP flow with its C\&C server. It can be seen that in this case, the server sends some data (not just acknowledgments) to the internal host. In each direction, 5 packets are exchanged between the end-host and the server over this flow with a duration of about 4600 ms. The three-way handshake (SYN:$P_{out1}$ $\rightarrow$ SYN-ACK:$P_{in1}$ $\rightarrow$ ACK:$P_{out2}$) completes within the first 160 ms. Following the establishment of this connection, the end-host sends an {\myverb{HTTP GET}} request ($P_{out3}$ with the size 420 bytes) to the server at the time about 4100 ms. The server responds with an ACK ($P_{in2}$) followed by 1001 bytes {\myverb{HTTP OK}} ($P_{in3}$) -- we were unable to inspect the packet payload since in our PCAP trace packets are truncated to their first 96 bytes. Right after that, the server initiates termination of this TCP flow by sending a FIN-ACK ($P_{in4}$) -- this packet is followed by ACK ($P_{out4}$) and FIN-ACK ($P_{out5}$) packets from the host, and the final ACK ($P_{in5}$) from the server.
			
			We also found two incidents of possible coordination \cite{gu2008botsniffer} across suspicious hosts while contacting their C\&C servers at around 2:10pm and 2:20pm. In the first instance (at 2:10pm), two suspicious end-hosts simultaneously initiated HTTP flows: one host initiated four flows. The other one initiated eight flows both with a C\&C server from Suppobox family with IP address ``{\myverb{184.168.131.241}} (corresponding to domain name  ``{\myverb{strengthstorm.net}}'' with the response TTL value of 600 sec) -- we verified that this IP address is blacklisted \cite{urlSuppo}.
			In the second instance (at 2:20pm), a WiFi NAT gateway and an end-host respectively initiated one and two HTTP flows with a C\&C server from the Suppobox family with the same IP address ``{\myverb{184.168.131.241}}".  %{\color{red}Are you able to mention the TTL value of the corresponding DNS reply here?}%The NAT gateway established one flow and the end-host established two flows simultaneously.

			\textbf{Compute Cost:} Our insights into network activities of malware were only obtained by manually analyzing a short but representative PCAP trace of the campus network. It is important to note that it becomes costly (practically infeasible) to capture and analyze packets of all network flows at high rates (10Gbps or more). We also note that malicious traffic (specifically for malware and botnet\\
			) often contributes to a tiny fraction of the total network traffic -- the majority of packets are benign. Therefore, it is needed to employ a systematic and scalable method to capture only suspicious traffic (corresponding to servers that are resolved as a result of DGA queries) and check whether it is malicious or not.  
			{In the next section, we will leverage the ability of SDN to dynamically mirror suspicious traffic flows and develop learning-based models (based on insights obtained in this section) to automatically detect malicious flows associated to their infected hosts.}

			\section{Modeling and Mirroring Traffic of Suspicious Malware Servers} \label{sec:SystemDesign}
			{
				In this section, we begin by developing our protocol-specialist models (one corresponding to each of HTTP, HTTPS, and UDP protocols) using CTU-13 network traces (discussed in \S\ref{subsec:ourData}). We, next, develop a system to select, mirror automatically, and diagnose traffic flows corresponding to suspicious malware servers. We employ SDN reactive rules to select and mirror suspicious traffic to our packet processing engine. The engine feeds our trained models by a set of flow attributes for diagnosis, determining whether these selected TCP/UDP flows are malicious or not.
			}

			\subsection{Modeling Traffic Behavior of Malware} \label{subsec:CTU-modeling}
			We use malware PCAP traces of the CTU-13 dataset \cite{garcia2014empirical}. 
			Authors of \cite{garcia2014empirical} primarily aimed to detect malicious ``hosts'' by developing clustering-based models from their dataset. They employed host-level attributes, including the count of remote IP addresses, count of remote/local transport port numbers, average packet size, and the average count of packets transmitted over windows every two-minute. We note that computing these attributes near real-time for every internal host (tracking metadata of all packets) will be computationally prohibitive, especially at scale. Also, traffic modeling at the host-level becomes slightly coarse-grained (aggregate of benign and malicious flows), resulting in reduced visibility into the activity of individual malware flows. { Our approach aims to characterize the behavior of malware activity on a per-flow basis and diagnoses a fraction of network flows, only those suspicious TCP/UDP flows pertinent to a DGA-related DNS query. We develop a set of machine learning models (a model per protocol) to determine if a flow is malicious or not. Many cybersecurity researchers \cite{ allard2011tunneling, garcia2014empirical, buczak2016detection} employed multi-class decision trees to distinguish malicious and benign traffic. However, balancing the training dataset to avoid overfitting remains a nontrivial challenge \cite{multiClass2008}.
				It has been shown \cite{ahmed2019monitoring,sivanathan2020detecting} that one-class classifiers or anomaly detection models can learn the distribution of training data (malicious flows in the context of this chapter) and detect any deviations (benign flows) during the testing phase. Our protocol-specialist models will generate ``negative'' output for malicious instances and ``positive'' output otherwise. This use of one-class models means that each model can be re-trained/updated (in case of extending the malware dataset), independent of the other models. 
			}
			
			%{\color{red}The purpose of opting {for }the flow-based attributes instead of host-based attributes is just to smartly aggregate the traffic with a minimal amount of cost at the packet processing server.}

			\subsubsection{Attributes and Classifiers for Malware Flows}
			Inspired by \cite{garcia2014empirical}, we identify eight attributes on a per-flow basis for malware traffic -- 4 per each direction (in/out). Our attributes of a malware flow are as follow.
			
			\begin{itemize}
				\item \textit{flow volume} in bytes (in/out).
				\item \textit{flow duration}, \ie the gap between the arrival time of the first packet and the last packet (in/out).
				\item \textit{number of packets} (in/out).
				\item \textit{average packet size} (in/out).
				%\item \textit{\color{red}\st{standard-deviation of packets size (in/out)}}.
			\end{itemize}

			\begin{table*}[t!]
				\centering
				\caption{Distribution ($\mu$ and $\sigma$) of attributes value for malicious flows in CTU-13 dataset.}
				\vspace{-3mm}
				\label{tab:featuresInsightsCTU}
				\begin{adjustbox}{max width=0.99\textwidth}   
					\renewcommand{\arraystretch}{1.5}
					\begin{tabular}{|lc|c|c|c|c|c|c|c|c|c|}
						\hline
						& & \multicolumn{4}{c}{\textbf{Outgoing}} & \multicolumn{4}{|c|}{\textbf{Incoming}} \\ \cline{3-10}
						& &  flow volume (B) & flow duration (s) & \# pkts & Avg. pkts size (B)  & flow volume (B) & flow duration (s) & \# pkts & Avg. pkts size (B)  \\ \hline 
						\textbf{HTTP} &   &  \cellcolor{yellow!25}(980, 666)  & \cellcolor{yellow!25}    (7.3, 29.5)  & {\cellcolor{yellow!25}(5, 1)} & (172.5, 94.1) &       \cellcolor{yellow!25}(1470, 1589) &     \cellcolor{yellow!25}(1.5, 7.6)  &  \cellcolor{yellow!25}(4, 1) &    (270.5, 235.4)  \\ \hline
						\textbf{HTTPS} &  &    \cellcolor{yellow!25}(1597, 4392)   &    \cellcolor{yellow!25}(761.3, 2832.3) & \cellcolor{yellow!25}(12, 41)  & (81.1, 59.4)     &  \cellcolor{yellow!25}(5805, 47043)   &     \cellcolor{yellow!25}(760.6, 2832.1) &  \cellcolor{yellow!25}(7, 20)  & (187.4, 390.9) \\   \hline        
						\textbf{UDP} & & \cellcolor{yellow!25}(5968, 100239) &    \cellcolor{yellow!25}(708.8, 2326.2)    &  \cellcolor{yellow!25}(14, 165)  & (201.6, 164.6)     &  \cellcolor{yellow!25}(5968, 100239)   &     \cellcolor{yellow!25}(712.3, 12451.2) &  \cellcolor{yellow!25}(14, 165)  & (201.6, 164.6) 
						\\ \hline
					\end{tabular}
				\end{adjustbox}
				\vspace{-4mm}
			\end{table*}
			
			We computed the above attributes for all (labeled) malicious flows in the CTU-13 dataset. { We show in Table~\ref{tab:featuresInsightsCTU} the distribution ($\mu$ and $\sigma$) of raw values of attributes across the entire dataset.} Note that the malware flows in the CTU-13 dataset are from three protocols, namely HTTP, HTTPS, and UDP. By analyzing these values, we observe that flow volume,  flow duration, and packet count for both incoming and outgoing directions are key attributes in characterizing the three categories of malware flows. As an example, considering the outgoing direction, the mean ($\mu$) volume of flow in HTTP, HTTPS, and UDP malware is about $1000$, $1600$, and $6000$ bytes, respectively (first column of Table~\ref{tab:featuresInsightsCTU}).

			{Let us make some high-level observations on the range of attributes across the three types of malicious flows.}
			Note that the variation of  flow volume is much larger in HTTPS (with $\sigma\approx4400$) and UDP (with $\sigma\approx100,000$) flows than in HTTP flows (with $\sigma\approx700$). A slightly similar pattern is observed in the flow duration and packet count attributes. UDP and HTTPS flows, compared to HTTP flows, are generally longer in duration (mean $700$s versus mean $7$s), and carry a larger number of packets (mean $12-14$ packets versus mean $5$ packets). Such a clear distinction between the three categories (HTTP, HTTPS, and UDP) can also be seen in the values of attributes for the incoming direction. Therefore, we train three separate models (each specific to a protocol), increasing the accuracy of detecting malware flows.
			{ We split the malicious data of each protocol-specific model into 60\% (for training and validation) and 40\% (for testing only). }

			\subsubsection{{Model Training}}
			%\textbf{Model Training:}
			We used {\fontsize{10}{48}\usefont{OT1}{lmtt}{b}{n}\noindent scikit-learn} and its APIs, an open-source machine-learning package written in Python, to train and test our models. 
			{
				Our prediction models are one-class classifiers trained by four popular algorithms, namely Isolation Forest (iForest) \cite{liu2008isolation}, Extended iForest (EiF) \cite{EiF2019}, K-means \cite{ding2004k}, and one-class support vector machines (OC-SVM) \cite{scholkopf2001estimating} using attributes of malicious flows obtained from the CTU-13 dataset.
				The models classify a flow: if the subject flow is tested negative by its corresponding model (\ie HTTP, HTTP, or UDP), then it is classified as ``malicious"; otherwise, it is ``benign". Later in this subsection, we will compare the performance of one-class models against that of multi-class classifiers.}
			
			The iForest algorithm works based on the concept of isolation without employing any distance or density measure. The algorithm divides instances into sub-samples to construct a binary tree structure -- by randomly selecting the attribute, and then randomly selecting the split values from a range (within min and max obtained from training) for that particular attribute -- { splitting values is always done by an ``axis-parallel'' hyperplane (\eg rectangular shape in 2D space of attributes).} 
			If the value of a given instance is less than the split value, the point is directed to the left branch of the tree structure otherwise, it goes to the right side branch. This branching is recursively performed until either a predefined height limit is approached or a single point is isolated in the dataset. The algorithm then marks the instances that travel less into the tree structure as an anomaly, while those that travel deeper into the tree structure are classified as benign. To avoid issues due to randomness, the process is repeated several times, and the average path length is calculated and normalized. 
			
			For the training phase of the iForest models, we consider three tuning parameters, namely the number of trees (\textit{n\_estimators}), height limit of trees (\textit{max\_samples}), and contamination rate. For each of the three models, we tune the value of each parameter while fixing the other two parameters and validate the accuracy of our specialized models for the malicious flows in the CTU-13 dataset. The default value for the number of trees is 100, the height limit of trees is set to ``auto'', and the contamination rate is 10\%. After tuning the individual three models, we found the optimal value of these tuning parameters: the number of trees equal to 10, the height limit of trees equal to 8, and the contamination rate of 1\% for all three models.  
			
			{
				It has been recently shown \cite{EiF2019} that while iForest is a computationally efficient algorithm, it suffers from a bias (affecting the anomaly score) arisen by its use of axis-parallel hyperplanes. Authors of \cite{EiF2019}, therefore, enhance the standard iForest algorithm by developing Extended Isolation Forest (EiF), which performs data splitting with ``random-slope'' hyperplanes.  
				Our EiF models are tuned in the same way as iForest.

				\begin{figure}[t!]
					\begin{minipage}{0.7\columnwidth} 
						\centering
						\vspace{0mm}
						\includegraphics[width=0.98\textwidth]{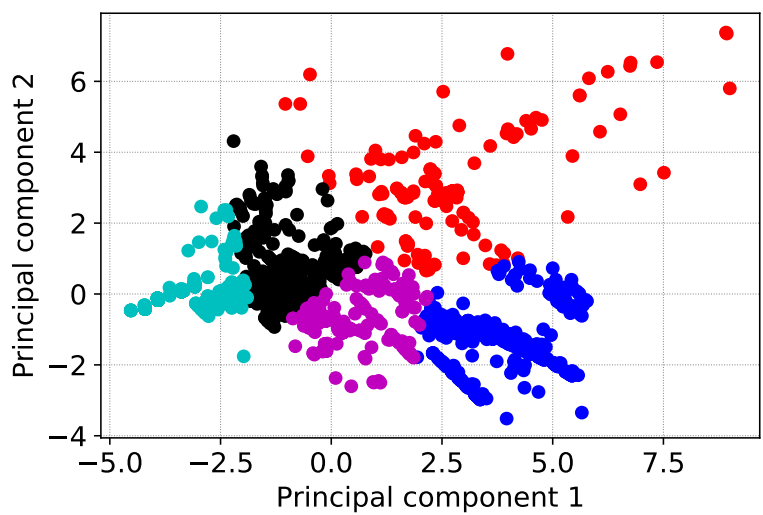}
						\vspace{-3mm}
						\caption{ Clusters of K-means model for HTTP flows.}
						\label{fig:httpKmeans}
					\end{minipage}
					\begin{minipage}{0.88\linewidth}
						\centering	
						\tabcaption{{ Distribution ($\mu$ and $\sigma$) of attributes for K-means HTTP clusters.}}	
						\begin{adjustbox}{max width=0.98\columnwidth}   
							\renewcommand{\arraystretch}{1.5}
							\begin{tabular}{|l|l|l|l|l|l|l|l|l|}		
								\hline
								& \multicolumn{4}{c|}{\textbf{Outgoing}}                                   & \multicolumn{4}{c|}{\textbf{Incoming}}                                 \\ \hline
								\textbf{}        & \textbf{\# pkts} & \textbf{flow vol.}      & \textbf{Pkt size} & \textbf{flow dur.}  & \textbf{\# Pkts} & \textbf{flow vol.}     & \textbf{Pkt size} & \textbf{flow dur.} \\ \hline
								\cellcolor[HTML]{FE0000} & (5, 0.3)         & (1103, 1645) & (214, 26)      & (2, 7)    & (4, 0.5)         & (995, 325)    & (218, 77)       & (0.42, 1.4)   \\ \hline
								\cellcolor[HTML]{333333} & (7, 0.8)         & (838, 216)     & (124, 30)      & (7, 19)   & (7, 1.1)         & (4993, 1222)  & (767, 222)     & (4.0, 15.1) \\ \hline
								\cellcolor[HTML]{3531FF} & (4, 1.1)         & (358, 138)   & (83, 19)     & (7, 32)  & (3, 0.7)         & (200, 95)   & (71, 2)        & (0.6, 2.6) \\ \hline
								\cellcolor[HTML]{00FDFF} & (7, 0.6)         & (2457, 420)     & (366, 55)   & (1, 6)  & (6, 0.9)         & (2770, 992) & (449, 104)  & (0.3, 0.8) \\ \hline
								\cellcolor[HTML]{FF40FF} & (5, 0.9)         & (663, 118)    & (125, 27)    & (15, 44) & (5, 0.8)      & (764, 295)   & (172, 80)    & (2.3, 9.7)  \\ \hline
							\end{tabular}
							\label{tab:clusterAttributes}	
							%			\vspace{2mm}
						\end{adjustbox}
					\end{minipage}\hfill
					\vspace{-5mm}
				\end{figure}
				
				The K-means algorithm finds groups of instances (aka clusters) for a given class similar to one another. Its centroid identifies every cluster, and an instance is associated with a cluster if the instance is closer to the centroid of that cluster than any other cluster centroids. For better performance of K-means models, it is essential to pre-process our data and tune certain parameters. We begin by recording each attribute's Z-score (\ie computing mean $\mu$ and standard deviation $\sigma$). We then normalize our dataset instances by calculating the deviation from the mean divided by the standard deviation for each attribute. %As discussed above that our instances are multidimensional (\ie 10 attributes) - we employ Principal Component Analysis (PCA) \cite{ding2004k} to project the instances to lower-dimensional space. 
				To tune a K-means model,  we need to compute the optimal number of clusters that are obtained by the elbow method \cite{ketchen1996application}. By applying this method, we found the optimal number of clusters for all three models to be equal to $5$. We show in Fig.~\ref{fig:httpKmeans} the resulting (color-coded) clusters  of training instances for HTTP flows. Note that our instances are multidimensional (\ie each instance contains 8 attributes), and thus cannot be easily visualized. Therefore, we employ Principal Component Analysis (PCA) to project the data instances onto two dimensions for illustration purposes. 
				
				Additionally, Table~\ref{tab:clusterAttributes} provides further insights into attributes of these five clusters. We make a few observations: these clusters cannot be distinctly identified by their packet count attribute -- almost similar across all clusters; cyan and black seem to represent top-heavy clusters (cyan by total flow volume in both directions and black by the average size of incoming packets); comparing red (top row) and purple (bottom row) clusters we find that the average size of packets in the red cluster is about 20\% (incoming) to 70\% (outgoing) larger than that of the purple cluster -- also, duration of flows in the red cluster is one order of magnitude shorter than that in the purple cluster.

				{
					OC-SVM is an algorithm that identifies anomalous instances by constructing a hyperplane boundary around expected training instances. 
					It comes with three main tuning parameters, namely \textit{Kernel},  \textit{gamma}, and \textit{nu} with default values, respectively equal to ``radial basis function (rbf)", ``scale" (inverse of product of attributes  count and attributes  variance) and ``0.5". 
					We tune OC-SVM similar to iForest and EiF where a parameter is fixed, and others are varied to find the best prediction. The optimal tuning parameters are found to be as follows: ``rbf'' kernel, gamma equals $0.125$, and nu equals $0.05$.}
				
				{
					\subsubsection{{Model Validation}}
					%\textbf{Model Validation:}  
					\label{subsubsec:ModelValidation}
					
					We validate the performance of our trained models against training instances (the only 60\% of the malicious dataset since benign instances are not used for training our one-class models). Validation results are shown by top row in Tables~\ref{tab:HTTPValidation} (HTTP model),~\ref{tab:HTTPSValidation} (HTTPS model), and~\ref{tab:UDPValidation} (UDP model).
					It is observed that three algorithms, namely iForest, EiF, and K-means, perform reasonably well (giving consistently high accuracy of more than $97\%$ across the three protocol-specialist models) during the validation phase. However, the OC-SVM algorithm performs relatively poorly even for validation, with the best malicious detection rate of less than $73\%$ given by the UDP model.}
				
				\begin{table}[t!]
					\centering
					\caption{Accuracy of HTTP models in correctly detecting malicious and benign flows.}
					\label{tab:HTTPValidation}
					\vspace{-2mm}
					\begin{adjustbox}{max width=0.99\columnwidth}   
						\renewcommand{\arraystretch}{1.5}
						\begin{tabular}{|c|l|l|l|l|l|l}
							\cline{1-6}
							\hline
							\multicolumn{1}{|l|}{}                                       & \multicolumn{1}{l|}{}                    & \multicolumn{1}{l|}{\textbf{True Positive}} & \multicolumn{1}{l|}{\textbf{False Positive}} & \multicolumn{1}{l|}{\textbf{True Negative}} & \multicolumn{1}{l|}{\textbf{False Negative}} \\ \hline
							\multicolumn{1}{|l|}{\multirow{2}{*}{\textbf{iForest}}}      & \multicolumn{1}{l|}{\textbf{Validation}} & \multicolumn{1}{l|}{98.71\%}                & \multicolumn{1}{l|}{-}                       & \multicolumn{1}{l|}{-}                      & \multicolumn{1}{l|}{1.29\%}                  \\ \cline{2-6} 
							\multicolumn{1}{|l|}{}                                       & \multicolumn{1}{l|}{\textbf{Testing}}    & \multicolumn{1}{l|}{97.8\%}                 & \multicolumn{1}{l|}{6.45\%}                  & \multicolumn{1}{l|}{93.55\%}                & \multicolumn{1}{l|}{2.2\%}                   \\ \hline
							\multicolumn{1}{|l|}{\multirow{2}{*}{\textbf{EiF}}}          & \multicolumn{1}{l|}{\textbf{Validation}} & \multicolumn{1}{l|}{99.05\%}                & \multicolumn{1}{l|}{-}                       & \multicolumn{1}{l|}{-}                      & \multicolumn{1}{l|}{0.95\%}                  \\ \cline{2-6} 
							\multicolumn{1}{|l|}{}                                       & \multicolumn{1}{l|}{\textbf{Testing}}    & \multicolumn{1}{l|}{98.85\%}                & \multicolumn{1}{l|}{6.09\%}                  & \multicolumn{1}{l|}{93.91\%}                & \multicolumn{1}{l|}{1.15\%}                  \\ \hline
							\multicolumn{1}{|l|}{\multirow{2}{*}{\textbf{K-means}}}      & \multicolumn{1}{l|}{\textbf{Validation}} & \multicolumn{1}{l|}{98.27\%}                & \multicolumn{1}{l|}{-}                       & \multicolumn{1}{l|}{-}                      & \multicolumn{1}{l|}{1.73\%}                  \\ \cline{2-6} 
							\multicolumn{1}{|l|}{}                                       & \multicolumn{1}{l|}{\textbf{Testing}}    & \multicolumn{1}{l|}{98.11\%}                & \multicolumn{1}{l|}{8.36\%}                  & \multicolumn{1}{l|}{91.64\%}                & \multicolumn{1}{l|}{1.89\%}                  \\ \hline
							\multicolumn{1}{|l|}{\multirow{2}{*}{\textbf{OC-SVM}}}      & \multicolumn{1}{l|}{\textbf{Validation}} & \multicolumn{1}{l|}{63.18\%}                & \multicolumn{1}{l|}{-}                       & \multicolumn{1}{l|}{-}                      & \multicolumn{1}{l|}{36.82\%}                 \\ \cline{2-6} 
							\multicolumn{1}{|l|}{}                                       & \multicolumn{1}{l|}{\textbf{Testing}}    & \multicolumn{1}{l|}{61.09\%}                & \multicolumn{1}{l|}{40.57\%}                 & \multicolumn{1}{l|}{59.43\%}                & \multicolumn{1}{l|}{38.91\%}                 \\ \hline
							\multicolumn{1}{|l|}{\multirow{2}{*}{\textbf{RandomForest}}} & \multicolumn{1}{l|}{\textbf{Validation}} & \multicolumn{1}{l|}{94.13\%}                & \multicolumn{1}{l|}{4.32\%}                  & \multicolumn{1}{l|}{95.68\%}                & \multicolumn{1}{l|}{5.87\%}                  \\ \cline{2-6} 
							\multicolumn{1}{|l|}{}                                       & \multicolumn{1}{l|}{\textbf{Testing}}    & \multicolumn{1}{l|}{81.32\%}                & \multicolumn{1}{l|}{20.82\%}                 & \multicolumn{1}{l|}{79.18\%}                & \multicolumn{1}{l|}{18.68\%}                 \\ \hline

							\cline{1-6}

						\end{tabular}
					\end{adjustbox}
				\end{table}

				\begin{table}[t!]
					\centering
					\caption{Accuracy of HTTPS models in correctly detecting malicious and benign flows.}
					\label{tab:HTTPSValidation}
					%\vspace{-2mm}
					\begin{adjustbox}{max width=0.99\columnwidth}   
						\renewcommand{\arraystretch}{1.5}
						\begin{tabular}{|c|l|l|l|l|l|l}
							\cline{1-6}
							\hline
							\multicolumn{1}{|l|}{}                                       & \multicolumn{1}{l|}{}                    & \multicolumn{1}{l|}{\textbf{True Positive}} & \multicolumn{1}{l|}{\textbf{False Positive}} & \multicolumn{1}{l|}{\textbf{True Negative}} & \multicolumn{1}{l|}{\textbf{False Negative}} \\ \hline
							\multicolumn{1}{|l|}{\multirow{2}{*}{\textbf{iForest}}}      & \multicolumn{1}{l|}{\textbf{Validation}} & \multicolumn{1}{l|}{98.96\%}                & \multicolumn{1}{l|}{-}                       & \multicolumn{1}{l|}{-}                      & \multicolumn{1}{l|}{1.04\%}                  \\ \cline{2-6} 
							\multicolumn{1}{|l|}{}                                       & \multicolumn{1}{l|}{\textbf{Testing}}    & \multicolumn{1}{l|}{97.16\%}                & \multicolumn{1}{l|}{5.49\%}                  & \multicolumn{1}{l|}{94.51\%}                & \multicolumn{1}{l|}{2.84\%}                  \\ \hline
							\multicolumn{1}{|l|}{\multirow{2}{*}{\textbf{EiF}}}          & \multicolumn{1}{l|}{\textbf{Validation}} & \multicolumn{1}{l|}{99.13\%}                & \multicolumn{1}{l|}{-}                       & \multicolumn{1}{l|}{-}                      & \multicolumn{1}{l|}{0.87\%}                  \\ \cline{2-6} 
							\multicolumn{1}{|l|}{}                                       & \multicolumn{1}{l|}{\textbf{Testing}}    & \multicolumn{1}{l|}{98.92\%}                & \multicolumn{1}{l|}{5.41\%}                  & \multicolumn{1}{l|}{94.59\%}                & \multicolumn{1}{l|}{1.08\%}                  \\ \hline
							\multicolumn{1}{|l|}{\multirow{2}{*}{\textbf{K-means}}}      & \multicolumn{1}{l|}{\textbf{Validation}} & \multicolumn{1}{l|}{98.42\%}                & \multicolumn{1}{l|}{-}                       & \multicolumn{1}{l|}{-}                      & \multicolumn{1}{l|}{1.73\%}                  \\ \cline{2-6} 
							\multicolumn{1}{|l|}{}                                       & \multicolumn{1}{l|}{\textbf{Testing}}    & \multicolumn{1}{l|}{98.45\%}                & \multicolumn{1}{l|}{8.08\%}                  & \multicolumn{1}{l|}{91.92\%}                & \multicolumn{1}{l|}{1.55\%}                  \\ \hline
							\multicolumn{1}{|l|}{\multirow{2}{*}{\textbf{OC-SVM}}}      & \multicolumn{1}{l|}{\textbf{Validation}} & \multicolumn{1}{l|}{65.59\%}                & \multicolumn{1}{l|}{-}                       & \multicolumn{1}{l|}{-}                      & \multicolumn{1}{l|}{34.41\%}                 \\ \cline{2-6} 
							\multicolumn{1}{|l|}{}                                       & \multicolumn{1}{l|}{\textbf{Testing}}    & \multicolumn{1}{l|}{64.90\%}                & \multicolumn{1}{l|}{37.06\%}                 & \multicolumn{1}{l|}{62.94\%}                & \multicolumn{1}{l|}{35.10\%}                 \\ \hline
							\multicolumn{1}{|l|}{\multirow{2}{*}{\textbf{RandomForest}}} & \multicolumn{1}{l|}{\textbf{Validation}} & \multicolumn{1}{l|}{93.89\%}                & \multicolumn{1}{l|}{3.27\%}                  & \multicolumn{1}{l|}{96.73\%}                & \multicolumn{1}{l|}{6.11\%}                  \\ \cline{2-6} 
							\multicolumn{1}{|l|}{}                                       & \multicolumn{1}{l|}{\textbf{Testing}}    & \multicolumn{1}{l|}{80.27\%}                & \multicolumn{1}{l|}{17.48\%}                 & \multicolumn{1}{l|}{82.52\%}                & \multicolumn{1}{l|}{19.73\%}                 \\ \hline 
							\cline{1-6}	
						\end{tabular}
					\end{adjustbox}
					%\vspace{-3mm}
				\end{table}

				\begin{table}[t!]
					\centering
					\caption{Accuracy of UDP models in correctly detecting malicious and benign flows.}
					\label{tab:UDPValidation}
					%\vspace{-2mm}
					\begin{adjustbox}{max width=0.99\columnwidth}   
						\renewcommand{\arraystretch}{1.5}
						\begin{tabular}{|c|l|l|l|l|l|l}
							\cline{1-6}
							
							\hline
							&                     & \textbf{True Positive} & \textbf{False Positive} & \textbf{True Negative} & \textbf{False Negative} \\ \hline
							\multirow{2}{*}{\textbf{iForest}}      & \textbf{Validation} & 96.96\%                & -                       & -                      & 3.04\%                  \\ \cline{2-6} 
							& \textbf{Testing}    & 96.28\%                & 7.86\%                  & 92.14\%                & 3.72\%                  \\ \hline
							\multirow{2}{*}{\textbf{EiF}}          & \textbf{Validation} & 97.60\%                & -                       & -                      & 2.40\%                  \\ \cline{2-6} 
							& \textbf{Testing}    & 97.03\%                & 7.01\%                  & 92.99\%                & 2.97\%                  \\ \hline
							\multirow{2}{*}{\textbf{K-means}}      & \textbf{Validation} & 98.76\%                & -                       & -                      & 1.24\%                  \\ \cline{2-6} 
							& \textbf{Testing}    & 97.58\%                & 7.97\%                  & 92.03\%                & 2.42\%                  \\ \hline
							\multirow{2}{*}{\textbf{OC-SVM}}      & \textbf{Validation} & 72.91\%                & -                       & -                      & 27.09\%                 \\ \cline{2-6} 
							& \textbf{Testing}    & 73.35\%                & 31.48\%                 & 68.52\%                & 26.65\%                 \\ \hline
							\multirow{2}{*}{\textbf{RandomForest}} & \textbf{Validation} & 95.32\%                & 3.79\%                  & 96.21\%                & 4.68\%                  \\ \cline{2-6} 
							& \textbf{Testing}    & 82.92\%                & 19.82\%                 & 80.18\%                & 17.08\%                 \\ \hline

							\cline{1-6}
							
						\end{tabular}
					\end{adjustbox}
					%	\vspace{-3mm}
				\end{table}

				\begin{figure*}[h!]
					\begin{center}
						%\vspace{-3mm}
						\mbox{
							%\hspace{-3mm}
							\subfloat[HTTP (iForest \& EiF)]{ \includegraphics[width=0.3\textwidth,height=0.20\textwidth]{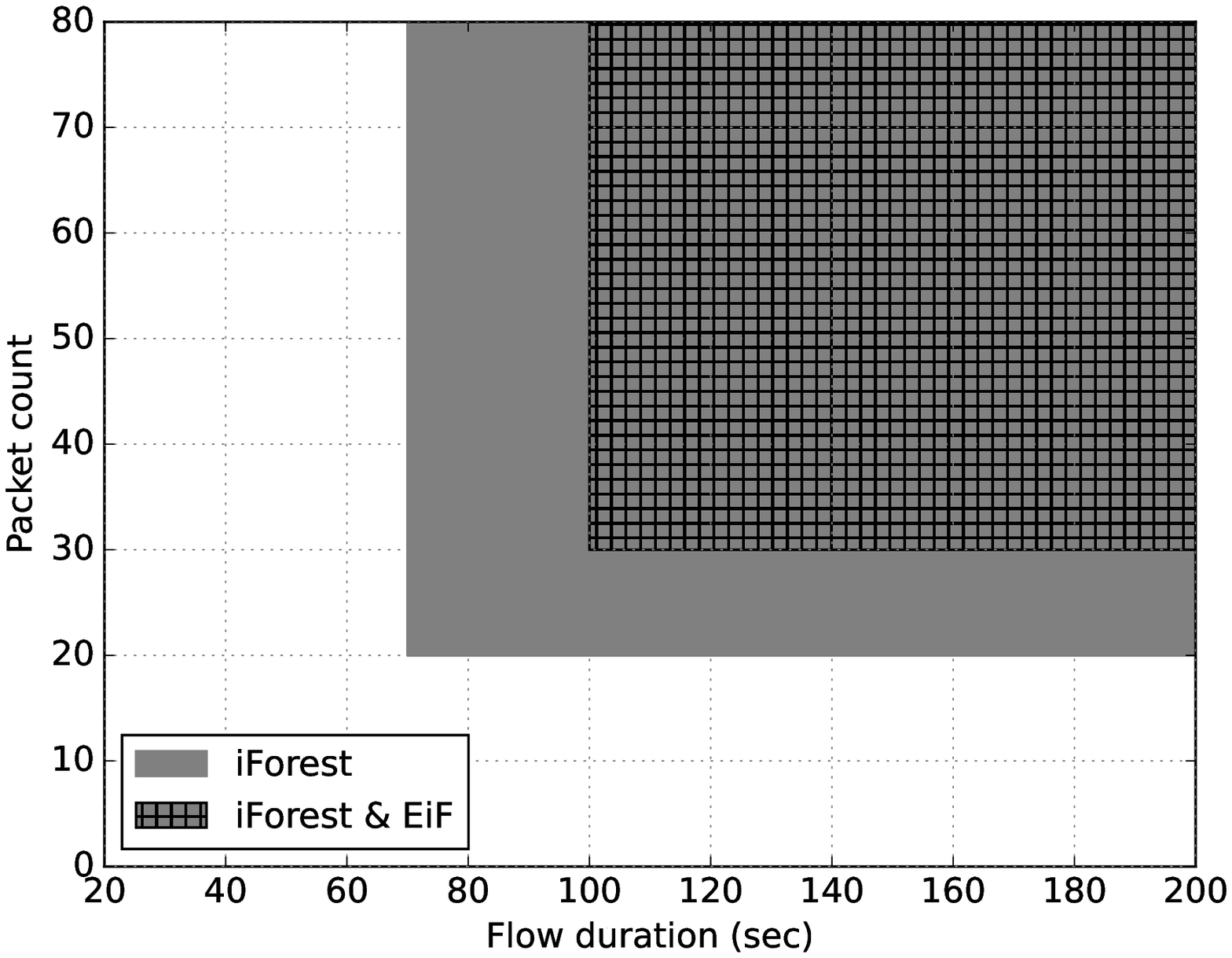}\label{fig:HTTP-misclassified-1}}\quad
							%\hspace{-5mm}
							
							\subfloat[HTTPS (iForest \& EiF)]{ \includegraphics[width=0.3\textwidth,height=0.20\textwidth]{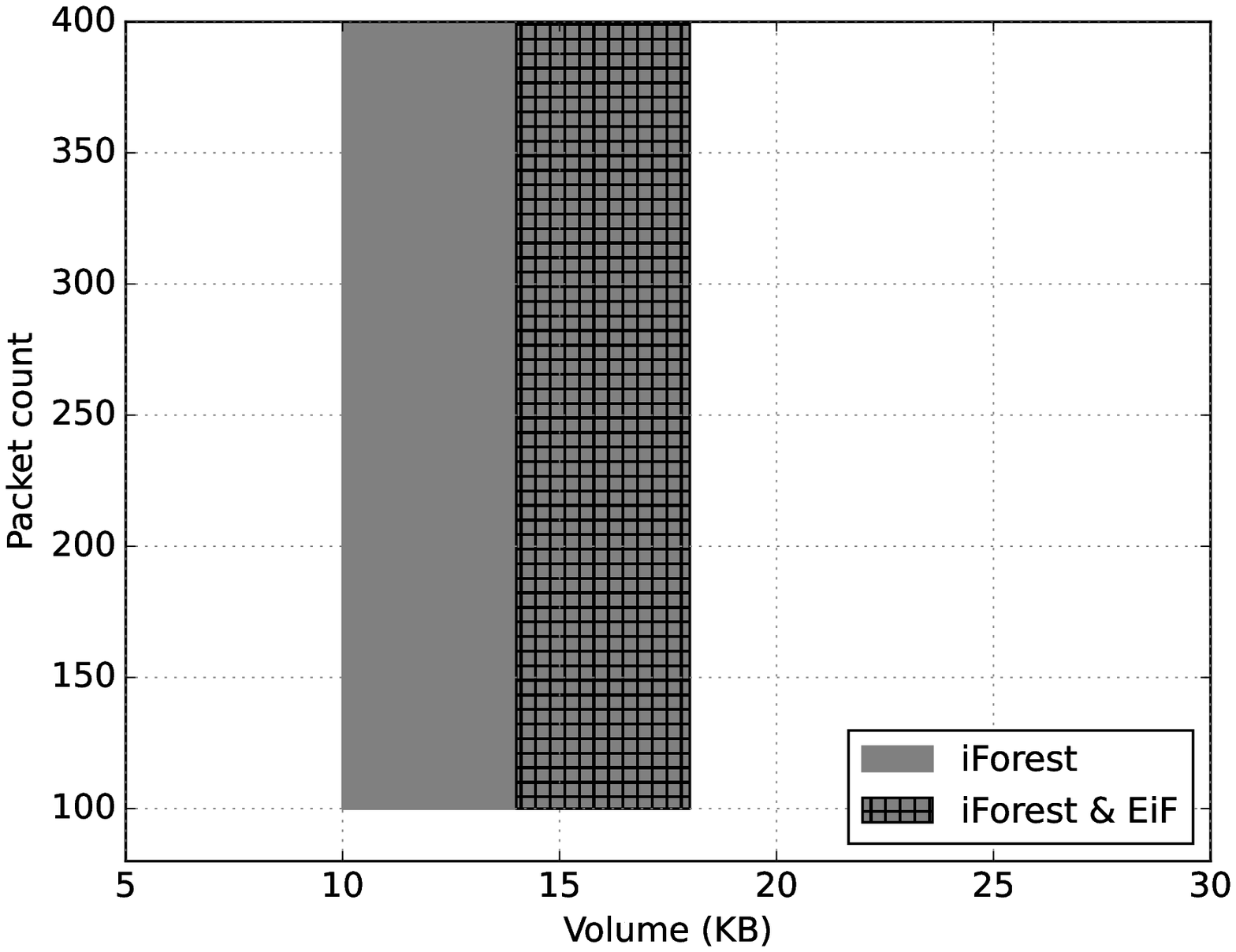}\label{fig:HTTPS-misclassified-1}}\quad
							%\hspace{-5mm}
							
							\subfloat[UDP (iForest \& EiF)]{ \includegraphics[width=0.3\textwidth,height=0.20\textwidth]{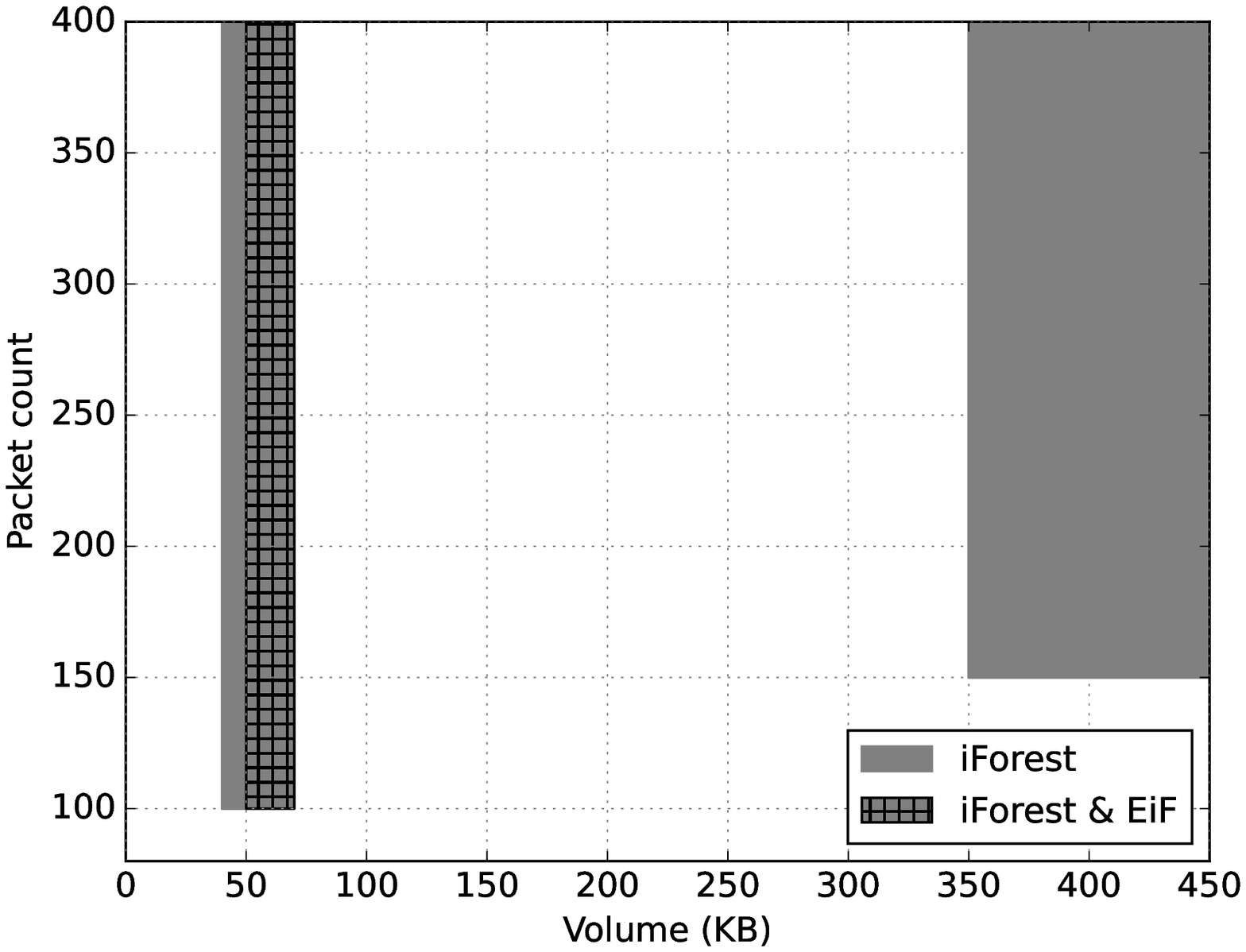}\label{fig:UDP-misclassified-1}}\quad
							%\hspace{-5mm}
							
						}
						
						\mbox{
							\subfloat[HTTP (K-means)]{ \includegraphics[width=0.3\textwidth,height=0.20\textwidth]{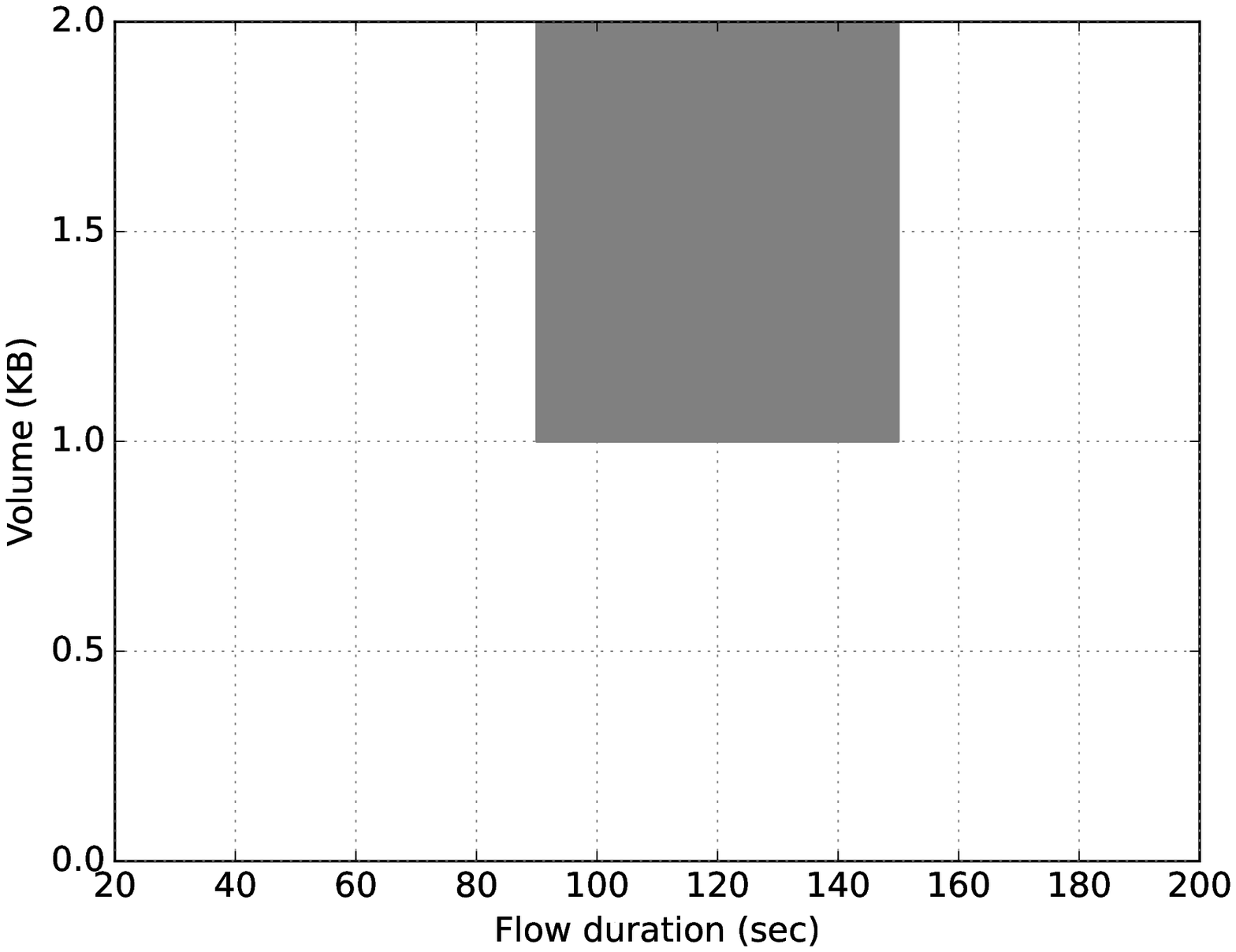}\label{fig:HTTP-misclassified-2}}\quad
							
							\subfloat[HTTPS (K-means)]{ \includegraphics[width=0.3\textwidth,height=0.20\textwidth]{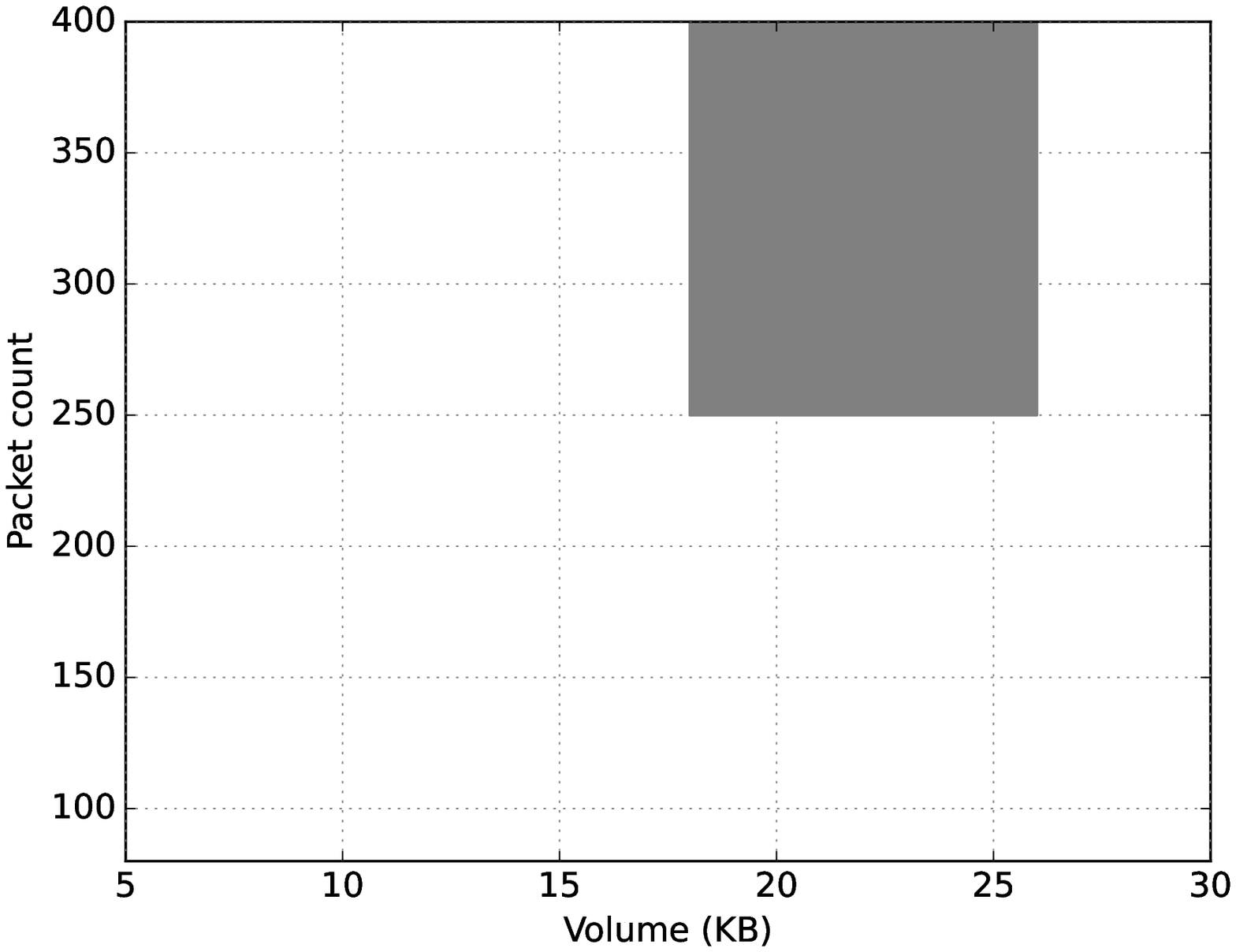}\label{fig:HTTPS-misclassified-2}}\quad
							
							\subfloat[UDP (K-means)]{ \includegraphics[width=0.3\textwidth,height=0.20\textwidth]{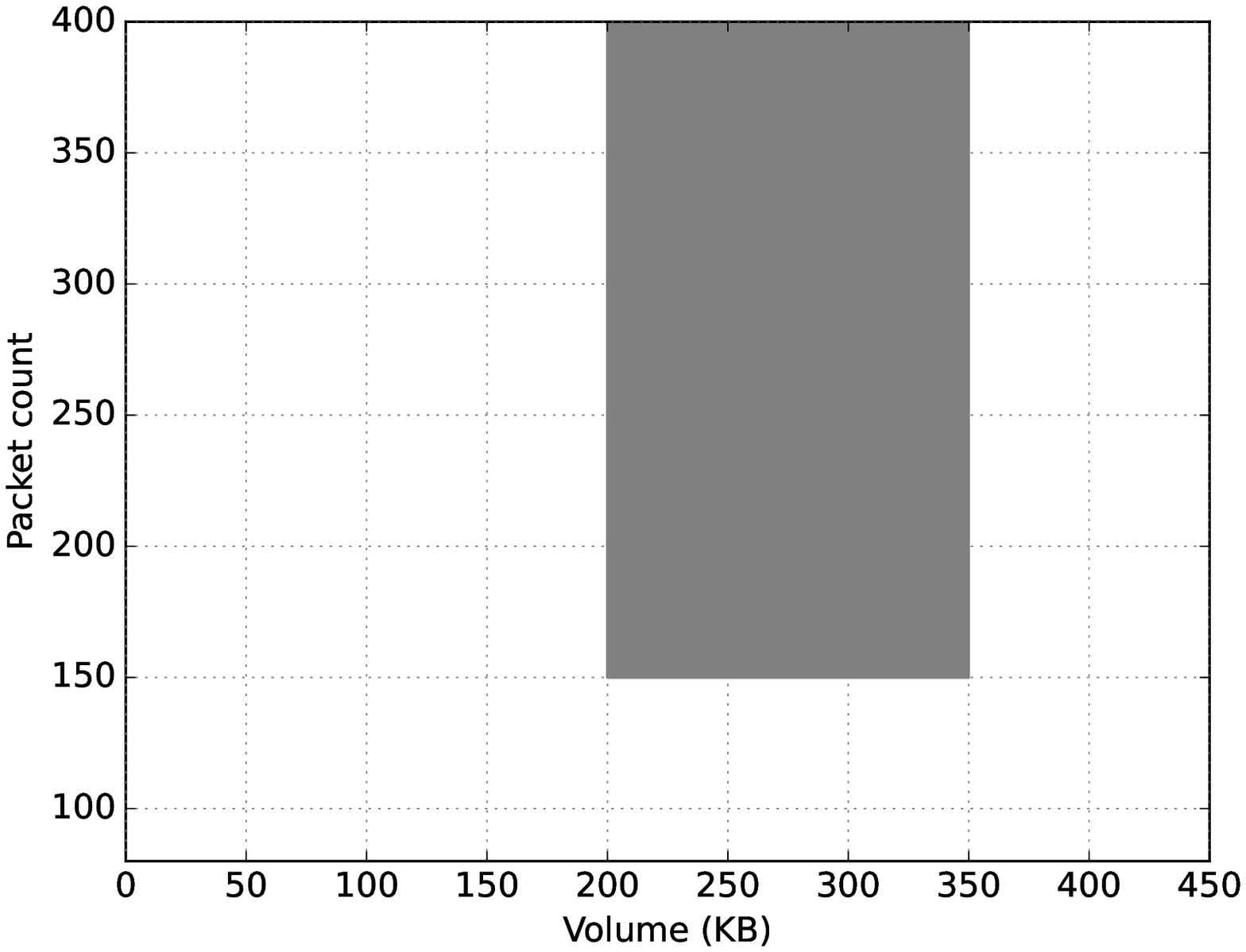}\label{fig:UDP-misclassified-2}}\quad
							
						}
						\vspace{-2mm}
						\caption{Attributes of misclassified flows: (a,d) HTTP, (b, e) HTTPS, and (c,f) UDP, across top performing classifiers.}
						%	\vspace{-4mm}
						\label{fig:MisclassifiedInstances}
					\end{center}
					%	\vspace{-2mm}
				\end{figure*}

				%{\color{magenta}
				{
					To better understand the inferencing capability of top-performing algorithms (\ie iForest, EiF, and K-means) we now focus on misclassified flows across these models. Fig.~\ref{fig:MisclassifiedInstances} visualizes an approximation of two-dimensional regions for key attributes of misclassified flows. We found that a diagnosed flow is misclassified (with a probability of more than $90$\%) by respective models if its attributes fall in the highlighted regions of Fig.~\ref{fig:MisclassifiedInstances}.
					Let us start with iForest and EiF models on the top row. First, it is seen that they misclassify those HTTP flows which are large in packet count and long in duration, as shown in Fig.~\ref{fig:HTTP-misclassified-1}, while for HTTPS and UDP flows, having large packet count (more than 100 packets) will probably lead to misclassification, as shown in Figures~\ref{fig:HTTPS-misclassified-1} and~\ref{fig:UDP-misclassified-1}.  
					K-means models, on the other hand, tend to misclassify HTTP flows with smaller volume but medium-length  (Fig.~\ref{fig:HTTP-misclassified-2}), and HTTPS and UDP flows with larger packet count (Fig.~\ref{fig:HTTPS-misclassified-2} and~\ref{fig:UDP-misclassified-2}).
					
					Taking these observations into account, none of these models seem distinct except by their overall accuracy, and hence we choose EiF models for our trial evaluation in \S\ref{sec:results}. 
					
				}

				%}
				
			}

			{
				\subsubsection{Models Testing}
				%\textbf{Model Testing:}
				Following validation, we quantify the performance of our trained one-class models against testing malicious instances (the remaining 40\% of malicious flows) as well as the entire set of benign instances.
				Testing results are shown by bottom row in Tables~\ref{tab:HTTPValidation} (HTTP model),~\ref{tab:HTTPSValidation} (HTTPS model), and~\ref{tab:UDPValidation} (UDP model). We observe that EiF consistently gives the best accuracy for both malicious and benign testing flows, across the three models -- true negatives of more than $97$\% and true positives of at least $93$\%. Unsurprisingly, OC-SVM is found to perform very poorly (compared with iForest, EiF, and K-means) during the testing phase with (malicious and benign) detection rates of mostly less than $70$\%. 
			}

			{
				\textbf{One-class versus Multi-class Models:}
				Lastly, to highlight the shortcomings of multi-class models in diagnosing the health of network traffic, we consider a two-class (malicious and benign) Random Forest classifier. It is trained with 60\% of the entire CTU-13 dataset, and its performance is tested with the remaining 40\% of instances. Validation and testing results are shown by the last column of Tables~\ref{tab:HTTPValidation},~\ref{tab:HTTPSValidation}, and~\ref{tab:UDPValidation}.
				It can be seen that Random Forest's detecting rates (true positives and true negatives) are around $80$\%, which is lower than those of iForest, EiF, and K-means particularly during the testing phase, though it gives acceptable detection rates ($\approx95$\%) during the validation phase.

				\subsection{Dynamic Traffic Selection using SDN} \label{subsec: SDN-SYSTEM}

				Fig.~\ref{fig:ProposedArch}  shows the functional blocks in our system architecture applied to a typical enterprise network. Enterprise users are on the left and can be on an access network  (wired and/or wireless). The Internet is on the right. Our solution is designed to be a ``bump-in-the-wire'' on the link at which traffic monitoring/management is desired (active management) -- an alternative approach is to feed our system a mirror of all network traffic (passive monitoring). Our system is therefore transparent to the network and does not modify packets in any way. {Further, no packet is sent to the SDN controller (for dynamic management of flow rules); instead, selected traffic (DNS packets, C\&C flows) that need inspection or diagnosis are sent as copies on separate interfaces of the switch, to which specialized traffic analytics engines (software inspection) are attached.} This protects the controller from overload from the data-plane, allowing it to scale to high rates and to serve other SDN applications. %Moreover, since incoming data packets are sent onwards by the switch immediately, the data-plane benefits in having minimal latency overhead and is protected from controller failures.
				Our solution comprises a DGA query finder (top right of Fig.~\ref{fig:ProposedArch}), which is fed by real-time incoming DNS responses, an SDN switch whose flow-table rules will be managed dynamically by API calls from the DGA finder to the SDN controller (center of Fig.~\ref{fig:ProposedArch}), a packet processing engine that extracts flow attributes (of suspicious network traffic only) feeding machine learning-based models (top left of Fig.~\ref{fig:ProposedArch}).

				\begin{figure}[t!]
					\centering
					\includegraphics[width=0.88\textwidth]{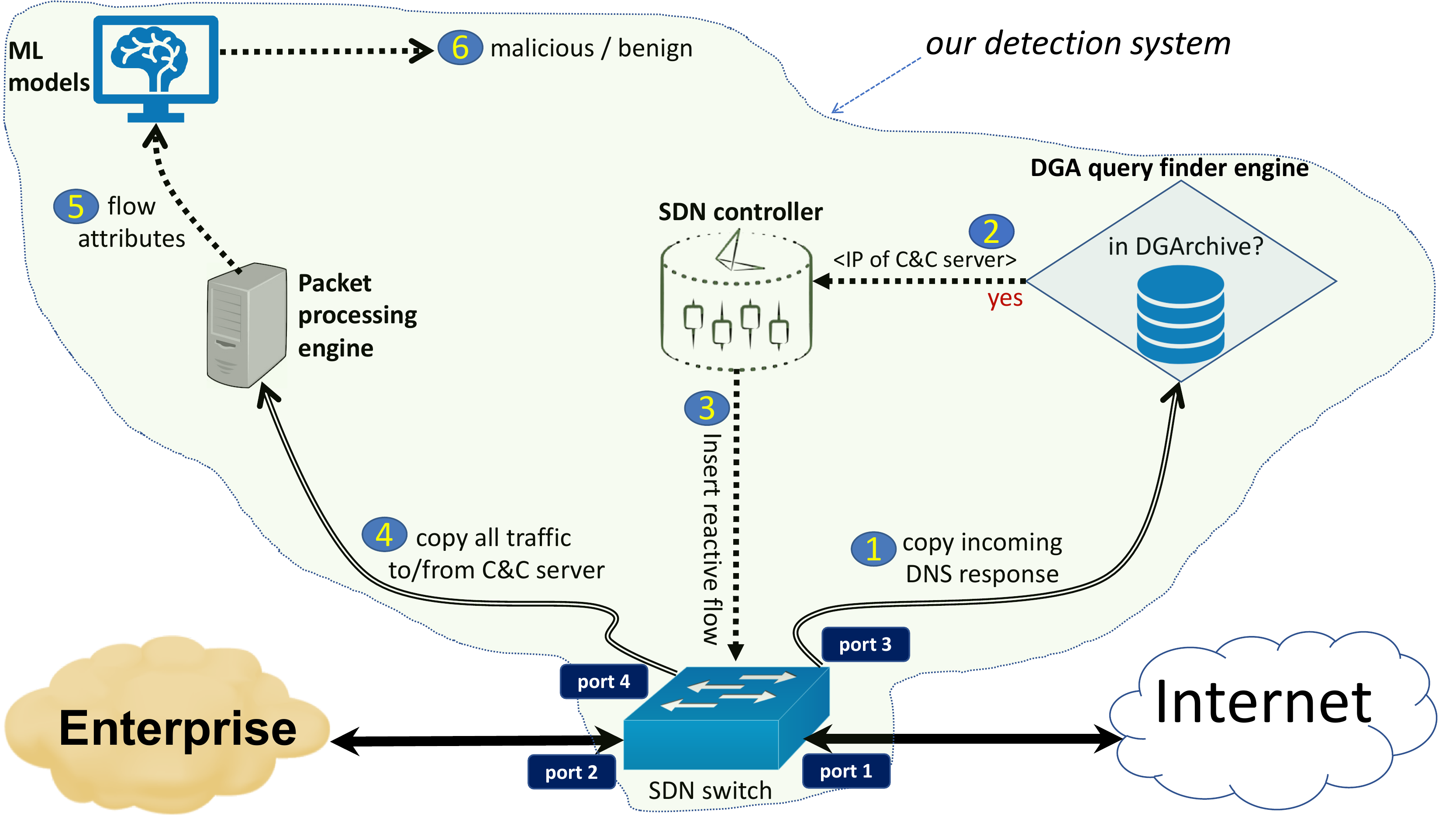}
					\caption{System architecture of our detection system.}
					\label{fig:ProposedArch}
				\end{figure}

				Note that our DGA query finder engine (handling DNS traffic of our university Internet link with a peak load of about 10 Gbps) runs on a virtual machine with 6 CPU cores, 8GB of memory, and storage of 500GB.
				We believe that cost and complexity of processing DNS packets in software can be reasonably managed at scale. In terms of processing costs, empirical results of our previous research studies \cite{PAM19,jawad2019DNS} on traffic analysis of our university campus network show that DNS constitutes a tiny fraction (less than 0.1\%) of total network traffic by volume. This corroborates with the analysis of the one-hour full campus traffic trace collected during the peak hour (\S\ref{subsec:1hourPCAP}), revealing that out of the total 3.2B packets only 2M are DNS -- about $0.06$\%. In terms of complexity of measurement, it is relatively easy to capture DNS traffic with only a few flow entries (\ie mirroring IPv4 and IPv6 UDP packets to/from port 53) in an SDN switch.

				We believe that the key contribution of this chapter is our detection method that dynamically (using SDN) and confidently (using specialized classifiers) identifies malicious flows established after DGA queries. Our system infers suspicious hosts by employing a broad signature from the public database DGArchive. It verifies its initial inference by applying specialized one-class classifiers. The key advantage of our method is its scalability and low rate of false alarms.
				Note that our detection method primarily relies on the real-time mapping (dynamic and/or static) of DGA-enabled malware domains to their IP addresses. The IP address of Internet-based (C\&C) servers is dynamically obtained from the respective DNS responses.

				For the SDN switch, we use a fully Openflow 1.3 compliant NoviSwitch 2122 \cite{noviflow} which is controlled by the Ryu SDN controller \cite{ryu}. The switch provides 240 Gbps of throughput, up to one million TCAM flow-entries, and millions of exact-match flow-entries in DRAM, and we found it to amply cater to the requirements of this project. Furthermore, we use a combination of proactive and reactive entries in the switch flow table. A proactive entry is statically pushed by the controller so that all DNS response packets (\ie UDP source port 53) received from the Internet are forwarded (on port 2) and mirrored (on port 3) to the DGA finder, as shown by step \ciao{1} in Fig.~\ref{fig:ProposedArch}. 
				{
					The finder looks up the queried domain name against DGArchive. If found, it extracts the IP address of the server resolved by DNS responses and subsequently calls the SDN controller (step \ciao{2}) that results in the insertion of a reactive flow-entry, shown by step \ciao{3}. 
					Note that our malware detection method is limited by the knowledge-base provided by the DGArchive repository. This means that to detect new families of DGA-enabled malware flows, whose DNS query name is outside of this database, one needs to update the DGArchive -- it has been consistently maintained over the past five years.
					
					One may also want to examine the registration date of domain names before mirroring the traffic, given the overlap between some DGA domains and legitimate domains (discussed in \S\ref{subsec:ourData}). We believe that examining the registration date of domains in ``real-time'' can be challenging since many TCP/UDP flows (between internal host and external C\&C server) often commence shortly (less than 100ms) after their DNS resolution (will be discussed later in \S\ref{sec:results}, Fig.~\ref{fig:ccdfDelay}), and hence the delay of registration lookup may cause missing the C\&C communication flow. In addition, mirroring ``selected flows'' which carry a fairly small number of packets and are relatively short (will be discussed later in \S\ref{sec:results}, Fig.~\ref{fig:ccdfflow}) would not incur significant cost of software processing or switch TCAM entries (will be discussed later in \S\ref{sec:results}, Fig.~\ref{fig:sdnFlows}).}
				%{ This step will ensure that our proposed model is scalable and computationally inexpensive as we are not inspecting all network packets.}

				Reactive rules that match the server's IP address (two rules per server: one matching source and one matching the destination IP address) are of the highest priority and get installed as a consequence of DGA queries detected by the DGA finder. To protect the SDN switch from TCAM exhaustion (scalable management of TCAM usage), reactive rules are automatically timed out after a period equals to the TTL obtained from their corresponding response. The reactive flow entries provide filtered packets (to/from potential C\&C servers) to the packet processing engine on port 4 in step \ciao{4}. Our packet processing engine (run on a generic server configured with Ubuntu version 16.04.4) analyzes suspicious traffic filtered and mirrored by the SDN switch. It constructs flow-level attributes that are fed, in step \ciao{5}, to the machine learning (ML) models for prediction. The models are one-class classifiers (discussed in \S\ref{subsec:CTU-modeling}) distinguishing, step \ciao{6}, malicious flows from benign ones.  We will describe in \S\ref{sec:results} the performance of our prototype under real traffic of our campus network.  
				Note that the proposed approach is generic and can be readily applied to any context, given that specialized ML-models are trained by the malware dataset of those domains (e.g., cloud-based or IoT-based setups).

				\section{Evaluation Results}\label{sec:results}

				\begin{table}[t!]
					\centering
					\caption{Distribution of malware families among suspicious flows, selected and mirrored by our SDN system.}
					\label{tab:distro_maliciousFlows45Days}
					\begin{adjustbox}{max width=0.88\textwidth}
						\begin{tabular}{@{}lr@{}}
							\toprule
							\textbf{DGA-enabled malware (\# C\&C servers)} & \multicolumn{1}{r}{\textbf{\# flows [\%]}}      \\ \midrule
							ModPack ($7$)    & $35941$    [$63.10$\%]   \\ \midrule
							Matsnu ($2$)    & $20806$    [$36.53$\%]
							\\ \midrule
							Ramnit ($2$)    & $169$ [$0.30$\%]
							\\ \midrule
							Suppobox ($8$)    & $37$ [$0.06$\%]
							\\ \midrule
							Bamital    ($1$) & $4$ [$0.01$\%]   \\ \midrule
						\end{tabular}
						%    \vspace{-3mm}
					\end{adjustbox}
				\end{table}

				{
					We have implemented a fully functional system (shown in Fig.~\ref{fig:ProposedArch}) and operated it during a 50-day trial (3-Dec-2019 to 21-Jan-2020) under entire campus network traffic. During this trial, our system automatically selected, mirrored, and recorded suspicious DGA-based flows (after DGA-based DNS responses) in real-time at the line rate of up to 10 Gbps using reactive SDN rules. In addition to suspicious flows, the entire DNS traffic was recorded during this trial to correlate DGA queries with their corresponding suspicious flows. In this section, we evaluate the efficacy of our trained models by applying them (in offline mode) to recorded suspicious flows from the trial, diagnosing their health (malicious or benign?). 
				}
				
				{\subsection{SDN-selected DGA Flows} %We now evaluate the efficacy of our approach during a 50-day trial by applying it to suspicious traffic (of the University campus network) that is reactively selected and mirrored by our SDN system. 
					We perform a correlation between SDN-selected suspicious flows and DNS responses of DGA families.} Table \ref{tab:distro_maliciousFlows45Days} summarizes the distribution of suspicious flows (a total of 56,957) that are associated with (DNS queries of) five DGA families listed in rows. It is seen that ModPack family dominates with $63.10$\% of flows, followed by the Matsnu family with $36.53$\% of flows. Note that these flows are exchanged between enterprise hosts and a small number (a total of 20) of C\&C servers on the Internet -- the number of unique servers (IP addresses) is listed in the bracket in front of their respective DGA family. Furthermore, we observe that the activity of these servers varies across families. For example, seven servers associated with the ModPack family handle about 36K flows while 8 Suppobox servers exchange 37 flows over the 50-day period of our experiment.

				During our trial, we found that C\&C servers of 14 DGA-enabled malware families were successfully resolved (476K DNS responses). However, the C\&C servers of only 5 families were contacted by internal hosts, following their DNS resolution. We observe that those 9 malware families which generated no C\&C communications, contribute to only  $0.1$\% of the total resolved DGA queries. 
				We note that the probability of communicating with an intended C\&C  server following a successful DNS resolution varies across families. For example, the top two families display a completely different pattern. ModPack, which dominates by making 35.9K flows, had 475K DNS queries resolved. Apparently, most of these DNS queries (with an average TTL value of $\approx$5 minutes) are made purely to keep their local cache updated by the latest IP address of their intended server. On the other hand, Matsnu generated 20.8K flows with only 22 DNS responses (with much longer TTL values averaged at $\approx$5 hours) during our 50-day trial. In this case, DNS queries are only made when certain communications are desired. Of the remaining three families, Ramnit behaves similar to Matsnu by exchanging 169 C\&C flows with 14 DNS responses (average TTL $\approx$4 hours), while Suppobox (37 flows, 97 DNS responses, average TTL $\approx$1.5 hours) and Bamital (4 flows, 11 DNS responses, average TTL $\approx$10 minutes) display a pattern like ModPack. %{\color{magenta}[HASSAN] can you verify that long TTL families make DNS query whenever they want to send data?[JAWAD: I have verified that after the successful DNS response of these long TTL families, they generate the malicious flows (data).]}     

				\subsection{Diagnosing  DGA Flows} 
				Of the total of 56,957 suspicious flows (mirrored by our system during this trial period), 35645 are HTTP, 19674 are HTTPS, and 1638 are UDP. { Recall from the previous section that we trained three EiF models (HTTP, HTTPS, and UDP), each with their respective malicious flows extracted from the CTU-13 dataset.}

				%We begin with the HTTP model, we passed all the suspicious flows (35.6K) from 50 days-worth of data.
				{
					Table~\ref{tab:evalResults} shows the testing results of suspicious flows against their corresponding EiF model. 
					It can be seen that more than $90$\% of suspicious flows across the three types (HTTP, HTTPS, and UDP) have been classified as malicious. These high detection rates verify the efficacy of our trained models in diagnosing suspicious DGA flows.
					
					In the absence of ground-truth data whether suspicious flows are indeed malicious or not, we further analyzed these selected flows and their attributes. It is seen that $99.94$\% of suspicious HTTP flows are predicted as malicious. We found that a vast majority of suspicious HTTP flows ($33.5$K) consist of the only three-way handshake (initiated by internal hosts) followed by a TCP RST (reset) packet sent by the initiating host. Almost all of these $33.5$K HTTP flows are classified as malicious -- the CTU-13 dataset also had 810 flows ($17$\% of total HTTP flows) of this kind. Excluding these specific $33.5$K HTTP flows, again a vast majority ($98.33$\%) of the remaining $2046$ suspicious  HTTP flows are predicted to be malicious, highlighting the fact that their traffic behavior conforms to the norms of known malware (\ie the CTU-13 dataset). Only 34 HTTP flows are classified as benign that carry a large number of packets (about 35 packets) compared to the malware norms (5 packets) -- this corroborates to a great extent with our observations from misclassified flows during model validation, shown in Fig.~\ref{fig:HTTP-misclassified-1}.
					Similarly, we investigated the attributes of suspicious HTTPS and UDP flows that are classified as benign during trial evaluation, and found that benign-predicated: HTTPS flows contain an average of more than 200 packets, carrying relatively high volume ($\approx$$25$ KB) of traffic (conforming to Fig.~\ref{fig:HTTPS-misclassified-1}); and UDP flows contain 320 to 390 packets, resulting in flow volume of average $350$ KB (conforming to Fig.~\ref{fig:UDP-misclassified-1}). In summary, suspicious flows which are classified as benign are probably misclassified by the EiF models. This means that traffic flows subsequent to a DGA-based query are likely to be malicious.   
				}

				\begin{table}[t!]
					\centering
					\caption{Results of testing suspicious flows against their corresponding EiF models.}
					\label{tab:evalResults}
					\begin{tabular}{lll}
						\begin{adjustbox}{max width=0.88\textwidth}
							\begin{tabular}{@{}lrrrr@{}}
								\toprule
								%\textbf{} & \multicolumn{1}{r}{\textbf{}}     \\ \midrule
								& \textbf{HTTP} & \textbf{HTTPS} & \textbf{UDP} & \textbf{Aggregate}\\ \midrule
								\textbf{\# suspicious flows}   & $35645$& $19674$& $1638$ & $56957$\\ \midrule
								~~~~\% malicious & $99.94$\%   & $93.92$\% & $92.71$\% & $97.63$\%\\ \midrule
								~~~~\% benign     & $0.06$\%        & $6.08$\%     &$7.29$\% & $2.37$\%\\ \midrule 
								\textbf{\# internal hosts making suspicious flows}     & $262$ & $2367$ & $45$ & $2488$\\ \midrule
								~~~~{\# hosts with all flows malicious}               & $239$ & $1731$  & $20$ & $1818$\\ \midrule
								~~~~{\# hosts with malicious and benign flows} & $17$    & $533$   & $25$ & $567$\\ \midrule
								~~~~{\# hosts with all flows benign} & $6$ & $103$& $0$ & $103$\\ \midrule
							\end{tabular}
							
						\end{adjustbox}
					\end{tabular}
				\end{table}

				\subsection{Infected Hosts Initiating Malicious DGA Flows}
				We have identified malicious flows succeeding DGA queries, but network operators are more interested in identifying hosts infected by malware. We mapped all of the suspicious flows (dynamically selected and mirrored by our system) to their corresponding hosts (inside the campus network) that initiated those flows. These hosts are in three categories, as shown by the bottom rows in Table~\ref{tab:evalResults}: (a) hosts with all of their suspicious flows are predicted as malicious (``pure-malicious''), (b) hosts with some of their suspicious flows are predicted as malicious, and some as benign (``mix-malicious-benign''), and (c) hosts with all of their suspicious flows are predicted as benign (``pure-benign'').

				\begin{figure}[t!]
					\centering
					\includegraphics[width=0.88\textwidth,height=0.47\textwidth]{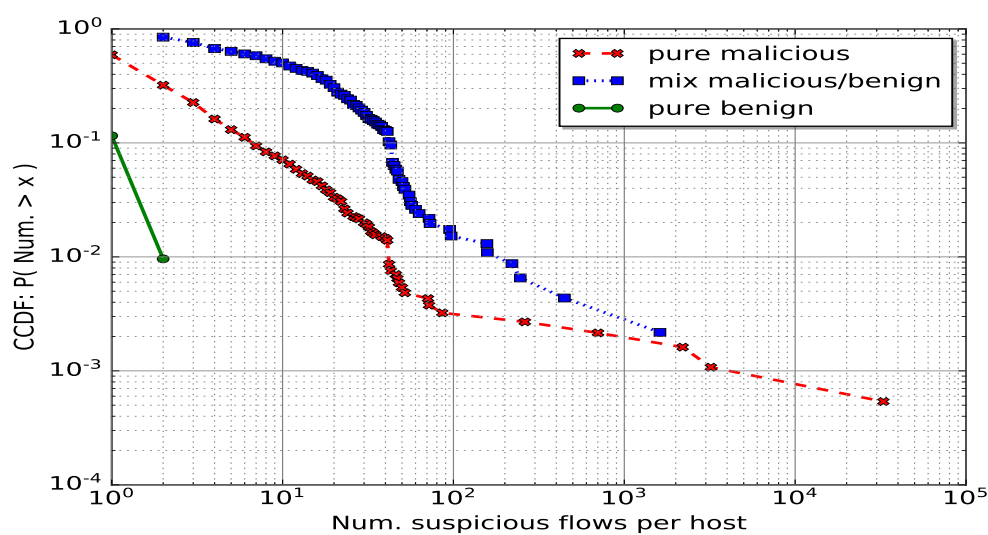}
					\vspace{-2mm}
					\caption{CCDF of number of suspicious flows per host.}
					\label{fig:CCDFsuspiciousFlows}
				\end{figure}

				\begin{figure}[t!]
					\centering
					\includegraphics[width=0.88\textwidth,height=0.49\textwidth]{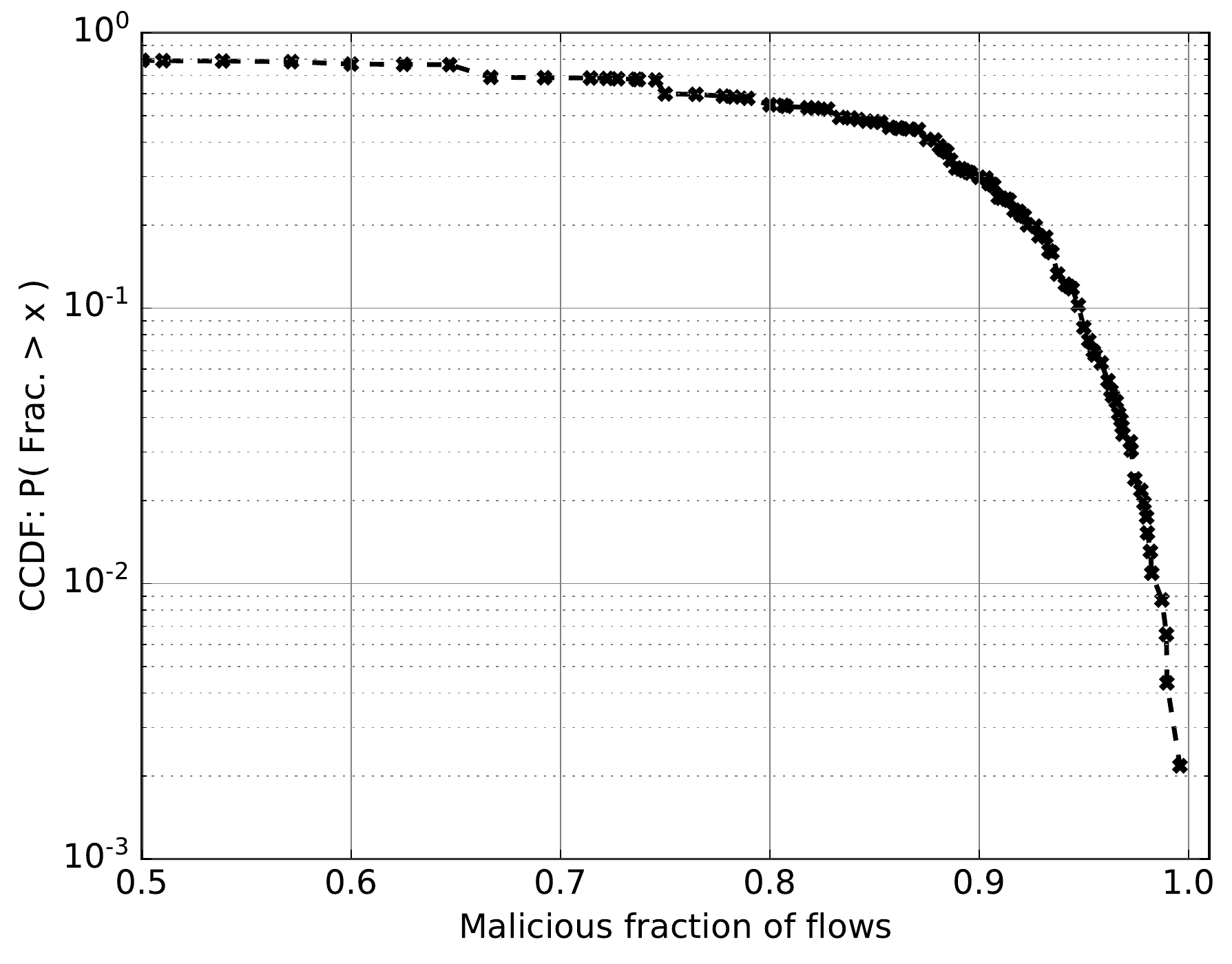}
					\caption{CCDF of malicious fraction of flows per host in mix-malicious-benign category.}
					\label{fig:CCDFFractionMaliciousFlows}
				\end{figure}

				It can be seen that at the aggregate level, shown in the last column of Table~\ref{tab:evalResults}, the pure-malicious category dominates with 1818 hosts, followed by mix-malicious-benign and pure-benign, respectively with 567 and 103 hosts -- across the campus network, there are a total of 2488 internal hosts that generate some suspicious flows (HTTP and/or HTTP and/or UDP). 
				Note that 302 of these hosts are NAT gateways (each representing several wireless hosts), and the remaining 2186 are actual end-hosts (clients or servers which are not NATed).

				Focusing on the infected hosts, we found that they belong to 226 different subnets of size /24. Of these subnets, $34$\% have more than ten infected hosts, and $25$\% have more than 20 infected hosts. These insights help network operators who want to tighten the security posture of certain subnets, those that have some degree of infection. % -- the distribution of infected host count per subnet is shown in Fig.~\ref{fig:HostPerSubetCCDF}. 

				Next, we performed reverse lookups to infer the nature of infected hosts, and found that: 302 hosts are WiFi NAT gateways (from two dedicated subnets) with names as ``{\myverb{SSID-pat-pool-a-b-c-d.gw.univ-primay-domain}}'' where ``{\myverb{a.b.c.d}}'' is the public IP address of the NAT gateway, and SSID is the WiFi SSID for the University campus network; 34 hosts are the Mac devices (spread across 23 subnets) with names containing strings like ``{\myverb{mbp}}'' or ``{\myverb{imac}}''; 616 hosts are returned with names including strings like `{\myverb{desktop}}''  that they indicate are a user desktop machine (Windows/Linux); and the remaining 1433 hosts (spread across 186 subnets) are returned with no name -- they are also typically regular end-hosts.

				We plot in Fig.~\ref{fig:CCDFsuspiciousFlows} the CCDF of suspicious flow count per host within each of the three categories mentioned above. It can be seen that $99\%$ of pure-benign hosts have at most two suspicious flows -- rarely active. In the other two categories, instead, far more suspicious flows are observed per host (average of 22 flows)  -- several hosts in both of these very active categories have more than 1000 flows. Finally, focusing on the mix-malicious-benign category, we see that $50$\% of hosts have more than ten suspicious flows, represented by the tail of their CCDFs. Also, more than $80$\% of hosts in the mix category have more than half of their suspicious flows classified as malicious -- the CCDF plot in Fig.\ref{fig:CCDFFractionMaliciousFlows} particularly illustrates the distribution of the malicious fraction of flows across all hosts of this category. For these reasons, we deem the two active categories, consisting of 2385 hosts, to be likely ``infected'' by malware.

				\begin{figure*}[t!]
					\begin{center}
						%\vspace{-3mm}
						\mbox{
							%\hspace{-3mm}
							\subfloat[Aggregate of all hosts.]{ \includegraphics[width=0.48\textwidth]{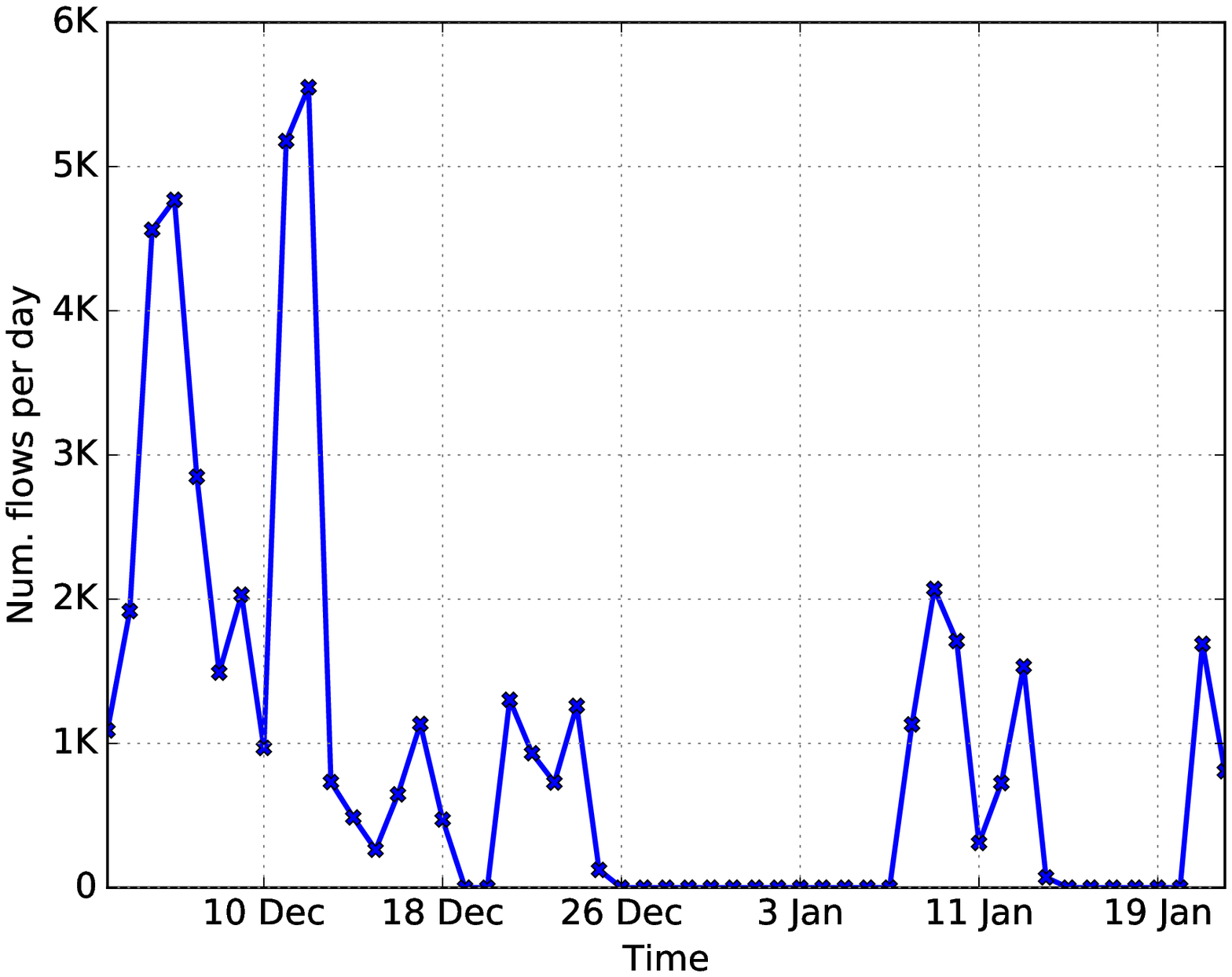}\label{fig:50DaysAllHostFlows-1}}\quad
							%\hspace{-5mm}
							\subfloat[Most active infected end-host (top) and a NAT gateway (bottom).]{
								\includegraphics[width=0.48\textwidth]{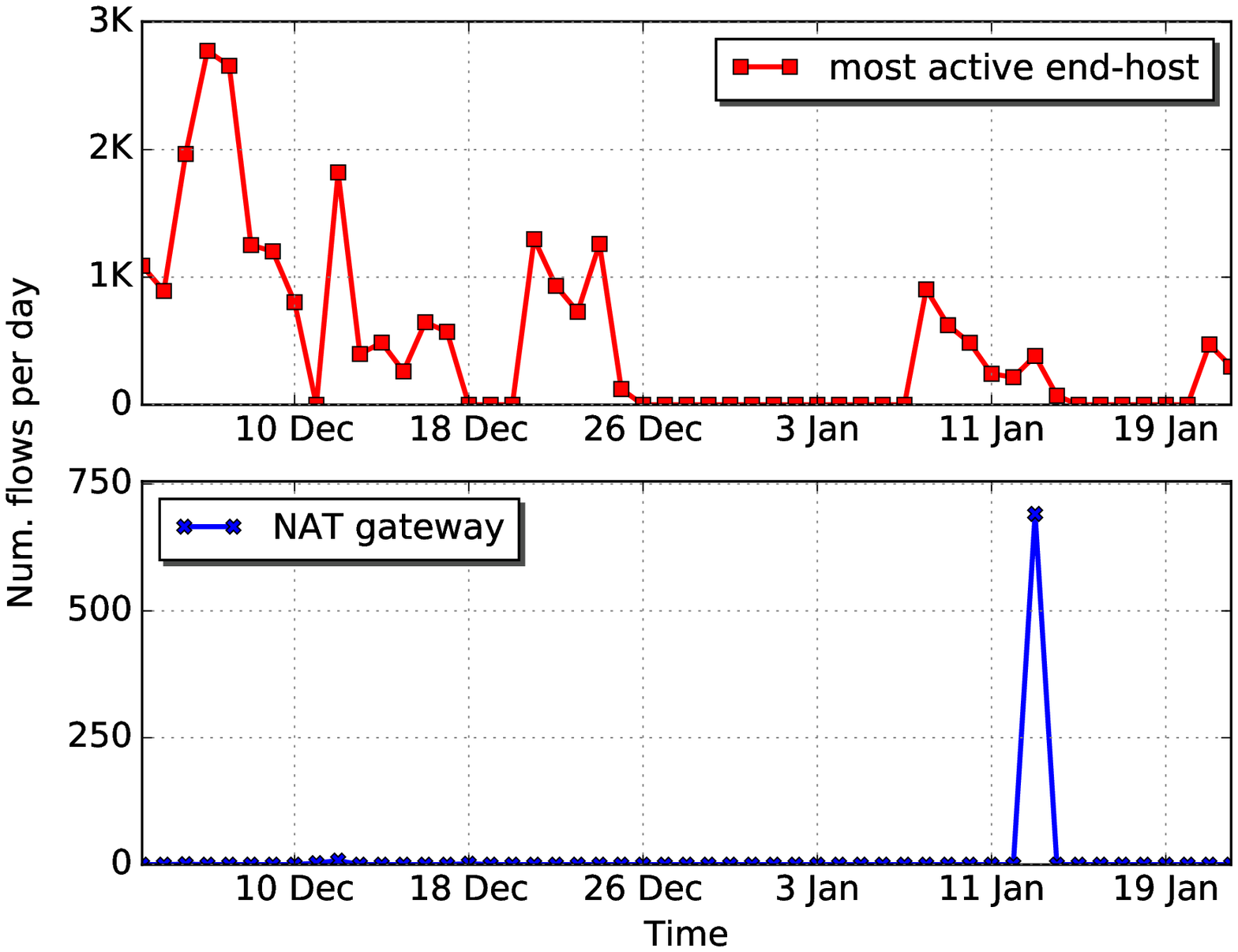}    \label{fig:worstHost-1}}\quad
						}
						
						\vspace{-2mm}
						\caption{Time trace of active malicious flows (between infected internal hosts and malware servers) during the 50-day trial for: (a) aggregate of all campus hosts, and (b) most active infected end-host (top) and a NAT gateway (bottom).}
						\vspace{-6mm}
						\label{fig:ActiveFlows-1}
					\end{center}
					\vspace{-2mm}
				\end{figure*}

				\begin{figure}[t!]
					\centering
					\includegraphics[width=0.88\textwidth,height=0.47\textwidth]{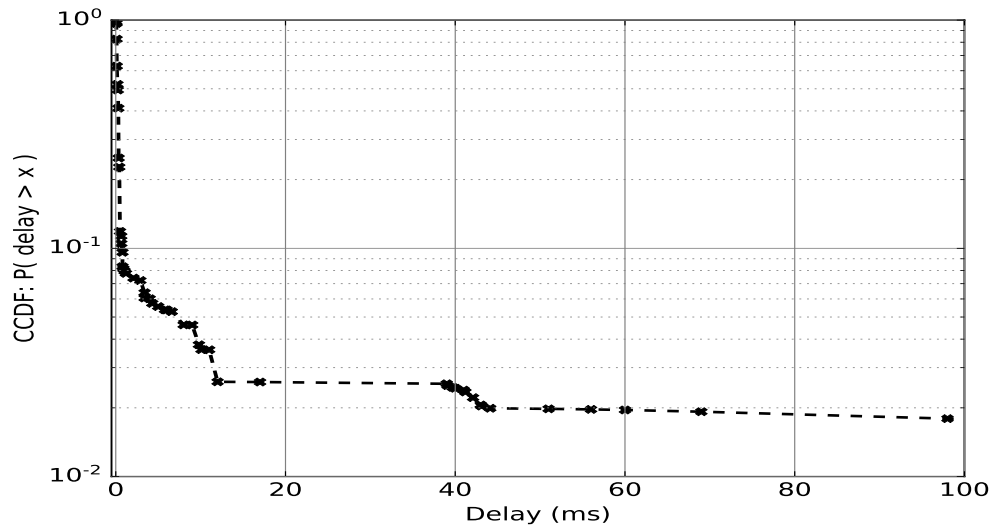}
					\vspace{-3mm}
					\caption{{Delay between DGA-based DNS responses and commencement of their subsequent TCP/UDP flow.}}
					\label{fig:ccdfDelay}
					\vspace{-6mm}
				\end{figure}

				We plot in Fig.~\ref{fig:ActiveFlows-1} the daily time trace of active malicious flows detected during our trial. Each data-point represents the number of active flows over a 24-hour window. 
				Looking at Fig.~\ref{fig:50DaysAllHostFlows-1}, we observe that the activity of DGA-enabled C\&C communications across the aggregate of all campus hosts was relatively high during the first half of December (peaking at a total of $\approx$5500 flows on 12th Dec), gradually fell reaching to a complete no-show during holiday shutdown periods (between 25-Dec-2019 and 8-Jan-2020), and afterward revived slowly. This pattern of malware activity correlates (to a great extent) with the number of active users on the network, suggesting infected regular hosts. 
				Moving on to Fig.~\ref{fig:worstHost-1}, we see malicious flow activity of two hosts. The top subplot corresponds to the most active infected end-host that generated a total of 24.8K malicious flows (all preceded by DGA responses of ModPack family). This host was active at the beginning of the trial (first three weeks in December) by making on average more than 1000 malicious flows per day. Still, its hyperactivity slowly diminished in the second half of our trial.   
				The bottom subplot is a NAT gateway that displays a different pattern of malware activity. It made a total of 14 malicious flows (pertinent to Matsnu family) during December, went silent for three weeks, and then suddenly became heavily active in the middle of January by making about 700 malicious flows from the Suppobox family. It is expected to see a diversity of families in the NAT gateway since they make flows on behalf of a group of end-hosts.

				{
					We show in Fig.~\ref{fig:ccdfDelay} the CCDF plot of delay between DGA responses (resolution of DGA query for C\&C server) and the commencement of the first associated TCP/UDP flow. It can be seen that more than $90$\% flows start in less than $2$ ms (very shortly) after their DNS resolution. Since flow arrival delays do not go beyond $100$ ms, it is crucial for our detection system to immediately insert a corresponding reactive rule into the SDN switch, capturing and diagnosing the communication between the internal host and the external C\&C server.
					%(1st week of trial period 3rd Dec - 9th Dec 2019) where each data point represents the delay after successful DGA-enabled queries and commencement of TCP/UDP flows. We observed that 90\% of flows have a delay of less than 2ms. This means that most of the TCP/UDP flows between the internal host and their C\&C servers commence shortly (less than 100ms) after a successful DGA-enabled response. 

					\begin{figure}[t!]
						\begin{center}
							%\vspace{-3mm}
							\mbox{
								%\hspace{-3mm}
								\subfloat[{Packet count.}]{
									\includegraphics[width=0.48\textwidth]{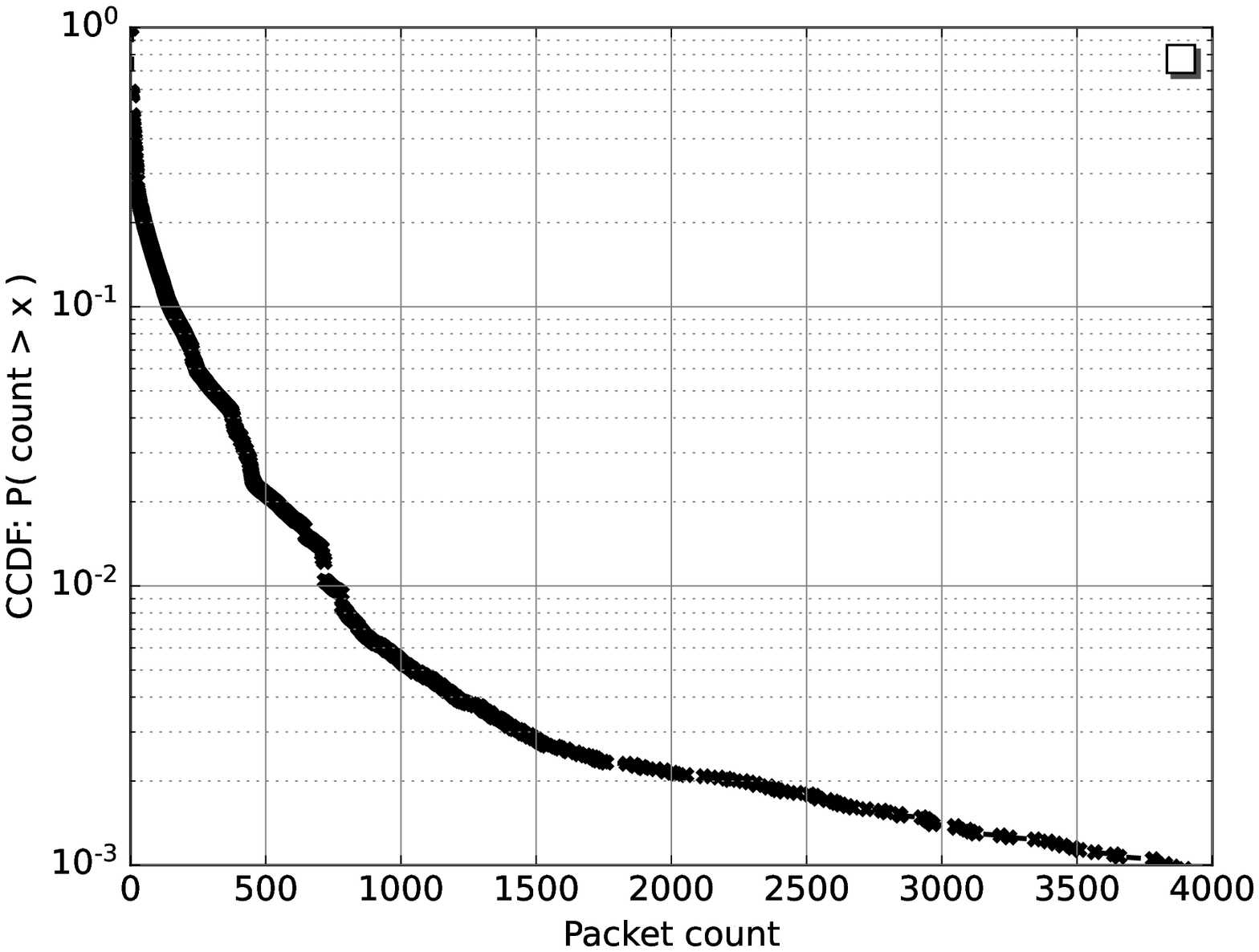}    \label{fig:ccdfflowPkt}}\quad
								
								%            \hspace{-7mm}
								%			\subfloat[\color{magenta}Volume.]{
								%				\includegraphics[width=0.32\textwidth,height=0.18\textwidth]{./flowVol.eps}    \label{fig:ccdfflowVol}}\quad
								
								\subfloat[Duration.]{
									\includegraphics[width=0.48\textwidth]{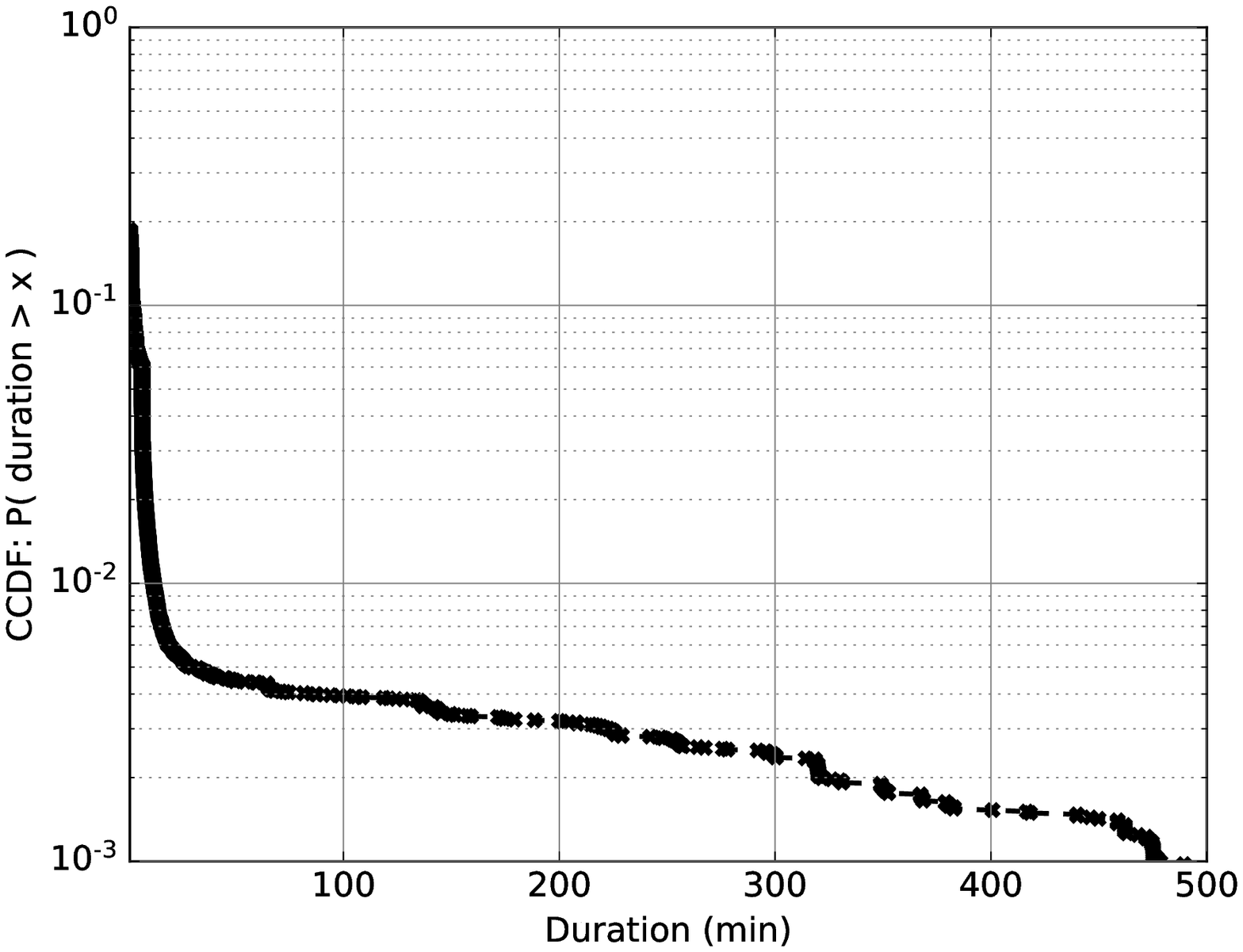}    \label{fig:ccdfDur}}\quad
								
							}
							\caption{CCDF of attributes: (a) {packet count}, and (b) duration, of suspicious flows.}
							\label{fig:ccdfflow}
						\end{center}
					\end{figure}

					Next, we analyze the size of mirrored traffic flows which need to be analyzed in software -- suggesting the computing cost. It can be seen in Fig.~\ref{fig:ccdfflowPkt} that reflected suspicious flows often carry a small number of packets -- more than $86$\% of flows have less than $100$ packets. Interestingly, these flows are relatively short  -- as shown in Fig.~\ref{fig:ccdfDur} more than $90$\% of flows last less than $2$ minutes, hence get timed out quickly from the switch TCAM table. This measure is essential since TCAM  is one of the precious resources in our system.  
					%This shows that the mirrored selected flows carry a small number of packets and are short as well.
					Considering the time trace of reactive rules (during our 50-day trial) in Fig.~\ref{fig:sdnFlows}, we see that no more than 400 entries per day are installed into the SDN switch by the controller.
					These metrics collectively serve as clear evidence of the cost-effectiveness of our solution.
				}

				\subsection{Comparing Our Diagnosis Models with Zeek IDS}
				Lastly, we validate our results against logs of an open-source IDS Zeek (formerly Bro) -- which is a powerful network analysis tool, widely used, and consistently maintained by the community for more than 20 years now \cite{zeek}.
				Our main intention is to compare our ML-based flow-level approach with a decent rule-based packet-level method in detecting malicious traffic.

				\begin{figure}[t!]
					\centering
					\includegraphics[width=0.88\textwidth,height=0.47\textwidth]{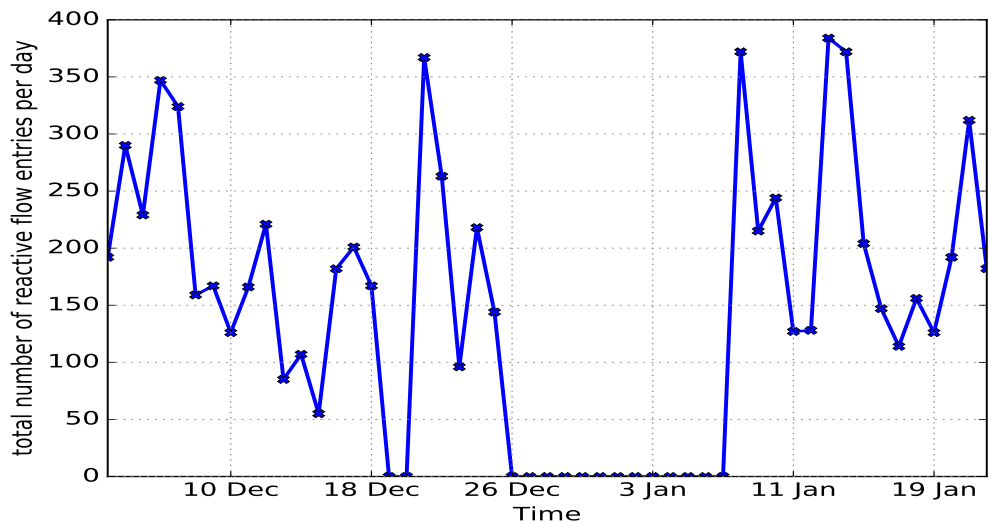}
					\caption{{Daily number of reactive flow entries installed by the SDN controller into the SDN switch, during the 50-day trial.}}
					\label{fig:sdnFlows}
				\end{figure}
				
				Zeek performs a packet-based analysis and raises alarms if a known malicious signature is found in a packet.  We replayed the 50-day worth of selected traffic (of suspicious flows) onto Zeek to check how it flags malicious activities. Of a total of 5.5 M packets, Zeek raised alarms for 23.7 K ($0.4$\%). To compare with our flow-level analysis, we aggregate those packets flagged by Zeek into flows. It turns out 14,455 flows (\ie $25.3$\% of suspicious flows), belonging to all of the five malware families (Table~\ref{tab:distro_maliciousFlows45Days}), are detected as malicious by Zeek.

				We found that all of Zeek flagged flows are a subset of malicious flows classified by our models.
				Starting from the HTTP flows, we found a small overlap (only 1087 flows, about $3$\%) in the outputs of our HTTP model and Zeek. This is mainly because those 33.5K HTTP flows ($94$\% of total) that only carry three-way handshakes with a reset do not cause Zeek to raise any alarms when inspecting their packets.
				Instead, the overlap was far better for HTTPS flows -- Zeek corroborates our HTTPS model by flagging 13368 flows ($\approx 68$\%) that we classified as malicious.
				Lastly, none of the UDP flows are flagged by Zeek, probably because individual UDP packets did not display any malicious pattern matching Zeek's known signatures. Zeek raises several different alert types for malicious HTTP and HTTPS packets --  about half of malicious HTTP and HTTPS receive more than one alert type. The top four alerts are:  ``{\myverb{above\_hole\_data\_without\_any\_acks}}'', ``{\myverb{bad\_TCP\_checksum}}'', ``{\myverb{possible\_split\_routing}}'', and ``{\myverb{data\_before\_established}}''. 

				\section{Conclusion}\label{conclusion}

				We have developed and validated a method for real-time selective mirroring network flows for diagnosis by trained models. We analyzed 75 days worth of DNS traffic (2.4B records), highlighted the prevalence of more than twenty DGA-enabled malware families across internal hosts, and obtained insights into their behavioral patterns while communicating with their corresponding C\&C server by analysis of a large PCAP collected during peak hour. We identified their traffic attributes and trained three one-class classifier models specialized in HTTP, HTTPS, and UDP protocols using public PCAP traces of known malware families. We then developed a system that continuously monitors DNS traffic and automatically (using SDN reactive rules) selects and mirror communications between internal hosts and malware C\&C servers pertinent to DGA queries. We then evaluated the efficacy of our models by testing suspicious traffic flows against our trained models, identified infected hosts from suspicious flows. Finally, we compared our detection approach with software IDS Zeek. In the next chapter, we will identify the malicious hosts of an enterprise network by analyzing non-existent DNS responses. 
				
				\chapter[Learning-Based Detection of Malicious Hosts by Analyzing Non-Existent DNS Responses ]{Learning-Based Detection of Malicious Hosts by Analyzing Non-Existent DNS Responses}
				\label{chap:ch5}
				\minitoc
				
				The previous chapter highlighted the prevalence of DGA-enabled malware in our campus network and identified infected hosts by selectively mirroring the suspicious network traffic using SDN. This chapter explores the use of DNS for service disruption (through random subdomain attack) by analyzing the incoming DNS responses. Parts of this chapter are under review at IEEE Globecom 2022.
				
				The DNS Water Torture attack (aka Slow Drip or Random Subdomain attack) is a type of DDoS attack on authoritative DNS servers and/or open resolvers, whereby the victim is bombarded with random non-existent domains (NXDs) DNS requests, exhausting their entire resources. A famous example of this attack was launched by Mirai botnet on Dyn DNS architecture in 2016. Researchers have proposed solutions to detect these attacks; however, they predominantly apply certain static thresholds to the count of NXD responses. This method can result in high false positives and needs to be customized to the traffic pattern of victim DNS servers, making it practically challenging for network operators to adopt them at the source of potential attacks. 
				
				This chapter aims to detect possibly infected hosts of a university campus network that take part in this specific type of DNS-based attacks. Our contributions are threefold: (1) We analyze 120 days’ worth of DNS traffic collected from the border of a large university campus network to draw insights into the characteristics of non-existent domain (NXD) responses of incoming DNS packets. We discuss how malicious NXDs differ from benign ones and highlight two attack scenarios based on their requested domain names; (2) We develop a method using multi-staged iForest models to detect malicious internal hosts based on the attributes of their DNS activity; (3) We evaluate the efficacy of our proposed method by applying it to live DNS data streams in our university campus network. We show how our models can detect infected hosts that generate high-volume and low-volume distributed non-existent DNS queries with more than 99\% accuracy of correctly classifying legitimate hosts.

				\section{Introduction}	
				\label{sec:intro}
				%\cite{ahmed2019monitoring}
				
				Over the last two decades, there has been tremendous growth in cyber-attacks and malicious activities exploiting DNS protocol as the number of network devices grows daily. %Domain Name Systems (DNS) is a fundamental protocol that maps the domain name with its corresponding IP address and vice versa. DNS has a decentralized, distributed, and hierarchical database having three levels of hierarchy \ie when a client types "www.google.com" on the browser, the DNS request is initiated by the client's computer which goes to a root server (which maintains the IP addresses of top-level domains (TLDs)), the root server returns the IP address of TLD. Next, the request goes to the TLD which gives the IP address of the authoritative name server of "google.com". Finally, the request goes to the authoritative server that returns the IP address of the requested website. 
				DNS is a mission-critical service but open by design and rarely monitored by the firewall compared to email, FTP, or HTTP.  According to a security firm \cite{efficientIP2017}, in 2017, the cost of damages caused by a DNS attack for a large organization (3,000+ employees) was estimated as \$2.2m. 
				Attack on DNS infrastructure of Dyn (providing DNS services to big companies such as AirBnB, Spotify, and Twitter) caused by a malware known as Mirai 2016 is a famous example to explain water torture attack in which thousands of vulnerable IoT devices took part in sending queries for random domains to the companies whose DNS infrastructure is operated by Dyn \cite{dyn2016}. According to the FBI, the attackers used random subdomain attacks to target US-based state-level voter registration and information website in 2020 \cite{fbi2020}. 
				
				DNS works in such a way that if a query is being asked from the DNS authoritative name server or open resolver, it is an obligation on them to answer it even if the query is non-existent in their ecosystem. Non-existent domains are of two types: (a) benign: popular search engines and anti-viruses utilize random-looking domains to convey a one-time signal to their servers known as disposable domains (\eg {\myverb{elb.amazonaws.com.cn}}, {\myverb{cloudfront.net}}, and {\myverb{avts.mcafee.com}}). Benign domains may also contain typo mistakes. For example, a user accidentally writes ``{\myverb{googel.com}}'' instead of ``{\myverb{google.com}}''; and (b) malicious: launch a type of DDoS attack, DNS water torture attack \cite{water2014} also known as random subdomain attack by dynamically generating random strings as the prefix of a victim domain. The DNS Water Torture Attack is a type of DDoS attack on DNS servers. This attack affects both authoritative servers and open/recursive resolvers, but mainly it targets the former.

				Cyber actors use bots (compromised devices) to send many randomly generated domain names on their victim servers. The queried domain names relate to the primary domain that is governed by its authoritative name server to return the IP address of that particular domain. During the attack, due to the high number of requests, the victim authoritative servers and/or the recursive resolvers may have slow response to the queries being asked or potentially become unavailable. Although the problem has been well understood over the last decade, it is mostly dealt with from the perspective of the victim server by identifying the malicious queries. We see this as an opportunity to detect potentially infected hosts of an enterprise that initiate non-existent queries to the outside world. It is important for enterprises to implement certain cyber hygiene practices that aid in boosting an organization’s overall security posture as well as miniating their reputation on the Internet community by preventing their internal hosts from attacking others.
				
				Researchers have proposed a range of countermeasures to detect NXD attacks; however their proposed methods are too simple (threshold-based) and hence fall short when it comes to sophisticated attacks - with a different objective being the detection of victim primary domains. Instead, we develop a monitoring system at the source that can detect enterprise hosts that generate high volumes of NXD attacks and low and distributed NXD attacks.
				
				In this chapter, we make the following three contributions: (1) We analyze 120 days' worth of DNS traffic collected from the border of a large university campus network to highlight and draw insights into the high volume of incoming Non-Existent Domains (NXD) responses and to identify the difference between two scenarios of water torture attack (attack on authoritative DNS server and/or open resolver) ; (2) Based on the behavioral attributes, we develop a multi-staged iForest model to classify the internal hosts (those receiving benign NX responses versus those taking part in water torture attack) and (3) We evaluate the efficacy of our proposed approach on live DNS data from the network border of a large university campus with an accuracy of over 99\% incorrectly classifying legitimate hosts.

				The rest of the chapter is organized as follows: \S 2 discusses existing literature on DNS security related to water torture attack. \S3 describes the analysis of incoming NXD responses in our university campus network. We describe our proposed scheme in \S4. We describe the efficacy of our scheme in §5. Lastly, we conclude this Chapter in \S6.

				\section{Analyzing Two Variants of DNS Random Subdomain Attacks in Our Campus Network}
				\label{sec:analysis}
				
				In this section, we first analyze the prevalence of NX domains in our campus network. We then look at the two variants of water torture attack - the first case is when the victim is an authoritative name server, and the second case is when the victim is an open/recursive resolver. 
				The study here considers data collected over four months from  31st Oct 2019 to 28th Feb 2020. %We collected daily DNS PCAP traces from the border of the University campus network. Each PCAP has a size of about 15 GB on average. %The IT department of the campus network provisioned a full mirror (both inbound and outbound) of its Internet traffic (on a 10 Gbps interface each) to our data collection system from its border router (outside of the firewall). We obtained appropriate ethics clearance (Human Research Ethics Advisory Panel approval number HC17499) for this study. % (approval number will be disclosed when this chapter is de-anonymized). % 
				Details of our data collection can be found in \S\ref{sec:3.2}. 
				%We extracted DNS packets from the enterprise Internet traffic streams in real-time by configuring rules to match incoming/outgoing IPv4 and IPv6 UDP packets on port 53 in an OpenFlow switch. 

				\begin{figure}[t!]
					\centering
					\includegraphics[width=0.85\textwidth,height=0.47\textwidth]{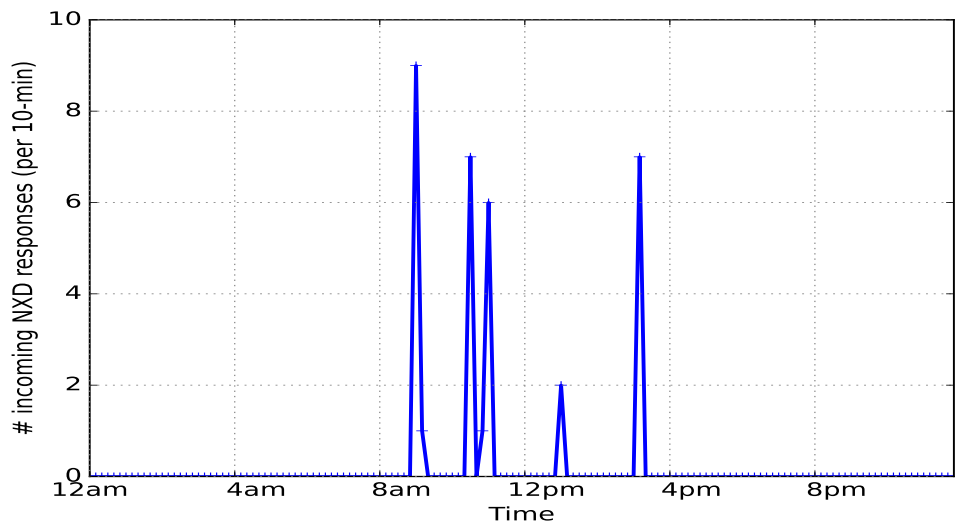}
					\caption{Daily trace of incoming DNS response with errors (typo mistakes: benign.)}
					\label{fig: benignActivityTypos}
				\end{figure}

				\begin{figure}[t!]
					\centering
					\includegraphics[width=0.85\textwidth,height=0.47\textwidth]{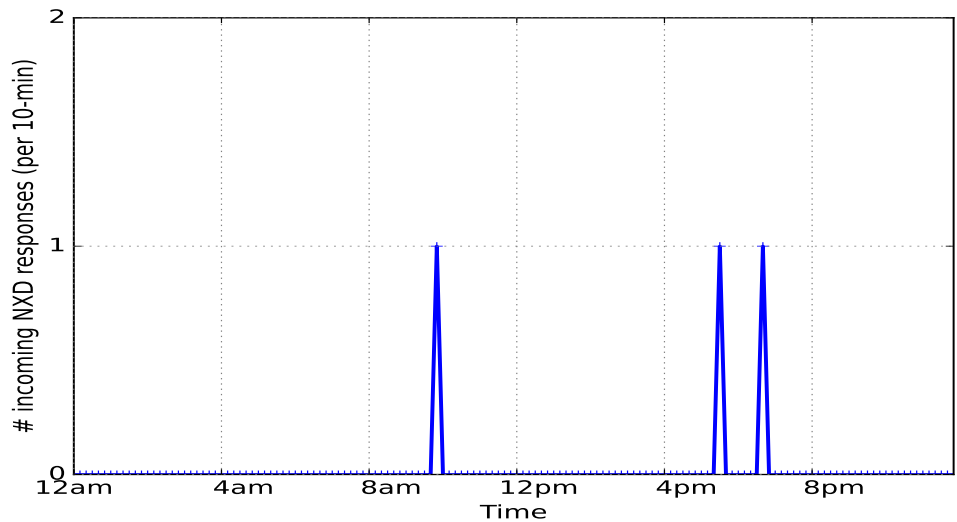}
					\caption{Daily trace of incoming DNS response with errors (Disposable domains from Sophos antivirus: benign.)}
					\label{fig: benignActivitySophos}
				\end{figure}
				
				\textbf{Benign incoming responses:}	Before exploring the anatomy of NXD attacks, we discuss the two possible cases of incoming benign NXDs responses in an enterprise network: (i) Typing mistakes, and (ii) Disposable domains used by antivirus tools (benign data exfiltration as described in Chapter \ref{chap:ch3}). Fig.~\ref{fig: benignActivityTypos} shows a full day trace in which a benign host just received 31 NXD responses due to typos in a day. Similarly, Fig.~\ref{fig: benignActivitySophos} shows a host which received just 3 NXD responses all due to disposable domains coming from the antivirus tools. Thus, it can be seen that the benign hosts do not receive a massive number of NXDs.

				\textbf{Illustration of Our Data Collection:} Let us understand the anatomy of water torture attack by taking a closer look at our data from a day. As shown in Fig.~\ref{fig: taxonomyWaterTorture}, we show the visual illustration of our data collection for the incoming NX responses. As we collect the DNS data from the border router, the identity of some of the internal hosts gets hidden behind the UNSW recursive resolvers (RRs) and NAT gateways. Hence the scope of this work is limited to those unhidden internal hosts receiving NXD responses. Out of 1.4 M incoming NXDs on this day, 700K responses were destined to 393 hosts. The responses are sourced from either open resolvers and/or authoritative name servers. In what follows, we will highlight how internal enterprise hosts behave during attack scenarios.

				\begin{figure}[t!]
					\centering
					\includegraphics[width=0.88\textwidth,height=0.47\textwidth]{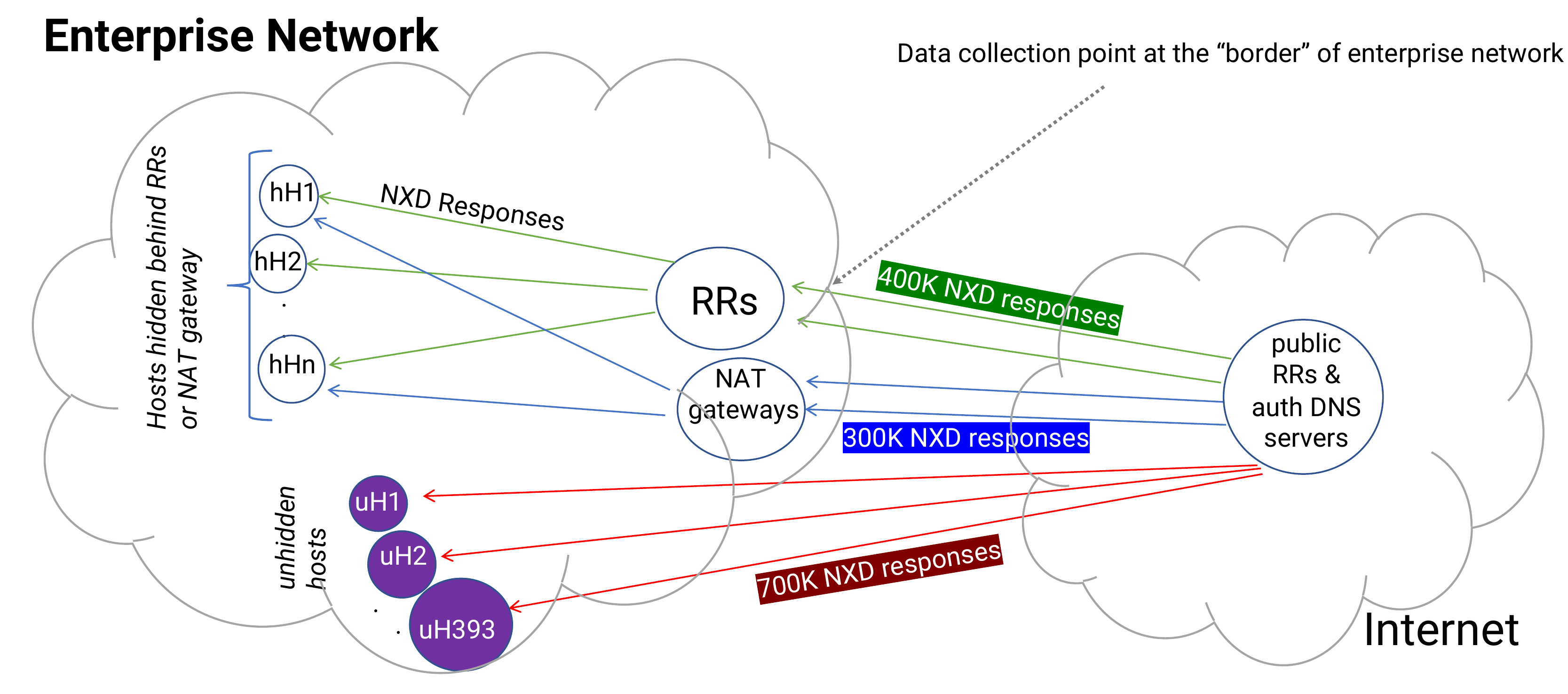}
					\caption{Visual illustration of our data collection and various entities identified.}
					\label{fig: taxonomyWaterTorture}
				\end{figure}

				\begin{figure}[t!]
					\centering
					\includegraphics[width=0.88\textwidth,height=0.47\textwidth]{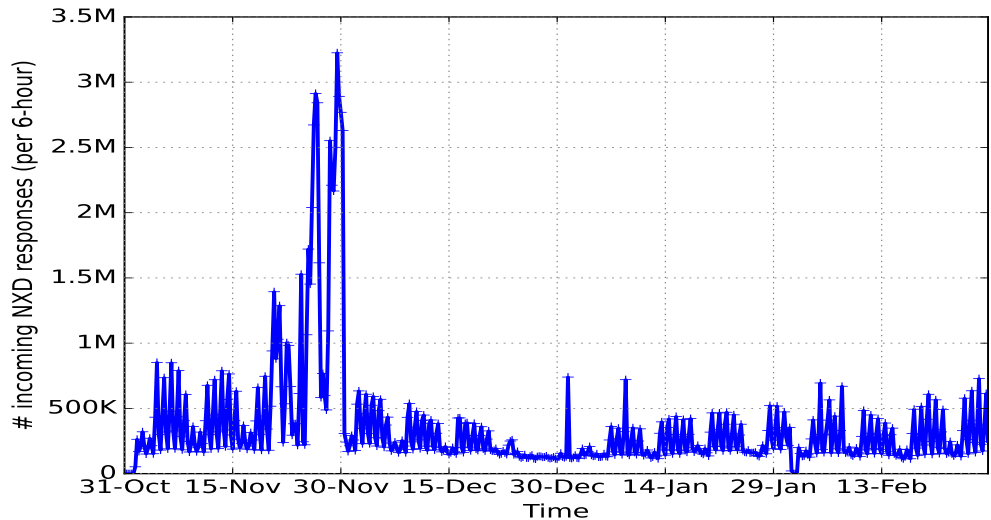}
					\caption{Timetrace of incoming  NX DNS responses.}
					\label{fig: 4monthTimeTrace}
				\end{figure}

				\subsection{Attack Scenarios}
				
				This section discusses the attack scenarios \ie attack on the authoritative name server of the victim domain, and an attack on the open resolver. We first plot in Fig~\ref{fig: 4monthTimeTrace} the time trace of incoming NXD responses in our dataset. 
				Each data-point in this plot represents the number of NXD responses over a 6-hour window. We observe that the typical values of the number of incoming NXDs are less than 500K domains. These domains are mostly benign (either typos or disposable domains). However, some spikes can be seen in the plot highlighting some abnormal activities. We further analyzed those days with spikes in the count of incoming NXD responses and found out the two different scenarios of NXD attacks as described in the following subsection. 
				
				\begin{figure}[t!]
					\centering
					\includegraphics[width=0.88\textwidth,height=0.47\textwidth]{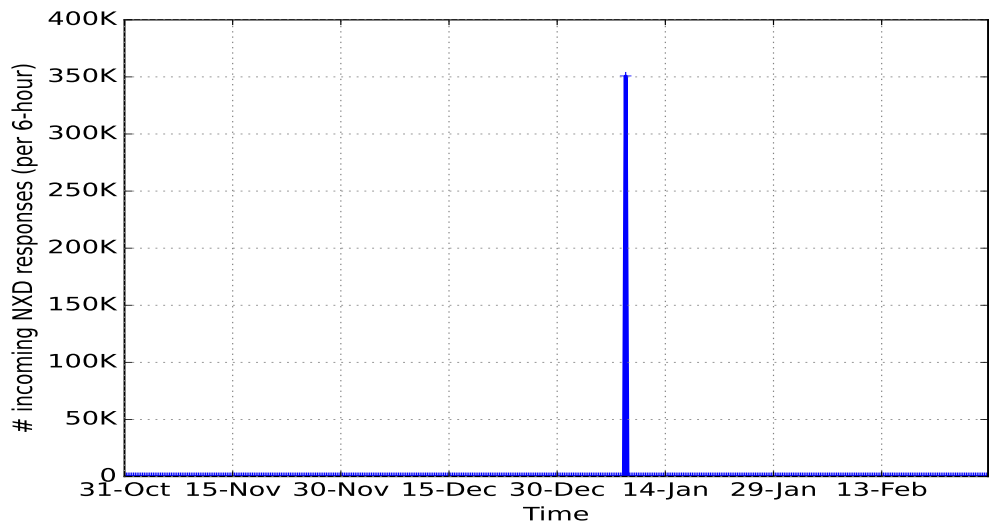}
					\caption{Time trace of an infected host  in our dataset - a sudden rise in NXD responses observed with in an hour only.}
					\label{fig: AuthDNSAttackTimeTrace}
				\end{figure}
				
				\begin{figure}[t!]
					\centering
					\includegraphics[width=0.88\textwidth,height=0.47\textwidth]{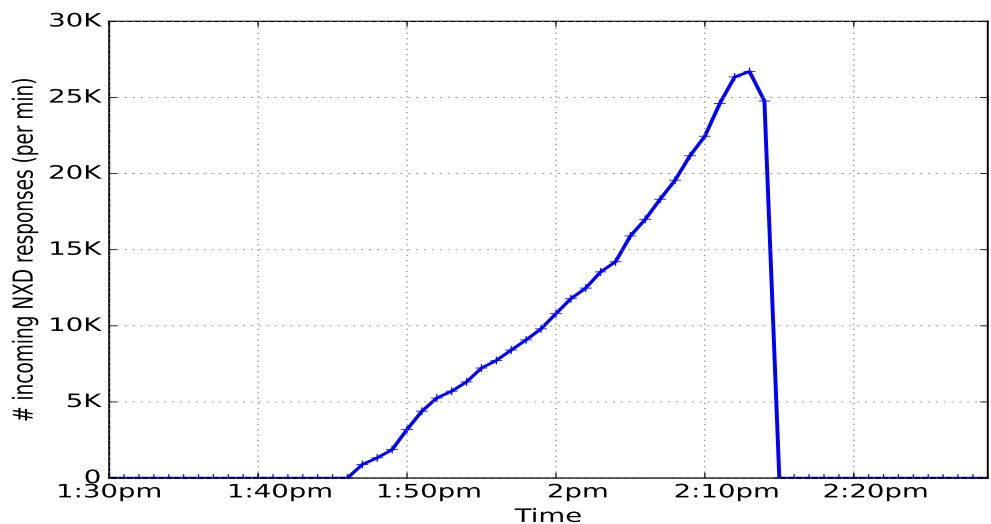}
					\caption{Zoomed-in time trace of the suspicious host during the hour of interest (8th January 2020 1:30 - 2:30pm).}
					\label{fig: AuthDNSAttackTimeTrace1hour}
				\end{figure}
				
				\subsubsection{Attack on Authoritative DNS Servers}
				
				Analyzing the spike on 8th January 2020, we found out that most of the incoming NXD responses  (350K) are destined to a regular host. To better understand the behavior of that host in terms of incoming NXD responses, we plot a time trace of activity for the entire duration of our dataset. Interestingly, the host displays unexpected (suspicious) behavior on 8th January 2020 as shown in Fig. ~\ref{fig: AuthDNSAttackTimeTrace}. We see no NXD activity other than one massive spike within an hour or two on 8th January.

				\begin{table}[t]
					\centering
					\caption{Sample of FQDNs used by suspicious host.}
					\label{tab:sampleAttack}
					\begin{adjustbox}{max width=1.5\textwidth}   
						\renewcommand{\arraystretch}{1.5}
						\begin{tabular}{|l|l|}
							\toprule
							\hline
							\textbf{S.No} & \textbf{FQDNs}            \\ \hline
							1                & {\myverb{3rd4.ahrtv.cn}}
							\\ \hline
							2                & {\myverb{6nq0.ahrtv.cn}}
							\\ \hline
							3                & {\myverb{3guc.ahrtv.cn}}
							\\ \hline
							4                & 	{\myverb{a28.ahrtv.cn}}
							\\ \hline
							5                & {\myverb{w2l9v.ahrtv.cn}}               \\ \hline
							6                & {\myverb{undertake.ahrtv.cn}}         \\ \hline
							7                & {\myverb{disposal.ahrtv.cn}}                   \\ \hline
							8                & {\myverb{mengla.ahrtv.cn}}     \\ \hline
							9                & {\myverb{launcher.ahrtv.cn}}                 \\ \hline
							10               & {\myverb{deliberately.ahrtv.cn}}       \\ \hline 
							\bottomrule
						\end{tabular}
					\end{adjustbox}
				\end{table}

				Let us zoom into the host activity on a per-minute basis for those particular hours to see a fine-grained pattern of NXD arrivals (whether spread across hours uniformly or bursty). As shown in Fig. \ref{fig: AuthDNSAttackTimeTrace1hour}, we observe that the number of NXD responses starts to grow linearly at around 1:45pm, peaks at 26K and sharply becomes zero at 2:15pm.  
				Analyzing their query names revealed that they all target a single primary domain name {\myverb{ahrtv.cn}} - this highlights an attack on the corresponding authoritative name server. Table \ref{tab:sampleAttack} lists some FQDNs found in those queries where some of them contain random subdomains \eg{``{\myverb{3guc.ahrtv.cn}}''}, and contain dictionary words \eg{``{\myverb{disposal.ahrtv.cn}}''} in the subdomain part. In general, we find out that the volume of incoming NXDs from water torture attack is significantly much greater than disposable and benign NXD domains as shown in Fig. \ref{fig: benignActivityTypos} and \ref{fig: benignActivitySophos}.

				We then plotted the CCDF of the total number of occurrences of unique FQDNs for this suspicious host, which can be seen in Fig.~\ref{fig: CCDFUnique NXDs}. In total, we found 350,695 NXD responses from ``{\myverb{ahrtv.cn}}'' out of which 350,581 responses use unique FQDNs (used just once as shown in Fig.~\ref{fig: CCDFUnique NXDs} - more than 99.99\% of FQDNs occurred just once, which is a very unusual behavior).

				\begin{figure}[t!]
					\centering
					\includegraphics[width=0.88\textwidth,height=0.47\textwidth]{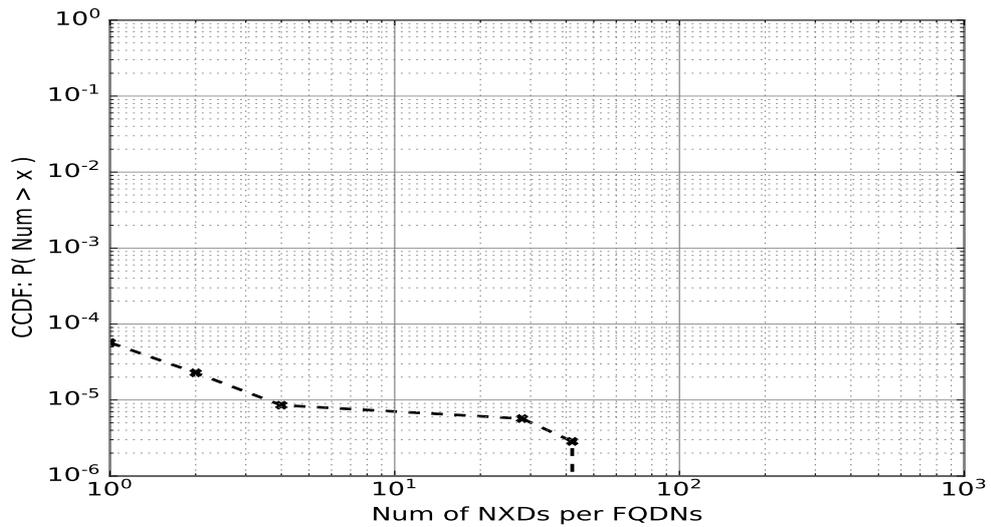}
					\caption{CCDF of number of occurrences of unique NXDs per FQDN for the suspicious host.}
					\label{fig: CCDFUnique NXDs}
				\end{figure}

				%\begin{figure}[t!]
				%	\centering
				%	\includegraphics[width=0.88\textwidth,height=0.47\textwidth]{./CCDF_PerInternalHostNXDs.eps}
				%	\caption{CCDF of number of incoming  NXDs per host in a day (26th Nov 2019)}
				%	\label{fig: CCDFNXDsPerHost}
				%\end{figure}

				\subsubsection{Attack on Open Resolvers}

				We found out another form of NXD attacks in which an open resolver was the target. After observing unusual spikes during the last week of Nov 2019 (shown in Fig.~\ref{fig: 4monthTimeTrace}), we focused on 26th Nov 2019 to better analyze the behavior of the involved host as well as top queried FQDNs and primary domains. We found out that an internal host of our campus made unusual queries to Google's public DNS resolver 8.8.8.8 - the query names were ``{\myverb{shu-Aspire-V3-572}}''
				and ``{\myverb{mtrnlab5}}'' and hence not fully qualified since they did not conform to a standard structure.
				We observed that the volume of those non-standard queries goes to 400K NXDs per hour, highlighting a very abnormal behavior by this internal host. 
				Fig.~\ref{fig: SuspiciousHostOpenResolver} shows the time trace of incoming NXD responses for this host in our dataset that shows a heavy NXD activity at the end of Nov 2019 with peak incoming responses around 2.5M over a 6-hour window, and another suspicious activity at the start of Feb 2020 with the peak number of incoming responses crossing 300K per 6-hour.
				
				\begin{figure}[t!]
					\centering
					\includegraphics[width=0.88\textwidth,height=0.47\textwidth]{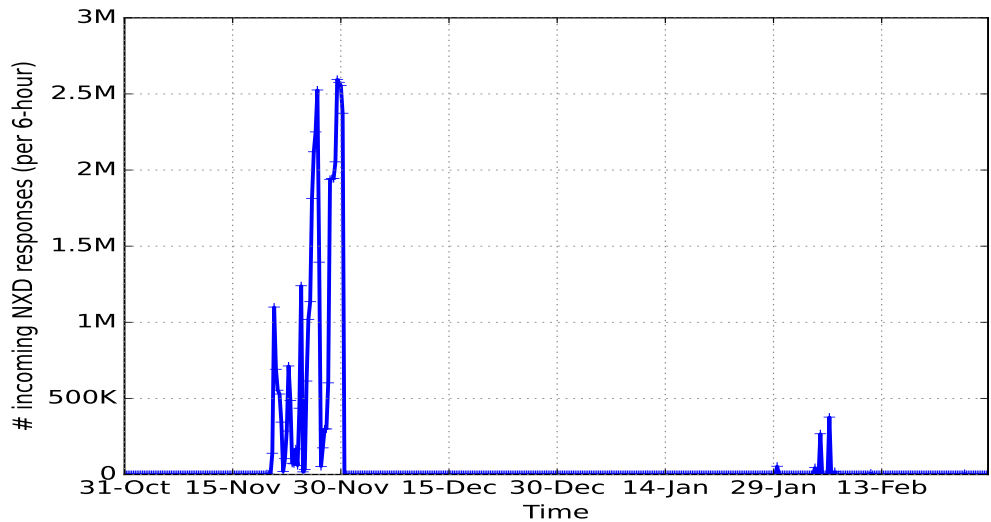}
					\caption{Time trace of  an infected host.}
					\label{fig: SuspiciousHostOpenResolver}
				\end{figure}

				\begin{figure*}[t!]
					\begin{center}
						%\vspace{-3mm}
						\mbox{
							%\hspace{-3mm}
							\subfloat[H1 targeting an authoritative server per hour.]{ \includegraphics[width=0.29\textwidth,height=0.18\textwidth]{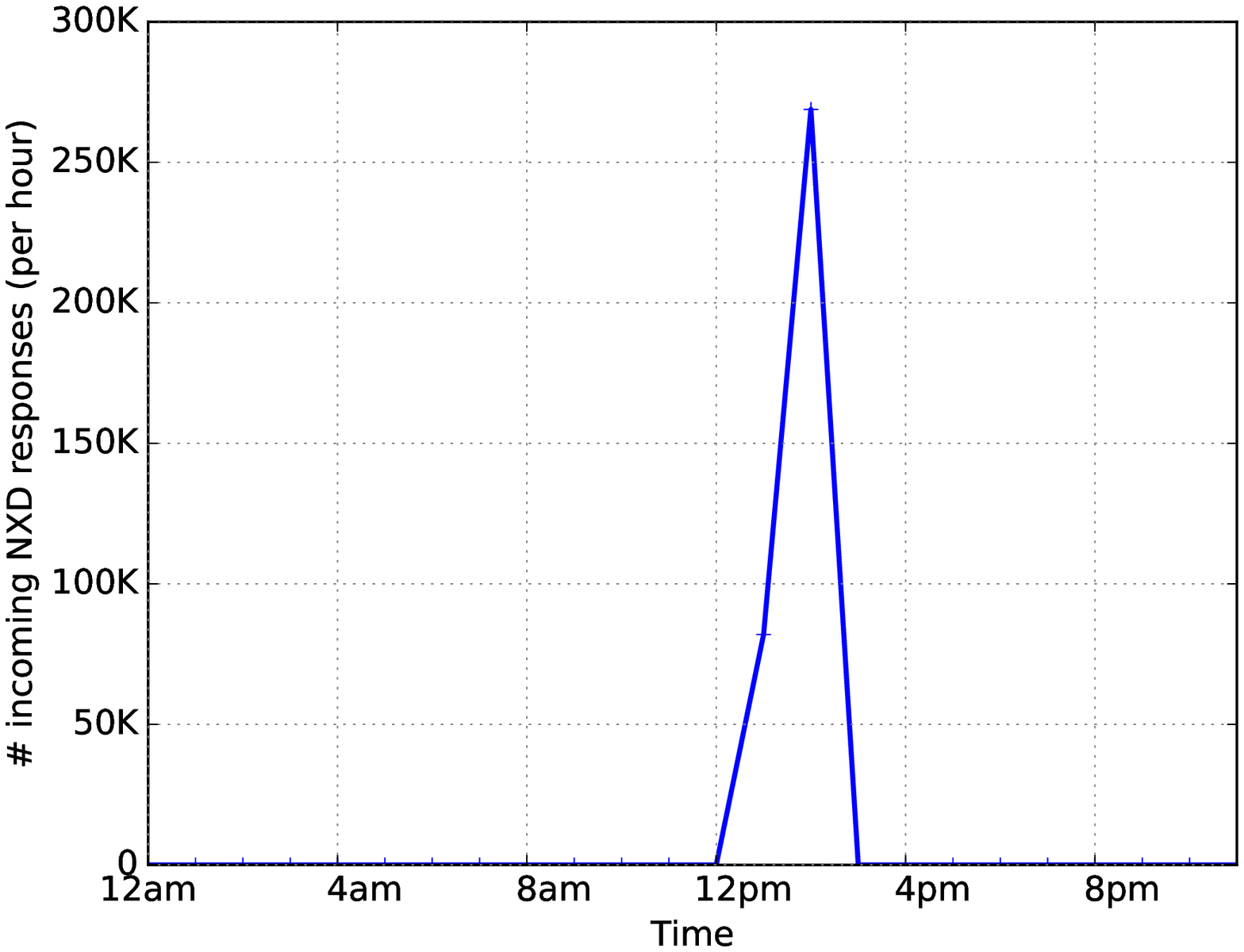}
								\quad
								\label{fig:inf1a}
							}

							\subfloat[H2 targeting a recursive server per hour.]{ \includegraphics[width=0.29\textwidth,height=0.18\textwidth]{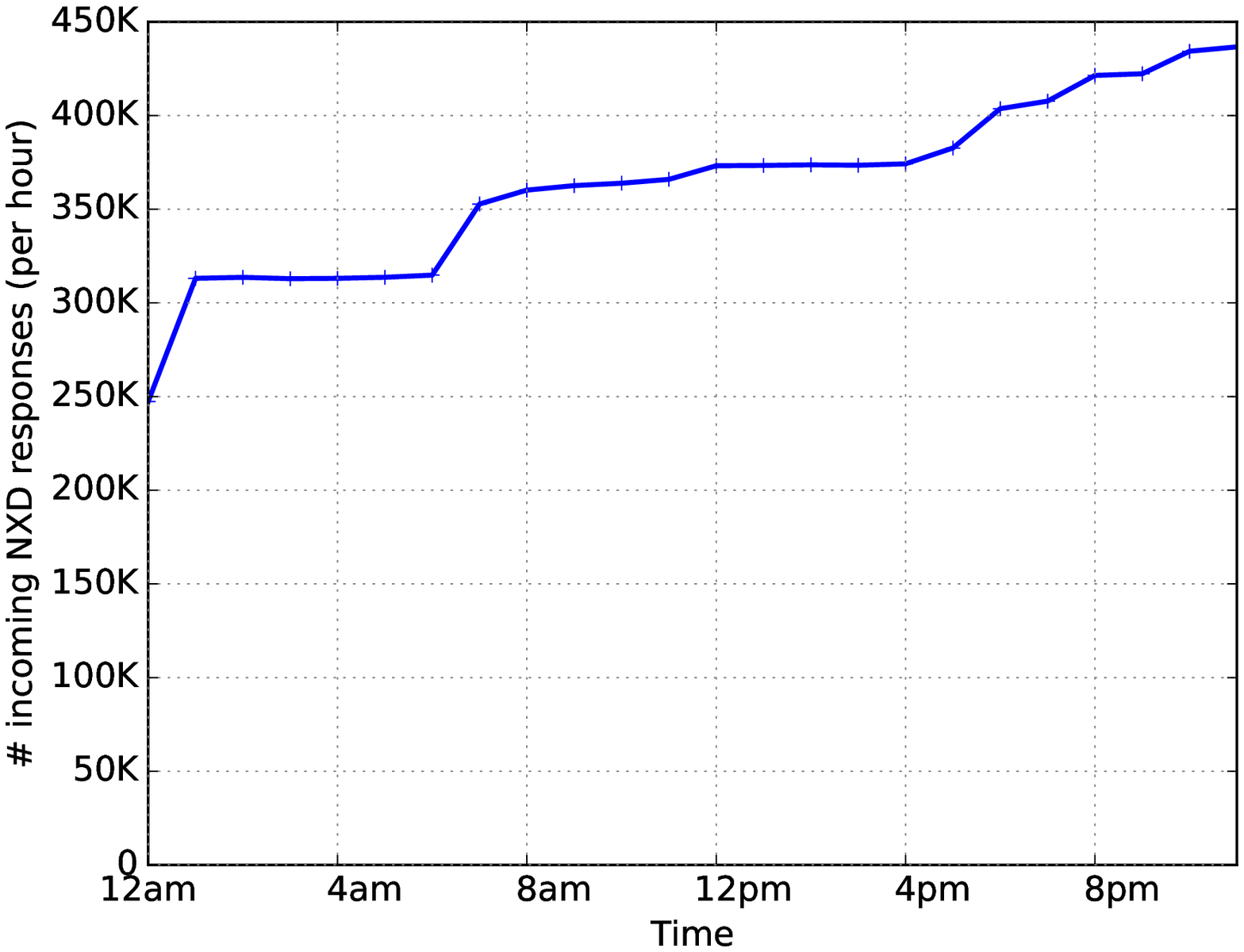}
								\quad
								\label{fig:inf2a}
							}

							\subfloat[H3 targeting an authoritative server per hour.]{ \includegraphics[width=0.29\textwidth,height=0.18\textwidth]{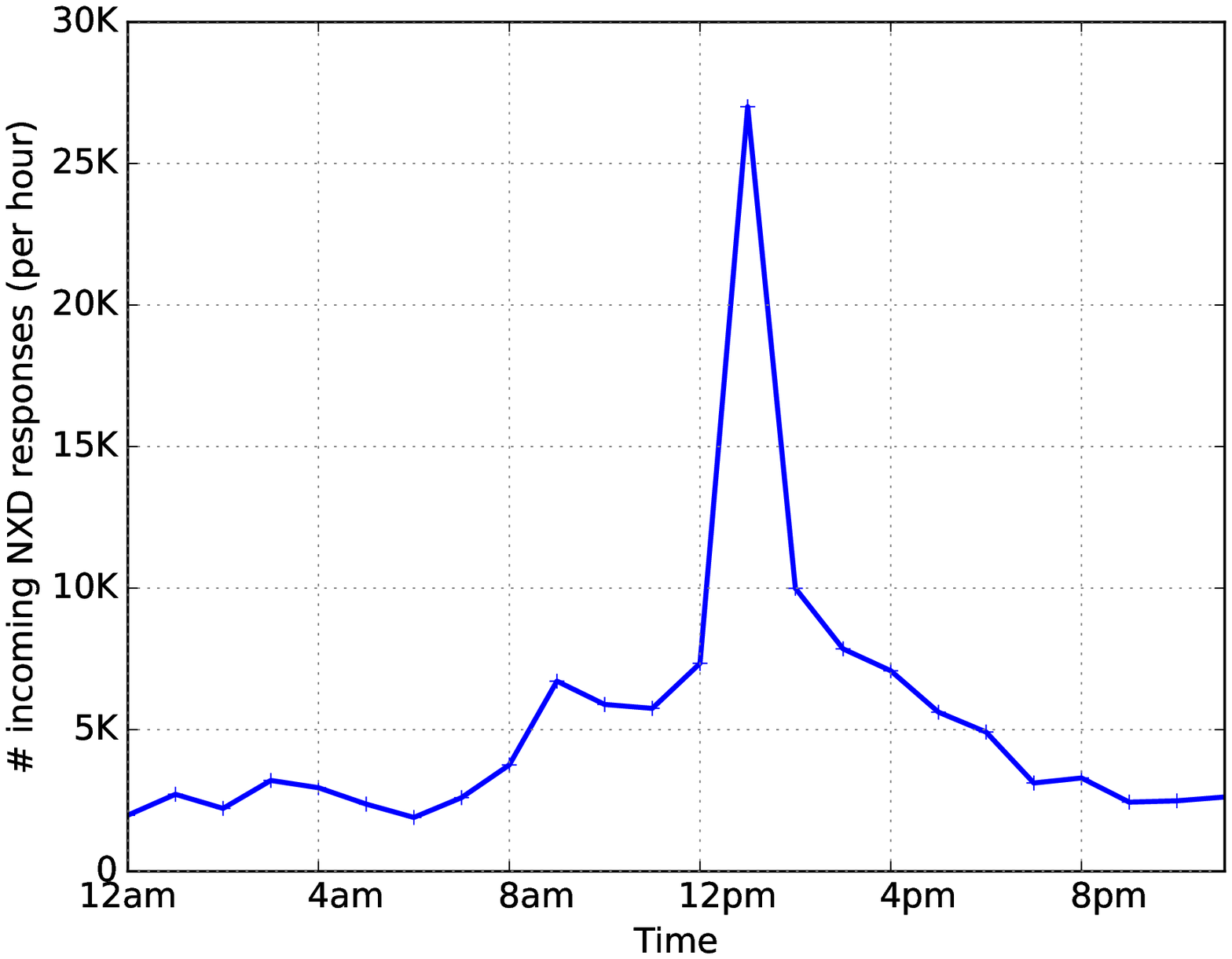}
								\quad
								\label{fig:inf3a}
							}
						}
						
						\mbox{
							%\hspace{-3mm}
							\subfloat[H1 targeting an authoritative server per minute.]{ \includegraphics[width=0.29\textwidth,height=0.18\textwidth]{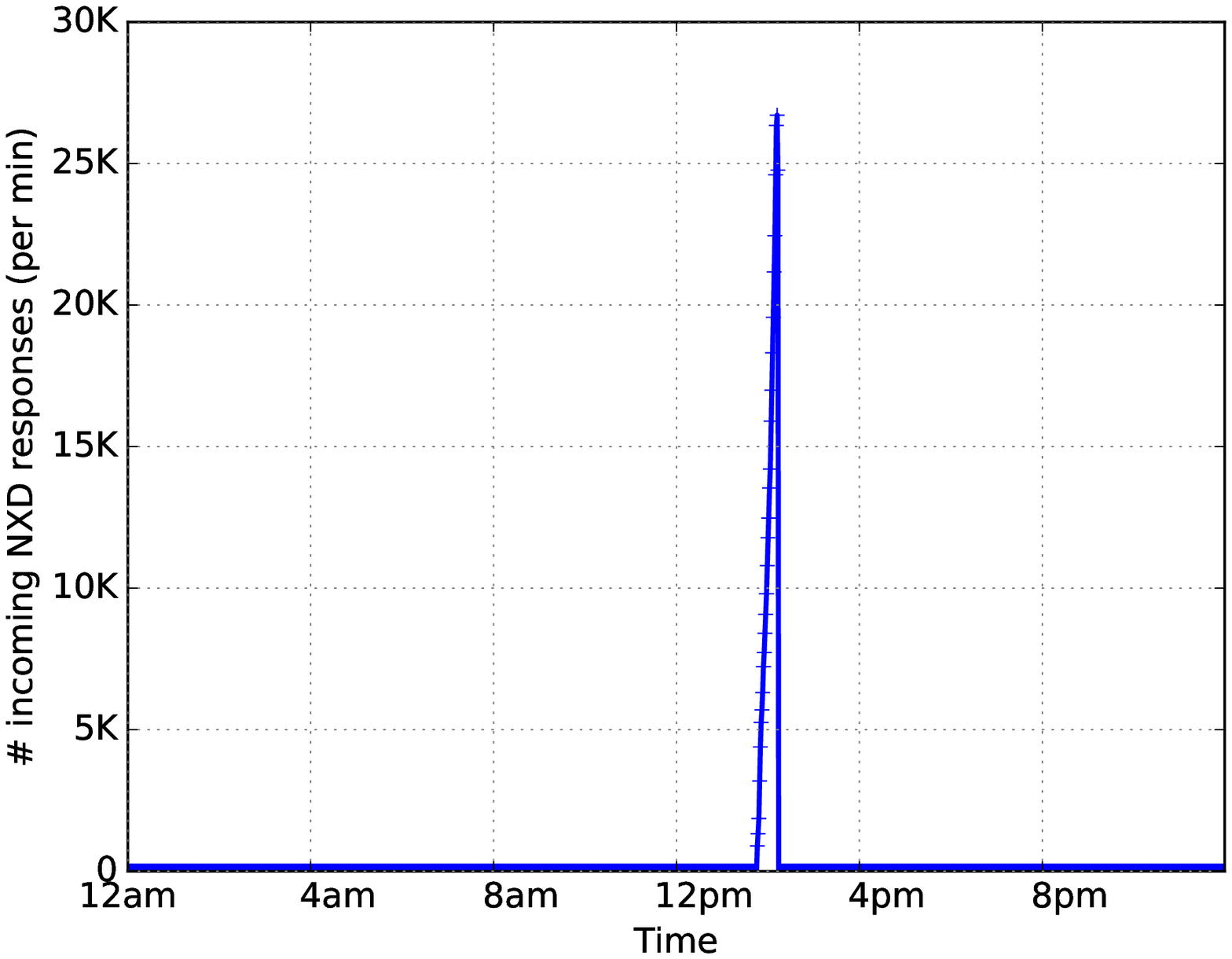}
								\quad
								\label{fig:inf1b}
							}

							\subfloat[H2 targeting a recursive server per minute.]{ \includegraphics[width=0.29\textwidth,height=0.18\textwidth]{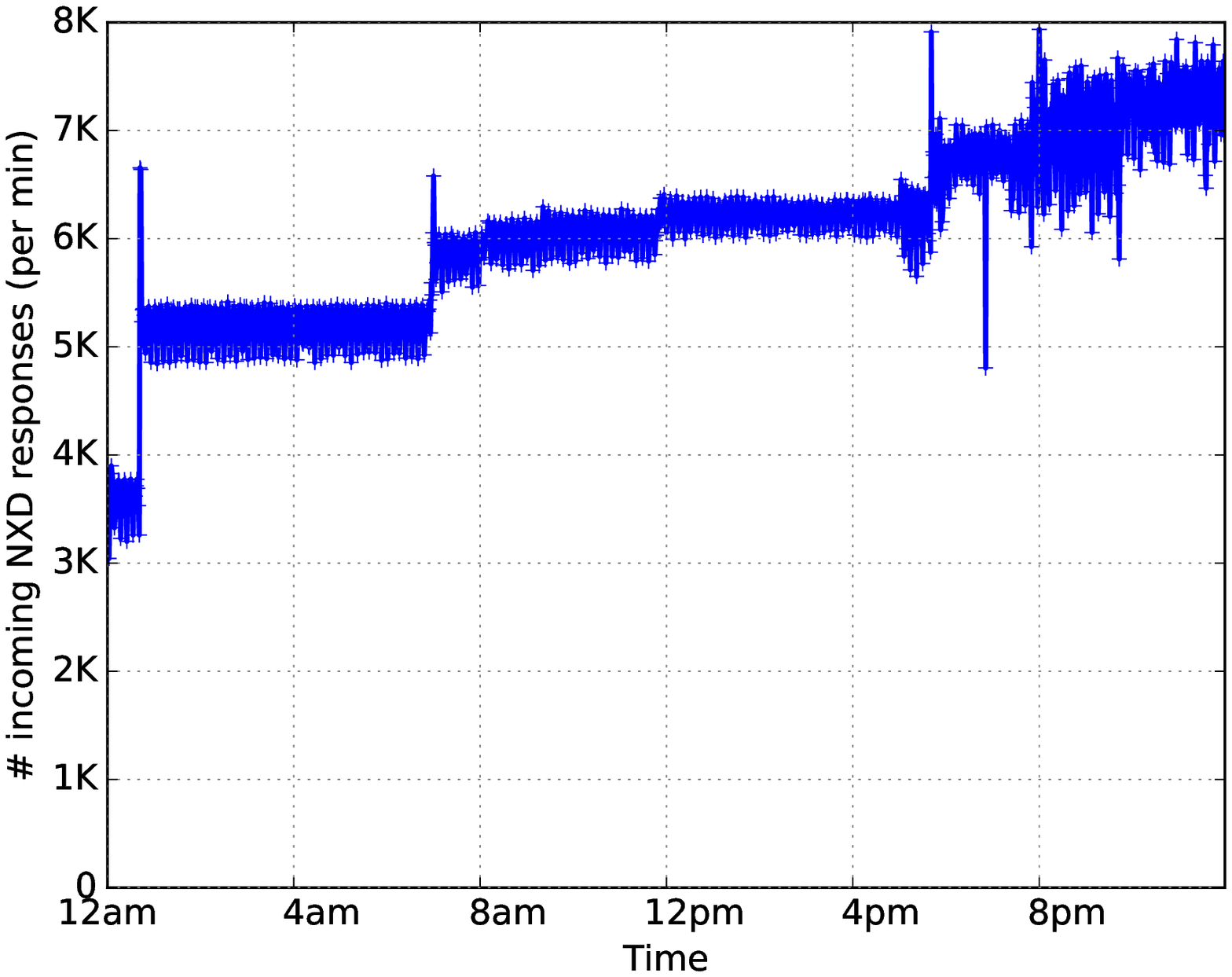}
								\quad
								\label{fig:inf2b}
							}
							
							\subfloat[H3 targeting an authoritative server per minute.]{ \includegraphics[width=0.29\textwidth,height=0.18\textwidth]{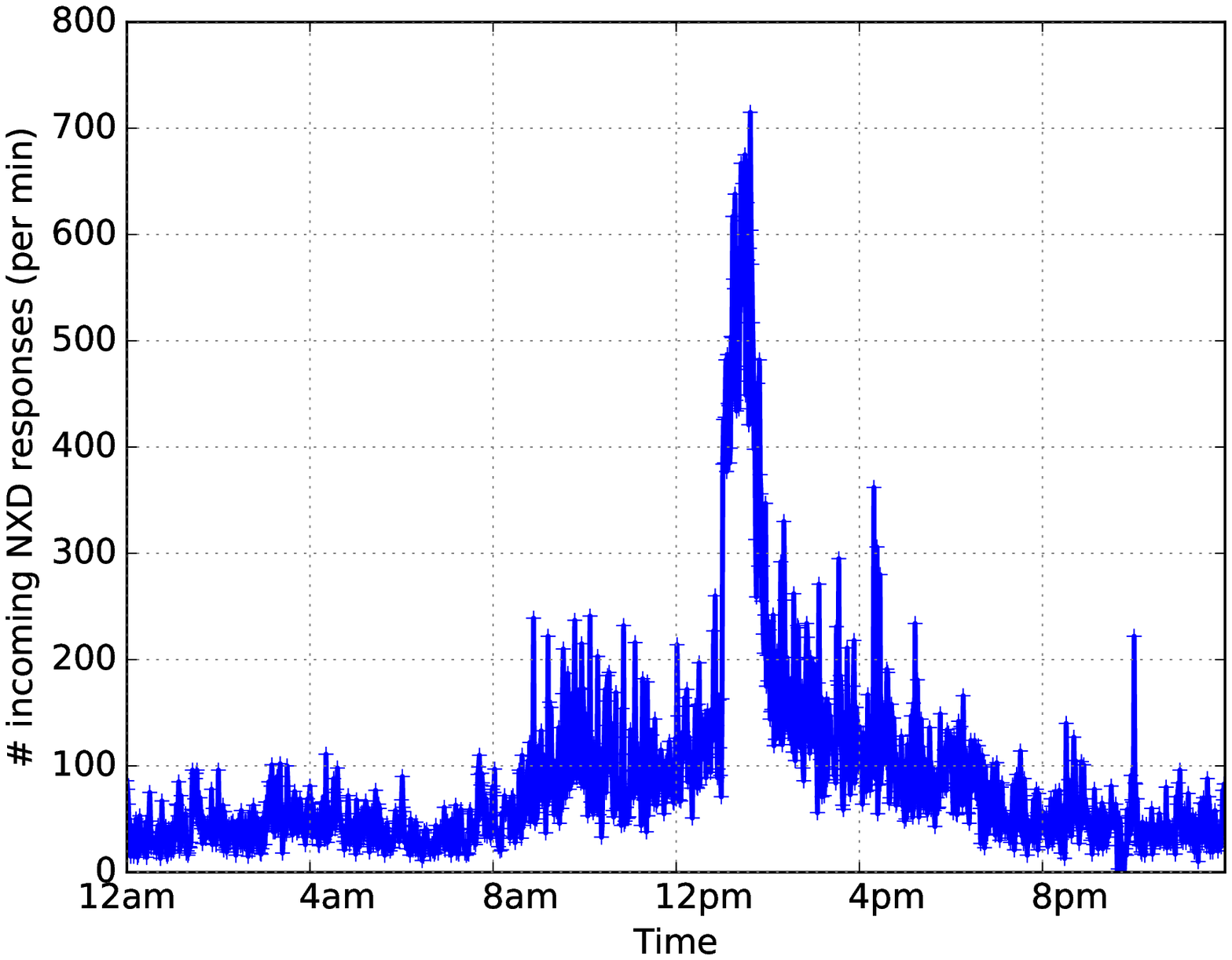}
								\quad
								\label{fig:inf3b}
							}
						}
						
						\mbox{
							%\hspace{-3mm}
							\subfloat[H1 targeting an authoritative server per second.]{ \includegraphics[width=0.29\textwidth,height=0.18\textwidth]{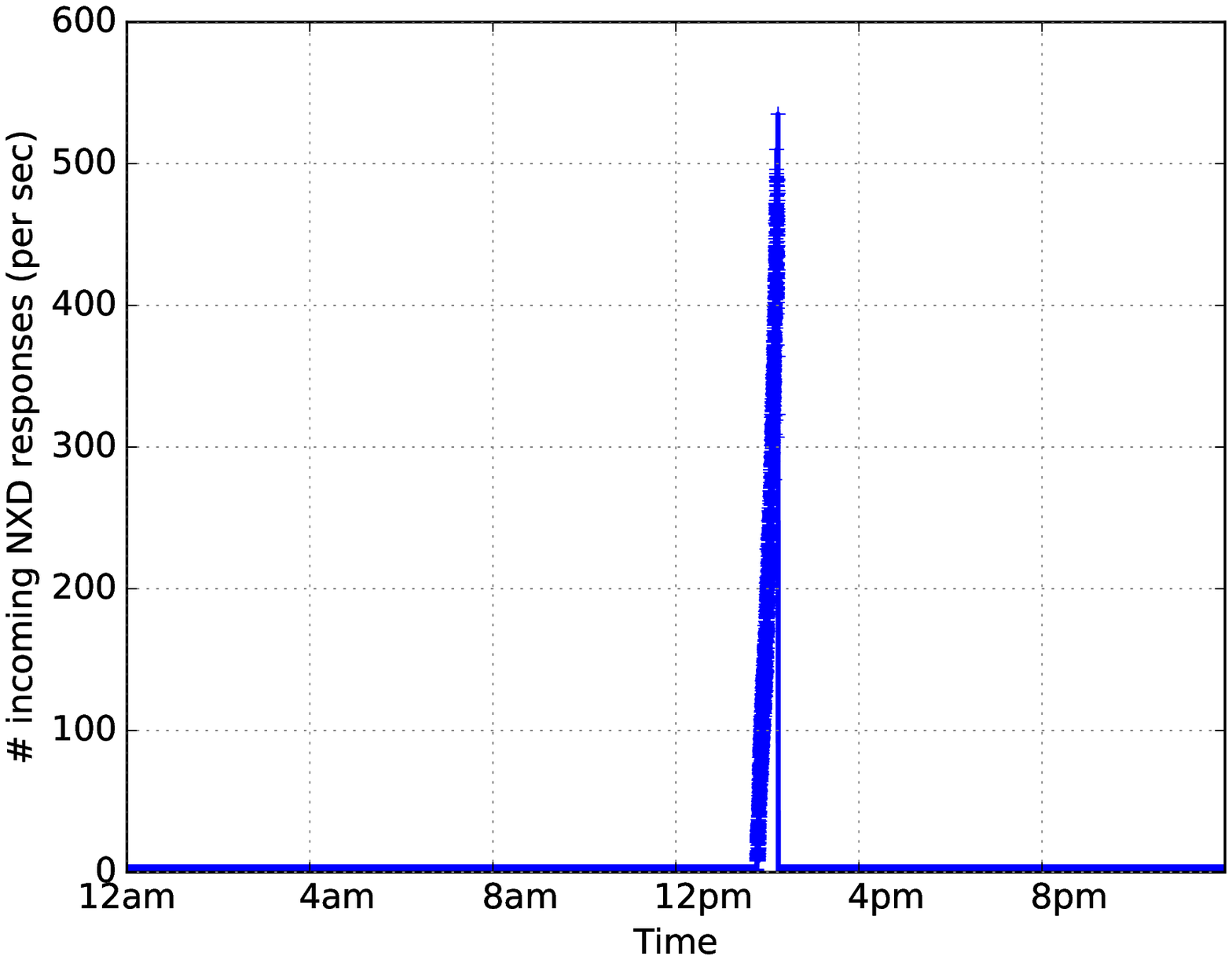}
								\quad
								\label{fig:inf1c}
							}

							\subfloat[H2 targeting a recursive server per second.]{ \includegraphics[width=0.29\textwidth,height=0.18\textwidth]{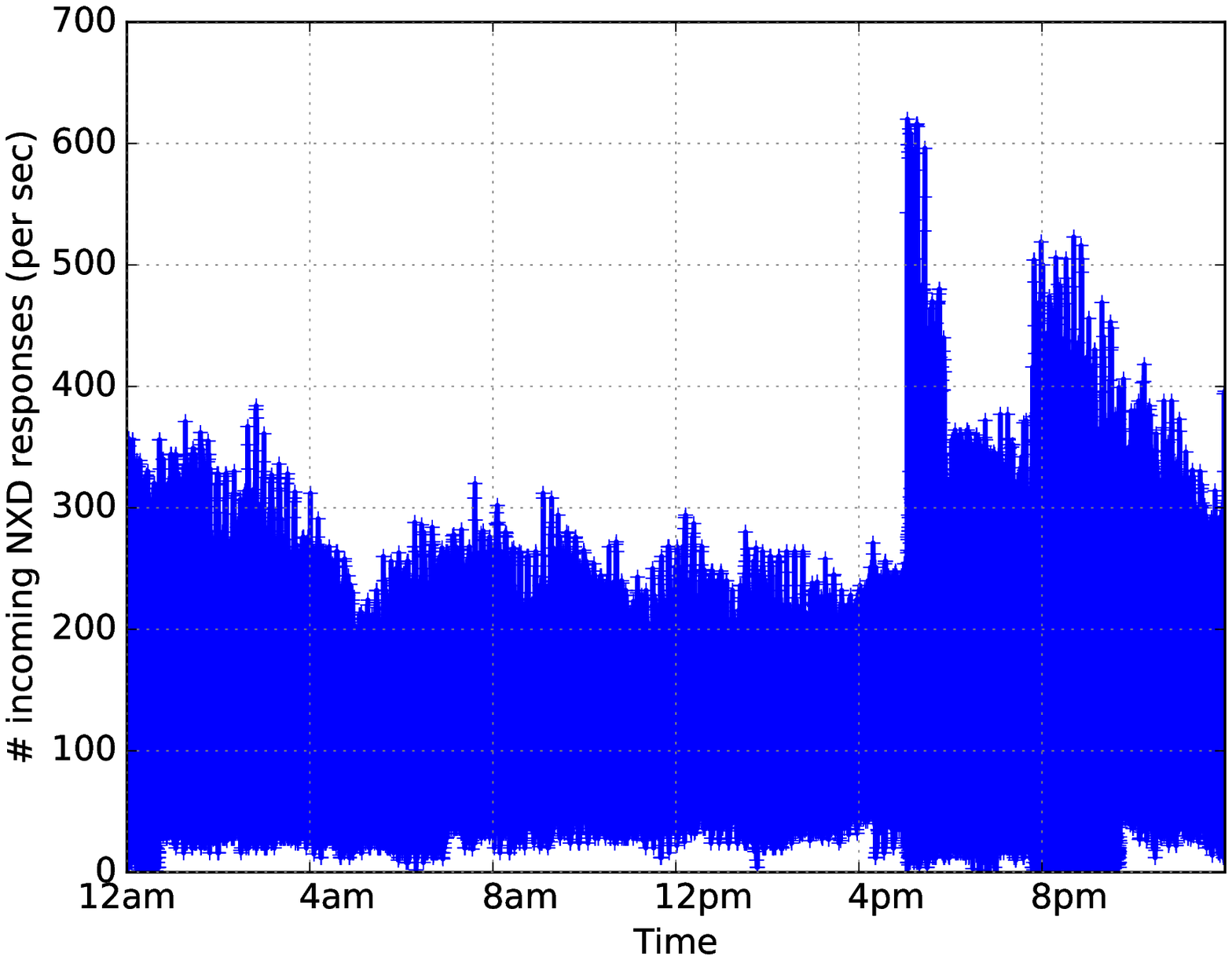}
								\quad
								\label{fig:inf2c}
							}

							\subfloat[H3 targeting an authoritative server per second.]{ \includegraphics[width=0.29\textwidth,height=0.18\textwidth]{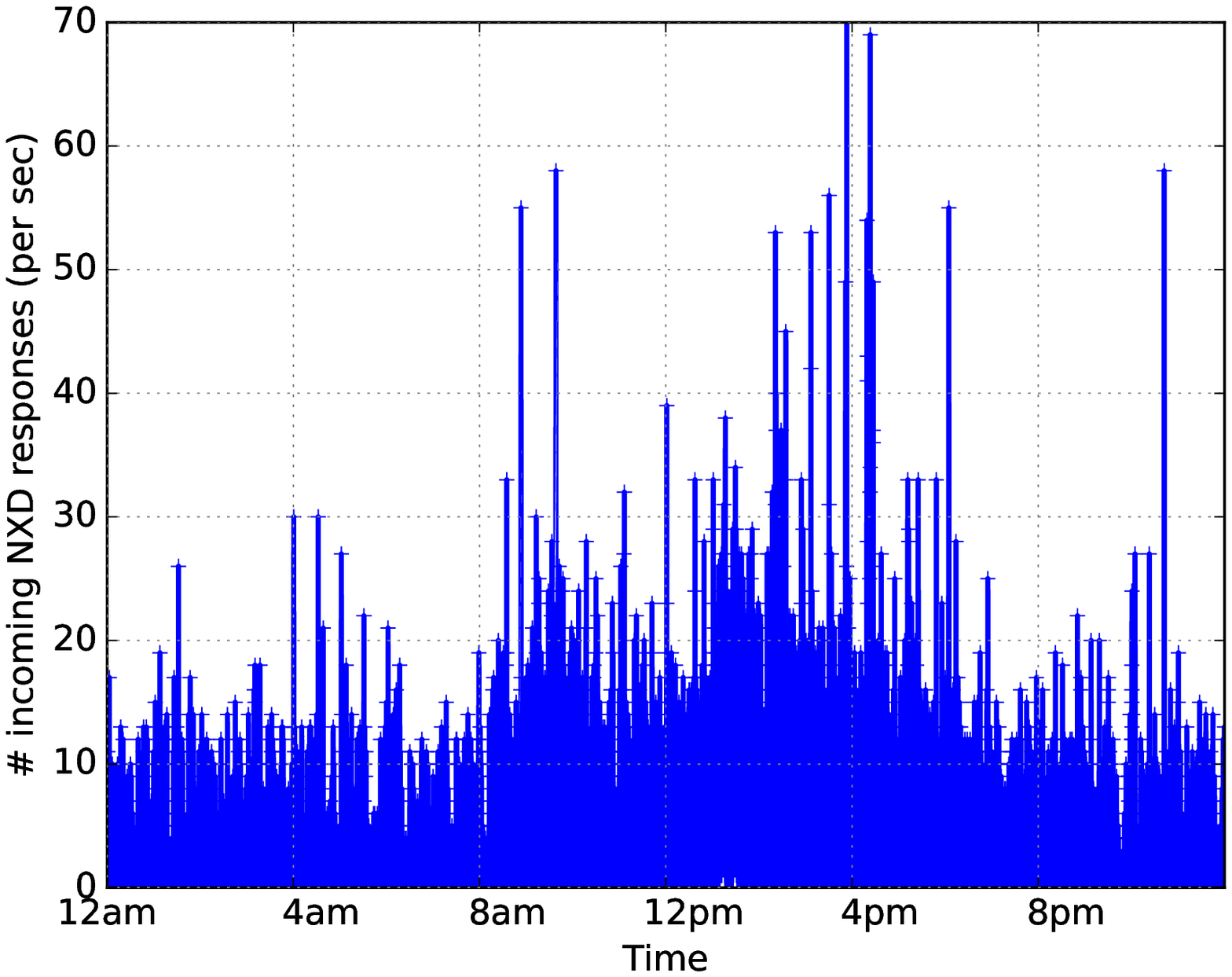}
								\quad
								\label{fig:inf3c}
							}
						}
						\caption{Time-trace of count of NXD responses for various infected hosts at various time granularity (per hour, per min and per sec): (a)(d)(g) Infected H1 (b)(e)(h) Infected H2, and (c) (f)(i) Infected H3.}
						%        \vspace{-5mm}
						\label{fig:different_infectedHosts}
					\end{center}
					\vspace{-6mm}
				\end{figure*}

				\subsection{Drawback of Using a Threshold for NXDs Attack Detection}
				
				In this section, we discuss the drawback of using threshold-based attack detection. The first point we want to make here is that various hosts behave differently on a campus network. Therefore, setting a threshold for hosts would be challenging, given the variation in the attack profile.
				% as we discussed in the previous subsection that some attacks are heavy in terms of the number of NXDs count per host whereas others are light due to the attack behavior in a distributed manner. 
				Fig.~\ref{fig:different_infectedHosts} illustrates the incoming NXD responses during a day for three representative infected hosts at different time granularity (per hour, per minute, and per second). Let us first compare the hourly count of incoming NXDs across these possibly infected hosts. Fig.~\ref{fig:inf1a} depicts the time-trace for the infected H1, the host is bursty that only became active at around 1pm in the day, peaking at 250K incoming NXD responses during a day. Comparing H1 with another infected host in Fig.~\ref{fig:inf2a} which is super active during the whole day with 300K incoming NXD responses on average. Based on the above two infected hosts, one may choose a threshold of more than 200K NXD responses per hour. However, for our third representative infected host (shown in Fig~\ref{fig:inf3a}), this threshold will not be triggered and this infected host (H3) will go undetected. Also, we test the thresholding method by changing the timescale to a per-minute basis. Fig.~\ref{fig:inf1b} provides the time trace of infected host 1 with incoming NXDs from 5K to 26K in a minute and comparing it with infected host 2 which ranges from 4K to 8K per minute. We can set a threshold of incoming NXDs to 5K (any hosts receiving more than 5K NXDs will be flagged as malicious). The infected host 3 will pass from this threshold criteria undetected as in Fig.~\ref{fig:inf3b}, we can see that incoming NXDs of infected host 3 range from 100 to 700 NXDs. Similarly, in Fig.~\ref{fig:inf1c}, \ref{fig:inf2c}, and \ref{fig:inf3c}, we compare the infected hosts on a per-second basis where the infected host 1 and 2 have the peak value of 500 and 600 respectively, whereas the infected host 3 has a peak incoming NXDs count of 70.

				\section{Multi-Staged Machine Learning Architecture}
				
				\begin{figure}[t!]
					\centering
					\includegraphics[width=0.88\textwidth,height=0.47\textwidth]{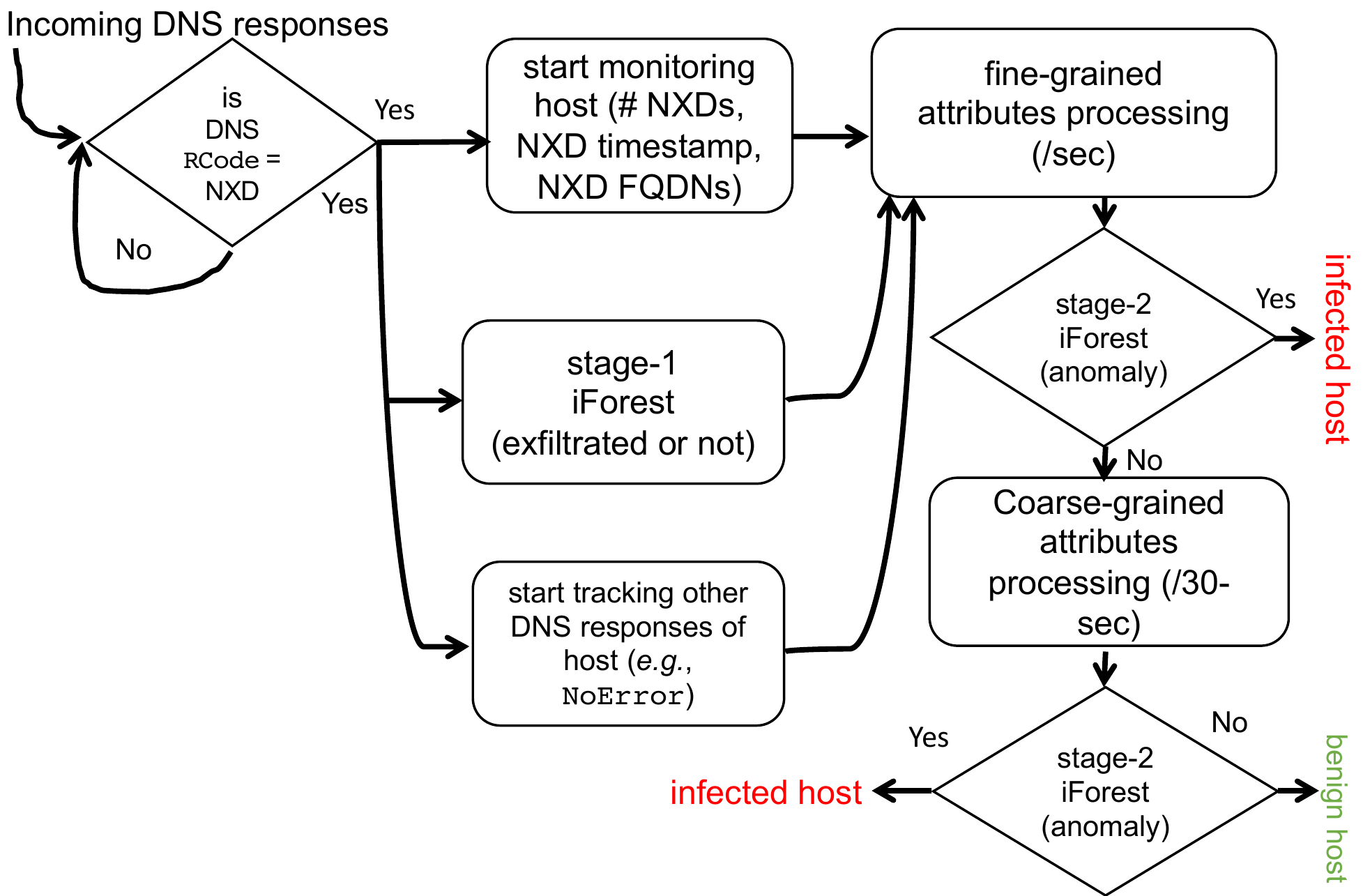}
					\caption{Overview of our proposed scheme.}
					\label{fig: Achitecture}
				\end{figure}
				
				In this section, we present the overall architecture of our method, and then we discuss the details of our multi-stage machine learning algorithm.

				\subsection{System Design}
				
				Fig.~\ref{fig: Achitecture} shows the structure of our detection system. An incoming NXD DNS response triggers our system. First, we start monitoring the behavior of the internal host which receives the NXD response. We track the number of NXD responses, timestamp of NXD responses, and FQDNs of individual NXD responses. Upon receiving the NXD response destined to the internal host, we start tracking all the responses to that specific host to get the ratio of NXD responses versus all the other response types. We have devised a new mechanism that uses cascaded machine learning-based models. In stage-1, we use an iForest model to detect whether the incoming NX response is exfiltrated or not (the exfiltrated response represents the benign disposable domains generated by antivirus tools). We design our system to detect volumetric NXD attack as well as distributed NXD attack. We devised two approaches \ie fine-grained approach and coarse-grained approach. 
				We then process the attributes on a per-second basis (fine-grained) and pass it to the stage 2 iForest model if the model identified that host as benign, we pass it to our coarse-grained model to detect the distributed NXD attack over time. For that, we process the attributes per 30 seconds interval (coarse-grained approach) and pass it to the iForest model to classify the host as malicious or benign.
				
				Our multi-staged ML is based on ``Isolation Forest (\textit{iForest})'' \cite{liu2008isolation}.%which is an effective algorithm in detecting anomalous instances in high-dimensional datasets with minimal memory and time complexities. %The iForest algorithm {\cite{liu2008isolation}} works based on the concept of isolation without employing any distance or density measure. This algorithm aims to isolate test instances by randomly selecting a feature and then selecting a split value from a range (within min and max obtained from training) values of the selected feature. Then, the score is calculated as the number of conditions (path length) to check for isolating a test instance. Note that isolating normal instances require more conditions. The process is repeated several times to avoid issues due to randomness, and the average path length is calculated and normalized. 
				In the first stage, we pass the FQDNs per host to the model that we used to detect DNS exfiltration in Chapter \ref{chap:ch3}. However, our focus here is to detect the non-exfiltrated domains. 
				The reason to choose non-exfiltrated domains is that the water torture attack queries do not conform to the attributes of exfiltrated domains. %As described earlier in  \S \ref{sec:intro}, we might have legitimate NXDs as well due to typo mistakes or disposable domains (single time used by antivirus).
				We have utilized our previously trained ML model from Chapter \ref{chap:ch3} to extract an attribute \ie {\textbf{fraction of non-exfiltrated domains}. We extract eight attributes (from the query name section of each incoming NX DNS response packet) that collectively have strong predictive power in determining whether the query name is exfiltrated or not (output of stage-1). The attributes include: 
					\begin{itemize}
						\item \textit{Total count of characters in FQDN}
						\item \textit{Total count of characters in sub-domain}
						\item \textit{Total count of uppercase characters}
						\item \textit{Total count of numerical characters}
						\item \textit{Entropy}
						\item \textit{Number of labels}
						\item \textit{Maximum label length}
						\item \textit{Average label length}
					\end{itemize}
					
					We compute the fraction of non-exfiltrated domains for each internal host as an attribute for our next stage ML-based model.
					
					For the second stage, we train two iForest models: (a) one is fine-grained for detecting volumetric water torture attack (in terms of volume of DNS requests), and (b) another is coarse-grained for detecting distributed water torture attack.  We consider the attributes discretely on a per-second basis for each host, feeding the iForest model for the fine-grained model. For the coarse-grained model, we compute the attributes discretely on 30 seconds basis for each host. 
					Based on the NXD analysis from our campus network, we define the following attributes (for each internal host) to detect the water torture attack in our second stage for both fine and coarse-grained iForest models (stage-2).

					\begin{itemize}
						
						\item \textit{Ratio of NXD response to other response types.}
						\item \textit{Average inter-arrival time between NXD responses. }
						\item \textit{Standard deviation of inter-arrival time between NXD responses. }
						\item \textit{Fraction of non-exfiltrated domains.}
						\item \textit{Average number of labels in query names of NXD responses.}
						%\item Average number of numerical characters in a subdomain
						%\item Average entropy of subdomain
						%\item Number of distinct TLDs used in a domain name
						%\item Number of distinct response types in a domain name
						%\item Number of distinct external IP addresses
					\end{itemize}
					
					\begin{figure}[t!]
						\centering
						\includegraphics[width=0.88\textwidth,height=0.57\textwidth]{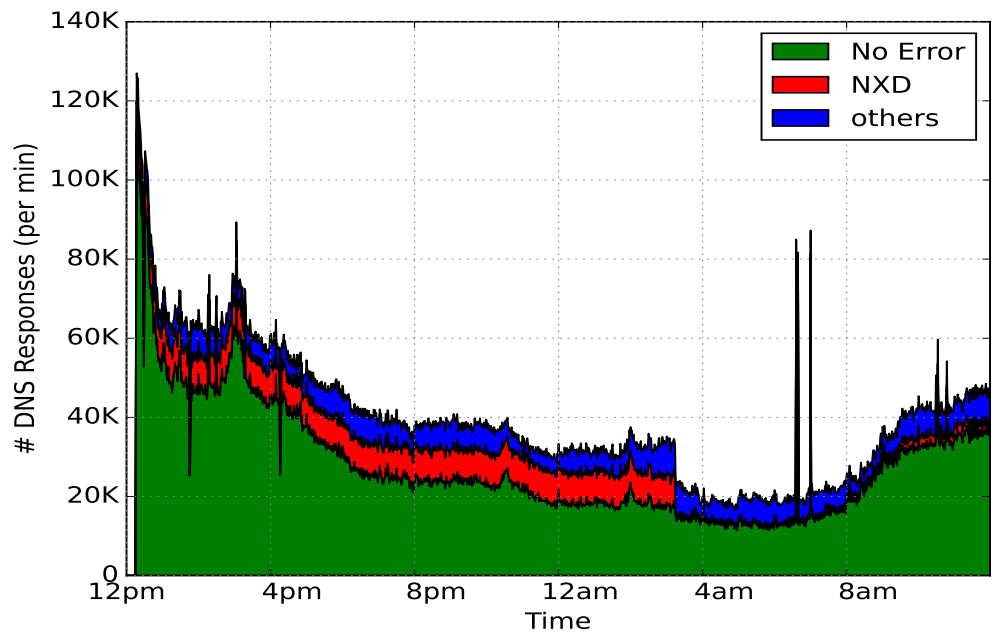}
						\caption{Stack plot of number of incoming DNS responses in our campus network over a day (26th Nov 2019).}
						\label{fig: stackPlot}
					\end{figure}

					\begin{figure}[t!]
						\centering
						\includegraphics[width=0.88\textwidth,height=0.57\textwidth]{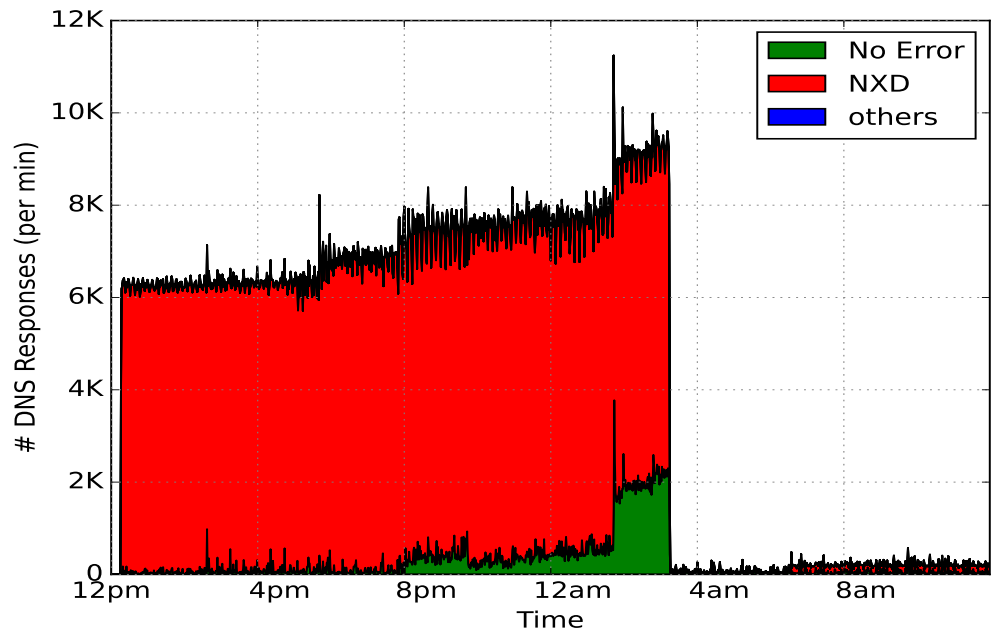}
						\caption{Stack plot of number of incoming DNS responses of a suspicious host over a day (26th Nov 2019).}
						\label{fig: stackPlotWeird}
					\end{figure}

					We will now explain the motivation behind choosing the above attributes for our ML models. The {\textbf{ratio of NXD response to other response types}} is the most important attribute for us to distinguish a malicious host from a benign one. To explain this, let us take a look at Fig.~\ref{fig: stackPlot} that shows the activity of all hosts of our campus network on a day. The stackplot depicts that the number of No Error responses is significantly higher than that of NXDs, with the peak value of more than 100K responses in a minute. In contrast, the NXD responses are less than 10K for overall hosts. We then plot in Fig.~\ref{fig: stackPlotWeird} the number of DNS responses (per min) for a suspicious host during that day. We can see that No Error responses peaked at 2K per minute, whereas NXD responses (shown in red) peaked at 6K. Therefore, the ratio of NXD to all the other responses becomes much greater than 1 from 12pm till 3am. We identify {\textbf{average and standard deviation inter-arrival time between NXD responses}} as other two main attributes for classification based on the analysis of benign and suspicious host as discussed in \S \ref{sec:analysis} where we showed that benign NXDs are distributed in time and do not occur very often over a day as opposed to malicious NXDs involved in a water torture attack. Similarly, {\textbf{fraction of non-exfiltrated domains} } gives the percentage of domains used other than disposable domains and {\textbf{average number of labels of NXD responses} captures whether a domain is a disposable domain or a water torture attack domain. We note that disposable domains are often long as shown in Table \ref{tab:sampleDNS} for {\myverb{sophosxl.net}} and contain more than 5 or 6 subdomains whereas the random NXD domains typically have one subdomain \cite{luo2018large}.

						\subsection{Model Training}

						In this subsection, we provide details of our ML model training used in each stage.

						\begin{figure}[t!]
							\centering
							\includegraphics[width=0.88\textwidth,height=0.57\textwidth]{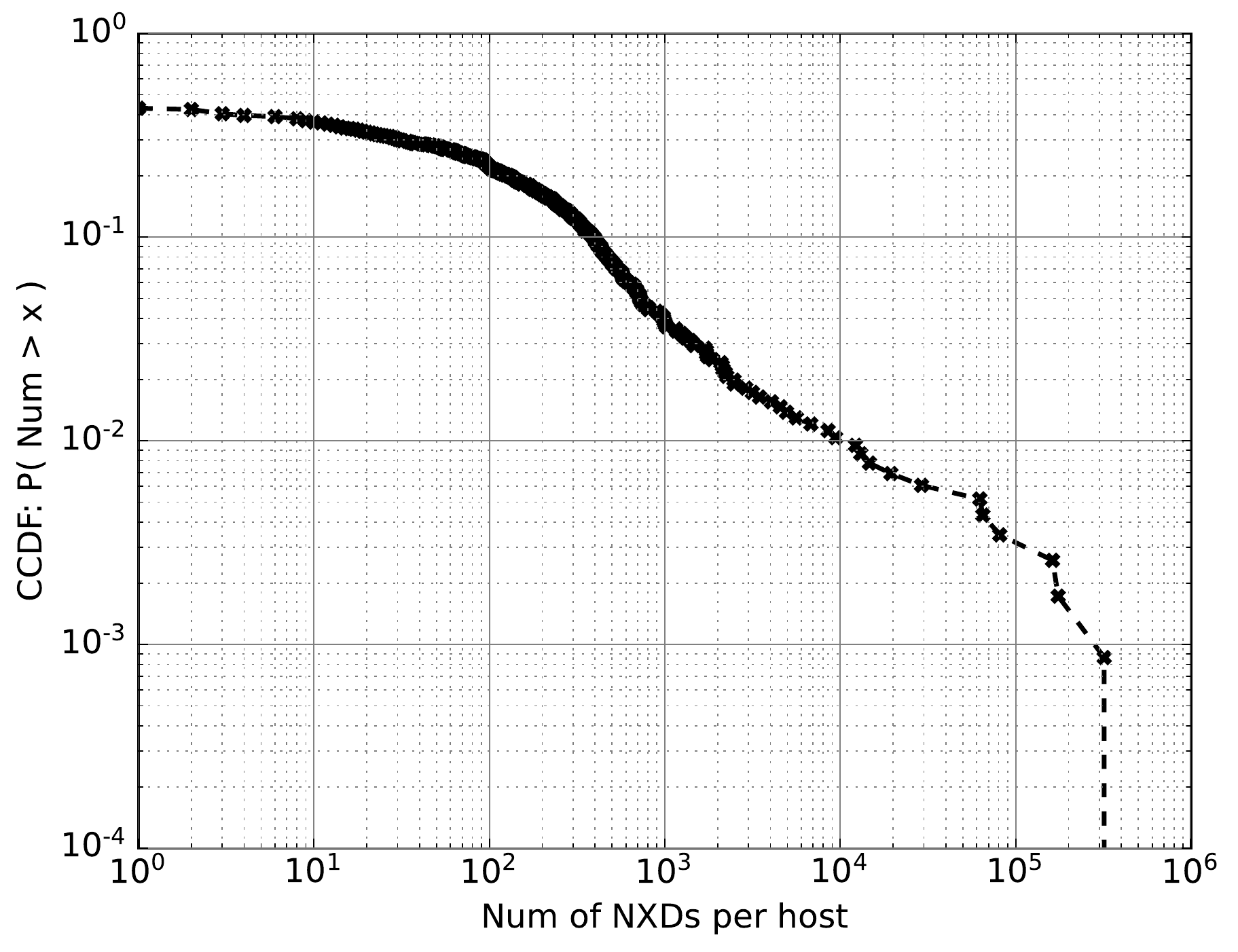}
							\caption{CCDF of number of occurrences of NXDs per host.}
							\label{fig: CCDFNXDs}
						\end{figure}
						
						{\textbf{Stage-1 Model:}} For the stage-1 model, we train an iForest model with benign data from four days of our DNS dataset. For ground truth of benign domains we use the same Majestic Million list that we used in Chapter \ref{chap:ch3}. We use the same approach that we used in \S\ref{sec:ML} for training \& tuning.

						{\textbf{Stage-2 Model:}} To train the stage-2 model, we need a dataset of benign NXDs that is quite challenging as there is no public dataset from enterprise hosts. Therefore, we construct our own dataset. To populate benign domains, we start with 4 days’ worth of NXD responses from our original DNS dataset. Fig.~\ref{fig: CCDFNXDs} shows the CCDF of the total count of NXD responses per host. Interestingly, more than 55\% of hosts just receive one NXD response. We believed these hosts accidentally mistyped a domain name, resulting in an NXD response. Therefore, we assumed those hosts to be benign. Also, if a host just receives NXD responses for disposable domains (generated by antivirus tools), we assume it as benign too (although the host may or may not involve in other malicious activities, our primary focus is on NXD attack). Note that this threshold value can be configured by the network administrator based on the requirement for their network.

						To achieve our objective of detecting NXD attacks at two levels of granularity (fine-grained and coarse-grained), we train two iForest models at stage-2, the first model is trained to detect the heavy volume of NXDs based on attributes computed on a per-second basis. Similarly, the second model is trained with attributes computer on a per 30-sec basis for a distributed NXD attack.

						\section{Performance Evaluation}
						
						In this section, we evaluate the efficacy of our scheme by cross-validating and testing the accuracy of the trained models for benign instances and quantifying their performance on a full campus traffic stream. 
						
						\subsection{Performance of Fine-Grained Model}
						
						\begin{table}[t!]
							\centering
							{
								\caption{Anomaly detection by fine-grained model.}
								\label{table:perfFine}
								\begin{adjustbox}{max width=0.49\textwidth}  
									\renewcommand{\arraystretch}{1.3}    
									\begin{tabular}{lccc}
										\toprule
										{\textbf{Input}}  & {\textbf{Output}}    & \textbf{Days 1-4} & \textbf{Days 5-7} \\ \hline\hline
										\multirow{2}{*}{Benign hosts} & normal    & $99.6$\%     & $98.5$\% \\
										& anomalous & {0.4\%}       & {1.5\%} \\ \cline{1-4}
										\multirow{2}{*}{Remaining hosts}         & normal    & $94.1$\%      & $93.3$\% \\
										& anomalous & $5.1$\%       & $6.7$\% \\  
										\bottomrule   
									\end{tabular}
								\end{adjustbox}
							}
						\end{table}
						
						In this subsection, we evaluate the performance of our fine-grained ML model. Table \ref{table:perfFine} shows that benign hosts are classified as normal with an accuracy of more than 98\% for both training and testing with false-positive rates of less than 2\%. Our objective here is to detect anomalous hosts when the models are applied to remaining hosts. We find that 5.1\% of the hosts are classified as anomalous. Further analysis revealed that this model captures all the heavy volume NXD attacks with some false positives. Some of the hosts classified as malicious are NAT gateways and recursive resolvers due to their heavy activity of receiving NXD responses (actual end hosts are hidden behind NAT gateways and resolvers, and hence, the detection of those actual end hosts in this case is beyond the scope of this work). 
						
						\subsection{Performance of Coarse-Grained Model}
						
						We next evaluate our coarse-grained model, which was trained by the iForest algorithm. Table \ref{table:perfCoarse} summarizes the results.  It can be seen that trained internal hosts are correctly classified as benign with an accuracy of more than 99\% during validation. Similarly, the benign hosts are correctly classified as benign with an accuracy of 98\% with a false-positive rate of less than 2\%. These results are similar to those of the fine-grained model. However, our intent of using a coarse-grained model was to detect distributed NX attacks. We can see here for the remaining hosts, the number of hosts classified as normal is decreased drastically to 90\%, whereas the anomalous hosts are nearly 10\%. When we analyzed the hosts classified as anomalous, we found out that some hosts were taking part in water torture attack with a very low volume of NXDs, but the requests were distributed in time. 
						
						%Moreover, we compare the infected hosts that are missed by our fine-grained ML model and picked by the coarse-grained ML. We have 
						
						\begin{table}[t!]
							\centering
							{
								\caption{Anomaly detection by coarse-grained model.}
								\label{table:perfCoarse}
								\begin{adjustbox}{max width=0.49\textwidth}  
									\renewcommand{\arraystretch}{1.3}    
									\begin{tabular}{lccc}
										\toprule
										{\textbf{Input}}  & {\textbf{Output}}    & \textbf{Days 1-4} & \textbf{Days 5-7} \\ \hline\hline
										\multirow{2}{*}{Benign hosts} & normal    & $99.4$\%     & $98.1$\% \\
										& anomalous & {0.6\%}       & {1.9\%} \\ \cline{1-4}
										\multirow{2}{*}{Remaining hosts}         & normal    & $90.6$\%      & $90.1$\% \\
										& anomalous & $9.4$\%       & $9.9$\% \\  
										\bottomrule   
									\end{tabular}
								\end{adjustbox}
							}
							\vspace{-4mm}
						\end{table}
						
						\subsection{Discussion} To better understand these findings, we have further analyzed hosts that are detected as anomalous in ``remaining hosts'' category. %-- note that DNS responses are exclusively used in this section for drawing further insights into anomalous queries. 
						First, we look at the anomalous hosts for our fine-grained model. In total, we have $45$ unique anomalous hosts. By analyzing these IP addresses, we found (by reverse lookup) that $8$ of them are NAT gateways (the actual number of hosts are hidden behind these NAT gateways). Other $37$ are all regular end hosts coming from $5$ different subnets of size /24 of our university campus. Interestingly, out of these five subnets, $18$ are from the same subnet, indicating that this particular subnet might be infected by malware.
						
						Moreover, we compare these numbers with the results of our coarse-grained model. We found out that the anomalous hosts after adding our coarse-grained model increased to $87$, which shows that it also flagged hosts involved in distributed random subdomain attacks. Furthermore, out of these $87$ unique anomalous hosts, by reverse lookup, we found out that $11$ of them are from NAT gateways and the remaining $76$ belong to regular end hosts coming from $9$ different subnet of size /24. Out of those $76$ hosts, $26$ hosts fall under one subnet during the entire week. Upon investigation, we found out the subnet is the same that our fine-grained model flags. This shows that some hosts are possibly involved in high volume random subdomain attack while other involved in distributed low-volume random subdomain attacks.

						\section{Conclusion}\label{conclusion} 
						
						Enterprise networks are a potential target of cyber-attackers, specifically those exploiting DNS to perform various attacks. We have developed a multi-stage machine learning-based solution to detect NXD attacks. First, by analyzing incoming NXD responses from DNS traffic of our campus network, we identified the difference between two scenarios of random subdomain attacks on authoritative DNS servers and open resolver. We developed a method using multi-staged iForest models to classify the malicious internal hosts that take part in water torture attack based on their behavioral attributes. Lastly, we evaluated the efficacy of our proposed approach on live DNS data from the network border of a large university campus.

\chapter{Conclusions and Future Work}
\label{chap:ch6}

\minitoc

\section{Conclusions}

Enterprise networks are under enormous threat to cyber attacks. DNS is an essential protocol used by every device connected with the network to map the domain name with an IP address and vice versa. Unfortunately, attackers use DNS as a covert channel as it is hardly policed by most enterprise networks' firewalls and IDS systems.
Existing mechanisms to detect the malicious activities in the network mostly contain a knowledge-based approach that is fruitful for the known attacks but fails when detecting zero-day attacks. This thesis studied three DNS-based cyber-attacks, specifically data exfiltration, DGA-based malware C\&C communication, and service disruption through NXDs. 
We analyzed the network traffic of our campus network and a Government research organization over six months and showed the prevalence of these attacks, devised the ML-based models to detect them, and evaluated the efficacy of our detection mechanism in the wild.

We summarize the essential contributions of this thesis towards achieving DNS security.
\begin{itemize}
	\item  In our first contribution, we tackled data exfiltration using DNS. We analyzed outgoing DNS queries and identified stateless attributes such as the number of characters, the number of labels, and the entropy of the domain name to distinguish malicious data exfiltration queries from legitimate ones. We trained our machines using ground-truth obtained from a public list of top 10K legitimate domains. We then empirically validated and tuned our models to achieve over 98\% accuracy in correctly distinguish legitimate DNS queries from malicious queries. 
	
	\item Our second contribution tackled malware C\&C communication using DNS. We analyzed DNS outgoing queries to identify more than twenty families of DGA-enabled malware when communicating with their C\&C servers. We have identified the characteristics of malicious network traffic and have trained three protocol-specific one-class classification models for HTTP, HTTPS, and UDP flows using public packet traces of known malware. In addition, we have developed a monitoring system to automatically and selectively reflect TCP / UDP flows related to DGA queries for diagnosis from trained models. We deployed our system in the field and evaluated its performance to show that it had potentially infected more than 2,000 internal assets, causing over a million suspicious flows, 97\% of which were off-the-shelf intrusion detection systems.
	
	\item For the third contribution, we studied the use of DNS for service disruption. We analyzed incoming DNS messages to distinguish between benign and malicious NXD, focusing on NXD DNS responses. We emphasized two attack scenarios based on their requested domain name. We developed multi-staged iForest classification models at various time windows (such as per second and 30 seconds) to detect internal hosts launching a service disruption attack using the internal host's NXD behavior attributes. We showed our models were able to capture high-volume and low-volume NXD based attacks with an accuracy of 99\% while classifying legitimate hosts.
	
\end{itemize}

\section{Future Work}
Our work is a significant milestone in achieving a practical solution for the DNS security of enterprise networks. However, our proposed methods can be further improved and extended to make way for exciting future breakthroughs. We outlined a few of them below.

\begin{itemize}
	\item As seen in Chapter \ref{chap:ch3}, we employed stateless attributes of the domain name to detect the data exfiltration. We drew insights into the practical considerations of using our detection scheme. This scheme can be extended by collecting states from the DNS traffic only for those hosts that generate anomalous queries and ultimately mitigate malicious DNS tunneling/exfiltration. %Another possible direction can be the use deep neural network to detect the DNS exfiltration. Some researchers have already started exploring this area
	
	\item In Chapter \ref{chap:ch4}, we focused on detecting the infected hosts by selectively mirroring the C\&C communication based on a well-known DGA domains database ``DGArchive'' containing more than 80 DGA malware families. Although we can add new DGA-enabled malware domains in the database, the process is still limited to the known malware. The module of the DGA query finder can be replaced by detecting the DGA domains instead of comparing with just the database to capture all the infected hosts in the enterprise network and improve the overall efficacy. 
	
	\item We demonstrated in Chapter \ref{chap:ch5} how analysis of incoming NXD responses could help identify infected hosts that launch volumetric attacks on Internet-based servers. However, we observed a lack of research in identifying the infected hosts generating the random subdomain attack that focuses on identifying the victim, such as authoritative name server and recursive resolver. Thus, although our study provides reasonable accuracy by identifying the source of random subdomain attacks, we still believe this can be further enhanced by using ML and deep learning-based models.
	
	\item Another future direction that can be explored includes the use of encrypted DNS such as DNSSEC \cite{arends2005dns}, DNS over TLS (DoT) \cite{rfc7858} and DNS over HTTPS (DoH) \cite{rfc8484}. Other researchers can use our ML detection models and encrypted DNS methods to enhance the security of existing DNS architecture, keeping in mind the increased overhead.
	
\end{itemize}

We hope other researchers will explore the future directions identified above.

%------------------------------------------------------------------
%	BIBLIOGRAPHY
%------------------------------------------------------------------
\backmatter
\renewcommand{\refname}{References} % Change the default bibliography title

% Run biber command for first time
%-------- IMP: DO NOT UPDATE THE LIBRARY DIRECTLY -----
%--------- IT IS LINKED TO ZOTERO----
%\bibliography{References} % Input your bibliography file
\appto{\bibsetup}{\sloppy}
\apptocmd{\sloppy}{\hbadness 10000\relax}{}{}
\begin{singlespace}
	\setlength\bibitemsep{10pt}   % length between two different entries
	\printbibliography[heading=bibintoc,title={document}]
\end{singlespace}

%------------------------------------------------------------------

% \input{./appendix_1.tex}

\end{document}